\documentclass[final,5p,times,twocolumn,numbers,sort&compress]{elsarticle}

\usepackage[T1]{fontenc}
\usepackage{amssymb}
\usepackage{amsmath}
\usepackage{url}
\usepackage{sidecap}
\usepackage{booktabs}
\usepackage{longtable}
\usepackage{lscape}
\usepackage{xcolor}
\usepackage{verbatim}
\usepackage{natbib}

\definecolor{RoyalBlue}{RGB}{65,105,225}
\newcommand{\mdd}[1]{\textcolor{black}{#1}}
\definecolor{olivegreen}{rgb}{0,0.6,0}
\newcommand{\gc}[1]{\textcolor{black}{#1}}

\usepackage[normalem]{ulem}
\let\oldsout\sout                   
\renewcommand{\sout}[1]{
  \textcolor{red}{\oldsout{#1}}     
}
\let\oldsout\sout                   
\renewcommand{\sout}[1]{
  \vphantom{\hphantom{#1}}     
}

\usepackage{bm} 
\usepackage[colorlinks=true]{hyperref}

\journal{Physics Reports}

\begin{document}

\begin{frontmatter}

\title{The Physics of News, Rumors, and Opinions}

\author[1,2,3]{Guido Caldarelli}
\affiliation[1]{Istituto dei Sistemi Complessi CNR-ISC Rome, Italy}
\affiliation[2]{DSMN and ECLT, Ca'Foscari University of Venice, Italy}
\affiliation[3]{London Institute for Mathematical Science, Royal Institution, London UK}
\author[4,5] {Oriol Artime}
\affiliation[4] {Departament de Fisica de la Materia Condensada, Universitat de Barcelona, 08028 Barcelona, Spain}
\affiliation[5]{Universitat de Barcelona Institute of Complex Systems (UBICS), Universitat de Barcelona, 08028 Barcelona, Spain}
\author[2,6]{Giulia Fischetti}
\affiliation[6] {Zurcher Hochschule fuer Angewandte Wissenschaften (ZHAW), Zurich, Switzerland}
\author[7]{Stefano Guarino}
\affiliation[7]{CNR-IAC Istituto per le Applicazioni del Calcolo, Roma Italy}
\author[8,9]{Andrzej Nowak}
\affiliation[8] {University of Warsaw, Poland}
\affiliation[9] {Florida Atlantic University, USA}
\author[10,11,12]{Fabio Saracco}
\affiliation[10]{Centro Studi e Ricerche e Museo della Fisica ``Enrico Fermi'' Rome, Italy}
\affiliation[11]{Istituto dei Sistemi Complessi CNR-IAC Florence, Italy}
\affiliation[12]{IMT School For Advanced Studies Lucca, Italy}
\author[13]{Petter Holme}
\affiliation[13]{Department of Computer Science, Aalto University, Finland}
\author[14,15]{Manlio De Domenico}
\affiliation[14] {Department of Physics, University of Padova, Padua, Italy}
\affiliation[15] {INFN Sez. Padova, Padua, Italy}

\begin{abstract}

The boundaries between physical and social networks have narrowed with the advent of the Internet and its pervasive platforms. This has given rise to a complex adaptive information ecosystem where individuals and machines compete for attention, leading to emergent collective phenomena. 
The flow of information in this ecosystem is often non-trivial and involves complex user strategies—from the forging or strategic amplification of manipulative content to large-scale coordinated behavior—that trigger misinformation cascades, echo-chamber reinforcement, and opinion polarization.
We argue that statistical physics provides a suitable and necessary framework for analyzing the unfolding of these complex dynamics on socio-technological systems. This review systematically covers the foundational and applied aspects of this framework.
The review is structured to first establish the theoretical foundation for analyzing these complex systems, examining both structural models of complex networks and physical models of social dynamics (e.g., epidemic and spin models). We then ground these concepts by describing the modern media ecosystem where these dynamics currently unfold, including a comparative analysis of platforms and the challenge of information disorders. The central sections proceed to apply this framework to two central phenomena: first, by analyzing the collective dynamics of information spreading, with a dedicated focus on the models, the main empirical insights, and the unique traits characterizing misinformation; and second, by reviewing current models of opinion dynamics, spanning discrete, continuous, and coevolutionary approaches.
In summary, we review both empirical findings based on massive data analytics and theoretical advances, highlighting the valuable insights obtained from physics-based efforts to investigate these phenomena of high societal impact.
\end{abstract}

\begin{keyword}
Disinformation \sep Spreading \sep Complex Networks \sep Communities and Polarization

\end{keyword}

\end{frontmatter}

\tableofcontents

\section{Introduction}
The study of social systems has long been approached by the definition of frameworks from economics, psychology, and sociology, which often focus on individual decision-making or qualitative interactions. However, many large-scale social phenomena such as opinion dynamics, crowd behavior, market fluctuations, and migration patterns exhibit emergent properties that bear striking resemblance to the collective behavior of complex  physical systems~\cite{artime2022origin}. This suggests a deeper analogy: some feature in the behaviour of human populations may be modeled as many-body systems governed by statistical mechanics,  where macroscopic social patterns arise from microscopic interactions between individuals \gc{(often not represented by an ordered lattice structure, but rather by complex graphs~\cite{caldarelli2002scale})}, much like how temperature and pressure emerge from molecular collisions.
In this sense, the boundaries between physical and social networks have narrowed since the advent of the Internet, with platforms such as social media and portable devices allowing real-time access to news everywhere on Earth. \sout{However, not all information is relevant or reliable, leading to a complex adaptive information ecosystem where individuals and machines produce content and compete for attention.}

In the last decades, the massive use of information and communication technologies (ICT) is producing a continuous large flow of traceable data that can be used to describe and understand the changes in our society, from an individual level to collectives~\cite{barabasi2005origin,lazer2009social,simini2012universal,pappalardo2015returners,lazer2021meaningful}. The complexity of this analysis calls for an approach that, on the example of statistical physics, can lead to a properly defined physics of society~\cite{caldarelli2018physics}, describing society through its interacting individuals. In this respect, the quantities of interest come from sociometric and behavioral data~\cite{nadal2012manifesto}, covering several aspects of human dynamics, from mobility to communication. 
Nevertheless, it is unlikely that such quantities can be statistically described using the same approaches adopted for large sets of particles--like in a box filled with gas--to obtain the equivalent of a Maxwell-Boltzmann distribution.
This is because individuals are more similar to active matter than gas particles and that groups of individuals--with their interactions--are best described by fat-tail distributions~\cite{pareto1935mind,zipf1949human,onnela2007analysis,song2010modelling,jiang2013calling,schlapfer2014scaling,alessandretti2020scales,xu2021emergence} and out-of-equilibrium dynamics~\cite{haken1975cooperative,castellano2009statistical,montanari2010spread,dedomenico2013anatomy,gallotti2016stochastic,hauser2019social}. 
On this rugged landscape, we must, therefore, stick to the few observational regularities--such as scaling laws--that we can uncover on this extremely heterogeneous structure given by individuals and their social networks. 

Statistical physics provides a natural toolkit for understanding such systems. By treating individuals as interacting particles subject to social forces (e.g., peer influence, cultural norms, or economic incentives), we can apply concepts like phase transitions, entropy, and stochastic dynamics to explain phenomena such as polarization, consensus formation, or sudden shifts in public opinion. For instance:
\begin{itemize}
\item Phase transitions may describe abrupt societal changes (e.g., revolutions or market crashes) as critical points where small perturbations trigger system-wide transformations.

\item Diffusion processes can model the spread of information or behaviors through social networks, analogous to heat propagation in materials.

\item Entropic forces might quantify the role of randomness in decision-making, where \mdd{a suitable definition of} "social temperature" reflects the variability of individual choices.
\end{itemize}
This paper explores how methods from statistical physics including agent-based models, mean-field theories, and non-equilibrium thermodynamics can shed light on social dynamics. We argue that such an approach offers two key advantages:
\begin{description}
    \item {\bf Quantitative predictability} By identifying \gc{appropriate parameters and laws}, we can forecast large-scale trends from local interaction rules. \gc{That is, we can identify order parameters (e.g., the average belief polarization of a population) and scaling laws that govern phase transitions. This allows us to move beyond anecdotal observation and forecast large-scale trends, such as the viral tipping point of a false narrative, directly from the local rules of user engagement, algorithmic amplification, and homophily.}
    \item {\bf Universality} Social systems may belong to the same universality classes as physical systems, enabling cross-disciplinary insights. \gc{The patterns of misinformation emergence, persistence, and decay may not be unique to social systems. Instead, they might belong to the same universality classes as established physical systems, such as the spontaneous magnetization in ferromagnetic materials or the activation dynamics in chemical reaction networks. This universality enables powerful cross-disciplinary insights, allowing us to adapt well-understood physical models to predict how misinformation will behave under different network topologies or moderation policies. Ultimately, by treating misinformation as a collective phenomenon emerging from micro-level interactions, statistical physics provides a rigorous, mathematically grounded toolkit to diagnose vulnerabilities, simulate intervention strategies, and potentially control the "epidemiology" of false information in digital ecosystems.}
\end{description}

Challenges remain, of course. Human behavior introduces complexities like memory, strategic intent, and heterogeneous interactions, which lack direct physical analogs. Yet, by adapting tools from active matter physics and disordered systems~\cite{cimini2019statistical} \mdd{with evolutionary game theory}, we may bridge these gaps. Ultimately, this framework aims to unify disparate social phenomena under the lens of statistical mechanics, revealing the "thermodynamics of society" and offering new levers for policy design~\cite{caldarelli2018physics}.

In this review, we shall describe the efforts to analyze this process through a complex network-based approach, since this seems to be a natural choice for modeling the structure and dynamics of a complex society \mdd{which is interconnected at both individual and group levels}. We justify this statement based on the definition of complexity, that is, the emergence of a new behavior whenever the number of ingredients (the particles, the agents, whatever describes the system under consideration) becomes very large~\cite{anderson1972more,ladyman2013what,buckley2017society,caldarelli2020perspective,artime2022origin}. Since complex systems are defined by system composed by many interacting parts, not surprisingly complex networks~\cite{newman2018networks} appear as the simplest mathematical model able to capture a variety of typical  features~\cite{cimini2019statistical} quantitatively. 
Complex networks have gained an increasing amount of consideration to be an optimal integration and investigation tool for real-world systems of diverse fields, from biology~\cite{kitano2004biological,gosak2018network}, to finance~\cite{barucca2018tackling,bardoscia2021physics}, from technology~\cite{caldarelli2000fractal,pastor-satorras2001dynamical} to sociology~\cite{castellano2009statistical}. Many of these systems show particular behaviors that distinguish them from other traditional systems, such as their scale-free nature, the emergence of power-law degree distributions and small-world phenomenon~\cite{watts1998collective,barabasi1999emergence,caldarelli2007scale}, the co-existence of several contexts--usually named layers--of interaction~\cite{mucha2010community,de2013mathematical,nicosia2013growing,wang2015evolutionary} and interdependency~\cite{buldyrev2010catastrophic,gao2012networks} \mdd{(see~\cite{artime2022multilayer,de2023more} for a recent review)}. 
Even if incredibly complex, a representation based on static interactions is still insufficient to explain information spreading and evolution in an interconnected society. On the one hand, even the structure of social systems changes dynamically and exhibits time-varying behavior to be accounted for~\cite{robins2001random,holme2012temporal,perra2012activity,borge-holthoefer2016dynamics}, with non-trivial memory effects~\cite{matamalas2016assessing,williams2022shape}. On the other hand, one has to consider that the sites of the networks (e.g., users in a socio-technical system) are active, so they can rearrange their links, join new groups, unfollow people and decide what to share, how and when~\cite{dedomenico2013anatomy}, leading non-trivial collective behaviors~\cite{lehmann2012dynamical, mocanu2015collective,lorenz-spreen2019accelerating,dedomenico2020unraveling}. Given such a broad and heterogeneous phenomenology, the activity of a node is a game-changer even in simple spreading models. An emblematic example is given by the mobility of active matter when sites are left free to explore their environment~\cite{vicsek1995novel}, and diffusion is characterized by a smaller dynamical exponent~\cite{cavagna2019dynamical} than the one we could expect on a network~\cite{delvenne2015diffusion}.

\subsection{Aims and scope}\label{ssec:aims}

The core goal of this review is to illuminate the principles governing the collective social dynamics of information and opinions and demonstrate how models and techniques rooted in statistical physics and network science provide the essential framework to achieve this understanding. Specifically, this paper addresses the following intertwined objectives that span structure, dynamics, and societal impact.

\subsubsection*{Understanding the structure and dynamics of socio-technical systems}

Online social platforms have become pervasive communication channels, where decentralized individual interactions drive complex, out-of-equilibrium dynamics. The process of news creation, diffusion, and opinion formation now occurs on highly disordered, active media where individuals and automated accounts (active units) initiate, terminate, or modify the flow of information.

The challenge lies in quantifying the impact of this systemic shift. We aim to clarify:
\begin{enumerate}
    \item \textbf{How and how fast do news, rumors, and opinions spread?} We seek to establish quantitative models that capture the dynamics of complex contagion and non-linear interactions characteristic of these systems.
    \item \textbf{What are the fundamental physical and network mechanisms governing spread?} This involves investigating the principles at the basis of spreading patterns in disordered structures and quantifying the pervasiveness of different narratives, from simple viral cascades to complex behavioral adoption.
\end{enumerate}

\subsubsection*{Characterizing the role of network topology and user behavior}

Social platforms are effectively complex networks where users (nodes) and their interactions (edges) encode the pathways for information flow. These networks exhibit highly heterogeneous connectivity, often characterized by fat-tailed degree distributions and significant mesoscale organization (clustering, communities, and homophily).

To understand reach and virality, we must characterize:
\begin{enumerate}
    \setcounter{enumi}{2}
    \item \textbf{What network features and user behaviors govern reach, virality, and threshold effects?} We will explore how topological features (e.g., directionality, weighting, community structure) interact with microscopic user-level dynamics (e.g., cognitive biases, attention allocation, reinforcement) to determine macroscopic outcomes. This calls for the use of rigorous methods, such as null models, to isolate the structural properties driving observed dynamics.
\end{enumerate}

\subsubsection*{Modeling collective phenomena: consensus, polarization, and fragmentation}

At the collective level, the interplay between structure and individual behavior gives rise to critical emergent phenomena. Interactions driven by homophily often create tightly-knit, ideologically insulated communities like echo chambers and polarized communities.

A central objective is to model and explain these collective states:
\begin{enumerate}
    \setcounter{enumi}{3}
    \item \textbf{Under what conditions does a population reach consensus, polarization, or fragmentation?} We will review computational opinion dynamics models--including bounded-confidence frameworks and adaptive network theories--that describe how network structure can amplify confirmation bias, filter information diversity, and lead to ideological extremization, providing critical insights into the mechanisms driving today's polarized discourse.
\end{enumerate}

\subsubsection*{Developing strategies to mitigate information disorder}

The complexity of the modern information ecosystem is intrinsically linked to the rise of information disorder, which encompasses misinformation, disinformation, and mal-information. The widespread dissemination of false narratives, often amplified by coordinated campaigns involving automated accounts, represents a major societal risk.

The final aim of this review is focused on practical applications:
\begin{enumerate}
    \setcounter{enumi}{4}
    \item \textbf{Can we design effective, physics-informed interventions to mitigate misinformation diffusion?} By fully understanding the dynamics--including the roles of simple vs. complex contagion, threshold effects, and temporal properties like burstiness and non-Markovianity--we can lay the groundwork for developing counter-strategies capable of disrupting malicious cascades and fostering a healthier informational environment.
\end{enumerate}

\paragraph{Limitations} 
As most of the results presented draw on evidence from online social platforms, it is worth recalling a major limitation recently highlighted in the literature to properly contextualise these findings~\cite{budak2024misunderstanding,falkenberg2025towards}. Most research focuses on a narrow group of high-income democracies, with the United States dominating the field. However, the organisation of political communication is correlated with the electoral system, which influences the extent of cooperation among political actors~\cite{urman2020context,vanvliet2021political}. In this respect, the U.S. plurality system represents a specific institutional setting that tends to foster polarisation. On the other hand, countries with proportional electoral systems tend to show lower levels of polarisation and more frequent interactions across opposing political camps~\cite{urman2020context,vanvliet2021political}. As a consequence, generalising findings from one country to others is risky, as different political environments can substantially affect the outcomes of studies and limit their applicability.

\subsection{Organization}

The remainder of this review proceeds as follows. We begin by establishing the necessary theoretical framework. Section~\ref{sec:networks} introduces the essential models of complex networks, covering structural properties of real-world socio-technical graphs and the probabilistic, mechanistic, and null models used for their analysis. This is followed by Section~\ref{sec:dynamics}, which reviews the core physical models of social dynamics, including classical frameworks like percolation, epidemic-like compartmental models, and spin models, which form the basis for understanding collective behavior.

We then ground these concepts in the modern context. Section~\ref{sec:media} details the main elements forming the present-day media ecosystem, offering a comparative discussion of major platforms (e.g., X, Facebook, Reddit) and summarizing the novel information disorders that have emerged in this context.

The core of the review applies this toolkit to specific collective phenomena. Section~\ref{sec:info_spreading} focuses on information spreading. This section first details the models of dynamics (from simple to complex contagion, including cognitive factors) and then presents empirical insights on topological and temporal characteristics. Crucially, it includes an in-depth analysis of misinformation dynamics, models, and proposed mitigation strategies. Following this, Section~\ref{sec:opinion_models} addresses models of opinion dynamics. We review major modeling paradigms, including discrete opinion models (like the Voter model) and continuous models (like Deffuant-Weisbuch), concluding with the critical challenge of modeling the coevolution of opinions and social structure.

Finally, Section~\ref{sec:conclusions} offers a summary and draws conclusions, highlighting key findings and future research directions in this rapidly evolving field.

\section{Models of Complex Networks}
\label{sec:networks}
Modeling complex social systems as complex networks is crucial for understanding the intricate relationships and emergent behaviors that characterize human interactions.
Traditional approaches often simplify these systems, overlooking the non-linear effects of connectivity.
By representing individuals or entities as nodes and their interactions as links, network models allow us to capture the underlying structure of social connections, which profoundly influences phenomena like opinion formation, disease spread, and collective action.
This network perspective provides a powerful framework to analyze how local interactions scale up to global patterns, revealing insights into the resilience, efficiency, and dynamics of social structures that would otherwise remain hidden.

\subsection{Properties of real-world networks}

Over the past two decades, the study of real-world networks---social, technological, biological, and information-based---has revealed a set of recurring structural features. These empirical properties are crucial for understanding processes such as diffusion, contagion, and percolation, and they serve as benchmarks for generative models of networks. 

One of the most studied properties of real-world networks is the distribution of degrees across nodes. Many networks exhibit highly skewed degree distributions, where most nodes have only a few connections while a small number of nodes act as hubs with very high degree. Early studies suggested that such distributions often follow a power-law form, $P(k) \sim k^{-\alpha}$ with $2 < \alpha < 3$ \cite{barabasi1999emergence}. This led to the characterization of many networks as ``scale-free.''

However, subsequent large-scale statistical evaluations found that pure power laws are less common than initially believed. Instead, many networks are better described by truncated power laws, log-normal distributions, or exponential cutoffs \cite{broido2019scale,clauset2009power}. Deviations from the idealized scale-free pattern, especially in the low-degree regime, have significant implications for dynamical processes such as information diffusion or epidemic spread. For example, while hubs can dramatically lower epidemic thresholds, the scarcity of low-degree nodes may reduce the potential for long-tail spreading cascades.

Another hallmark of real networks is the ``small-world'' property: average path lengths are short, typically scaling logarithmically with the number of nodes \cite{watts1998collective}. This ensures that information or contagion can propagate efficiently across large systems. At the same time, real-world networks exhibit unusually high clustering coefficients compared to random graphs, reflecting the strong tendency for two neighbors of a node to also be connected.

Clustering is often heterogeneous, with a clustering spectrum that varies by degree: high-degree nodes typically exhibit lower clustering, while low-degree nodes belong to more tightly knit communities \cite{ravasz2003hierarchical}. Such patterns suggest hierarchical or modular organization.

A pervasive property of real-world networks is the presence of community or modular structure: nodes cluster into groups with dense internal connections and sparse external links \cite{fortunato2010community}. Such mesoscopic organization underpins functional specialization in biological systems, topic groups in citation networks, and social circles in human interaction graphs.

Communities can overlap, evolve over time, and exist at multiple scales. Their presence strongly influences diffusion dynamics: spreading processes can be slowed by modular boundaries, but inter-community bridges often serve as critical conduits for global cascades.

Networks also differ in how nodes of similar degree connect. Social networks often show assortative mixing, where high-degree nodes preferentially connect to other high-degree nodes \cite{newman2018networks}. In contrast, technological and biological networks are frequently disassortative, with hubs tending to connect to low-degree nodes.

Degree correlations significantly impact diffusion processes: assortative networks can facilitate robust spreading within highly connected cores, while disassortative structures enable rapid reach across peripheral nodes. Generative models often incorporate assortativity through rewiring mechanisms or correlation-preserving constraints.

The properties outlined above---heavy-tailed degree distributions, small-world effects, clustering, community structure, and assortativity---constitute the empirical foundation for modern network science. They highlight the limitations of classical random graph models and motivate the development of richer generative frameworks, such as preferential attachment, small-world models, stochastic block models, and fitness-based approaches. Any comprehensive survey of diffusion models must first situate them against these structural features, since they directly shape the pathways and thresholds of information spread.

\subsection{Probabilistic network models}
\label{sec:probabilistic_models}

The study of complex networks has been profoundly shaped by a number of foundational modeling frameworks, each of which highlights different mechanisms by which connectivity patterns may arise.  Among the most influential contributions are random graph theory~\cite{erdos1959on,erdos1960on,bollobas2011random}, the Watts--Strogatz small-world model~\cite{watts1998collective}, models based on suitable definitions of fitness~\cite{caldarelli2002scale,bianconi2001competition}, exponential random graph models (ERGMs)~\cite{holland1981exponential,frank1986markov}, stochastic block modeling~\cite{holland1983stochastic,peixoto2014hierarchical}, and latent geometry approaches~\cite{papadopoulos2012popularity}.  Together, these frameworks form a diverse toolkit that can be used to illuminate distinct empirical regularities observed in social, biological, and technological networks.

Random graph theory, as formalized by Erd\H{o}s and R\'enyi~\cite{erdos1959on}, considers the ensemble of graphs in which edges between node pairs are placed independently with equal probability.  This simple generative prescription has the striking consequence that the typical distance between two randomly chosen nodes grows only logarithmically with the number of nodes: in this narrow sense, Erd\H{o}s--R\'enyi (ER) graphs already display the ``small-world'' property of short path lengths.  However, ER graphs fail to reproduce at least two ubiquitous empirical signatures, making these graphs unsuitable as faithful models of real complex systems, such as social networks.
On the one hand, the model yields networks with Poisson-like degree distributions, as opposed to the heavy-tailed distributions of real networks.
On the other hand, in ER ensembles the clustering coefficient (the tendency of neighbors of a node to be connected to one another) vanishes as the system size increases, whereas empirical networks present high local clustering. 

The configuration model extends the ER model proposing a fundamental random graph ensemble that generates networks with an arbitrary, user-specified degree sequence \cite{molloy1995critical,newman2018networks}. Each node is assigned a number of half-edges (stubs) according to the desired degree distribution, and edges are formed by randomly pairing stubs until none remain. This produces graphs that exactly preserve the degree sequence while remaining maximally random in all other respects. 

A distinct but complementary class of models was proposed to address the more general problem of reproducing a set of desired network properties. Exponential random graph models (ERGMs)~\cite{holland1981exponential,frank1986markov} extend the random graph paradigm by endowing the ensemble of networks with a Gibbs-like probability distribution that explicitly biases graph occurrence according to chosen structural features.  Formally, ERGMs assign to each graph a probability proportional to $\exp(-\sum_i \theta_i C_i(G))$, where the $C_i(G)$ are structural statistics (e.g., total number of edges, degree counts, numbers of triangles) and the $\theta_i$ are parameters (Lagrange multipliers) that control their expected values in the ensemble.  In this way ERGMs provide a principled, maximum-entropy style route to generate networks that match empirical constraints while remaining otherwise as random as possible.  Their flexibility makes ERGMs particularly attractive for social network analysis, because one can tune the model to reproduce effects such as degree heterogeneity, homophily, and triadic closure.  At the same time, this flexibility comes at a cost: parameter estimation and sampling in ERGMs can be computationally demanding, prone to issues of model degeneracy and poor mixing, and often requires careful model specification and diagnostics for reliable inference.

The Watts--Strogatz (WS) model~\cite{watts1998collective} addressed a complementary limitation of simple random constructions by proposing an extremely parsimonious mechanism that interpolates between regular order and randomness.  The WS recipe begins with a regular ring lattice in which each node is connected to its $k$ nearest neighbors, a structure that yields very high clustering but long average path lengths.  A single parameter, the rewiring probability $p$, controls the transformation: with probability $p$ each edge is detached from its original neighbor and reconnected to a uniformly chosen random node, avoiding duplicates and self-loops.  Two extremes are notable: when $p=0$ the network remains the original lattice (maximal clustering, large distances), while when $p=1$ the network becomes essentially a random graph with short distances but negligible clustering.  Crucially, for small but nonzero values of $p$ the WS construction produces networks that simultaneously exhibit high clustering and short average path length: a few randomly rewired edges act as shortcuts across the lattice and drastically reduce global distances while the dense local neighborhood structure responsible for clustering is preserved.  This two-effect behaviour—shortcuts + retained local structure emerging by tuning a single parameter—was one of the first parsimonious explanations for how empirical networks can combine global navigability with strong local cohesion, and it highlighted the importance of wiring heterogeneity (shortcuts) in producing the small-world phenomenon in a clustered substrate.

A different modeling perspective focuses on intrinsic node heterogeneity rather than on topology-preserving rewiring.  In fitness-based or hidden-variable models~\cite{caldarelli2002scale,bianconi2001competition}, each node is endowed with an intrinsic attractiveness or fitness drawn from a distribution; the probability that two nodes connect is then a function of their fitnesses.  This mechanism can generate heavy-tailed degree distributions and reproduce degree heterogeneity observed in many empirical systems without invoking explicit growth dynamics.  Extensions incorporate evolving fitnesses, multiple fitness dimensions controlling different types of ties, or coupling between fitness and spatial/semantic proximity; fitness models therefore provide an appealing route to explain degree variability as an outcome of latent, node-level attributes.

Despite the value of these paradigms, no single class of models captures all empirical regularities simultaneously.  ER graphs capture short path lengths but lack clustering; WS graphs explain clustering and short paths but do not naturally produce broad degree heterogeneity; fitness models generate heavy-tailed degrees but do not by themselves explain modular structure or strong clustering at multiple scales.  To bridge these gaps, subsequent modeling efforts have incorporated similarity or latent structure as an additional organizing principle.  Latent space models~\cite{hoff2002latent} posit that nodes occupy positions in an abstract similarity space and that connection probabilities decay with distance; stochastic block models (SBMs)~\cite{holland1983stochastic} encode group-based affinities and modularity explicitly, and degree-corrected variants~\cite{karrer2011stochastic} add node-specific parameters to account for heterogeneous degree while preserving community structure.  Hierarchical and Bayesian extensions of SBMs~\cite{peixoto2014hierarchical,peel2017ground} enable principled multiscale inference and model selection.  From a complementary statistical-physics perspective, maximum-entropy ensembles~\cite{park2004statistical} formalize null models constrained by empirical observables and have been generalized to incorporate correlated topologies, clustering constraints, and higher-order structures~\cite{squartini2011analytical,bianconi2009entropy} (more details on this subject can be found in Subsection~\ref{ssec:maxent}).

Finally, latent geometric formulations—most notably Random Hyperbolic Graphs (RHGs) and related constructions—provide a parsimonious unifying picture in which popularity (expected degree) and similarity are encoded as radial and angular coordinates in a negatively curved space.  This geometric embedding yields heavy-tailed degree distributions, strong clustering, and navigable small-world topologies within a compact and interpretable generative mechanism~\cite{krioukov2010hyperbolic,boguna2010sustaining,papadopoulos2012popularity,muscoloni2018machine,boguna2009navigability}, and has proven useful both as an explanatory model and as a practical tool for network synthesis and analysis.

\subsection{Generative network models}

A central strand in the development of network science has been the proposal of 
\emph{generative models} in which the network grows or evolves 
through explicit rules that specify how new nodes and edges are added, rewired, or removed. 
Unlike purely statistical ensembles, these models are designed to capture the actual 
processes that drive the formation of links in real systems. 
In many cases, they were directly motivated by social networks, 
where ties emerge from concrete interaction mechanisms such as popularity, similarity, or repeated encounters. 
As such, mechanistic models can be seen as stylized but insightful hypotheses 
about the micro-level dynamics underlying observed macroscopic structures 
\cite{caldarelli2007scale,newman2018networks}.

The most influential example is the Barab\'asi--Albert (BA) model~\cite{barabasi1999emergence}, 
which introduced the principle of \emph{preferential attachment}. 
In this framework, networks grow by the sequential addition of new nodes, 
each of which connects to existing nodes with a probability proportional to their degree. 
The resulting ``rich-get-richer'' mechanism naturally generates scale-free degree distributions, 
a hallmark of many real-world systems, including the Web and scientific citation networks. 
Although simple, the BA model provided a mechanistic explanation for 
heterogeneous connectivity patterns and inspired a large family of extensions: 
nonlinear preferential attachment rules, aging effects (where older nodes become less attractive over time), 
and fitness-based modifications where the propensity of a node to attract links depends on intrinsic qualities 
\cite{krapivsky2000connectivity,bianconi2001bose,caldarelli2002scale,medo2011temporal}. 
These refinements retain the core idea of cumulative advantage while incorporating 
more realistic assumptions about competition, obsolescence, or node heterogeneity.

Another class of mechanistic models was explicitly tailored to social networks. 
A well-known example is the \emph{networked seceder model} by Gr\"onlund and Holme~\cite{gronlund2004networking}. 
Here the central mechanism is identity formation: individuals seek to distinguish themselves from the average 
while still remaining within a recognizable community. 
The algorithm operationalizes this by letting nodes compare their positions to others and 
rewire links toward neighbors of nodes that are farther from the population center. 
The emergent outcome is the spontaneous formation of community structure, 
a robust feature of empirical social networks that the BA model alone cannot capture. 

Similarly, the Jin--Girvan--Newman model~\cite{jin2001structure} 
was motivated by empirical observations of friendship networks among students. 
It combines three behavioral rules: 
(i) new friendships are more likely between individuals who already share mutual acquaintances 
(triadic closure), 
(ii) there is a cognitive or social cap on the number of friendships a person can maintain, 
and (iii) ties decay over time unless reinforced by repeated interaction. 
These simple but realistic assumptions lead to networks with clustering, turnover, and community-like structures, 
all of which are absent in purely degree-driven models. 
The explicit use of triadic closure in particular has been recognized as a key mechanism 
in shaping the topology of social networks, complementing the role of preferential attachment.

Further refinements include models that incorporate \emph{homophily}, 
the tendency of individuals to form ties with others who are similar in attributes such as opinions, 
demographics, or geography. 
Homophily-driven attachment rules can produce modular networks and echo chambers, 
providing mechanistic insight into how social cleavages translate into network segregation 
\cite{mcpherson2001birds}. 
Other models have focused on \emph{temporal dynamics of link formation}, 
highlighting bursty and heterogeneous activity patterns: 
for instance, Myers and Leskovec~\cite{myers2014bursty} 
analyzed follower dynamics on Twitter and proposed mechanisms 
where exposure, novelty, and individual activity rates shape the formation and dissolution of ties.

In summary, mechanistic models offer a complementary perspective to purely statistical or geometric ones. 
They provide explicit hypotheses about how microscopic processes—preferential attachment, 
triadic closure, identity formation, homophily, or activity bursts—combine to generate 
the large-scale structural features observed in empirical networks. 
While our review is not primarily focused on network formation, 
it is important to stress that the spreading of news, rumors, and opinions 
takes place on social substrates whose very structure is shaped by such mechanisms, 
often with strong feedback loops between information dynamics and network evolution.

\subsection{Testing the right model: null models \& maximum-entropy ensembles}\label{ssec:maxent}
In Physics, it is standard practice to develop null models to separate genuine signals from random noise. These models partially encode the structure of the real system and are used to assess the statistical significance of empirical measurements. For example, null models were instrumental in the characterization of the Higgs boson’s properties~\cite{aad2013evidence}. If a null model fails to reproduce an observed phenomenon—e.g., if the associated p-value falls below a pre-defined significance threshold—then the observation is considered informative, indicating that the model lacks key explanatory elements. Conversely, if the model reproduces the observation, the result can be attributed to the mechanisms already embedded in the model and is thus not considered to provide additional insight.

Selecting an appropriate null model is, to some extent, an art. Such models should reflect the essential characteristics of the empirical system while remaining sufficiently flexible, as overly restrictive models may obscure the distinct contributions of various structural ingredients to the observed data.
In the absence of specific model assumptions, a well-founded null hypothesis is given by the maximum-entropy random graph constrained by known empirical quantities. By construction, maximum-entropy null models are maximally random—based on the maximization of Shannon information entropy—and thus serve as unbiased statistical benchmarks.
Although rooted in Information Theory, maximum-entropy null models have a direct analogue in Statistical Physics through the canonical ensemble. In this context, the constraints used in entropy maximization play a role analogous to energy in the canonical ensemble. The formal equivalence between Statistical Physics and Information Theory was first established by Jaynes in 1957~\cite{jaynes1957information}; a comprehensive application of this framework to complex networks is discussed in~\cite{cimini2019statistical}. This approach has been successfully applied to systems such as international trade networks, financial networks~\cite{bardoscia2021physics}, and, more recently, to Online Social Media.

The definition of maximum entropy null models is essentially structured in three steps, beginning with the empirical network to be analyzed, hereafter denoted as $G^*$\footnote{All quantities marked with an asterisk $*$ refer to empirical values.}.

First, one defines the ensemble $\mathcal{G}$, i.e., the set of all possible graphs with the same number of nodes, ranging from the empty graph to the fully connected graph~\cite{park2004statistical}.
Next, each graph in the ensemble is assigned a probability by performing a constrained maximization of the Shannon entropy over the ensemble, namely  
\begin{equation*}
    S = -\sum_{G \in \mathcal{G}} P(G)\,\ln P(G).
\end{equation*}

If $\vec{C}(G)$ is a vector of quantities measured on a generic graph $G \in \mathcal{G}$—for instance, the total number of links or the degree sequence—one may require the null model to be maximally random (i.e., to maximize its entropy) while fixing $\langle \vec{C} \rangle$, the ensemble-average of $\vec{C}$, to a prescribed value~\cite{park2004statistical}. As in statistical ensembles, it follows that the probability satisfying these constraints has the form  
\begin{equation}\label{eq:ERG_p}
    P(G) \;=\; \frac{e^{-\vec{\theta} \cdot \vec{C}(G)}}{Z(\vec{\theta})},
\end{equation}
where $Z(\vec{\theta})$ is the partition function and $\vec{\theta}$ is the vector of Lagrange multipliers, analogous to the inverse temperature in the canonical ensemble.

Finally, to tailor the model to empirical data, one typically sets $\langle \vec{C} \rangle = \vec{C}(G^*)$, which determines the numerical values of the multipliers $\vec{\theta}$~\cite{garlaschelli2007self,garlaschelli2008maximum,squartini2011analytical}. Remarkably, for maximum entropy null models, this constrained entropy maximization is equivalent to likelihood maximization; other constructions may exhibit inconsistencies between their definitional constraints and the likelihood, thereby introducing nontrivial, hard-to-control biases in the analysis~\cite{garlaschelli2008maximum}.

Let us conclude this paragraph with some remarks. 

First, Erd\H{o}s--R\'enyi random graphs, as well as Stochastic Block models, can be obtained through a maximum entropy approach, just constraining, respectively, the total number of links, and the number of links inside each block and between each couple of blocks. Furthermore, the reader may have noticed the similarity between the probability per graph in Eq.~(\ref{eq:ERG_p}) and the definition of Exponential Random Graph models. Indeed, Park and Newman first showed how ERGMs can be derived from a maximum entropy approach~\cite{park2004statistical}. Nevertheless, there is a crucial difference in the way those models are used: in Section~\ref{sec:probabilistic_models}, models were presented to explain the network structures observed in empirical systems. In the present Subsection, the logic is reversed: instead of describing the phenomenon, the null model acts as a benchmark. By comparing the empirical system with a model that incorporates properties of the real system, one can identify non-trivial features of the empirical system that the constraints cannot explain. In principle, any model can serve as a null model; however, choosing maximum-entropy null models helps reduce the bias introduced in the statistical benchmark.

Second, in the canonical ensembles of Statistical Mechanics, the quantity determining each microstate’s probability is global --namely, the total energy. In contrast, when applying the same framework to complex networks, the constraints are often local, defined at the node level. Consequently, the model incorporates more detailed information about the empirical data, akin to knowing the average energy of each particle in a gas. While in Statistical Mechanics the per-particle energy is less relevant since all particles have the same properties~\cite{huang2009introduction}, in many online systems agent behavior follows fat-tailed distributions. For example, user activity metrics such as content creation, reposting frequency, and content virality often exhibit power-law–like behavior. Therefore, specifying the local “energy” at the node level becomes particularly informative~\cite{bianconi2007entropy,bianconi2009entropy}.

Third, although one can in principle constrain any network-based quantity, in practice certain choices are preferred. In many cases, the degree and strength sequences are especially informative~\cite{bianconi2007entropy}. Moreover, linear constraints on the adjacency matrix—such as those enforcing degree or strength sequences—allow the factorization of each graph’s probability into edge-specific probabilities~\cite{park2004statistical}. Hence, these constraints are among the most popular.  

Finally, within the maximum entropy framework and in the absence of additional information, links are distributed as uniformly as possible, subject to the imposed constraints. In other words, a maximum entropy null model allocates weights and edges uniformly, consistently with those constraints. In this sense, maximum entropy null models provide an ideal statistical benchmark for detecting an excess of connections among node groups, a feature of particular relevance for online social platforms, which often exhibit strongly modular structures~\cite{conover2011political,wu2011who}. Indeed, multiple studies have shown that users cluster according to shared interests, sensibilities, or political leanings. Accordingly, maximum entropy null models are especially effective for capturing such information.

\section{Physical Models of Social Dynamics}
\label{sec:dynamics}

The study of complex social phenomena necessitates moving beyond static structural descriptions to embrace the dynamics that unfold upon these networks. While Section~\ref{sec:networks} provided the foundational graph-theoretical tools for characterizing the topology of socio-technical systems, this section introduces the physical models crucial for understanding time-dependent processes like spreading, influence, and collective behavior. Models rooted in statistical physics--originally developed to explain phase transitions, critical phenomena, and non-equilibrium dynamics in materials and biological systems--offer a powerful, parsimonious framework to capture the complex, active, and out-of-equilibrium nature of human interactions. Specifically, we survey approaches that model the propagation of states (e.g., adoption of information or infection), the spatial and temporal exploration of the network, and the collective alignment or polarization of opinions. These models allow for the quantitative investigation of how microscopic interaction rules translate into macroscopic societal outcomes.

\subsection{Percolation and branching processes}
\label{sec:percolation_branching}

Percolation and branching processes are two fundamental stochastic frameworks that offer complementary perspectives on propagation phenomena. Percolation theory considers the emergence of large-scale connectivity in a network when edges or nodes are randomly occupied with probability $p$ \cite{grimmett2012percolation,stauffer2018introduction}. At the percolation threshold $p_c$, the system undergoes a phase transition: below $p_c$, occupied sites form only small clusters, whereas above $p_c$, a giant connected component emerges that enables system-wide propagation. On random networks with arbitrary degree distributions, percolation can be analyzed using generating functions. If $P(k)$ is the degree distribution and
\begin{equation}
    G_0(x) = \sum_{k=0}^\infty P(k)x^k, \quad
    G_1(x) = \frac{G_0'(x)}{G_0'(1)},
\end{equation}
then the condition for the emergence of a giant connected component is
\begin{equation}
    G_1'(1) = \frac{\langle k^2 \rangle - \langle k \rangle}{\langle k \rangle} > \frac{1}{p},
\end{equation}
which yields the critical occupation probability $p_c$ \cite{newman2001random,callaway2000network}.

Branching processes, in contrast, model reproduction dynamics generation by generation. In the classical Galton--Watson process, each individual produces a random number of offspring according to distribution $\{p_k\}$, with generating function
\begin{equation}
    f(s) = \sum_{k=0}^\infty p_k s^k.
\end{equation}
The extinction probability $q$ is the smallest nonnegative solution of
\begin{equation}
    q = f(q).
\end{equation}
If the mean number of offspring
\begin{equation}
    m = f'(1) = \sum_{k=0}^\infty k p_k
\end{equation}
satisfies $m \leq 1$, then $q=1$ and extinction occurs almost surely; if $m>1$, then $q<1$ and there is a positive probability of survival \cite{watson1875probability,harris1963theory,athreya2012branching}. In epidemiological and information-diffusion settings, $m$ plays the role of a basic reproduction number $R_0$: when $R_0>1$, widespread propagation becomes possible.

These two perspectives are deeply linked. On locally tree-like networks, the percolation threshold coincides with the branching-process criticality condition: above the threshold, the Galton--Watson approximation admits survival, and simultaneously the network exhibits a giant connected component \cite{newman2001random,callaway2000network}.

Branching-process intuition also extends naturally to continuous-time settings via Hawkes processes. A Hawkes process has intensity
\begin{equation}
    \lambda(t) = \mu + \sum_{t_i < t} g(t - t_i),
\end{equation}
where $\mu$ is a baseline rate and $g(\cdot)$ is a triggering kernel. Each event can generate offspring events in time, and the average branching ratio is
\begin{equation}
    n = \int_0^\infty g(s)\,ds,
\end{equation}
which parallels the mean offspring number in Galton--Watson processes. If $n<1$ the process is subcritical and event cascades die out quickly, while $n\geq 1$ implies potential unbounded growth \cite{hawkes1971spectra,crane2008robust}.

In practice, percolation theory provides insight into global connectivity and final outbreak sizes, while branching processes capture early stochastic dynamics and extinction probabilities. Temporal models such as Hawkes processes enrich the picture by incorporating realistic timing of events. Modern studies of information diffusion increasingly integrate these perspectives, combining percolation for critical thresholds, branching-process approximations for early dynamics, and point-process models for empirical temporal patterns \cite{watts2002simple,kempe2003maximizing,crane2008robust}.

\subsection{Epidemic-like compartmental models}
\label{sec:epidemic_models}

Network epidemiology models assume that network edges serve as the principal mechanisms of disease transmission.
They generally abstract away many intricate biological facets to instead concentrate exclusively on population dynamics, focusing on how the disease spreads through the population as a whole rather than on detailed biological processes within an individual.
Each individual within a defined population is categorized into one of a limited number of possible, mutually exclusive states, often referred to as compartments.

There are three particularly notable compartmental models that are fundamental to understanding epidemic dynamics:
\begin{itemize}
    \item In the S $\xrightarrow{\beta}$ I Model (Susceptible-Infected Model) the only permissible transition for an individual is from the "Susceptible" (S) compartment to the "Infected" (I) compartment. Once an individual becomes infected, they remain in the infected state indefinitely, meaning there is no recovery or removal from the infectious pool.
    \item The S $\xrightarrow{\beta}$ I $\xrightarrow{\gamma}$ S Model (Susceptible-Infected-Susceptible Model) introduces a recovery mechanism, positing that infected individuals, after a certain period, eventually return to the susceptible state. This implies that they regain susceptibility and can be reinfected multiple times.
    \item Finally, in the S $\xrightarrow{\beta}$ I $\xrightarrow{\gamma}$ R Model (Susceptible-Infected-Recovered Model) infected individuals eventually recover from the disease (or, in some contexts, die) and, crucially, gain permanent immunity, meaning they cannot be reinfected. 
\end{itemize}

In the deterministic formulation of these epidemic models, the transition rules governing the movement of individuals between compartments are quantitatively expressed in terms of constant rates, with $\beta$ generally used for the infection rate and $\gamma$ for the recovery rate. 
Consequently, the macroscopic dynamic behavior of the epidemic, detailing how the numbers of individuals in each compartment change over time, is meticulously described by a system of coupled differential equations.
These equations represent the average behavior of the population, assuming large numbers of individuals and well-mixed interactions.

The solution to the differential equations offers a valuable mean-field picture for large, perfectly well-mixed populations.
In the SI model, all individuals eventually catch the disease, and the mean-field solution shows that the fraction of infected individuals increases sigmoidally, approaching 1 (the entire population) as time goes to infinity.
The SIS and SIR models, instead, exhibit a crucial threshold behavior determined by the basic reproduction number, $R_0 = \beta/\gamma$. This dimensionless value quantifies the average number of secondary infections produced by a single infected individual when introduced into a completely susceptible population.
When $R_0 \le 1$, the epidemic will die out, because the recovery rate is high enough relative to the transmission rate to prevent a large-scale epidemic from sustaining itself.
When $R_0 > 1$, on the other hand, an epidemic occurs: in the SIS model, the disease will persist in the population, reaching a non-zero stationary state $i^* = 1 - 1/R_0$; in the SIR model, the disease will spread through a significant portion of the population (the epidemic size) before eventually dying out, primarily due to the depletion of susceptible individuals.

When transitioning from the simplifying assumption of fully-mixed deterministic models to the more realistic realm of stochastic models explicitly defined on networks, the underlying network structure becomes of paramount importance.
The specific arrangement of contacts and relationships within the network significantly and fundamentally influences the epidemic's progression, its final extent, and its overall outcome.
In particular, the homogeneous mean-field approximation works poorly for networks with a heterogeneous degree distribution. 

To address this issue, Pastor-Satorras and Vespignani~\cite{pastor2002epidemic} proposed the heterogeneous mean-field (HMF) approximation, based on formulating a specific equation for the relative density $i_k(t)$ of infected nodes of degree $k$.
Under the assumption that all links are equally likely--i.e., that the network follows the so-called configuration model--the dynamical equation for the infected fraction \(i_k(t)\) reads:
\[
\frac{di_k}{dt} = -\gamma i_k + \beta k [1 - i_k - r_k]\;\Theta(t),
\]
where \(\Theta(t)=\sum_{k'} \frac{k' P(k')}{\langle k\rangle} i_{k'}(t)\).
Solving these self-consistency equations yields the epidemic threshold $\beta/\gamma > \langle k \rangle/\langle k^2 \rangle$.

The HMF has two main drawbacks: it neglects the actual structure and assumes that all nodes of degree $k$ are equivalent, and it predicts a vanishing threshold for scale-free networks in the thermodynamic limit~\cite{pastor2001epidemic}. 
A further approximation, independently obtained by Wang, Chakrabarti et al.~\cite{wang2003epidemic,chakrabarti2008epidemic} and by G\'omez et al.~\cite{gomez2010discrete}, is referred to as quenched mean-field (QMF) (as opposed to the HMF being annealed).
The QMF is also called individual-based mean-field and it is based on studying the temporal evolution of the probability that a node is infected, considering the actual structure of the network.
One approximates each node's infection probability $x_i(t)$ by
\[
x_i'(t) \approx -\gamma x_i + \beta \sum_j A_{ij} (1 - x_i)x_j,
\]
treating neighbor states as independent.
Linear stability analysis then shows an epidemic threshold at $\beta/\gamma > 1/\Lambda_{\max}$, where $\Lambda_{\max}$ is the largest eigenvalue of the adjacency matrix $A$ of the network.
This QMF result has been rigorously derived and verified for SIS dynamics~\cite{pastor-satorras2015epidemic}.
Intuitively, the most connected core of the network sets the scale for spreading.

More general approximate methods that try to take various network structural considerations into account do exist~\cite{pastor-satorras2015epidemic,kiss2017mathematics,britton2019stochastic}, as well as more detailed epidemic models where, for instance, there can be intermediate states accounting for exposure periods where individuals are not yet infected~\cite{pastor-satorras2015epidemic,kiss2017mathematics,britton2019stochastic}.
More generally, for more complex epidemic models and arbitrary networks, the system evolution can be simulated, even on temporal~\cite{masuda2013predicting} or higher-order~\cite{wang2024epidemic} networks.

\subsection{Diffusion, random walks, navigation and routing}
\label{sec:diffusion_navigation}

The physics concept of diffusion has also been used as a model of social phenomena, albeit the term ``diffusion'' is often used in the social sciences for a different phenomenon, where whatever spreads does not need to have a conserved mass.
Physical diffusion~\cite{masuda2017random} is the most basic dynamic process that allows information to flow between distinct parts of an online social network. In physics, the standard diffusion equation for a field $u(\Vec{r},t)$ can be written as 
\begin{equation}
    \frac{\partial u(\Vec{r},t)}{\partial t}=D\nabla^2 u(\Vec{r},t). 
\end{equation}
when the time derivative is equal to zero, the stationary solution corresponds to evaluating the Laplacian operator $\nabla^2 u(\Vec{r},t)=0$. 
On a complex network with adjacency matrix $\mathbf{A}$, such as an online social system, we must consider a discretized field $\Vec{u}(t)$, where $u_{i}(t)$ indicates the value of the field on node $i$ at time $t$. Accordingly, passing to finite differences to describe the continuous derivatives of the field, the diffusion equation takes the form 
\begin{equation}
    \frac{\partial \Vec{u}(t)}{\partial t}=(\mathbf{K}-\mathbf{A}) \Vec{u}(t)=\mathbf{L}\Vec{u}(t)
    \label{eq:lapl},
\end{equation}
which in the case of a stationary field $\Vec{u}^{\star}$ reduces to
\begin{equation}
    \mathbf{L}\Vec{u}^{\star}=0
    \label{eq:lapl2},
\end{equation}
where $\mathbf{K}$ is the degree matrix---a diagonal matrix where $K_{ii}=k_{i}$ is the degree, or the number of neighbors, of the $i$-th node---and $L$ is the combinatorial Laplacian matrix.

Physical diffusion is the macroscopic counterpart of a random walk process. While it is a general powerful model to address the phenomenon, it is nonetheless a poor model of any kind of information propagation: (i) (Dis)information typically doesn't leave the spreader. (ii) The broadcast from a source node to its neighbors is not deterministic: either some neighbors might not be reached for some reason, or they are reached and do not want to spread further that piece of information, with or without modification. Accordingly, it is possible to consider other spreading processes that can capture various details observed in empirical systems. 
On the other hand, since we have decided to describe the system in the form of a network, standard diffusion phenomena clarify what will be the role of Laplacian matrix in the study of information disorder.

For instance, random walk dynamics might more accurately capture information diffusion if the information is spread from a node to a single neighbor due to some form of contact at time $t$. In this case, the governing equation is very similar to Eq.~(\ref{eq:lapl2}), where instead of the combinatorial Laplacian, the normalized Laplacian matrix defined by $\tilde{\mathbf{L}}=\mathbf{K}^{-1}\mathbf{L}$ appears.

Regardless of the specific spreading process considered, once the formalism is set up, estimating how long the information takes to reach a destination at a distance $d$ from the source is possible. In physical diffusion, the mean square displacement, usually adopted for such a purpose, grows linearly with time. For active matter, and then presumably for temporal networks that rearrange their links, the presence of topological shortcuts, such as in small-world~\cite{watts1998collective} and ultra-small-world networks~\cite{cohen2003scale}, might lead to superdiffusion~\cite{allen-perkins2019markov,ji2023signal}. 
Nevertheless, specific other topological configurations, such as the ones characterized by strong topological clusters, where the similarity with echo chambers is evident, might trap the random walker into specific parts of a network for a long time, longer than expected from networks with less marked groups.

While diffusion and random walks offer valuable models for understanding how information spreads in an unguided, exploratory manner, the study of navigability and routing shifts the focus to how that spread can be deliberately optimized by following specific, efficient pathways. 
Refs.~\cite{watts2002identity,kleinberg2000navigation} highlight the significance of Stanley Milgram's famous ``small-world'' experiment, in which participants were asked to forward a package through their personal acquaintance networks~\cite{milgram1967small,travers1969experimental}.
This experiment served as a foundational case study and motivated a wide range of research in complex networks, notably through Ref.~\cite{watts1998collective}.
Beyond analyzing conditions for local clustering and the scaling of global path lengths, several studies emphasized that the experiment's main contribution lies in showing that individuals can effectively navigate social networks.
This insight supported the broader view that humans are adept at managing their social ties~\cite{boissevain1974friends,granovetter1973strength}.
Building on this idea, Ref.~\cite{watts2002identity} proposed that people rely on a hierarchical mental representation of their social groups to perform such navigation.

The navigation problem becomes academically interesting when one has some partial information to guide the search for a path from the source to the target node~\cite{lee2011pathlength,benevenuto2012characterizing}.
In the limit of little information, in fact, the problem approaches random walks, whereas in the case of full information, it could be solved by standard shortest-path algorithms.
The navigability of real-world complex networks can be significantly improved and explained by considering an underlying latent geometry, which allows for the efficient use of local information to guide the search for a path from a source to a target node~\cite{boguna2009navigability,krioukov2010hyperbolic}. 

A problem strictly related to navigability and information diffusion in complex systems is \textit{routing}, that is, finding efficient message passing strategies to guarantee the delivery of some information piece to one or many target nodes in a network~\cite{yan2006efficient,danila2006optimal}.
Key challenges include minimizing the number of hops, avoiding congestion, and ensuring reliable delivery with only local information available to each node.

Greedy Routing~\cite{fraigniaud2014greedy} is a localized routing strategy where, at each step, a message is forwarded to the neighbor that appears "closest" to the destination based on a predefined metric, such as geographic distance, a calculated shortest-path estimate, or some abstract "distance" in a latent space. The rule is simple: if node $u$ wants to send a message to node $t$, it passes the message to its neighbor $v$ such that $d(v,t) < d(u,t)$, where $d(\cdot,\cdot)$ is the distance metric, and $v$ is chosen to minimize $d(v,t)$ among all neighbors. A significant limitation, however, is that greedy routing does not guarantee message delivery in all network topologies. Messages can become trapped in local minima where all neighbors are "further" from the destination than the current node, even if a global path exists.

Kleinberg's model, also known as the Decentralized Search Model~\cite{kleinberg2000small,kleinberg2004small}, specifically addresses the "small-world problem" and how short paths can be found using only local information, a challenge that simple greedy routing often fails at. The model proposes a network construction where nodes possess both local links, connecting them to geometrically or socially close neighbors, and a few long-range links, connecting them to distant nodes. The key insight is that for efficient decentralized search, where greedy routing proves effective, these long-range links must be chosen with a specific probability distribution. Specifically, a node at position $u$ creates a long-range link to node $v$ with a probability $P(u \to v) \propto d(u,v)^{-r}$, where $d(u,v)$ represents the distance between them, and $r$ is an exponent. Kleinberg demonstrated that if the exponent $r$ is precisely equal to the dimensionality of the underlying space $D$ (i.e., $r=D$), then a greedy routing strategy can locate the target in a poly-logarithmic number of steps, approximately $(\log N)^2$ for $D=2$, where $N$ denotes the network size. Conversely, if $r \neq D$, the search process becomes inefficient. This model thus offers a theoretical explanation for the navigability of social networks and the empirically observed "six degrees of separation" phenomenon, suggesting that efficient navigability arises from a precise balance in the formation of long-range ties.

To manage the inherent complexity of routing in very large networks, hierarchical routing strategies are employed. These strategies organize the network into a layered structure, thereby reducing the amount of information required at each node for making routing decisions~\cite{kleinrock1977hierarchical,corominas2002hierarchical}. Hierarchical structures achieve this by organizing the network's nodes into logically defined clusters or hierarchies. Routing decisions are not made on a flat, global scale but rather at multiple levels of abstraction. These levels typically include global routing, which pertains to the process of routing messages between different clusters or higher-level hierarchical domains, where nodes at this level only need to know how to reach other clusters, not every individual node within them. Once a message has reached the appropriate high-level cluster, local routing mechanisms take over to guide the message within that specific cluster, and potentially its sub-clusters, until it arrives at the destination node.

For hierarchical routing, the total cost of routing a message is the sum of the costs incurred at each level of the hierarchy. This can be expressed as $C = \sum_{l=1}^L C_l$, where $L$ represents the total number of hierarchical levels in the network structure, and $C_l$ denotes the cost of routing a message at a specific level $l$ within the hierarchy. This cost can be quantified in various terms, such as the number of hops, time elapsed, or computational effort expended.

\subsection{Linear thresholds, independent cascades and influence maximization}
\label{sec:influence_tools}

A central framework for modeling the diffusion of information, ideas, and opinions in social networks is the \emph{Linear Threshold Model} (LTM). In this model, each node $v$ has an associated threshold $\theta_v \in [0,1]$ and is influenced by its neighbors through weighted edges $w_{uv}$ such that $\sum_{u} w_{uv} \leq 1$ \cite{granovetter1978threshold,schelling2006micromotives,kempe2003maximizing}. A node becomes active once the cumulative influence of its already active neighbors exceeds its threshold, i.e.,
\[
\sum_{u \in A \cap N(v)} w_{uv} \geq \theta_v,
\]
where $A$ is the set of currently active nodes. Starting from an initial seed set $S$, the process unfolds in discrete steps until no more activations are possible. The influence function $f(S)$ is defined as the expected number of eventually active nodes when $S$ is the initial seed set.

The threshold model naturally captures the phenomenon of \emph{information cascades}, where individuals sequentially adopt behaviors or beliefs based on observations of their neighbors' actions \cite{bikhchandani1992theory}. This connects to the broader literature on social learning, where agents update their beliefs through observation of others' decisions, potentially leading to efficient information aggregation but also to fragile mass behaviors and learning blockages \cite{banerjee1992simple,acemoglu2011opinion}.

A relevant question is how to choose the $k$ nodes that maximize the influence function $f(S)$. This \emph{influence maximization problem} is NP-complete under general thresholds \cite{kempe2003maximizing}. A natural restriction that makes the problem analytically treatable is requiring $f$ to have a diminishing-return behavior. A typical example are submodular functions, for which, given a pair of sets $A \subseteq B$ and any element $v$, 
\[
f(A \cup \{v\}) - f(A) \geq f(B \cup \{v\}) - f(B).
\]
The submodular model is obtained from the general threshold framework by requiring that each local function $f_v$ is submodular.

An alternative but closely related approach is the \emph{Independent Cascade Model} (ICM) and its variants \cite{kempe2003maximizing}. In this model, every edge $(u,v)$ is assigned an activation probability $p_{uv}$. When a node $u$ becomes active at time $t$, it has a single chance to activate each inactive neighbor $v$ at time $t+1$, with probability $p_{uv}$. Each attempt is independent of the others and of the past history. More general cascade models allow the activation probability $p_{uv}$ to depend on the set $A$ of neighbors that have already tried (and failed) to activate $v$, i.e.\ $p_{uv}(A)$. The classical ICM corresponds to the special case $p_{uv}(A) = p_{uv}$, constant and independent of $A$. 

The cascade framework has been extended to capture more sophisticated dynamics, including heterogeneous agent behaviors \cite{watts2002simple}, temporal network effects \cite{holme2012temporal}, and competitive diffusion scenarios where multiple pieces of information or products compete for adoption \cite{bharathi2007competitive}. Recent work has also explored how network structural properties and correlations affect cascade dynamics, revealing that traditional mean-field assumptions may fail in realistic social network topologies \cite{gleeson2007seed}.

Influence maximization under the general cascade model is also NP-complete \cite{kempe2003maximizing}. However, as in the threshold case, tractability can be restored by imposing a diminishing-return condition. This leads to the \emph{Decreasing Cascade Model}, in which $p_{uv}(S) \geq p_{uv}(T)$ whenever $S \subseteq T$: the more failed attempts are made to activate $v$, the harder it becomes to influence it. 

The computational challenges of influence maximization have spurred considerable algorithmic innovation. Beyond the seminal greedy approximation algorithm of Kempe et al., researchers have developed more scalable approaches including the Cost-Effective Lazy Forward (CELF) algorithm \cite{leskovec2007cost}, sketch-based methods for large networks \cite{cohen2014sketch}, and reverse influence sampling techniques \cite{borgs2014maximizing}. These advances have enabled practical applications to viral marketing, social media campaigns, and public health interventions in networks with millions of nodes \cite{chen2010scalable,tang2014influence}.

Together, the LTM and ICM constitute the two canonical frameworks for diffusion processes in networks. Their submodular variants are particularly important because they enable efficient approximation algorithms for influence maximization, with provable performance guarantees \cite{kempe2003maximizing}. The rich interplay between information cascades, social learning theory, and algorithmic approaches to influence maximization continues to drive advances in understanding how information, behaviors, and innovations spread through complex social systems \cite{jackson2010social,easley2010networks}.

\subsection{Spin models, oscillators and synchronization}
\label{sec:ising_models}

The Ising model was originally introduced in the 1920s by Wilhelm Lenz and Ernst Ising as a simple model of ferromagnetism \cite{ising1925beitrag}. In its most basic form, the model consists of binary variables (``spins'') $s_i = \pm 1$ located on the nodes of a lattice. Each spin interacts with its neighbors through pairwise couplings $J$, and may also experience the effect of an external field $h$. The energy of a configuration is given by
\begin{equation}
E = - J \sum_{\langle i,j \rangle} s_i s_j - h \sum_i s_i,
\end{equation}
where the first term encodes local alignment and the second represents global bias. Despite its simplicity, the Ising model captures the emergence of collective order (magnetization) and the existence of sharp transitions between ordered and disordered phases \cite{stanley1971phase,baxter2016exactly}.

The central importance of the Ising model lies in its role as a paradigmatic example of \emph{critical phenomena}. Near the critical temperature $T_c$, the system exhibits long-range correlations, power-law fluctuations, and scale invariance. These features are not specific to ferromagnets, but are shared by a wide variety of systems, from liquid-gas transitions to neural avalanches \cite{stanley1999scaling,beggs2003neuronal}. This observation gave rise to the concept of \emph{universality}: very different microscopic systems can exhibit identical macroscopic critical behavior, characterized by the same set of critical exponents and scaling functions \cite{kadanoff1966scaling,fisher1974renormalization}.

In the context of complex systems beyond physics, the Ising framework provides a minimal representation of collective dynamics driven by local interactions. Individuals in a society, for instance, may be represented as binary spins indicating whether they adopt or reject a piece of news. Neighbor-to-neighbor influence corresponds to the coupling $J$, while external drivers such as mass media can be modeled by a field $h$. Just as in magnets, such simple local rules can give rise to macroscopic consensus, polarization, or sudden phase transitions in collective opinion \cite{galam1991towards,weidlich2006sociodynamics,castellano2009statistical}. Recent work has extended Ising-like approaches to model the spread of information and memes in online social networks, highlighting how social reinforcement and external fields (media, platforms) can push systems close to criticality, amplifying small fluctuations into large-scale cascades \cite{suchecki2005voter,gleeson2013binary}.

\textit{Synchronization} is a ubiquitous phenomenon in coupled dynamic systems, where individual components adjust their rhythms or states to align with each other.
In social systems, this can manifest as the alignment of opinions, behaviors, or even emotional states among individuals. 
The Kuramoto model~\cite{strogatz2000kuramoto} is a classic and widely used mathematical framework to describe the dynamics of synchronization.
It models a collection of $N$ coupled oscillators, each possessing a specific phase $\phi_i$ and a natural frequency $\omega_i$.
The interaction between any two oscillators $i$ and $j$ on the network is typically represented by a sinusoidal coupling term, which tends to pull their phases closer together.
The evolution of the phase of each oscillator is given by the differential equation:
$$\frac{d\phi_i}{dt} = \omega_i + \frac{K}{N} \sum_{j=1}^N a_{ij} \sin(\phi_j - \phi_i)$$
Where $K$ is a global coupling strength, determining the strength of interaction between oscillators, and $a_{ij}$ are the elements of the adjacency matrix.
Despite its simplicity, the model reveals rich collective dynamics, including the emergence of phase synchronization, where a macroscopic fraction of oscillators lock into a common rhythm.

A primary consideration for Kuramoto-like models for social dynamics is the underlying social network structure.
The topology of the network, whether it be scale-free, small-world, or random, profoundly influences the synchronization process and, consequently, the patterns of diffusion and opinion formation.
Researchers meticulously incorporate these intricate network structures into their models to reflect real-world social connections \cite{rodrigues2016kuramoto}, and have utilized variations of the Kuramoto model to investigate how opinions polarize or converge within a society, particularly under different network structures, ranging from highly connected to more fragmented communities \cite{hong2011kuramoto}.
Within this paradigm, each individual in a social network is conceptualized as an oscillator, and their interest for a spreading topic is represented by the phase of this oscillator.
The interactions between individuals, such as conversations, debates, or exposure to others' viewpoints, are then modeled as the coupling strength between these oscillators.
Modifications include bounded-confidence couplings and high-dimensional opinion spaces on unit spheres, leading to stable consensus or bipolarized states.

\section{Modern Media Ecosystem}
\label{sec:media}

Contemporary information diffusion unfolds across an intricate ecosystem of channels, including traditional broadcast media, their digital successors, and a diverse array of online social networks (OSNs). Each platform generates rich digital traces—followers lists, repost logs, hyperlinks, reaction metadata—providing raw material for constructing graph-based models. 
In this section, we systematically detail the workflow for extracting network representations from primary diffusion tools, examine why each platform has attracted scientific scrutiny, summarize key empirical patterns uncovered to date, and discuss practical considerations such as API limitations, data biases, and ethical constraints. 
We also briefly review emerging properties of these networks at all scales.
Finally, we report on the main issues introduced by this novel information paradigm.

\subsection{X (Twitter)}

X (formerly Twitter—a name still prevalent in scholarly literature) continues to serve as a paradigmatic case study for at least two principal reasons. 

First, the platform’s constraint on message length rendered it particularly well-suited for the rapid dissemination of news and political messaging. From its inception, Twitter soon became one of the preferred platforms by journalists and political figures, who demonstrated high levels of activity (see, for instance, a report about Italian journalism~\cite{agcom2020journalism}). In this regard, although Twitter users have never been demographically representative of the general population due to the platform’s relatively limited diffusion compared to other social media, it has nonetheless managed, in many national contexts, to more accurately reflect and shape public and political discourse. The present situation, following the platform’s 2022 change of ownership, marks a significant departure from this earlier role. The removal of content moderation, the algorithmic promotion of particular types of content, and the overall increase in polarization have, in recent months, led to a worldwide outflow of users and diminished the platform’s relevance to public debate~\cite{reuters2025dnr}.

Second, the availability of data for academic research has historically made Twitter an exceptionally valuable object of study. Over the years, it has arguably granted researchers greater access to data than any other social media platform. However, the change in ownership has substantially complicated this access, making it practically impossible to freely retrieve data using the official API for research projects~\cite{x_api_about}. One exception is the European Union. Under the Digital Services Act (DSA), social platforms are required to disclose how their systems work to regulators~\cite{EU_DSA_2022_Article40}. In practice, however, access to data on platform X is only granted to research projects that have been approved by the platform itself—and even then, access is limited to a short time window.

In practice, users author posts (or \emph{tweets}) of up to 280 characters, which can be reshared (or \emph{retweeted}), quoted, replied to, or liked.
Given the nature of this social network, different graph structures can be outlined.
While in almost every case, the vertices are given by the other users of the service, an edge (directed) can be defined either by a following relation (I am the follower of another account) or by retweeting somebody else text.
To study diffusion, researchers especially focus on the following complementary networks:

\begin{itemize}
  \item \emph{Follower network}: a directed graph where each edge $(u,v)$ indicates that user $u$ follows user $v$. This almost-static backbone is typically fetched via the REST API or academic ``full-archive'' datasets.
  \item \emph{Retweet network}: a directed, temporal graph built by parsing retweet events. Each retweet at time $t$ of an original tweet $m$ yields an edge from the retweeter to the original author, annotated with timestamp $t$. Remarkably, the information about how the retweeter met the content that later retweeted, i.e. if she saw the original post by following the author of the post or if she is following someone that reposted the original message, is not available, thus limiting the analysis.
  \item \emph{Mention and reply networks}: directed graphs in which edges represent explicit references: if user $u$ mentions $v$, an edge $(u,v)$ is added; if $u$ replies to $v$, a similar edge is created. These networks capture conversational pathways and can highlight tightly knit conversation clusters \cite{romero2011differences}.
\end{itemize}

X's layered networks of follows, retweets, and mentions enable disentangling potential versus realized influence.
One of the first comprehensive studies of the entire ``Twittersphere''~\cite{kwak2010what} showed that the follower-following network topology is characterized by a heavy-tail
follower distribution, a short effective diameter, and low reciprocity, which represents a deviation from known characteristics of human social networks. Another interesting observation is a certain level of homophily among reciprocated accounts. Network nodes were ranked using three procedures to identify potential accounts: by the number of followers, by PageRank, and by the total number of retweets. Findings suggested that ranking by number of followers and PageRank accounts for independent information concerning ranking by number of retweets. 
Early work showed that cascade sizes follow a power-law distribution with heavy tails: most tweets die quickly, but a few go viral, reaching millions within hours \cite{kwak2010what,goel2015structural}.
While the follower network dictates potential reach, the retweet network governs actual diffusion.
Temporal analyses reveal diurnal cycles and burstiness, captured by self-exciting Hawkes process models \cite{zhao2018trendburst,kobayashi2021tideh}, with high activity immediately after posting followed by a long-tail decay \cite{Lerman2009Computational}.

It is worth noting that the user experience on the platform has evolved over the years. Users are more or less exposed to certain topics according to their interest in them, as evaluated by suitable algorithms.
While this individuates different systems across time, most of the result presented here focus on the diffusion across channels that users decided themselves to create or conversely delete. These topics-based structures have already showed a reasonable stability. 
In particular, a non-directed edge can be created amongst different accounts if they participate in a specific discussion. 
The power of network theory allows us to define a metric in the space of opinions; for example, when considering politics, we can obtain a ``spectroscopy'' of politics, as in the activity of the website \url{https://politoscope.org}.

Before the 2022 change of ownership, data access was enabled by the Twitter REST and Streaming APIs, though public endpoints enforce rate limits (e.g., 900 requests per 15 minutes), leading some researchers to rely on academic licenses for full-archive access.
Sampling biases between the Streaming "gardenhose" and the Firehose have been quantified, revealing that partial feeds can underrepresent rare but impactful events \cite{Morstatter2013Sample}.
Additionally, tweet deletion and account suspension can create missing data.
As mentioned above, in the present day, public endpoints do not allow downloading data, and free access to data for researchers, when possible, is subject to the approval of the platform. 
Ethical considerations include user privacy, informed consent for research, and adherence to X’s terms of service.

\subsection{Facebook}
Facebook’s platform centers on bidirectional friendships, complemented by public pages and groups.
Facebook’s user base remains over 2.5 billion monthly active users.
However, unlike X, much user content is private, raising unique access and ethical issues.
Key network representations include:

\begin{itemize}
  \item \emph{Friendship network}: an undirected graph in which edges connect mutual friends. This network can be sampled via consenting participants using the Graph API, but full‑graph access is extremely limited and subject to platform policy changes and privacy regulations.
  
  \item \emph{Sharing network}: a directed graph where an edge $(u,v)$ indicates that user $v$ reshared a post originally published by $u$. Historically, this could be extracted via CrowdTangle or approved API apps, but CrowdTangle was discontinued on August 14, 2024. Its direct replacement, the Meta Content Library and API~\cite{meta_content_library_2025}, is available only to qualified academics or non‑profit researchers via the virtual clean room of ICPSR (\emph{Inter-university Consortium for Political and Social Research}~\cite{icpsr_2025,SocialMediaArchive_FAQ}) and lacks retroactive historical access and many export functionalities.

  \item \emph{Reaction/comment networks}: for public posts, graphs can be constructed from reactions or comments. The newer Content Library API now includes public comments and view counts, but extraction is restricted to in‑clean‑room analysis and lacks many of the interactive or bulk-export features present in CrowdTangle.

  \item \emph{Group membership bipartite network}: users and groups form two disjoint node sets; an edge connects a user to each group they belong to. Projection yields co‑membership graphs. Pre‑2024, some group metadata was accessible via CrowdTangle, but since its shutdown, access to group-level membership is even more restrictive and not publicly documented by Meta.
\end{itemize}

The amount of information stored in Facebook's posts and preferences is known to be incredibly efficient in the profiling of the users~\cite{bachrach2012personality,quercia2012personality,cadwalladr2017great}. 
Analysis of the complete Facebook friendship network~\cite{ugander2011anatomy} showed a broad, right-skewed degree distribution that falls off more rapidly than a power law, although hubs certainly exist. 
Facebook was almost a fully connected network, with short average path lengths and high clustering. The small-world effect was also confirmed by measuring the average vertex distance of $4.74$ \cite{backstrom2012four} on the giant component. Users and friends on Facebook were highly clustered, and their friendships possessed dense cores. Moreover, community structure emerged at the scale of friendships between and within countries.
In this respect, the social structure of Facebook-friendship networks was investigated considering a hundred American institutions~\cite{traud2012social}. Different dominant contributions were discovered by carrying out complementary measures of assortativity and regression model coefficients on observed ties to evaluate homophily at a local level and community detection to capture modularity at the macroscopic level.

Research on Facebook diffusion has emphasized complex contagion, where multiple independent exposures from different friends increase the probability of adoption, in contrast to simple contagion on Twitter \cite{centola2010spread}.
Empirical analyses reveal that the sharing network is highly clustered, leading to echo chambers in which homogeneous viewpoints reinforce themselves \cite{bakshy2015exposure}.
Facebook's structural and algorithmic features create a uniquely fertile environment for the proliferation of disinformation. 
As the world's largest social network, its core functionalities – including an engagement-optimized News Feed algorithm, viral sharing mechanics, and community-oriented Groups – systematically amplify controversial and emotionally charged content~\cite{preston2021detecting}.

Access to Facebook’s (Meta’s) public data requires app review, and limited public page, group, and event content is now accessible primarily via the Meta Content Library \& API.

The Graph API enforces strict permission checks through an onerous App Review process and business verification. Developers report high rejection rates and multiple attempts before gaining approval, even for basic permissions like page access or comments retrieval.  
Meanwhile, as mentioned earlier, CrowdTangle has been terminated (August 14, 2024), and its replacement is only available via ICPSR’s clean‑room environments under stringent access terms.

Ethical challenges include obtaining user consent, anonymizing sensitive data, and satisfying GDPR requirements for European users \cite{gonzalez-bailon2014social}. Thus, most academic research now focuses on publicly available pages, groups, events or verified profiles via the Content Library, avoiding private personal data and recognizing the constraints imposed by Meta’s governance structure.

\subsection{Reddit}
Although Reddit is highly relevant in the USA, its global reach remains concentrated, as US users comprise 45\% of all daily active users worldwide~\cite{reddit_q2_2025}.
With respect to the previously described online social platforms, where users can form relationship ties, such as friendships or followings, Reddit has a distinct structure. Following its own description, 
``Reddit is a vast network of communities that are created, run, and populated by you, the Reddit users''\footnote{\url{https://redditinc.com/policies/reddit-rules}}.
An effect of the self-organisation of thematic \emph{subreddits} is the presence of internal policies of behaviour and users' moderation of the debates. Users interact mostly with content created by others in the subreddit by upvoting, downvoting or replying to it. 
While much discourse is public, the platform does not expose follower-like social ties; instead, researchers derive networks from posting and commenting behaviors:

\begin{itemize}
  \item \emph{User–subreddit bipartite network}: captures which users post or comment in which communities. Projections onto users reveal shared-interest networks; projections onto subreddits map topical co-occurrence \cite{weninger2012exploration}.  
  \item \emph{Comment-reply trees}: for each post, the comment thread forms a rooted tree with edges from commenters to the users they reply to. Tree depth and branching factor measure engagement and spread \cite{Janetschek2021Understanding}.  
  \item \emph{Cross-post and hyperlink network}: edges represent posts linking to other Reddit content or external URLs, facilitating study of cross-community information flow.  
\end{itemize}

The platform’s compartmentalized "subreddits" act as ideological enclaves, where community-specific norms dictate information credibility. Research shows that while some subreddits (e.g., r/science) enforce strict moderation, others (e.g., conspiracy-themed communities) foster unchecked rumor proliferation through upvote-driven visibility. This creates a paradox: Reddit’s decentralized moderation can both suppress and amplify fake news, depending on subreddit culture.

Reddit diffusion research highlights how community norms shape content spread: niche subreddits can incubate content that later propagates to mainstream communities.
Analyses of comment trees reveal that emotionally charged and controversial topics produce deeper, more branching threads with longer lifespans \cite{Janetschek2021Understanding,Kumar2018Community}, whereas broadly agreeable content spreads across many subreddits but with shallower threads \cite{Kumar2018Community}. 
Voting dynamics also interplay with structural position, as highly upvoted comments gain visibility and catalyze further engagement.

Although Reddit API and third‑party archives once enabled historical reconstruction and large‑scale research, changes since mid‑2023 have severely limited both ability and ethical permissibility. Most contemporary researchers rely on limited real‑time data, moderator‑mediated access, or new curated datasets.

\subsection{Other media}

While X, Facebook and Reddit are the most widely studied media for information diffusion, many others have been considered in the scientific literature.
In the following, we briefly review the most important.

\paragraph{Instagram}

Instagram emphasizes visual content—photos and short videos—with captions and hashtags providing textual context.
Public profiles can be scraped or accessed via the Business Account API under stringent rate limits.
Network constructs include: a follower network analogous to X's, though data availability is more limited; a user–hashtag bipartite network, which may be projected onto users to detect topical communities and trend propagation~\cite{Hu2014What}; a mention/comment networks, extracted from post metadata, capturing direct interactions~\cite{zhao2018trendburst}.
Hashtag diffusion studies reveal that network position (e.g., central influencers vs.\ periphery users) significantly affects spread: influencers spark broad initial exposure, while community members sustain propagation. Content virality correlates with visual aesthetics metrics (e.g., color composition) and hashtag novelty \cite{Hu2014What,zhao2018trendburst}.
Homophily in follower ties is stronger than on X, leading to denser community clusters and slower but more targeted diffusion~\cite{highfield2016instagrammatics}.
However, API restrictions 
and lack of streaming endpoints hinder comprehensive temporal analysis. Data collection often relies on manual scraping scripts, which require constant maintenance against layout changes and carry risks of account bans.

\paragraph{Weibo}
Weibo parallels Twitter’s structure but emphasizes repost comments, multimedia, and verified user badges. 
As the leading Chinese microblogging service, Weibo offers follower, repost, comment, and like interactions.
Public data can be scraped or accessed via limited APIs.
The possible network representations essentially coincide with those used for X: follower networks, repost cascades~\cite{Zhou2017Rumor}, and comment and like networks.
Empirical studies of rumor diffusion on Weibo reveal that rapid reposting lowers the critical threshold for epidemic-like spread, but dedicated "rumor debunker" accounts can significantly reduce cascade size when they intervene early \cite{Yang2012Rumor}.
Cascades on Weibo often display chain-like propagation in grassroots networks and star-like diffusion around celebrity accounts. 
Government censorship and bot activity introduce noise and structural breaks in diffusion patterns. Scraping is impeded by frequent layout changes and anti-automation defenses; partnerships with Chinese institutions can ease access but raise geopolitical sensitivities.

\paragraph{WhatsApp/Telegram}
WhatsApp and Telegram’s end-to-end encryption precludes direct observation of message content or network ties.
Researchers thus rely on indirect methods: group membership hypergraph, where participants in a group chat form a hyperedge; forwarding chains, reconstructing cascades from metadata (e.g., unique message IDs) to build trees similar to retweet cascades.
Forwarding logs (with anonymized IDs) can be collected from consenting users' device backups, revealing who forwarded which message and when \cite{Kumar2020Characterizing}.
Studies find that misinformation spreads rapidly in high-trust private groups, with analysis showing that small, dense groups can generate wide but shallow forwarding trees, mismatching public platform dynamics \cite{Kumar2020Characterizing}. 
Data collection hinges on user recruitment and informed consent.
Legal obligations (e.g., GDPR) demand strict anonymization and secure storage.

\paragraph{Traditional and Online Mass Media}
Traditional outlets—newspapers, television, radio—have migrated online, creating additional data sources:
\begin{itemize}
  \item \emph{Media hyperlink network}: nodes are media outlets, edges are hyperlinks between articles. Crawling sitemaps or RSS feeds enables construction of this directed graph, revealing inter-outlet influence patterns \cite{lerman2010information}.  
  \item \emph{Article comment threads}: each thread forms a reply tree similar to Reddit, allowing sentiment and engagement analysis.  
  \item \emph{Social sharing logs}: embedded share buttons record snapshots of how often an article is shared on platforms like Facebook or Twitter, permitting cross-platform diffusion mapping \cite{Altay2018Media}.  
\end{itemize}
Agenda-setting research uses time-series on topic mentions across outlets to detect core-periphery structures, where a small set of agencies leads narrative diffusion and others follow with lag times.
Temporal cross-correlation of article publication times quantifies influence hierarchies \cite{lerman2010information}.
Comment-sentiment networks highlight polarization and echo chamber formation.
Data access is achieved via RSS, web scraping, and social APIs; challenges include rate-limited requests, paywalls, and copyright constraints.

\subsection{Comparative discussion of social media features}

Extracting network representations from online social networks' platforms requires aligning the research question with data affordances and practical constraints.
Twitter, Reddit, and other public OSNs have long provided rich, timestamped, user-level data, making them valuable resources for cascade reconstruction and temporal network analysis. But recent access restrictions have made retrieving even basic information increasingly difficult. Facebook and Instagram offer deeper insights into clustered communities and visual contagion, but impose stricter API restrictions.
Weibo’s large user base grants cross-cultural perspectives, though scraping challenges and censorship affect data fidelity.
Encrypted platforms like WhatsApp demand novel hypergraph techniques and participant-driven data collection.
Traditional media complement these insights by tracing narrative origins and intermedia transfer. Across all platforms, researchers must navigate rate limits, sampling biases, legal frameworks, and ethical imperatives, making the choice of network model and extraction pipeline as crucial as the analytical methods applied to the resulting graphs.

Building network representations for information diffusion requires careful orchestration of extraction pipelines, platform-specific data artifacts, and analytical models. Table~\ref{tab:platform-comparison} summarizes key features across platforms. 

\begin{table*}[h!]
\centering
\fontsize{7.5pt}{9pt}\selectfont
\setlength{\tabcolsep}{4pt}
\begin{tabular}{llllll}
\toprule
Platform & Public Data & API Access & Primary Graphs & Temporal Logs & Limitations \\
\midrule
X/Twitter   & Yes & Academic Research (EU‑only)/Business API & Follow, Retweet, Mention & Yes & API caps, DSA scope; Privacy, ToS \\
Facebook  & Part & Graph API/Content Library API via ICPSR & Friend, Share, React, Group & Yes & Restricted permissions; GDPR, Consent \\
Instagram & Part & Graph API/Content Library API via ICPSR & Follow, Hashtag, Comment & Partial & Restricted permissions; GDPR, Consent \\
Reddit    & Yes & API/Pushshift & User–Subreddit, Comment & Yes & Archival gaps; Sensitive content \\
Weibo     & Part & Limited API & Follow, Repost, Comment & Partial & Censorship, Scraping; Geopolitical ethics \\
WhatsApp  & No  & None (user-sourced) & Hypergraph groups & Yes & Recruitment bias; Privacy, Encryption \\
Media     & Yes & RSS/Scrape & Hyperlink, Comment, Share & Yes & Paywall, HTML changes; Copyright \\
\bottomrule
\end{tabular}
\caption{Comparison of information diffusion platforms and their network extraction characteristics.}
\label{tab:platform-comparison}
\end{table*}

In sum, each platform’s unique data model shapes the feasible network abstractions and subsequent analysis. The interplay between data accessibility, graph complexity (monoplex vs.\ multiplex vs.\ hypergraph), and ethical/legal constraints dictates methodological choices. Robust diffusion studies thus depend as much on meticulous data engineering pipelines as on advanced theoretical frameworks.

\subsection{Information disorders in social media}

\emph{Information disorder} is an umbrella term to cover the multifaceted aspects and undesirable effects of the interaction between individuals and information. 
Following the definitions in Ref.~\cite{wardle2017information}, the \emph{information disorder} can manifest in 3 main forms: \emph{mis-information}, i.e. when false information is shared without the intent of producing harm; \emph{dis-information}, i.e. when false information is intentionally shared to produce harm; and \emph{mal-information}, i.e. when ``genuine information is shared to cause harm''. From the very definition, it is clear that those terms are broader than other broadly used terms that will recur in the present review, such as \emph{fake news}, which according to Lazer {\em et al.}~\cite{lazer2018science} is ``{\em .. information that mimics news media content in form but not in organizational process or intent.}''. 

More generally, complex adaptive information ecosystems are susceptible to various forms of exploitation, ranging from the diffusion of harmful content to coordinated inauthentic behaviors that manipulate individual and collective attention. Table~\ref{tab:misinfo} summarizes the main problems related to information diffusion in complex socio-technical systems, highlighting the diversity of mechanisms and outcomes. Variations in definitions across studies can produce divergent interpretations: for example, Vosoughi et al.~\cite{vosoughi2018spread} found widespread penetration of misinformation, whereas Grinberg et al.~\cite{grinberg2019political} reported that only a small fraction of users interacted with misinformation sources. Subsequent work suggests that the primary differences between true and false news lie in their ``infectiousness'', rather than in the underlying diffusion mechanisms~\cite{juul2021comparing}. A recent report has monitored the presence of mis- and dis-information in various online social platforms (Facebook, Instagram, LinkedIn, TikTok, X/Twitter, YouTube) in some European countries (France, Poland, Slovakia, Spain)~\cite{vincent2025measuring}. Among multiple different measurements regarding the first half of 2025, TikTok resulted in the most exposed social platform to misinformation, with $\sim$20\% of
posts targeting misinformation. On the opposite side of the spectrum lies LinkedIn, with nearly $\sim$2\% posts delivering false or misleading information. Furthermore, the authors noted that accounts consistently spreading low-credibility information receive more engagement than high-credibility accounts on every platform except LinkedIn. The gap is particularly pronounced on YouTube and Facebook, where low-credibility accounts generate nearly eight and seven times, respectively, the engagement per post per 1,000 followers compared to high-credibility accounts.

\begin{table*}
\centering
\begin{tabular}{|p{3cm}|p{13cm}|}
\hline
\textbf{Concept} & \textbf{Description} \\ \hline
Affective Polarization & Dislike of the ``other side" \cite{lazer2018science} \\ \hline
Astroturfing & The practice of masking the sponsors of a message or organization to make it appear as though it originates from grassroots participants \cite{keller2020political} \\ \hline
Echo chambers & The result of selecting a set of friends and information that adhere to our system of beliefs, thereby forming polarized groups \cite{delvicario2016spreading} \\ \hline
False Amplifiers & Methods, such as false news, disinformation, or networks of fake accounts aimed at manipulating public opinion \cite{weedon2017information} \\ \hline
False News & News articles that purport to be factual, but which contain intentional misstatements of fact \cite{weedon2017information, scheufele2019science} \\ \hline
Information operations and coordinated inauthentic behavior & Actions taken by organized actors (governments or non-state actors) to distort domestic or foreign political sentiment \cite{weedon2017information} \\ \hline
Social bot, Sybil account & Computer algorithm that automatically produces content and interacts with humans on social media \cite{davis2016botornot} \\ \hline
Misinformation & The spread of inaccurate information without malicious intent \cite{weedon2017information} \\ \hline
Disinformation & Inaccurate or manipulated information/content that is spread intentionally \cite{weedon2017information} \\ \hline
Infodemics & Overflow of information of varying quality that surges across digital and physical environments during an acute public health event. It leads to confusion, risk-taking, and behaviors that can harm health and lead to erosion of trust in health authorities and public health responses \cite{tangcharoensathien2020framework, calleja2021public} \\ \hline
\end{tabular}
\caption{A short list of the most used terms to refers to the various levels of disinformation that is possible to encounter when navigating social platforms.}
\label{tab:misinfo}
\end{table*}

\section{Information Spreading}
\label{sec:info_spreading}

\begin{figure}
\includegraphics[width=\linewidth]{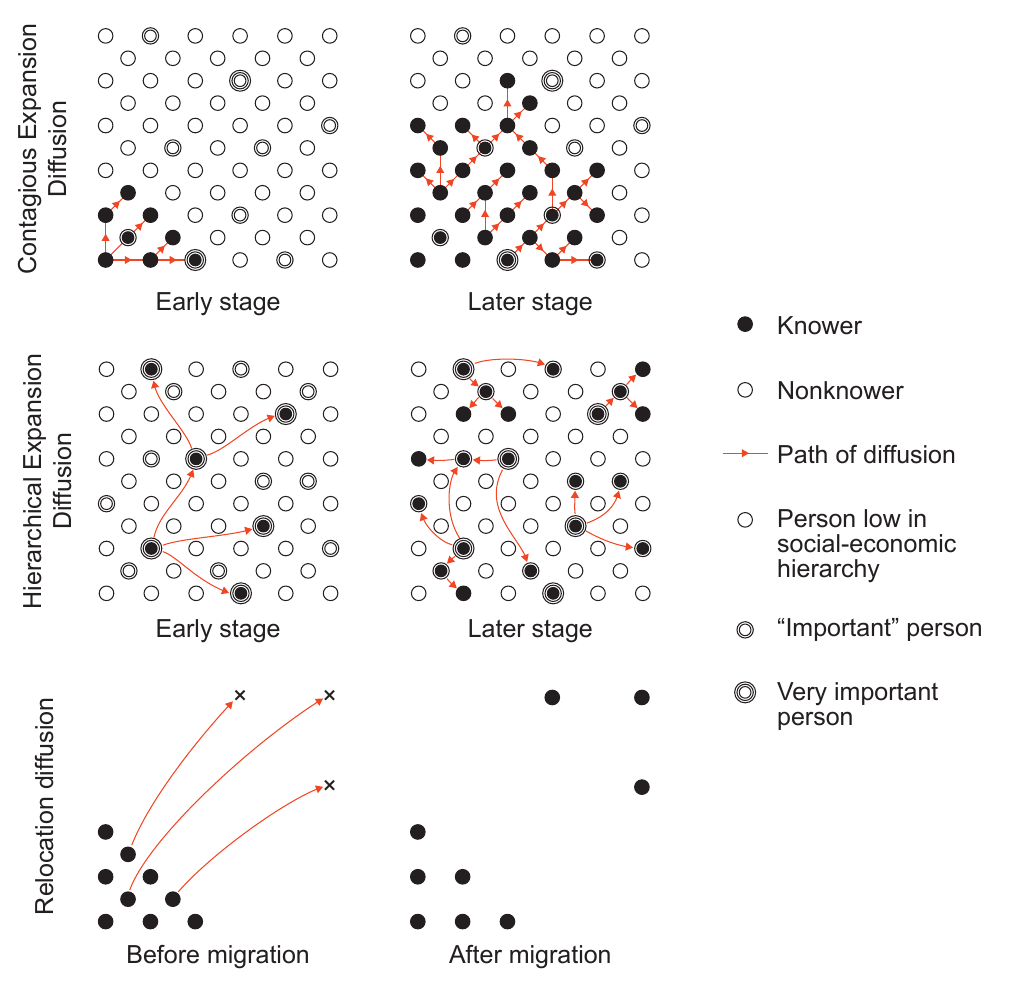}
\caption{Models of cultural diffusion---an illustration adapted from Jordan's 1982 \textit{The Human Mosaic}~\cite{jordan1982human}. (Slightly modified for consistency.) Note that ``diffusion'' in this figure is used in the sense of the social sciences and computer science: I.e., it does not preclude a change in the total mass of whatever spreads. Note also that the order of spreading is no longer thought to be determined by socioeconomic status~\cite{rosen2002anatomy}.
\label{fig:cultural}}
\end{figure}

Social and behavioral sciences have long emphasized that human cognition, perception, and social interaction shape the way information diffuses. 
Fig.~\ref{fig:cultural} shows three different theories of the spreading of ``ideas, innovations, and attitudes'' from a 1982 cultural geography textbook~\cite{jordan1982human}.
In the eyes of a physicist ``relocation diffusion'' resembles convection in the theory of heat transfer, whereas ``contagious expansion diffusion'' is more like conduction.
The other contagion model in Fig.~\ref{fig:cultural}, ``hierarchical expansion diffusion'', is reminiscent of Lazarsfeld's two-step model of communication~\cite{lazarsfeld1944people,katz1955personal}, originally intended to describe the influence of mass media on elections.
It proposes a picture of simultaneous influence by mass media and network peers that has been confirmed in social media studies~\cite{cha2012world}.
Its closest analogy among well-studied mathematical models (still too far to delve deeper into) might be reaction-diffusion systems~\cite{harrison1993kinetic}.

In the social sciences, the traditional type of study was centered around the S-curve of market share of a new technology taking over an old, with a focus on the timing of the adoption dividing the individuals into ``innovators,'' ``early adopters,'' ``early majority,'' ``late majority,'' and ``laggards''~\cite{rogers1983diffusion}.
Within this body of literature, there were also studies focusing on how the micro-dynamics of contagion translate to macro-patterns---like H\"agerstrands spatial diffusion models~\cite{hagerstrand1953innovationsf}, or network spreading approaches~\cite{valente1996network,valente1996social}.
Innovation diffusion remains a well-studied topic in the age of social media. See, e.g., Refs.~\cite{toole2012modeling,arieli2020speed}, that support a picture of innovation spreading being influenced by many factors such as the personality of the adopters, the product, network structure, and the platforms they spread over. Highly related to the spread of innovation is the spread of language. Researchers have, for example, studied the propagation of neologisms in social media~\cite{kooti2021emergence,eisenstein2014diffusion}.

From a sociological perspective, the study of information diffusion builds on foundational work in communication theory.
The aforementioned two-step flow model~\cite{lazarsfeld1944people,katz1955personal} proposed that mass media influence flows through opinion leaders to the broader public, a concept that finds modern expression in hierarchical diffusion models--as depicted in Fig.~\ref{fig:cultural}--and the role of influencers in social networks.
Granovetter's work on the strength of weak ties~\cite{granovetter1973strength} highlighted how information spreads through bridging connections between social clusters, while theories of social proof and conformity from psychology explain why individuals are more likely to adopt behaviors when they observe others doing so, particularly under uncertainty.
Psychological research further demonstrates that repeated exposure enhances persuasion \cite{zajonc1968attitudinal,hassan2021effects}, that cognitive biases (e.g., confirmation bias, motivated reasoning) strongly filter what individuals share \cite{nickerson1998confirmation}, and that emotional content, especially moral or arousing emotions, increases transmission likelihood \cite{berger2012what}. Together, these insights highlight that information spreading is not only a structural process on networks but also one shaped by thresholds of attention, memory, and trust. 

In modern times, social media platforms such as Twitter, Facebook, Weibo, and Telegram have fundamentally reshaped how information propagates through society.
Large-scale cascades of news or misinformation can unfold in minutes, and understanding why certain content "goes viral" has become a critical scientific and societal challenge.
Conventional data-driven methods, network statistics approaches or machine-learning classifiers often identify correlations without explaining structural origins of thresholds, critical points, or universal scaling laws. Descriptive social science frameworks catalog what happens but may not predict when a cascade becomes global or explain why specific network architectures amplify misinformation. To address these gaps, researchers have turned to statistical physics and complexity science. Models originally developed for ferromagnets, percolation in porous media, and epidemic outbreaks offer insights into how local interactions among users yield emergent, macroscopic patterns on social media.

The problem of news propagation can be cast into a problem of spreading on very disordered and discretized media, where sites -- e.g., newspapers, people, and mobile phone apps -- form an interconnected web of communication channels. Furthermore, every site is not a simple but an active unit: it can start, terminate, or modify the process. Accordingly, such sites are more similar to active matter~\cite{fodor2016how,shaebani2020computational} characterized by non-linearity, lack of detailed balance and of time-reversal symmetry, and out-of-equilibrium dynamics.

By treating social media as a time-evolving network—nodes represent users and edges denote follower or friend relations—one can apply various physics-inspired models. These include epidemic-like compartmental models (SIR, SIS), percolation theory (site and bond), mean-field approximations, branching processes, and maximum-entropy ensembles. These methods are instrumental in capturing phenomena such as thresholds separating confined chatter from global virality, power-law cascade-size distributions near criticality, and the significant role of network topology in shaping diffusion dynamics.

This section builds on the structural foundation outlined in Sec.~\ref{sec:networks} to understand how information, such as news, rumors and memes, actually spread across network topologies.
It explores various concepts, models, and techniques drawn from the field of physics and its applications, that offer a fundamental understanding of information diffusion in a broad sense.
It focuses into the underlying physical principles that govern spreading processes, including simple and complex contagion, diffusion, random walks, and message passing, providing a theoretical lens to analyze how information spreads, interacts, and evolves within complex systems.

\subsection{Modeling the dynamics of information spreading}

A central challenge in the study of information diffusion is to translate the diverse mechanisms of contagion into formal models that can generate, predict, and explain observed spreading patterns. Modeling approaches capture how microscopic processes—ranging from individual cognitive biases and behavioral rules to multilayer structures and higher-order interactions—aggregate into collective phenomena. This section reviews the principal modeling frameworks, from simple to complex contagion and from single-layer abstractions to multiplex and higher-order representations, highlighting how insights into human decision-making and social influence are incorporated into quantitative theories of diffusion.

\subsubsection{From simple to complex contagion}
\label{sec:concepts_definitions}


Research in information spreading typically focuses on understanding how local interactions yield emergent, macroscopic patterns. Key research targets include:

\begin{itemize}
    \item \textit{Individual influence}: Understanding how single users or small groups can trigger large cascades, including the identification of "hidden influentials" and optimal seeding strategies
    \item \textit{Community effects}: How network structure, clustering, and community organization affect diffusion pathways and cascade containment
    \item \textit{Topological effects}: The role of degree distribution, network heterogeneity, and structural properties like k-core organization
    \item \textit{Cognitive constraints}: Memory limitations, attention capacity, information processing bounds, and cognitive biases that filter information sharing
    \item \textit{Temporal dynamics}: Non-Markovian activity patterns, burstiness, and time-dependent interactions that affect spreading timescales
    \item \textit{Multilayer systems}: Interactions across multiple communication channels and platforms, including cross-layer amplification effects
    \item \textit{Systemic outcomes}: Emergence of collective attention, polarization, consensus formation, and large-scale behavioral coordination
\end{itemize}




The most fundamental type of social spreading process happens when something---an infection, an idea, the use of a product, etc.---spreads from one person to another. I.e., everyone who has the thing got it from exactly one other, but the total amount of what spreads is not necessarily conserved (as in random walks and physical diffusion discussed above). This type of spreading dynamics, where one spreading event is fully described by a unique transmission tree~\cite{liben-nowell2008tracing,cha2008characterizing}, is usually referred to as \textit{simple contagion}~\cite{centola2018how}.
Its simplicity, while allowing for a deeper mathematical analysis than other social processes~\cite{pastor-satorras2015epidemic,kiss2017mathematics,britton2019stochastic}, is sometimes seen as a limitation to their realism.

By far, most theoretical work on simple contagion has been done in the context of infectious disease modeling, which in turn builds on results from percolation theory and branching processes.
Applications to information spreading illustrate the utility of these frameworks. In independent cascade or SIR-type models, each informed node transmits to neighbors with independent probability $p$, so the final set of informed individuals corresponds to a percolation cluster \cite{newman2001random}. Threshold models such as Watts’ cascade model \cite{watts2002simple} generalize this by requiring multiple exposures: a node adopts only if a sufficient fraction of neighbors are already informed, producing global cascades under certain degree distributions and threshold heterogeneity.

\begin{figure}
\includegraphics[width=0.9\linewidth]{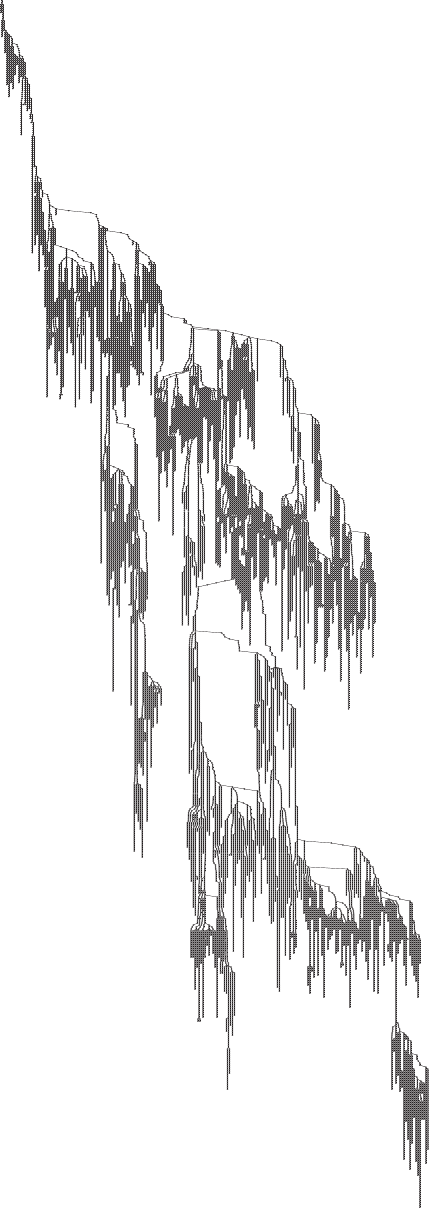}
\caption{A visualization from Ref.~\cite{liben-nowell2008tracing} showing the propagation of chain-letter e-mails petitioning against the 2002--2003 Iraq war. The generation number of the email increases downwards.}
\label{fig:e-mail}
\end{figure}

In Fig.~\ref{fig:e-mail} we see one information cascade from a 2002 e-mail chain letter~\cite{liben-nowell2008tracing}.
While one might be tempted to model this type of information spreading as a simple contagion, the shape of this tree is very different from what one can naively expect from an SIR model, where, on average, infected people would be much closer to the infection source.
This tree has 18,119 nodes, of which 17,079 (94\%) have exactly one child. The median node depth is 288, and the width of the tree is 82.

Despite being often modeled with similar frameworks, in fact, epidemic dynamics and information-spreading dynamics exhibit fundamental differences in their underlying mechanisms and transmission patterns. 
These were first demonstrated in the seminal work by Daley and Kendall~\cite{daley1964epidemics, daley1965stochastic}, followed shortly thereafter by the Maki–Thompson rumor model~\cite{maki1973mathematical}.
In both models, a closed population is partitioned into three classes: ignorants (who have not heard the rumor), spreaders (who actively tell it) and stiflers (who know the rumor but no longer spread it).
Spreaders contact others at random: when a spreader meets an ignorant the ignorant becomes a spreader; when a spreader contacts another spreader or a stifler the initiating spreader becomes a stifler (loses interest).

The pioneering Daley-Kendall~\cite{daley1964epidemics} and the Maki–Thompson~\cite{maki1973mathematical} models rely on the intuition that one key element that differentiates information from disease spreading is the decay of interest.
They do so through "stifling" interactions: once a spreader has encountered enough people who are already aware of the news piece, they stop sharing it.
In standard epidemic models, individuals who were previously infectious transition into the removed category through death, isolation, or recovery, at a rate proportional to the number of infectious individuals.
When it comes to news and rumors, however, individuals often stop spreading just because the information is no longer perceived as novel and potentially interesting.

These type of models have been widely used as baselines for analytical studies and network-based extensions of rumor/information propagation.
For the sake of simplicity, they often assume that a single encounter with either another active spreader or a former spreader is sufficient to stop an individual from continuing to spread the rumor. 

With the standard SIR notation--S being the ignorant, I the spreaders, R the stiflers--the probability for an individual to stop spreading information is proportional to the frequencies of meetings between members of the various classes. The list of the possible transitions in the time interval $(t, t+dt)$ with associated probabilities is the following:
\begin{align}
    &(S, I, R) \rightarrow (S-1, I+1, R), \quad SIdt + o(dt); \\
    &(S, I, R) \rightarrow (S, I-2, R+2), \quad \frac{1}{2}I(I-1)dt + o(dt); \\
    &(S, I, R) \rightarrow (S, I-1, R+1), \quad IRdt + o(dt).
\end{align}
The key aspect of this rumor-propagation mechanism is the lack of a threshold effect; the proportion of the population that eventually hears the rumor remains roughly constant, regardless of the population size, $N$.
On complex networks, the presence of hubs alters outcomes: hubs quickly become stiflers and thus reduce the final informed fraction, though they accelerate early dissemination. 


The Maki-Thompson model has been extensively analyzed and applied to understand rumor spreading in social networks through several key contributions. \cite{nekovee2007theory} pioneered the adaptation of the classical Maki-Thompson and Daley-Kendall frameworks to complex network structures, developing mean-field theoretical approaches that account for the heterogeneous connectivity patterns characteristic of real social networks. Building on this foundation, \cite{kandhway2014optimal} extended the Maki-Thompson framework to address practical concerns about information control, developing optimal strategies for rumor containment and mitigation in networked environments. From a mathematical perspective, \cite{gani2000maki} provided a comprehensive theoretical analysis of the classical discrete-time Maki-Thompson model, establishing rigorous analytical foundations for understanding rumor dynamics. This work was further refined by \cite{belen2011classical}, who developed continuous-time formulations of the classical model and introduced mathematical refinements that enhanced the model's analytical tractability. More recently, \cite{ferraz2022subcritical} demonstrated how network structural correlations fundamentally alter Maki-Thompson dynamics, revealing subcritical behavior and correlation-induced phase transitions that challenge traditional mean-field predictions, thereby advancing our understanding of how network topology influences information spreading processes.

A term often used to describe these extended contagion models is \textit{complex contagion}.
While some authors~\cite{aral2017exercise} use the term in a more specific sense, we will use it in a general sense, to refer to any mechanism where multiple exposures or reinforcing signals are required for transmission.
This type of spreading is very common in social contagions, such as the adoption of new technologies, political movements, or risky behaviors, where social validation or perceived lower risk from multiple sources is necessary for adoption.

Key mechanisms include:
\begin{itemize}
    \item \textit{Social reinforcement}: The probability of adoption increases with multiple exposures from different sources
    \item \textit{Peer pressure}: Social influence that depends on the proportion of one's social circle that has adopted  
    \item \textit{Stifling}: The process by which individuals stop spreading information once they encounter others who are already aware of it
    \item \textit{Threshold effects}: Where adoption requires a minimum number or proportion of activated neighbors
    \item \textit{Memory and repeated exposure}: Where cumulative exposure over time influences adoption probability
\end{itemize}

In an early model of complex contagion~\cite{granovetter1978threshold}, Granovetter noticed that for some social spreading the cost of changing behavior (like joining a riot~\cite{berk1974gaming}) decreases with the number of social neighbors who already did. Then Granovetter stipulated that everyone has an individual threshold for changing behavior to arrive at his celebrated \textit{threshold model} of collective behavior~\cite{granovetter1978threshold}.
Threshold models are a common framework for describing complex contagion processes on networks. They formalize the idea that an individual's adoption of a behavior or idea depends on the number or proportion of their contacts who have already adopted it.
In the case of binary states, Watts' model~\cite{watts2002simple} defines the following update rule for the state of a node $i$ at time $t+1$, denoted $x_i(t+1)$ (where $x_i=1$ if active/infected and $x_i=0$ otherwise):
$$x_i(t+1) = \begin{cases}
1, & \text{if } \sum_{j \in N(i)} a_{ij}x_j(t) \geq \phi_i \\
0, & \text{otherwise}
\end{cases}$$
Where $N(i)$ is the set of neighbors of node $i$, $a_{ij}$ are the elements of the adjacency matrix, and $\phi_i$ is the individual threshold for node $i$.
Possibly, only a subset of $k$ neighbors sampled independently for each agent might be considered.
In this framework, the condition for a cascade is that the largest cluster of most influenceable agents (with $\phi_i<1/k$) percolates.
If the underlying network is sufficiently sparse, cascade propagation is limited by the network's global connectivity; otherwise, if the network is sufficiently dense, it is limited by the stability of the nodes.

The model can be generalized to weighted networks, where the weight $w_{uv}\in[0,1]$ represents the influence the $u$ exerts on $v$~\cite{kempe2003maximizing}.
Given a set $S_0$ of active nodes at time $t = 0$, the cascade develops deterministically in discrete time.
At each time step $t$, each inactive node becomes active if the sum of weights on edges from activated neighbors reaches its threshold values.
This linear model can be easily generalized by combining the influence of the neighbors by any monotone function $f$ of the set of currently active nodes, such that $f(\emptyset) = 0$.

More generally, Watts's model triggered a gold rush of threshold-model studies, where physicists and other theoreticians modified it to: work on temporal networks~\cite{karimi2013threshold,takaguchi2013bursty}, varying the influence of individuals~\cite{watts2007influentials}, study the effects of bridge nodes~\cite{centola2007cascade}, to account for the age of the phenomenon spreading~\cite{gronlund2005network}, to allow for correlation between degree and threshold~\cite{lee2017social}, or distance to the source and threshold~\cite{nishioka2022cascading}. Gleeson~\cite{gleeson2008cascades} proposed an analytically tractable model of cascades. A general conclusion is that threshold models can reverse the insights from simple contagion models: in threshold models, bursty contact dynamics can speed up spreading~\cite{karimi2013threshold,takaguchi2013bursty}, or depend little on long-range links~\cite{centola2007complex}, in contrast with epidemic spreading.

\begin{SCfigure*}[0.5]
\includegraphics[width=1.3\linewidth]{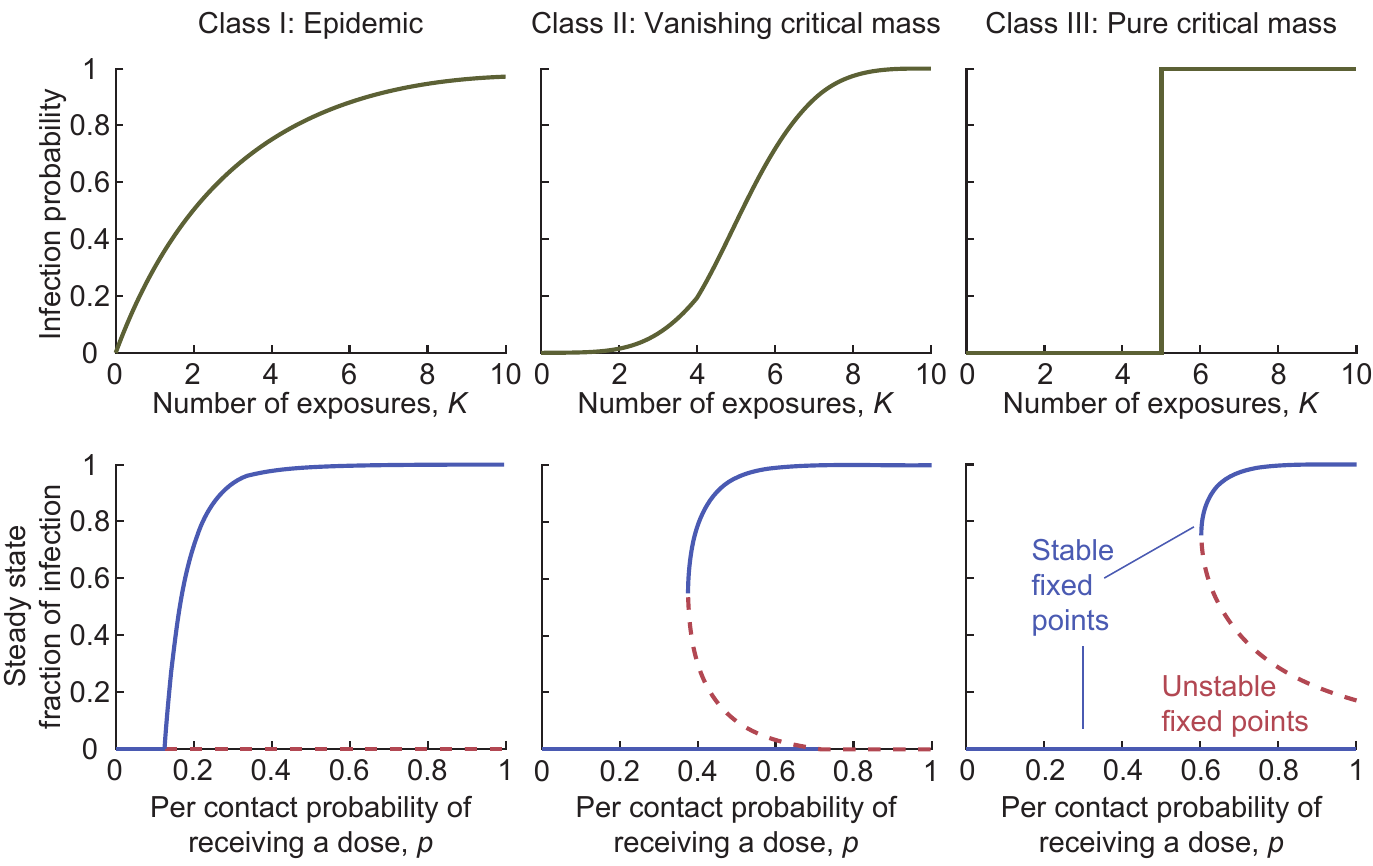}
\caption{The Dodds-Watts model of social contagion~\cite{dodds2004universal}. The top row shows the ``dose-response curve,'' i.e.\ the probability of an individual becoming infected given a number $K$ of neighbors having been infected during the last unit of time. The second row of figures shows the steady-state fraction of infected individuals as a function of the probability of receiving a dose (contributing to the chance of contagion) from an infected neighbor. 
\label{fig:dodds_watts}}
\end{SCfigure*}

In an important contagion model interpolating between simple and complex contagion, Dodds and Watts~\cite{dodds2004universal} introduced the idea that individuals retain memory of past exposures to a contagious influence.  At time $t$, each $i$ receives a dose $d_i(t)$ from another randomly chosen individual $j$ with probability $p$. The total amount of doses over the past $T$ time steps
\begin{equation}\label{eq:doddsuniversal}
    D_i(t)=\sum\limits_{t^{\prime}=t-T+1}^{t}d_i(t^{\prime})
\end{equation}
is kept stored: when $D_i(t)$ exceeds a given threshold $d_i^*$ a susceptible individual becomes infected. A contagion model can belong to three universal classes depending on the equilibrium behavior. Class I includes the epidemic threshold models, which exhibit SIR-like dynamics with an epidemic threshold $p_c$ corresponding to the level of infectiousness needed for an initial group of infected individuals to cause an epidemic. The other two classes require a critical mass to initiate and sustain the contagion. However, while in class II, the critical mass goes to zero for $p<1$ (vanishing critical mass models), in class III, the critical mass remains always finite (pure critical mass models). Dodds and Watts' study suggested that modifications to the individual's threshold cause a considerable impact on the possibility of triggering a contagion and that the presence of easily influenced individuals increases the contagion's chances more than the presence of highly influential individuals (a.k.a., opinion leaders). See Fig.~\ref{fig:dodds_watts} for an overview of the model's behavior.

Complex contagion is closely linked to higher-order network models~\cite{bianconi2021higher,majhi2022dynamics}. Higher-order networks typically express relations involving more than just two individuals, thus moving the topological constraints of the contagion mechanism to the representation of the underlying network~\cite{iacopini2019simplicial}. In this literature, Ref.~\cite{cencetti2023distinguishing} discusses model selection between various higher-order models of social spreading. Furthermore, complex contagion is often built into more intricate models of social phenomena, like the survival of minority opinions~\cite{lee2017modeling}, the formation of youth subcultures~\cite{holme2005modelling}, climate action~\cite{barrett2014sensitivity}, etc.

\subsubsection{Cognitive and behavioral mechanisms at the individual and collective level}
\label{sec:empirical_behaviour}

A central insight from the study of information diffusion is that spreading is not merely a structural process but is deeply shaped by human behavioral and cognitive mechanisms. The foundational work of Goffman and Newill \cite{goffman1964generalization} drew an influential parallel between infectious disease dynamics and the dissemination of ideas through the SIR model framework. In this analogy, individuals progress from being unaware or uninterested (susceptible, S) to adopting or engaging with an idea (infected, I), and eventually losing interest or becoming resistant to further influence (removed, R). While initially metaphorical, this framework established the foundation for a rich class of models that explicitly incorporate behavioral states such as forgetting, attention decay, and psychological immunity, demonstrating how cognitive factors fundamentally shape cascade trajectories.

Building on this epidemiological perspective, researchers have revealed how psychological resistance and peer dynamics interact to determine diffusion outcomes. Ruan et al. \cite{ruan2015kinetics} demonstrated through analytical and simulation studies that while even a small constant rate of spontaneous adopters suffices for global diffusion, the speed of equilibrium attainment is highly sensitive to the density of immune individuals. As immunity increases, diffusion transitions from rapid to slow propagation, reflecting the fundamental tension between peer-driven adoption cascades and psychological resistance to persuasion.

The predictive power of behaviorally-informed models has been validated in large-scale empirical studies. Castiello et al. \cite{castiello2023using} calibrated an Ignorant–Spreader–Recovered (ISR) model on over 40,000 Twitter cascades of COVID-19 hashtags, estimating infection and recovery rates ($\beta, \gamma$) via maximum likelihood. Their model achieved remarkable accuracy, predicting peak cascade volume with a mean absolute error of less than two hours, demonstrating that cognitive states such as attention and disinterest, when properly parameterized, enable both explanatory and predictive models of online behavior.

Individual heterogeneity in activity patterns represents another crucial cognitive-behavioral factor. Borge-Holthoefer et al. \cite{borgeholthoefer2012emergence} adapted the SIS framework by incorporating node-specific activity rates $\lambda_i$ that reflect differences in login frequency and engagement patterns. Empirical fits to Twitter subgraphs revealed that heavy-tailed distributions of $\lambda_i$—characterized by a few highly active users and many less active participants—substantially extend outbreak durations and fundamentally alter cascade-size distributions. This finding underscores that the heterogeneous rhythms of human activity are not merely technical details but defining features of collective dynamics.

\begin{figure}
\includegraphics[width=\linewidth]{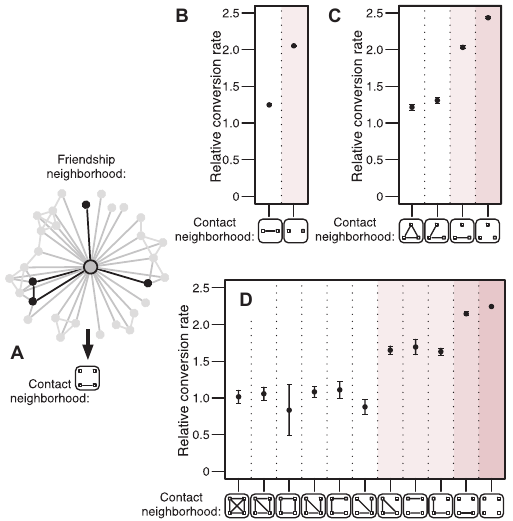}
\caption{A figure from Ref.~\cite{ugander2012structural} (studying the early growth of Facebook---the time when the platform primarily grew by email invitations) showing the paper's main finding. In (A), the black nodes are the focal node's friends who have sent the focal node an email invitation. This defines the ``contact neighborhood'' which the paper claims is the strongest determinant of the conversion rate (successful recruitment per time). Panels (B–D) show the relative conversion rates for two-node, three-node, and four-node contact neighborhood graphs. Shading indicates differences in component count. Invitation conversion rates are reported on a relative scale, where 1.0 signifies the conversion rate of one-node neighborhoods. Error bars represent 95\% confidence intervals and implicitly reveal the relative frequency of the different topologies.}
\label{fig:ugander}
\end{figure}

Empirical research has consistently highlighted the critical importance of reinforcement and threshold effects in adoption decisions. Multiple studies \cite{sprague2017evidence,hodas2014simple,ugander2012structural} demonstrate that diverse, multiple exposures significantly increase adoption likelihood compared to single or redundant contacts. Ugander et al. \cite{ugander2012structural} showed that Facebook adoption during early expansion was best predicted not by the total number of invitations received, but by their structural diversity—specifically, the number of disconnected social clusters represented among inviters (see Fig.~\ref{fig:ugander}). Similarly, Aral et al. \cite{aral2017exercise} identified reinforcement effects in a dataset of one million runners, while controlled experiments confirmed that adoption probability increases when influence comes from peers in distinct communities \cite{mason2012collaborative,centola2010spread}. These findings reveal that social influence operates less like simple contagion and more as a cognitively mediated process where diverse contextual cues enhance salience and credibility.

Formal models have been developed to capture these reinforcement mechanisms explicitly. Feng et al. \cite{feng2015competing} proposed a fractional-SIR (FSIR) model where susceptible users only become "infected" when the fraction of infected neighbors exceeds a threshold $\Gamma$, representing limited attention and information overload effects. When fitted to over one million 
Weibo cascades, the model identified empirical parameters $\beta_{\mathrm{emp}} \approx 0.03$ and $\Gamma_{\mathrm{emp}} \approx 0.2$ that successfully reproduced the observed truncated power-law distribution of cascade sizes.

Xie et al. \cite{xie2021detecting} advanced this framework with a percolation-based model incorporating exposure-dependent adoption probabilities $p_k$ that increase with the number of prior exposures. Analysis of large-scale Twitter and Weibo data confirmed this "social reinforcement" mechanism, showing that it effectively lowers percolation thresholds by 30–50\% compared to constant-probability models. Importantly, their work also revealed that automated bot clusters can exploit reinforcement dynamics to create spurious feedback loops that artificially amplify diffusion.

Cognitive biases play a fundamental role in shaping misinformation adoption and propagation. Physics-inspired models typically incorporate these biases as heterogeneous node-level parameters that modify susceptibility to misinformation or thresholds for belief adoption. Such heterogeneities introduce disorder analogous to quenched randomness in statistical mechanics \cite{sobkowicz2018opinion}.

Confirmation bias, where individuals preferentially process information aligning with prior beliefs, can be modeled through bounded-confidence mechanisms. In Deffuant-type models, agents only interact when their opinions differ by less than a tolerance threshold, representing cognitive openness \cite{battiston2020networks}. Low tolerance values lead to fragmentation and polarization—an emergent phase characterized by multiple stable opinion clusters that resist convergence.

Sobkowicz \cite{sobkowicz2018opinion} extends this approach by incorporating Bayesian belief updating filtered through personal biases, introducing nonlinearity and multistability where the system can settle into distinct fixed points depending on initial conditions and bias distributions. This resembles random-field Ising models with frustration and glassy dynamics, where disorder prevents global consensus. Moreover, biases introduce memory effects such as the continued influence effect, where misinformation persists even after correction, reflecting non-Markovian dynamics and hysteresis that require multi-compartment frameworks to capture accurately.

Beyond statistical regularities, higher-order structures such as narratives provide the semantic scaffolding through which information is processed and shared \cite{rimmon-kenan2003narrative,murray2015narrative,schank2013scripts,bruner2009actual}. Narratives impose coherent ordering on facts and events, linking them into storylines that fundamentally shape collective understanding. Well-structured stories exploit cognitive heuristics such as availability bias \cite{tversky1973availability}, enhancing processing fluency and perceived plausibility. This makes narrative-embedded disinformation particularly potent: fluency and coherence generate trust \cite{nowak2023integration}, while confirmation bias \cite{wason1960failure,nickerson1998confirmation} promotes selective adoption of content aligned with pre-existing beliefs. Consequently, narrative-driven disinformation spreads differently from epidemic contagion, forming polarized "narrative communities" or echo chambers \cite{zollo2018social,peruzzi2019confirmation} where reinforcing stories circulate with minimal opposition.

Emotional dynamics further amplify these higher-order effects. Research consistently demonstrates that emotionally charged content—whether positive (awe, joy) or negative (fear, disgust, outrage)—exhibits enhanced contagiousness \cite{berger2012what,stieglitz2013emotions,cotter2008influence,heath2001emotional}. Emotional arousal blurs fact-fiction distinctions, intensifying narrative "transportation" \cite{van2019storytelling} and increasing susceptibility to misinformation \cite{green2000role}. Crucially, corrective counter-narratives achieve maximum effectiveness when they themselves engage emotional responses \cite{sangalang2019potential}, highlighting that narratives and emotions, as higher-order organizing principles, modulate information flows in ways that exceed the explanatory scope of epidemic-like contagion models.

The dynamics of collective attention reveal additional layers of cognitive constraint on information diffusion. Lorenz-Spreen et al. \cite{lorenz-spreen2019accelerating} analyzed competition among cultural items for finite attention resources, demonstrating that content trajectories are accelerating with steeper gradients and shorter life cycles, reflecting saturation of human attention under increasing content production. Similarly, Eom et al. \cite{eom2015twitter} modeled daily political tweet volumes via geometric Brownian motion, showing that when averaged over optimal time windows, collective attention reliably predicts electoral outcomes.

Global-scale attention dynamics were illustrated in studies of Twitter activity surrounding the Higgs boson discovery announcement \cite{dedomenico2013anatomy}, where activation probabilities increased with repeated neighbor exposures, producing epidemic-like attention bursts. The "preferential attention" model \cite{dedomenico2020unraveling} incorporated heterogeneity in network connectivity and individual reactivity, with response probabilities scaling as $q_i(t)\sim k_i^{\alpha}$. This mechanism successfully reproduced sharp activity spikes, revealing that influential network hubs disproportionately drive collective attention through multilayer reinforcement effects that amplify sudden visibility surges.

Recent methodological innovations have integrated these behavioral and cognitive insights into increasingly sophisticated predictive frameworks. Random walk methods such as DeepCas \cite{li2017deepcas} and CasWarn \cite{gao2021public} model diffusion paths while capturing temporal and structural adoption features. Graph neural networks further enrich this perspective: DeepInf \cite{qiu2018deepinf} predicts adoption probabilities from local subgraph structure, CasGCN \cite{xu2020casgcn} models cascade growth via graph convolutions, and DyDiff-VAE \cite{wang2021dydiff} incorporates evolving user interests through dynamic latent variables.

Complementing these machine-learning approaches, differential-equation-based methods explicitly link adoption kinetics to spatio-temporal processes. Wang et al. \cite{wang2020modeling} developed ODID to model temporal–spatial diffusion on Digg, Cheng et al. \cite{cheng2024information} combined probabilistic diffusion with neural ODEs for popularity prediction, and Foroozani et al. \cite{foroozani2021nonlinear} modeled the anomalous diffusion patterns observed in real cascade data.

This comprehensive body of research demonstrates that cognitive and behavioral mechanisms—including attention limits, memory decay, structural diversity requirements, activity rhythms, and reinforcement thresholds—are not peripheral details but fundamental determinants of information diffusion. By embedding these mechanisms within epidemic, percolation, and machine-learning frameworks, researchers have established that collective information dynamics can only be understood through explicit consideration of how human cognition shapes the probabilities and pathways of social contagion.

\subsubsection{Cross-layer, multiplex, and higher-order models}
\label{sec:empirical_multiplex}

The study of information spreading in complex social systems increasingly requires frameworks that move beyond single-layer networks to account for multiplexity, cross-platform flows, and higher-order mechanisms of collective behavior. Modern information ecosystems are characterized by multiple interacting dimensions: users participate across different platforms simultaneously, content propagates through various modalities (text, images, video), and adoption decisions emerge from group interactions that cannot be reduced to pairwise influences. Understanding these multifaceted spreading processes demands sophisticated theoretical frameworks that capture the rich structural and dynamical complexity of contemporary digital communication.

Information propagation seldom occurs within isolated platforms. Users frequently inhabit multiple online environments (e.g., Twitter, Facebook, WhatsApp), with information flowing between these interconnected systems. From a network physics perspective, this scenario is optimally modeled using multilayer or multiplex networks, where each layer represents a distinct platform, and inter-layer edges capture cross-platform user activity \cite{battiston2020networks}.

In such multilayer frameworks, information spreads both within and across layers through fundamentally different mechanisms. A narrative originating on one platform may cross over to mainstream platforms via bridge nodes—users active on multiple platforms—who function as inter-layer links in the multiplex structure. Xian et al. \cite{xian2019misinformation} analyzed misinformation dynamics on correlated multilayer networks, demonstrating that inter-layer correlations and degree heterogeneity systematically lower the effective outbreak threshold $\beta_c$, thereby facilitating global cascade formation across the entire multiplex system.

Analytically, multilayer spreading can be modeled through coupled differential equations or message-passing frameworks that explicitly account for both intra-layer and inter-layer transmission processes. The effective reproduction number in such systems aggregates contributions from each individual layer and their mutual interactions. Critically, epidemic thresholds and phase transitions can shift dramatically due to inter-layer coupling: independent subcritical spreading processes within isolated layers can collectively produce supercritical cascades when coupled, analogous to the interdependent percolation transitions observed in infrastructure networks~\cite{liu2018epidemic,chang2021analytical,granell2013dynamical}.

From a structural perspective, multiplex networks introduce higher-order correlations and coupling effects that fundamentally modify percolation properties. The presence of overlapping hub nodes across multiple layers creates structural vulnerabilities that function as super-spreader bridges, dramatically accelerating cross-platform contagion. Targeted interventions on such critical bridge nodes—analogous to immunizing high-degree nodes in single-layer networks—can therefore significantly hinder cross-platform information propagation~\cite{cellai2013percolation,santoro2020optimal}.

Contemporary information frequently propagates through multimodal content that combines text, images, audio, and video elements. From a physics perspective, this introduces interacting contagions across distinct information channels, which can be naturally modeled using multiplex network structures or higher-order representations \cite{battiston2020networks}.

Each modality can be conceptualized as a distinct network layer, with information propagating separately yet interacting synergistically across these parallel channels. Wu and Chen \cite{wu2020spreading} developed a comprehensive analytical framework for such coupled spreading processes, revealing that while individual modality layers may remain subcritical in isolation, their mutual interaction can drive system-wide cascade formation. Simultaneous exposure via multiple modalities—such as text and images—can reinforce belief adoption through cognitive redundancy, effectively increasing the overall transmissibility $\beta_\mathrm{eff}$ of the composite contagion process.

This multimodal synergy can be formalized using coupled differential equations or message-passing algorithms across multiplex layers, incorporating explicit cross-modality reinforcement terms that capture the enhanced persuasive power of coordinated multimedia messaging. These multimodal interactions function as coupling constants in the effective Hamiltonian describing the system's spreading dynamics, systematically modifying the phase diagram and critical thresholds of information outbreak formation~\cite{chang2021analytical,granell2013dynamical}.

Beyond traditional multiplex models, multimodal content naturally calls for hypergraph representations, where group interactions—such as simultaneous exposure to video and text within group messaging contexts—are modeled as higher-order hyperedges connecting multiple nodes simultaneously. Battiston et al. \cite{battiston2020networks} demonstrated that such higher-order network structures can fundamentally alter spreading dynamics, enabling nontrivial collective effects and phase transitions that remain invisible in conventional pairwise interaction models.

From a percolation theory standpoint, multimodal content enhances information redundancy: when information fails to propagate successfully via one modality channel, it may still succeed through alternative pathways. This redundancy systematically lowers the effective percolation threshold, analogous to reinforcement mechanisms observed in interdependent network systems. For misinformation intervention strategies, recognizing modality-specific dominance becomes critically important. Physics-inspired models can parameterize channel-specific transmissibilities (e.g., $\beta_\mathrm{text}$, $\beta_\mathrm{video}$, $\beta_\mathrm{audio}$) and optimize mitigation efforts accordingly, with targeted interventions on the most effective channel—the dominant layer in multiplex terminology—potentially achieving disproportionate reductions in overall misinformation spread~\cite{cellai2013percolation,santoro2020optimal}.

The theoretical elegance of multiplex spreading models is exemplified by their capacity to produce universal scaling behaviors that transcend platform-specific details. O'Brien et al. \cite{obrien2019spreading} mapped meme propagation across multiplex networks to multitype branching processes, where each platform or communication channel forms a distinct layer contributing to the overall offspring distribution of information cascades. Within this mathematical framework, large-scale cascades emerge when the largest eigenvalue of the next-generation matrix reaches the critical value unity ($\Lambda_{\max}=1$), yielding a universal cascade-size distribution characterized by a robust power-law exponent of $-3/2$.

This critical scaling behavior, independently confirmed by Notarmuzi et al. \cite{notarmuzi2022universality}, underscores the fundamental robustness of branching-process analogies and highlights the central importance of cross-layer coupling mechanisms in shaping global propagation patterns across diverse information ecosystems. The theoretical universality of this scaling relationship suggests that despite the enormous diversity of platforms, content types, and user behaviors, certain statistical regularities emerge at large scales that reflect deep structural principles governing information flow in multiplex social systems.

However, the practical application of these elegant theoretical frameworks remains constrained by limited empirical access to comprehensive data on inter-platform dependencies and cross-layer coupling strengths, which are often proprietary and closely guarded by platform operators. This data scarcity represents a significant challenge for validating and calibrating multiplex models against real-world spreading phenomena~\cite{chen2018optimal}.

The convergence of multiplex network theory with cognitive and behavioral mechanisms reveals that information diffusion emerges from a complex interplay of structural coupling across platforms, competition for finite attentional resources, narrative framing effects, and emotional engagement processes. These multifaceted mechanisms collectively constitute a truly multilayered contagion process where mathematical universality coexists with strong cognitive and cultural contingencies.

Higher-order spreading processes thus represent far more than the simple additive outcome of individual link-by-link transmission events. Instead, they emerge from the dynamic interaction between network topology, content characteristics, user psychology, and cross-platform coupling, creating emergent collective behaviors that can only be understood through integrated theoretical frameworks that bridge network science, cognitive psychology, and information theory~\cite{granell2013dynamical,chen2018optimal}.

This integration of structural and cognitive mechanisms positions the study of higher-order spreading phenomena as a central frontier in contemporary network science, with profound implications for understanding and managing information flow in an increasingly complex and interconnected digital communication landscape.

\medskip

In summary, modeling work on information diffusion demonstrates how principles from contagion dynamics, cognitive science, and network theory can be translated into formal representations of social spreading processes. From simple compartmental models to multilayer and higher-order frameworks, these approaches highlight the mechanisms through which local behavioral rules and structural constraints generate emergent collective patterns. At the same time, models are necessarily selective abstractions, and their predictive power depends on the extent to which they capture the interplay between individual decision-making, network structure, and technological mediation. Continued progress in this area hinges on integrating insights from behavioral sciences, computational experiments, and empirical data analysis into unified modeling paradigms.

\subsection{Empirical insights from simulations and social data}

Beyond purely theoretical considerations, important progress has come from empirical analyses of large-scale datasets as well as controlled simulations of specific spreading dynamics. Together, these approaches provide complementary insights into how structural and temporal features shape diffusion outcomes, from the role of network topology to the emergence of communities and the signatures of critical behavior. This section reviews such findings, emphasizing how results from simulations and social data inform and constrain theoretical models, while also revealing mechanisms by which information diffusion reshapes the organization of social networks.

\subsubsection{Topological effects on diffusion dynamics}
\label{sec:empirical_influentials}

Real-world directed networks display hierarchical structures dominated by the emergence of leader nodes: in ecology, the hierarchies among individuals of animals or food webs; in control engineering, the master-slave coupling of oscillators; in social interactions, the source nodes that spread infection in contagion dynamics. O'Brien et al.~\cite{obrien2021hierarchical} quantified the network polarization through entropy rates and graph non-normality measures. They showed that when a given non-normality level was exceeded, the entropy rates underwent a first-order transition to zero abruptly. This was related to the emergence of leader nodes, a term used by the author to describe sink nodes which only have incoming edges. The distance from such leaders induces a hierarchical ranking of the other nodes. Sacco et al.~\cite{sacco2021emergence} have reported the emergence of leaders on the popular microblogging platform Twitter during the COVID-19 pandemic. They have shown that knowledge communities characterized global communication during the pandemic, with information cascades driven by a few users and scientific communication playing a marginal role in the digital information ecosystem.

Another study tracked the spread of political news on Twitter in 2016 and 2020 around US elections~\cite{flamino2023political}. Firstly, users who were highly relevant to disseminating information on Twitter (influencers) were identified by building the retweet network corresponding to different news media categories: fake, extreme bias right, right, right-leaning, center, left-leaning, and left. In this network, the out-degree is the number of users who have retweeted them at least once, and it is a measure of the influential power of the user. Then, influencers were connected into similarity networks based on the affinity of their retweeters. The similarity networks displayed a high degree of polarization and echo chamber structure, which became more prominent from 2016 to 2020.

In their work, Pei and Makse \cite{pei2013spreading} conducted a comparative analysis of various centrality measures, including degree, betweenness, closeness, and k-shell coreness, with the aim of identifying optimal spreaders on LiveJournal data. Their critical observation was that nodes situated in higher k-shells—that is, those residing in the dense "core" of the network—consistently triggered larger cascades than even the highest-degree nodes, as measured by average outbreak size. This provides strong empirical evidence that a node's structural position within the network's core-periphery organization holds more significance for widespread diffusion than its raw number of connections. This concept of "hidden influentials," who may not be traditional hubs but are strategically located to connect diverse parts of the network, represents a significant departure from purely degree-based spreading strategies and offers valuable insights for intervention strategies, suggesting that targeting core nodes might be more effective than exclusively targeting the most connected individuals.

Teng et al. \cite{teng2016collective} formalized the concept of identifying influential nodes beyond simple degree by introducing the Collective Influence (CI) metric. CI for a node $i$ is defined as $(k_i - 1)\sum_{j\in\partial B(i,\ell)}(k_j - 1)$, which effectively accounts for the influence of a node's multi-hop neighborhood. They demonstrated that seeds chosen based on the highest CI values on diverse networks, including APS, Facebook, and LiveJournal, consistently yielded 20-30\% larger cascades than seeds chosen by degree alone, particularly when the seeding budget was limited. Furthermore, removing top CI nodes from these networks caused the giant component to collapse at lower removal fractions compared to degree-based removal, indicating that CI better identifies structural bottlenecks crucial for network connectivity and diffusion. This work reinforces the idea that strategic placement, often at the boundaries connecting multiple communities, is key for maximizing cascade size, thereby providing a practical method for identifying these "hidden influentials."

The understanding of influence in multiplex systems was further extended by Bontorin and De Domenico \cite{bontorin2023multi}, using the BBC Pandemic face-to-face and digital layers. Their study revealed a counter-intuitive finding: hubs, traditionally considered effective spreaders, can sometimes act as "firewalls" if their attention is fragmented across multiple layers, leading them to absorb information rather than efficiently transmitting it across layers. They introduced a multi-pathway temporal distance metric to capture hidden diffusion paths, demonstrating that a non-hub with a short temporal distance across layers could be more influential than a high-degree hub confined to a single layer. This emphasizes that effective influence in multilayer systems depends not just on connectivity within a layer, but also on the ability to bridge information across different communication channels and on the temporal dynamics of these interactions.

Addressing the critical problem of optimal seeding, Altarelli et al. \cite{altarelli2013optimizing} investigated how to select an initial set of $q$ nodes to maximize the final spread of information. They framed this as a problem of minimizing a cost function over seed sets under an SIR dynamic. To solve this efficiently on large networks, they leveraged Belief Propagation (BP) on a factor graph representation, which enabled them to approximate the marginal probabilities of each node being activated. Applying their method to the Epinions trust network, they demonstrated that BP-derived seeds consistently outperformed greedy heuristics based purely on node degree by a margin of 10-20\%. This study is significant for providing a computationally tractable, physics-inspired approach to a practical problem in viral marketing and information dissemination, showcasing the power of message-passing algorithms in complex network optimization.

At the mesoscopic scale, social networks exhibit emergent structures--such as echo chambers and polarized communities--that shape collective opinion dynamics and information diffusion. These intermediate-scale patterns arise from homophily-driven interactions, where individuals preferentially engage with like-minded peers, creating densely connected clusters with limited cross-group communication. Computational models, including bounded-confidence frameworks and adaptive network theories, reveal how such topologies amplify confirmation bias and filter diversity, fostering ideological extremization. Empirical studies of online platforms further demonstrate that algorithmic recommendation systems often reinforce these dynamics by promoting content aligned with users' existing beliefs, effectively trapping them in self-reinforcing informational loops. The interplay between network structure and individual behavior at this scale explains why polarization persists despite abundant information access, highlighting the need for interventions—such as bridge nodes or neutral content injection—to mitigate fragmentation. By bridging micro-level cognitive biases with macro-level societal divides, mesoscopic analysis offers critical insights into the mechanisms driving today’s polarized discourse.

A study~\cite{bakshy2015exposure}, based on deidentified data from 10.1 million U.S. Facebook users, measured ideological homophily, that is, the tendency to connect preferentially with similar-minded individuals. The findings suggest that individuals' choices, rather than algorithmic ranking, have a more significant influence in restricting exposure to cross-cutting content, emphasizing the role of user behavior in shaping information exposure on social media.

The role of clustering and community structure has emerged as a central theme in understanding the dynamics of rumor and information diffusion. Early extensions of classical rumor models, such as the one by O’Sullivan et al. \cite{osullivan2015mathematical}, introduced a stifler (R) state to capture the tendency of individuals to cease spreading after losing interest. Their primary contribution was to examine how clustered networks alter the spreading process. Through cellular automata simulations on synthetic clustered graphs, they demonstrated that clustering accelerates rumor propagation by enabling multiple overlapping exposures within tightly knit groups. This “social reinforcement” effect mirrors complex contagion, where redundant signals from trusted peers push individuals past activation thresholds, leading to abrupt local surges. In this way, clustering acts as an amplifier of local diffusion, highlighting how mesoscale structures such as tightly connected neighborhoods shape spreading outcomes.

However, clustering does not universally promote global diffusion. Work on complex networks has shown that the presence of hubs can actually diminish the fraction of ultimately informed individuals compared to homogeneous graphs, because hubs quickly transition into stiflers and suppress further spread \cite{liu2003propagation, moreno2004dynamics}. Despite this reduction in reach, spreading efficiency—measured as the ratio of informed individuals to the overall traffic generated—tends to be higher in heterogeneous networks. This contrast points to a nuanced balance between local amplification and global reach, dependent on the interplay of degree heterogeneity and clustering.

The seminal studies by Zanette \cite{zanette2001critical, zanette2002dynamics} further highlighted how structural randomness modulates rumor dynamics. Using Watts–Strogatz networks with tunable rewiring, Zanette identified a critical threshold $p_c$: below this value, the fraction of informed individuals vanishes in the thermodynamic limit. Their framework, paralleling SIR epidemiological models, defined individuals as susceptible, infected, or refractory, with dynamics governed by local contacts and loss of interest. Simulations revealed that the rewiring parameter directly determines whether a rumor remains localized or percolates through the network, thereby linking small-world properties to large-scale spreading capacity.

More recent work has confirmed these theoretical predictions in empirical contexts. Davis et al. \cite{davis2020phase}, applying the Maki–Thompson rumor model to real-world systems, observed sharp phase transitions between localized and global spreading on networks with contrasting topologies: a sparsely clustered European railway network and a highly clustered DBLP co-authorship graph. While both exhibited critical thresholds consistent with mean-field approximations, the DBLP network required a higher spreading rate for global contagion. This finding underscores a paradox: clustering may intensify local cascades but simultaneously raise barriers to global reach, as information can become trapped within cohesive communities.

The importance of mesoscale structures has also been demonstrated in online social networks. Baños et al. \cite{banos2013role}, analyzing Twitter retweet cascades, showed that community organization strongly shapes diffusion patterns. While hubs could trigger rapid local bursts, $k$-shell coreness—reflecting a node’s embedding in the network core—was a more reliable predictor of final cascade size. Hubs tied to a single community often acted as “firewalls,” confining information, whereas core nodes bridging multiple communities enabled wider propagation. This reinforces the idea that diffusion depends less on raw connectivity than on strategic placement within modular structures.

Finally, Su et al. \cite{su2018optimal} formalized these insights by explicitly modeling the role of community mixing. Introducing a parameter $\mu$ to represent the fraction of inter-community edges, they derived mean-field equations that revealed three regimes of diffusion. For low $\mu$ values, cascades remained confined within communities. At intermediate values ($\mu_\ell < \mu < \mu_u$), global cascades became possible at minimal infection rates. Yet paradoxically, at high $\mu$ values, the network approached homogeneity and required higher infection rates to sustain spreading, as the reinforcing role of communities diminished. Their identification of an optimal mixing value $\mu^*$ that maximizes cascade size highlights the delicate balance between cohesion and connectivity. Monte Carlo simulations validated this phase diagram across the $(\mu, \beta)$ plane, offering robust evidence that modularity is not merely an obstacle but, when properly tuned, a driver of large-scale virality.

Taken together, these studies paint a consistent picture: clustering and community structures profoundly shape both the speed and extent of rumor diffusion. While dense local ties amplify reinforcement and accelerate spread, they may also hinder global contagion unless balanced by sufficient inter-community connectivity. Understanding this dual role is crucial for predicting, controlling, and leveraging information cascades in real-world networks.

\subsubsection{Discursive communities, opinion leaders and echo chambers} 
In the analysis of online social networks, the term \emph{discursive community} is used to identify groups of individuals contributing to the formation of a common discourse, sharing implicit rules and following a common goal~\cite{shaw2012politics,radicioni2021analysing}. Originally proposed in social science, the term has been extended to the context of online social media, although there is no complete agreement on a perfect translation of the original phenomenon into this novel framework~\cite{shaw2012politics}.

Using the definition above, a standard ``network theory'' community structure differs from a discursive community. The former, regardless of the community detection method employed, captures the excess of links inside a group of nodes, while the latter focuses on identifying groups of users gathering around a common set of concepts and ideas. Nevertheless, tools from network theory can be useful to detect discursive communities when appropriately adapted~\cite{guarino2024verified}.

To detect discursive communities on Twitter, an effective method proposed in the literature revolves around verified accounts~\cite{becatti2019extracting}. Before its change in ownership in 2022, on Twitter, the identity of the owners of accounts relevant for the public debate was certified autonomously by the platform~\cite{Wiki_verified,X2009,X2016}. While the introduction of `verified' users was perceived by some as the introduction of a VIP class of accounts, it was intended to avoid unauthorised impersonations of famous persons on the platform~\cite{X2009}. The verification of an account is graphically depicted as a blue checkmark close to the username. Nevertheless, since late November 2022, the verification of the account can be obtained upon payment as a feature of the Premium account~\cite{X2022}.
The main idea of Ref.~\cite{becatti2019extracting} is to detect similarities among verified users by looking at their audience composed of ``standard users'': two accounts contributing to the same discourse should be retweeted by the same standard users. Otherwise stated, the shared discourses are captured as they are perceived by the audience of standard users. 
With this aim, a bipartite network of verified and unverified users is defined; a link between a standard user and a verified one is present if the former has retweeted at least once the latter. The information about the interaction between these two classes of users is then projected into the layer of verified users: for each couple of verified user the so-obtained co-occurrences capture the number of `standard' users that interacted with both of them. Due to the heterogeneity in the behaviours of verified accounts, the noise removal is a necessary step; therefore, the co-occurrence network is compared with the one obtained by projecting a maximum entropy randomisation of the original (bipartite) system. Finally, a statistical validation returns a monopartite binary network among verified users in which a link connects two verified users if the number of common standard retweeters is statistically significant. Then, a community detection is run on the validated network, and the so-obtained labels are propagated to standard users via a label propagation algorithm on the entire retweet network.

Such a framework was applied in multiple cases, providing good results: in political online debates on Twitter, discursive communities align with political coalitions~\cite{becatti2019extracting,caldarelli2020role,caldarelli2021flow,mattei2022bow,radicioni2021analysing,pratelli2024entropy}. Furthermore, it was recently observed that verified users are much more efficient as seed for the detection of discursive communities than other class of users based exclusively on the activity~\cite{guarino2024verified}: in a sense, it seems that the prestige provided by the verification of an account is particularly relevant in the public debate. Such an observation generates numerous concerns about the opportunity of providing verification checkmarks upon payment.

According to several scholars, the heterogeneous behaviors exhibited by the different accounts may be the signal of an underlying phenomenon: the presence of a cohort of users who mediate the information from official sources to wider audiences, drawing on a novel adaptation of Lazarsfeld and Katz’s 1955 ``two-step flow'' theory~\cite{katz1955personal}. The essence of the two-step flow theory is that political information does not travel directly from primary authorities or news outlets to the general public; instead, it is predominantly filtered and interpreted by ``opinion leaders'' who then convey it to a broader audience.
At the time this idea was first introduced, the role of opinion leaders was primarily envisioned as being carried out through direct, in-person interactions—such as face-to-face meetings or public events. The central aim of the two-step flow model was to clarify how and when messages disseminated by mass media could reach individuals who lacked direct access to such sources. This theory was supported by various social experiments that demonstrated promising results~\cite{katz1955personal}. However, with the widespread adoption of television and the declining participation in social gatherings, individuals began to exert greater control over their own information environments. Consequently, both the predictive power and relevance of the two-step flow theory diminished.

\begin{figure*}[h!]
    \centering
\includegraphics[width=.7\textwidth]{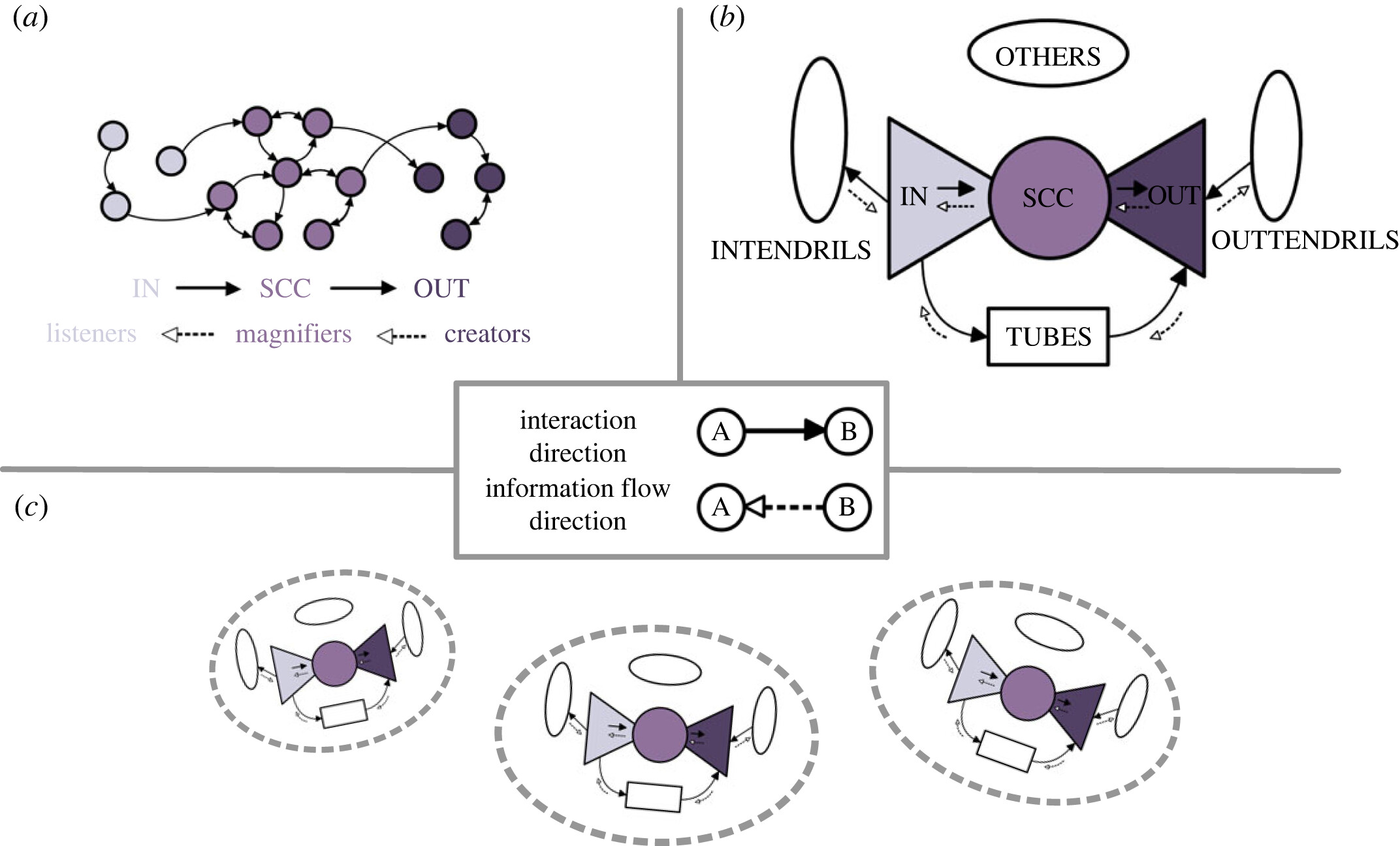}
    \caption{{\textbf{Mesoscale structures in online social networks.} 
Bow-tie structure in a complex network highlights the organization into a core, identified by a strongly connected component (SCC) where all users can reach each other through information diffusion and peripheral components. The IN component consists of users who can reach the SCC but not vice versa, whereas the OUT component consists of users from whom the SCC can reach but cannot. Tendrils attach to these sub-systems without connecting to the core. See the text for further details.} Figure reproduced from Ref.~\cite{han2024modelling}.}
    \label{fig:bowtie}
\end{figure*}

Numerous scholars have since attempted to adapt the concept of ``institutional'' message mediation by ``special'' individuals to the context of Online Social Networks, resulting in a variety of interpretations of the opinion leader’s role~\cite{wu2011who,weeks2015online, choi2015two, hilbert2017one,dubois2020who}. Nevertheless, there is no consensus among social scientists regarding the continued significance of the two-step flow in shaping opinions within social networks. For instance, Bennett and Manheim~\cite{bennett2006one} proposed the “one-step flow” theory, suggesting that the primary mechanism for opinion formation lies in platform recommendation systems. These systems leverage the digital footprints users leave on social networks to tailor content to individual preferences, thereby diminishing the influence of opinion leaders in favor of algorithmic curation.\\
Subsequent research has indicated that one-step, two-step, and even more complex multi-step flows—such as those involving influence among opinion leaders themselves—can coexist within the same communication ecosystem~\cite{hilbert2017one}.\\

Building on this perspective, a recent stream of research has identified a recurring mesoscale pattern in user interactions, known as the bow tie structure~\cite{broder2000graph,yang2011bow}. Broder et al.~\cite{broder2000graph} originally introduced the bow tie decomposition to characterize the World Wide Web’s architecture. Specifically, one first identifies the Largest Strongly Connected Component\footnote{The Largest Strongly Connected Component is the largest set of nodes within which each node can reach every other node by following directed edges.} (LSCC); nodes that can access to (but not part of) the LSCC are labeled IN, while those reachable from (but not part of) the LSCC are labeled OUT. Yang and colleagues later refined this classification~\cite{yang2011bow}, introducing other blocks in the decomposition. The bow-tie structure is displayed pictorially in Fig.~\ref{fig:bowtie}.\\ 
In their depiction of the Web, nodes represented websites and edges represented hyperlinks: the LSCC contained most websites, the IN segment was dominated by search engines, and the OUT segment comprised authoritative sources such as Wikipedia. This framework has since been applied to various systems, including the Tor network~\cite{bernaschi2022onion}, the control networks of transnational corporations~\cite{vitali2011network}, and more recently, online social platforms~\cite{mattei2022bow}.\\

In the context of social networks, Mattei et al.~\cite{mattei2022bow} observed that, when the debate is societal or political, nearly every discursive community exhibits an informative bow tie structure, encompassing nearly all participating accounts. It is natural, at this point to wonder if the observed structure appears by chance and is just due to the different behaviours of the accounts or if it is the effect of something deeper, as the two-step flow mentioned above. The empirical retweet network was then compared with a maximum entropy null model preserving, for each account, the number of different retweeted and retweeting users, i.e. the in- and the out-degree. Results show that the dimension of the blocks in terms of the number of nodes was statistically significant, i.e. it was either too small or too big to be explained by the activity of the users only. In this sense, only looking at the activity of users cannot explain the number of users mediating the information coming from standard sources.\\
Notably, the bow tie structure aligns naturally with the two-step flow framework: nodes within the LSCC correspond to opinion leaders, who synthesize messages originating from the authorities in IN and disseminate them to the wider audience in OUT. Motivated by these findings, diffusion models that exploit the bow tie organization of interactions have been proposed~\cite{han2024modelling}.

In political discourse, users with similar political leanings often form their opinions based on the same news sources~\cite{jamieson2008echo,garrett2009echo}. Although this phenomenon is unsurprising --many outlets openly disclose their political orientation-- it can become problematic when opinions become highly polarized and diverge markedly (even if there are some specific counterexamples~\cite{bail2018exposure}). In this scenario, {\em echo chambers} emerge: they are clusters of accounts sharing identical viewpoints, accepting only information that aligns with their preexisting beliefs and showing strong resistance to alternative perspectives~\cite{jamieson2008echo, garrett2009echo}. While the original concept was articulated qualitatively within social science, subsequent work has translated it into a quantitative framework and validated it across multiple online debates~\cite{delvicario2016spreading, zollo2018social,cinelli2021echo}. In fact, the presence of echo chambers is investigated in several different social platforms, i.e. Facebook~\cite{delvicario2016spreading, zollo2018social}, X/Twitter~\cite{cinelli2021echo, pratelli2024entropy} and Reddit~\cite{defrancisci2021noecho,monti2023evidence}.\\

The analysis of large-scale Facebook data by Del Vicario et al.~\cite{delvicario2016spreading} and Zollo et al.~\cite{zollo2018social} highlights how the circulation of both scientific information and conspiracy-like content is shaped by the emergence of echo chambers, i.e. sharply segregated, homogeneous communities, where users preferentially consume and propagate content consistent with their worldview. In this polarized environment, attention patterns within each community are remarkably similar regardless of the narrative, with heavy-tailed distributions of likes, shares, and comments, and comparable persistence of user engagement over time. However, the dynamics of information cascades differ: scientific news typically reaches peak diffusion rapidly, with cascade lifetime largely independent of size, whereas conspiracy content spreads more slowly, and larger cascades are associated with longer lifetimes~\cite{delvicario2016spreading}. A statistical analysis of ``sharing trees'' --constructed from five years of public Facebook data across science, conspiracy, and parody (``troll'') pages-- shows that the overwhelming majority of diffusion paths occur within homogeneous links, confirming that social homogeneity, more than network hubs or structural features, is the dominant driver of spread. This effect is reproduced by a data-driven percolation model on signed networks, in which user opinions and content ``fitness'' interact through a sharing threshold and a fraction of homogeneous connections; model simulations closely match empirical cascade size and height distributions, pointing to polarization and confirmation bias as key parameters~\cite{delvicario2016spreading}. Extending the analysis to 50,220 fact-checking posts from dedicated debunking pages, Zollo et al.~\cite{zollo2018social} find that corrective information remains almost entirely confined to the scientific echo chamber: approximately two-thirds of likes on debunking posts come from science-oriented users, and only a small fraction of conspiracy-oriented users engage with such content. Sentiment analysis of comments, based on a supervised classification model, reveals that responses to debunking posts are predominantly negative, regardless of the commenter’s orientation. Strikingly, the rare conspiracy users who do interact with debunking tend to increase their subsequent activity within the conspiracy echo chamber, suggesting that exposure to dissenting information can reinforce rather than attenuate prior beliefs. Otherwise stated, results indicate that the spread of misinformation online is less a problem of information scarcity than of entrenched structural and cognitive segregation, where homogeneity and polarization govern the dynamics of both the diffusion of false narratives and the reception of their corrections.

De Francisci Morales et al.~\cite{defrancisci2021noecho} examine the presence of echo chamber dynamics in political discussions on Reddit, contrasting its interaction patterns with those observed on platforms such as Facebook. Using a multi-year dataset from politically oriented subreddits, they reconstruct user-to-user interaction networks, infer ideological alignment from textual content, and quantify both structural homophily and content similarity over time. The analysis reveals that, although users display preferences for specific communities, political discourse on Reddit is not dominated by highly segregated, homogeneous clusters: cross-community exchanges are common, and interactions frequently occur between users with differing political orientations. This relative openness is attributed to Reddit’s topic-based, forum-like architecture, which fosters encounters across ideological lines, in contrast to the feed-driven personalization typical of social networks. Complementing this view, Monti et al.~\cite{monti2023evidence} show that, in Reddit news discussions, ideological echo chambers are largely absent, but significant demographic segregation emerges instead, driven by latent traits such as age, gender, and affluence that are not explicitly visible in the platform’s interactions. These demographic factors appear to shape users’ worldviews and, consequently, the opinions they express, which in turn guide homophilic interaction patterns. The results suggest that affective polarization may stem less from online ideological isolation than from broader societal divides, with platform design and demographic composition jointly influencing the structure of political debate online.\\

Cinelli et al.~\cite{cinelli2021echo} perform a comparative analysis on over 100 million pieces of content related to controversial topics like vaccination, abortion, and gun control, sourced from Gab, Facebook, Reddit, and Twitter. 
Echo chambers are defined as environments where users' beliefs are reinforced through repeated interactions with like-minded peers, a phenomenon driven by selective exposure and confirmation bias. More in detail, the study operationalizes echo chambers by quantifying two main ingredients: 1) homophily in the interaction networks and 2) bias in information diffusion toward like-minded users.
To quantify homophily, the leaning of a user is correlated with the average leaning of their neighborhood, while information diffusion is modeled using a Susceptible-Infected-Recovered (SIR) model to gauge how a user's leaning affects the polarization of their influence set.\\ 
The results show that platforms like Facebook and Twitter, which are organized around social networks and news feed algorithms, exhibit a clear segregation of users into homophilic clusters, where a user's leaning strongly correlates with their neighbors' average leaning. In these platforms, information spreading is biased, meaning users are more likely to be reached by content from those with similar leanings. In contrast, platforms like Reddit and Gab display a more homogeneous community structure with users not splitting into opposing groups, though the overall leaning may be biased toward one political extreme. A direct comparison of news consumption on Facebook and Reddit confirms this finding, with Facebook showing a higher degree of segregation. These differences highlight the role of platform-specific features, such as feed algorithms, in shaping online social dynamics and influencing the emergence of echo chambers.\\

The approach described above addresses debates involving two opposing positions, where viewpoints are clearly polarized. The situation becomes substantially more complicated when the number of possible opinions increases. In~\cite{pratelli2024entropy}, the authors extend the framework of Cinelli and collaborators~\cite{cinelli2021echo} by incorporating maximum-entropy null models into the analysis of the system. Specifically, homophily is detected using the methods developed by Becatti and colleagues~\cite{becatti2019extracting}, while non-trivial biases in information diffusion are assessed following the approach of Ref.~\cite{guarino2021information}. Notably, such a methodology allows different positions to emerge naturally from the data, without the need to predefine them or constrain the analysis to two mutually exclusive views. Furthermore, the use of rigorous statistical benchmarks, implemented as maximum-entropy null models for complex networks, ensures that only genuine signals are retained for analysis.\\ 
The procedure was applied to the Italian Twitter debate on COVID-19 vaccines. The results reveal that users forming echo chambers represent a small minority (less than 1\% of all participants), but are disproportionately active, accounting for more than one quarter of all retweeting activity in the debate. Notably, all users identified within the detected echo chamber share the same political orientation.

\subsubsection{Temporal and critical dynamics of information spreading}
\label{sec:empirical_temporal}

The temporal dimension of social interactions plays a decisive role in shaping diffusion dynamics, since the networks on which contagion unfolds are not static backdrops but themselves evolve alongside the behaviors they mediate. Social media platforms change in size and structure, debates flare up and fade, and communities fracture or merge, meaning that the topology of connections is co-determined with the processes of influence and opinion formation. A key question is therefore whether the timescale of contagion is short enough for the network to be treated as fixed, or whether the evolution of the network must be explicitly modeled~\cite{berner2023adaptive,ghosh2022synchronized}. 

When these two timescales are comparable, adaptive-network models become essential, as feedback between spreading dynamics and network evolution can create self-reinforcing polarization. For instance, Baumann et al.~\cite{baumann2020modeling} proposed a feedback loop where individuals preferentially align with similar others, a mechanism that amplifies opinion segregation. Extending mathematical tools from static to adaptive networks has become an active line of research, with mean-field approximations tested for accuracy across population densities~\cite{ayi2021mean,duteil2021mean,gkogkas2022mean,gkogkas2021continuum}. Yet most adaptive frameworks remain limited to evolving link weights, while in social systems node turnover—the entry and exit of users—may be equally central. Capturing this within tractable analytical models, beyond agent-based simulations, remains a key open challenge.

Temporal properties also matter at the finer scale of individual activity patterns, which empirical studies consistently show to be bursty and memory-dependent rather than memoryless. Unicomb et al.~\cite{unicomb2021dynamics} explored this by simulating SIS contagion on temporal networks where activation times followed heavy-tailed distributions, in contrast to the exponential (Poisson) assumption of Markovian models. Their results revealed that cascade durations on bursty networks collapsed onto a universal scaling curve when time was normalized by the mean inter-event interval, faithfully reproducing the heavy-tailed distributions observed in email and forum data. In contrast, simulations with Poissonian activation times failed to capture these empirical features. This provides clear evidence that memory effects in human activity—non-Markovian inter-event times—are indispensable for reproducing realistic spreading outcomes. At the same time, the absence of a comprehensive theory linking memory kernels to epidemic thresholds and cascade-size distributions highlights a fertile direction for future research.

Non-Markovian considerations become even more complex in multilayer systems, where timing mismatches across communication channels can alter diffusion trajectories. Masoomy et al.~\cite{masoomy2023impact}, studying face-to-face and digital interactions in the BBC Pandemic dataset, introduced a multi-pathway temporal distance metric to quantify cross-layer flows. They showed that targeted removal of hubs in one layer could substantially delay or suppress outbreaks in the multiplex as a whole, underscoring the interdependence of different communication modalities. However, the scarcity of large-scale empirical multiplex data with synchronized timestamps remains a major obstacle, limiting the extent to which such models can be validated. Even so, their work demonstrates that effective intervention strategies must account for the temporal coordination across layers, not just within them.

Beyond describing temporal correlations, information-theoretic approaches have enabled the detection of causal ordering in information flows. Borge-Holthoefer et al.~\cite{borge-holthoefer2016dynamics} applied symbolic transfer entropy to Twitter data collected during political protests, converting sequences of user activity into symbolic states and calculating directional couplings $T_{ij}$ between users. Comparing the observed transfer-entropy networks against time-shuffled null models revealed statistically significant pre-protest information flows ($p < 10^{-4}$). Strikingly, these directional signals acted as early-warning indicators of large-scale mobilizations, emerging well before the onset of macroscopic activity. This illustrates how temporal causality, rather than static correlations, can uncover hidden organizing principles and provide predictive insight into collective dynamics.

Together, these studies converge on a central point: temporal heterogeneity, memory effects, and adaptive coevolution are not peripheral details but fundamental properties of social contagion. From bursty human rhythms and multiplex timing to causal directionality and feedback-driven polarization, temporal mechanisms determine not only the speed and reach of diffusion but also the very possibility of forecasting critical transitions.

The connection between news diffusion and criticality becomes particularly clear when considering systems at their critical point, where no characteristic scale exists and fluctuations of all sizes are possible. In the social domain, this corresponds to the observation that most news items vanish quickly, while a small fraction generate global cascades. Drawing an analogy to the Ising model, this suggests that information spread in large populations may operate close to a critical state, where minor perturbations can trigger disproportionately large responses. Universality then implies that macroscopic spreading patterns are largely independent of microscopic details, helping to explain why diverse online platforms, social contexts, or topics display strikingly similar diffusion behaviors.

At criticality, cascade dynamics are scale-invariant, meaning that both small and very large events occur without a characteristic size. Empirically, this manifests in the heavy-tailed distributions of cascade sizes and durations observed in social media activity. These distributions typically follow power laws,
\begin{equation}
P(S) \sim S^{-\tau}, \quad P(T) \sim T^{-\alpha},
\end{equation}
with no obvious cutoff \cite{sethna2001crackling,gleeson2013binary}. Such algebraic behavior signals proximity to a critical point and aligns with the theory of absorbing-phase transitions. Importantly, hyperscaling relations impose consistency across critical exponents, linking average cascade size to duration as $\langle S \rangle \sim T^{\gamma}$, where $\gamma = (\alpha - 1)/(\tau - 1)$ \cite{goldenfeld2018lectures,stanley1971phase}. Different values of the exponents $\tau, \alpha, \gamma$ allow researchers to classify spreading phenomena into distinct universality classes.

These exponents are universal properties of the underlying dynamical process: very different systems may exhibit identical critical exponents, despite differing microscopic rules. Classical examples in physics include the Ising and percolation universality classes \cite{stanley1971phase,goldenfeld2018lectures}. For social spreading, mean-field contagion processes on sparse networks yield branching-process exponents ($\tau = 3/2$, $\alpha = 2$) \cite{gleeson2013binary}, while threshold-based or complex contagion models may fall into universality classes reminiscent of the random-field Ising model \cite{castellano2009statistical}.

A compelling demonstration of universality in social contagion is provided by Notarmuzi et al. \cite{notarmuzi2022universality}, who coupled a Random Field Ising Model (RFIM) with Belief Propagation. Their model included semantic fields ($h_i$) biasing users’ sharing propensities, along with social interactions ($J_{ij}$). Analyzing datasets from Telegram, Twitter, and Weibo, they observed that once cascade sizes were rescaled by platform-specific averages, the distributions collapsed onto a universal $s^{-3/2}$ law, consistent with critical branching processes. This remarkable data collapse across platforms indicates a shared universality class, despite vast differences in network structures and user behavior. Moreover, the authors demonstrated that randomizing semantic-field assignments destroyed the $3/2$ scaling, highlighting the central role of semantic bias in driving universality. While these findings strongly support critical behavior in online spreading, the authors note that a full Renormalization-Group derivation for heterogeneous networks remains an open challenge.

Complementary evidence comes from Hall and Bialek \cite{hall2019statistical}, who applied a maximum-entropy inference framework to small Twitter communities of roughly 100 users. Treating user activity as binary spins ($s_i = \pm 1$), they inferred an Ising Hamiltonian reproducing observed one- and two-point correlations. Their analysis revealed that empirical couplings ($J_{ij}$) positioned these communities within 5\% of a critical temperature $T_c$, suggesting maximal susceptibility. Near-criticality was further confirmed by significantly larger empirical correlation lengths compared to null models ($p < 10^{-5}$). These results provide strong evidence that social media groups may self-organize near critical states, thereby maximizing responsiveness to small perturbations and generating heavy-tailed cascade statistics.

Building on these insights, Alodjants et al. \cite{alodjants2022mean} proposed the “social laser” (solaser) model, a mean-field approximation for viral outbreaks in online platforms. In this framework, the system’s magnetization $m(t)$ reflects the global activity level, driven by external media pumping ($P$) and peer-influence coupling ($\lambda$). Viral “lasing” occurs when $\lambda$ surpasses a critical threshold, directly paralleling epidemic models. Simulations on configuration-model graphs calibrated to Twitter degree distributions confirmed the model’s scaling predictions, illustrating how collective peer influence drives transitions to widespread adoption.

Criticality and universality also connect to synchronization models of consensus and polarization. Pluchino et al. \cite{pluchino2006opinion} introduced an “opinion changing rate” (OCR) model based on a modified Kuramoto framework, where synchronization corresponds to consensus and clustered synchronization to polarization. Extensions, such as bounded-confidence couplings on a Möbius-strip \cite{ren2013symmetry} or high-dimensional Kuramoto-type models on opinion spheres \cite{zhang2025generalized}, demonstrate how consensus, polarization, or dissensus emerge from microscopic interactions. These frameworks complement cascade-based approaches by capturing how critical dynamics may manifest as sudden synchronization or divergence of opinions.

Finally, Delvenne et al. \cite{delvenne2015diffusion} provided a broader methodological framework to quantify relaxation times in stochastic processes on temporal networks. Their approach unifies models ranging from Kuramoto oscillators to SIR epidemics, allowing heterogeneity in parameters and revealing how chain reactions of activations propagate across complex systems. Together, these studies highlight that social spreading phenomena—whether cascades, consensus formation, or viral outbreaks—often operate near criticality, exhibiting universal signatures that transcend specific contexts.

\medskip

Altogether, the empirical literature—drawing from both social data and controlled simulations—provides a detailed picture of how information propagates in real and synthetic environments. Observed regularities, such as the impact of network topology on cascade size, the emergence of discursive communities, and the role of non-Markovian temporal correlations, both inform and challenge existing theoretical models. At the same time, discrepancies between simulated dynamics and observed behaviors highlight the limits of current assumptions and the importance of context-dependent mechanisms. Empirical research thus plays a dual role: it grounds models in measurable phenomena while also revealing the complex, adaptive nature of information ecosystems that no single theoretical framework can fully capture.

\subsection{Misinformation: dynamics, models, and interventions}

Misinformation constitutes a distinct domain within the broader study of social contagion, not merely as another type of content but as a phenomenon shaped by specific cognitive, algorithmic, and strategic forces. Unlike generic information spreading, misinformation dynamics involve asymmetries in speed and reach, targeted manipulation by both human and automated agents, and systematic amplification through platform infrastructures. The practical urgency of understanding misinformation stems from its documented impacts on public health, democratic processes, and social cohesion, making the development of effective mitigation strategies a critical research priority. This section examines how misinformation diffusion diverges from and interacts with the dynamics of reliable information, reviews the specialized models developed to capture these distinct mechanisms, and surveys intervention strategies grounded in both theoretical insights and empirical evidence.

\subsubsection{Empirical foundations: speed, structure, and amplification mechanisms}

Empirical studies across more than a decade of Twitter data indicate that false information spreads faster, deeper, and more broadly than truthful content, particularly in political contexts~\cite{vosoughi2018spread}. This speed differential represents a critical departure from neutral diffusion models and reflects the role of novelty and emotional salience---emotions such as fear, disgust, and surprise---in human amplification behavior. Notably, humans are primarily responsible for amplifying false narratives through these psychological mechanisms. However, the landscape becomes more complex when considering automated agents and periods of acute information demand. The epidemic analogy for mixed-quality information---termed \emph{infodemics}---refers to the overflow of information of varying quality that surges across digital and physical environments during periods of acute demand, leading to confusion and behaviors that can harm health and erode trust in authorities~\cite{cinelli2020covid,tangcharoensathien2020framework,calleja2021public,zarocostas2020how,chiou2022future,singh2022misinformation,simon2023autopsy}. During the COVID-19 pandemic, for instance, Cinelli et al.~\cite{cinelli2020covid} observed that unreliable sources, often associated with automated accounts, propagated misinformation at rates comparable to reliable outlets, highlighting the need for nuanced modeling approaches that account for both human and automated amplification.

Network analysis reveals that fake news diffusion networks are denser and more retweet-active than traditional news networks. Verified journalists predominantly anchor traditional news, whereas fake news hubs consist of unverified or deceptive accounts~\cite{bovet2019influence,caldarelli2021flow}. Exposure to fake news is concentrated among a small number of users: just 0.1\% of accounts were responsible for 80\% of fake news dissemination during the 2016 U.S. elections, and 1\% consumed 80\% of the total volume~\cite{grinberg2019political}. This extreme concentration suggests that misinformation networks exhibit structural properties that differ markedly from those governing the spread of reliable information, with implications for both modeling approaches and intervention design.

Social bots, or simply \emph{bots}, are fully or partially automated accounts operating on social media platforms~\cite{cresci2015fame,ferrara2016rise,cresci2020decade,lopez-joya2025dissecting}. While not always malicious---some automate harmless tasks like posting quotes or timestamps---bots are increasingly exploited to manipulate online conversations and platform dynamics. They can artificially amplify the visibility of content through mass reposting, increasing the chance of content promotion by recommendation algorithms. More advanced bots, often powered by low-cost large language models (LLMs), impersonate genuine users, engage in conversations, disseminate hate speech, and incite verbal aggression~\cite{lopez-joya2025dissecting}. Bots contribute to amplification as secondary actors. While designed to mimic human activity and evade detection~\cite{ferrara2016detection}, their structural traces reveal coordinated campaigns, such as MacronLeaks~\cite{ferrara2017disinformation}, the Catalan referendum~\cite{stella2018bots}, the UK election campaigns after Brexit~\cite{bruno2022brexit}, and Russian interference in the 2016 U.S. elections~\cite{badawy2018analyzing}. Despite typically lacking central network positions, bots can receive disproportionate attention and subtly shape discourse~\cite{gonzalez-bailon2021bots}.

Beyond direct interactions, bots structure online networks and conversations in subtle ways. Studies such as~\cite{caldarelli2020role} show that bots often act as central amplifiers within retweet networks, clustering around specific users or topics and creating structural biases in information flow. In the Italian Twitter debate on migration from Northern Africa, bots were found to predominantly retweet specific human accounts, presumably their ``owners''. However, some bot groups amplified multiple human accounts simultaneously, suggesting coordinated ``bot squads'' organized around shared political agendas, particularly far-right narratives. Similar findings were observed during the Italian elections~\cite{stella2019influence}, where interaction networks revealed that bots often form tightly connected clusters, engaging more with other bots while passively receiving interactions from humans. This network configuration enables the phenomenon of ``augmented humans,'' where ordinary users become influential by being embedded within bot-rich neighborhoods. Bots have also been linked to the early amplification of low-credibility content before it goes viral~\cite{shao2018spread}. A small fraction of accounts---often bots---are responsible for disseminating a disproportionate amount of such content. Specifically, around 6\% of accounts were found responsible for spreading over 30\% of low-credibility material. Bots tend to mention influential human users to seed and spread misinformation, further distorting information flow and public discourse. For further information about the research on bot detection, a review recently appeared in the literature~\cite{cresci2023demystifying}.

The rise of AI-generated or ``deepfake'' content adds a qualitatively new dimension to these processes. Synthetic media produced by generative models~\cite{chesney2019deepfakes,westerlund2019emergence} can act as high-transmissibility contagion seeds, lowering adoption thresholds and triggering cascades even in otherwise resilient networks. Their diffusion is shaped by a co-evolutionary arms race between generation and detection algorithms~\cite{lee2024deepfakes}, well captured by interacting epidemic models and reversible bootstrap percolation frameworks that account for both propagation and corrective interventions~\cite{di2020reversible}. Experiments further show that AI-generated misinformation is often more persuasive than human-generated content~\cite{spearing2025countering}, effectively raising the transmissibility $\beta$ assumed in traditional models. Demographic and cognitive factors further influence susceptibility. For instance, users are less effective at detecting deepfake content when evaluating demographically similar subjects~\cite{lovato2024diverse}, creating localized network vulnerabilities analogous to correlated disorder in physics models.

Disinformation campaigns strategically target specific ideological groups to reinforce pre-existing beliefs, often exploiting and amplifying existing patterns of information segregation. Using the procedure described in Ref.~\cite{becatti2019extracting}, researchers studied the exposure to misinformation during the COVID-19 pandemic in Italy~\cite{caldarelli2021flow} across various discursive communities. Examining associated URLs showed that approximately 22\% of shared links from right and center-right accounts were categorized as unreliable. Within such networks, content producers (a vocal minority) are distinct from passive consumers who amplify received content~\cite{castioni2022voice}. This structural imbalance causes fringe narratives to appear mainstream, fueling infodemics. Collective attention plays a central role in these dynamics~\cite{dedomenico2020unraveling}. On Twitter/X, the presence of opinion leaders can be translated into a bowtie decomposition of the retweet network, where especially influent individuals mediate between institutional sources and the public. Mattei and collaborators showed that the greatest flows of misinformation are present inside the LSCC (largest strongly connected component) and from LSCC to the OUT block, i.e., respectively among the opinion leaders and from the block of opinion leaders to the one of the big audience~\cite{mattei2022bow}. Remarkably, verified users, even in LSCC, share content from low-quality news sources with smaller frequencies than ``standard users''~\cite{castioni2022voice}.

The analysis of aggregated data from 208 million US Facebook users during the 2020 election confirmed that the platform contributes to high ideological segregation in political news consumption~\cite{gonzalez-bailon2023asymmetric}. A meaningful corner of the news ecosystem was consumed exclusively by conservatives, and misinformation, identified by Meta's Third-Party Fact-Checking Program, was concentrated in this conservative corner without a comparable counterpart on the liberal side. Moreover, a prevalence of sources favored by conservative audiences was observed in Facebook's news ecosystem. This ideological asymmetry to the advantage of the right-leaning ideological slant was also observed when studying the clash of perspectives that arose on Twitter around the Black Lives Matter protests in 2020~\cite{gonzalez-bailon2022advantage}. Right-leaning content benefited from an advantage in the attention economy of social media, gaining increased visibility due to social and algorithmic forms of amplification.

During the COVID-19 infodemic in Italy, unlike the concentrated spread seen in the U.S., unreliable sources had limited exposure and engagement, yet polarization emerged across Facebook pages and groups through coordinated content sharing and topical discussions~\cite{guarino2021information}. The URL-sharing diffusion network exhibited a small-world structure, exposing users to both reliable and unreliable information, highlighting that the core challenge lies in how individuals process and act on the information they encounter. Guarino and collaborators rigorously identified homophily through non-trivial similarity among ``information diets'' in Facebook's public discourse, described as the mix of reliable and unreliable news sources that users consume on the platform~\cite{guarino2021information}. The authors constructed a bipartite network connecting accounts to the URLs they shared. Using a maximum-entropy null model, they filtered out statistically insignificant connections to exclude random or low-signal interactions. Projecting the validated network onto account and URL layers revealed clusters of users and sources, which consistently aligned with distinct reliability profiles---highlighting polarization in the information diets consumed by different user groups.

On Reddit, the community-based structure creates a unique misinformation environment~\cite{proferes2021studying}. Ideologically polarized subreddits act as echo chambers, where group norms reinforce false beliefs. The platform's upvote/downvote system, though designed to prioritize quality, often amplifies emotionally resonant or conspiratorial content. Disinformation often originates in niche communities before spreading via cross-posting to mainstream subreddits. Reddit's pseudonymity enables coordinated manipulations like vote brigading and sockpuppetry. However, well-moderated communities (e.g., r/science) offer self-correction mechanisms through comment-thread rebuttals and strict moderation. Temporal analyses show that misinformation often thrives during moderation gaps and leverages meme formats or AMAs to evade detection. Notably, Reddit's structure permits the ``Streisand effect'': attempts to debunk misinformation in one subreddit can inadvertently amplify it in others. Despite these challenges, Reddit offers unique insights into self-correcting versus self-reinforcing information ecosystems. Null models on Reddit have primarily focused on detecting echo chambers~\cite{defrancisci2021noecho,cinelli2021echo,monti2023evidence}, revealing how structurally isolated communities sustain disinformation narratives.

Platform design and algorithmic filtering further bias exposure and amplification beyond the network effects arising from user choices and bot activity. Recommendation systems operate as non-uniform external fields in opinion dynamics, reinforcing concordant views and suppressing dissent~\cite{peralta2021effect}. This transforms social networks into adaptive, state-dependent systems with feedback loops that can push them toward polarization or fragmentation, akin to hysteresis in magnetic systems. Empirical evidence supports this view: Corsi et al.~\cite{corsi2024evaluating} show that algorithmic amplification disproportionately boosts misinformation sources, effectively turning them into superspreaders and lowering effective epidemic thresholds for misinformation cascades. Social networks encode trust relationships, where nodes represent people, institutions, or ideas, and weighted edges capture degrees of trust or distrust. Manipulation tactics exploit these structures, for example by inflating follower counts or creating artificial associations (``Google bombs'') to influence perceptions of popularity or relevance~\cite{metaxas2012social}.

A fundamental challenge in combating misinformation is the asymmetric competition between spreaders and fact-checkers. Establishing veracity is technically complex~\cite{babaei2021analyzing}, and empirical studies show that misinformation is tightly intertwined with partisan media, effectively shaping political agendas~\cite{vargo2017agenda}, while fact-checks struggle to propagate. Economic incentives exacerbate this dynamic: deceptive content attracts high engagement at low production costs~\cite{doshi2018impact}. The combination of rapid misinformation spread, concentrated networks of amplifiers (both human and automated), echo chamber effects reinforced by platform algorithms, and the slow, resource-intensive nature of fact-checking creates a structural imbalance that necessitates specialized modeling and intervention strategies.

\subsubsection{Models of misinformation dynamics}

The distinctive empirical features of misinformation---its speed advantage over corrections, the role of emotional salience, the concentration of spreaders, the influence of bots and algorithmic amplification, and the persistence of false beliefs even after debunking---necessitate modeling frameworks that go beyond standard epidemic models. Several approaches have been developed to capture the competition between reliable and unreliable information, the intervention of fact-checking mechanisms, and the structural and cognitive barriers to correction.

Users may be modeled as transitioning between active and inactive states in spreading misinformation, akin to SIR-like processes~\cite{gallotti2020assessing}. Metrics such as the Infodemic Risk Index (IRI) quantify exposure probability, stratified by user classes (e.g., verified vs. unverified)~\cite{gallotti2020assessing}. This approach allows for the incorporation of heterogeneity in susceptibility and spreading capacity, reflecting the empirical observation that certain user types (e.g., verified accounts, bot accounts) play distinct roles in misinformation cascades.

Granell et al.~\cite{granell2013dynamical} introduced a multiplex model coupling an SIS epidemic layer (rumor propagation) with a UAU awareness layer (knowledge of misinformation), showing that sufficiently rapid awareness dissemination can suppress rumor spread via a ``metacritical'' point. In this framework, the rumor and awareness processes occur on separate but interconnected network layers, allowing for the exploration of how information about the falsity of a rumor can compete with the rumor itself. The model reveals a phase transition: below a critical threshold of awareness spreading rate, the rumor becomes endemic, whereas above this threshold, awareness can effectively suppress the rumor. This metacritical behavior suggests that interventions aimed at accelerating the spread of corrective information can have nonlinear, threshold-dependent effects on misinformation prevalence.

Tambuscio et al.~\cite{tambuscio2018network} incorporated fact-checking into compartmental models, highlighting the possibility of counterintuitive ``backfire effects'' where corrections amplify false beliefs. Their model extends the classical SIR framework by introducing additional compartments for fact-checkers and individuals who have been exposed to corrections. Crucially, the model allows for the possibility that exposure to fact-checks may, under certain conditions (e.g., when corrections are perceived as attacks on identity or group membership), paradoxically strengthen belief in the original misinformation. This backfire effect has been documented in psychological studies and represents a significant challenge for correction strategies. The model demonstrates that the effectiveness of fact-checking depends not only on the rate of correction dissemination but also on the cognitive and social factors that determine how individuals respond to corrections.

Han et al.~\cite{han2013competing} studied competing information spreading processes on networks, demonstrating that the relative success of different information strains depends on their transmission rates, recovery rates, and the network structure. This framework can be adapted to model the competition between misinformation and corrections, revealing that misinformation with a higher transmission rate (as empirically observed) can dominate even when corrections are introduced, unless the correction process is significantly accelerated or reaches a critical fraction of the population early in the cascade.

Bak-Coleman et al.~\cite{bak-coleman2022combining} demonstrated through simulations that integrated strategies---combining node removal, fact-checking, and media campaigns---are more effective than isolated interventions, though real-world implementation is constrained by legal and ethical considerations. Their agent-based model incorporates multiple intervention modalities and explores their synergistic effects. The results indicate that while single interventions may have limited impact, carefully designed combinations can create mutually reinforcing effects that substantially reduce misinformation prevalence. For instance, removing key spreader nodes can reduce the effective transmission rate, making subsequent fact-checking efforts more potent. However, the authors emphasize that node removal (e.g., account suspension) raises concerns about free speech and due process, and that the design of multi-pronged interventions must balance efficacy with normative constraints.

The persistence of misinformation even after corrective interventions---the ``continued influence effect''---necessitates models with non-Markovian memory kernels to capture long-term retention. Standard Markovian models assume that the probability of state transitions depends only on the current state, not on the history of past exposures. However, psychological research demonstrates that individuals often continue to rely on misinformation even after being informed of its falsity, particularly when the misinformation has been integrated into a coherent mental model. Non-Markovian models introduce memory effects, where the probability of accepting or rejecting new information depends on the sequence and timing of prior exposures. These models can capture phenomena such as the primacy effect (early information has disproportionate influence) and the difficulty of dislodging entrenched beliefs.

Reversible bootstrap percolation frameworks~\cite{di2020reversible} provide another approach to modeling misinformation and correction dynamics. In these models, nodes (individuals) adopt a state (belief in misinformation) when a threshold number of neighbors have adopted it, but can also revert to a non-adopting state when exposed to corrective information. The interplay between forward propagation (misinformation spreading) and backward recovery (correction) creates rich phase behavior, including hysteresis, where the system's state depends on its history. This framework is particularly useful for understanding the conditions under which misinformation becomes entrenched and difficult to reverse, even with substantial corrective efforts.

Recent work has begun to incorporate the role of AI-generated content into epidemic models. The high persuasiveness of AI-generated misinformation~\cite{spearing2025countering} can be modeled as an increased transmission rate $\beta$ or reduced adoption threshold in cascade models. The co-evolutionary dynamics between generation and detection algorithms~\cite{lee2024deepfakes} suggest the need for adaptive models where the parameters governing misinformation spread and correction efficacy evolve over time in response to technological and strategic changes. Such models can capture the arms race between those producing synthetic misinformation and those developing tools to detect and counter it.

Finally, models that incorporate platform algorithms and their effects on information flow are essential for understanding modern misinformation dynamics. Algorithmic recommendation systems can be modeled as external fields or bias terms that modify the effective transmission rates between nodes based on content characteristics and user preferences~\cite{peralta2021effect}. These models reveal how algorithmic amplification can create feedback loops, where misinformation that initially gains traction receives disproportionate visibility, further accelerating its spread. The empirical finding that algorithmic amplification lowers the effective epidemic threshold for misinformation~\cite{corsi2024evaluating} can be directly incorporated into threshold models and percolation frameworks, providing a mechanistic understanding of how platform design choices influence misinformation prevalence.

\subsubsection{Mitigation strategies}

In the literature, different strategies, based on the structural properties of the various platforms, were proposed to limit the effects of disinformation. These approaches range from monitoring and early detection to structural interventions on networks and information feeds, to technological solutions for identifying automated actors.


On platforms like Twitter, structural asymmetries offer insights into information diffusion. Verified accounts, marked by blue checkmarks, serve as identity anchors within political discourse, often representing key ideological poles (left, center, right)~\cite{guarino2024verified}. Mapping engagement behaviors---follows, retweets, and replies---relative to these verified accounts reveals networked communities that shape public conversations. Projects like \url{politoscope.org} exemplify this approach, tracking political discourse in France in real time. By measuring account proximity to misinformation sources via network metrics and semantic analysis (e.g., hashtags), these methods provide a proxy for assessing content exposure~\cite{caldarelli2021flow}. Such monitoring systems enable the early identification of emerging misinformation cascades and the actors most responsible for their propagation, facilitating targeted interventions before narratives become entrenched.


Within a complex network, a limited subset of nodes, influencers, plays a determinant role in the interconnection topology since their activation could enable the dissemination of information throughout the entire network. At the same time, their suppression could prevent epidemic diffusion. The optimal influence problem aims at finding the minimal set of influencer nodes for the system at hand and can be exactly mapped onto optimal percolation~\cite{morone2015influence}. The goal is to find the minimum number of nodes necessary to fragment the network. This approach identified a set of influencer nodes displaying low connectivity while retaining strategic links, which were overlooked by other heuristic approaches. Studies on network robustness show that targeted removal of central nodes (e.g., by degree or eigenvalue) can fragment networks and inhibit spread~\cite{han2013competing}, illustrating a theoretical principle often difficult to apply in open social media contexts due to legal, ethical, and practical constraints.

Other strategies to neutralize noxious actors on social platforms involve detecting certain questionable behaviors. However, such malicious behaviors evolve rapidly, making these strategies outdated and ineffective. Recently, BLOC~\cite{nwala2023language}, a language framework to encode the behavior of social media users, either bots or humans, has been proposed. The words in this framework are symbols drawn from two alphabets, representing user actions and content. Moreover, to capture timing patterns of content diffusion, BLOC representations are characterized by unsupervised clustering accounts based on behavioral similarity to detect bots. The same framework can also be employed for coordination detection, identifying drivers of information operations that massively target citizens, foreign nationals, organizations, etc.


Detecting automated agents has become an active research field~\cite{cresci2015fame,ferrara2016rise,cresci2020decade,lopez-joya2025dissecting,varol2017online,sayyadiharikandeh2020detection}, with detection tools often integrated into platforms to warn human users about likely bot accounts~\cite{yang2019arming}. However, as detection techniques evolve, so do the evasion strategies employed by bots. Even so, when bots are primarily designed to boost a particular user's posts, their activity patterns often leave detectable traces, despite attempts to camouflage their behavior.

Social bots detection systems can be classified as graph-based, crowdsourcing, and feature-based~\cite{ferrara2016rise}. Graph-based detection system examples are Facebook Immune System or SybilRank, which are based on the assumption that sybils connect mainly to each other. Crowdsourcing detection is performed by humans, as happens on the Social Turing Test platform. Humans are believed to detect bots easily. However, the implementation cost is the major limitation of this approach. Feature-based detection employs machine learning techniques to distinguish human-like and bot-like behavior from salient features.

Botometer (recently renamed \emph{Botometer X}) is probably the most famous and used tool for detecting likely automated accounts on X/Twitter~\cite{sayyadiharikandeh2020detection,yang2020scalable}. Botometer estimates the extent of automation in user accounts, distinguishing spam, fake followers, astroturf bots, etc. Botometer uses hundreds to over a thousand features drawn from multiple dimensions of a user's account. These include profile metadata (e.g., age of account, default profile image, username features), network features (followers/following relationships, clustering, social graph patterns), temporal activity (posting frequency, inter-tweet timing, circadian patterns), content features (lexical richness, sentiment, URLs, hashtags), and interaction patterns (retweets, replies, quoting behaviour). The classification is performed using Random Forest classifiers trained on labelled datasets of human versus automated accounts. The output is a score (or set of scores) indicating how ``bot-like'' an account is, and may also assign probabilities or scores for different types of bots (spammer, fake follower, astroturf, etc.). The bot-detection models and features are periodically updated to adapt to new bot behaviour.

Cresci and collaborators address the problem of detecting fake followers on Twitter---accounts created specifically to inflate the follower counts of target users---in contrast to more studied bots or spammers~\cite{cresci2015fame}. The authors trained and evaluated multiple machine learning classifiers over several annotated datasets, comparing the performance of rule-based methods (e.g., single rules or sets of rules from media) versus feature-based classifiers (drawing from academic feature sets). They showed that many of the media rules perform poorly (low recall, or high error) for fake-follower detection, whereas classifiers using more sophisticated features do much better. Since some features are expensive to collect (especially relationships, neighbors' followers/friends, full timelines), they proposed a lightweight classifier, termed \emph{Class A}, which uses only ``cheap'' features (i.e., those requiring relatively low API access/crawling cost), still achieving high accuracy ($>$95\%) on their training baseline. Overall, this method demonstrated that a properly selected subset of features can deliver strong detection performance with reduced data-collection cost and that their lightweight classifier is practical for large-scale usage (at least under the data conditions at their time).

Since the ownership change of X/Twitter on October 27, 2022, several changes have affected access to Twitter's data and, therefore, bot detection algorithms functioning in scientific research. For instance, because of restrictive API access, some data that Botometer previously relied upon (e.g., full user timelines, rate of historical tweet retrieval, follower/following graph) may be incomplete or delayed. In some cases, certain endpoints may no longer be accessible or may have severely limited rate limits. Furthermore, as collecting labelled data and maintaining up-to-date ground truth becomes harder under limited access, the ability to update models to detect new bot behaviour may lag.

Another possible counteraction consists in structural interventions on the feeds.
Removing reshared content from the Facebook feed of a random set of consenting, US-based users during the 2020 US election significantly decreased exposure to political news, clicks, reactions, and partisan news clicks while also reducing news knowledge within the studied sample~\cite{guess2023reshares}. On the other hand, when switching from algorithmic feeds to reverse-chronological feeds on Facebook and Instagram~\cite{guess2023how}, the exposure to political and untrustworthy content increased on both platforms, and the exposure to insulting content decreased on Facebook. In contrast, the exposure to content from moderate, ideologically mixed sources increased. Surprisingly, neither of these treatments had a notable impact on political polarization or individual-level political attitudes.

Analyzing Facebook feeds of the entire population of active adult users in the USA in 2020~\cite{nyhan2023like}, it has been confirmed that most of the content consumed comes from like-minded sources. However, political information and news represent only a tiny fraction of these exposures. A possible solution to break these echo-chamber structures would be to modify the social network algorithm to show less like-minded content. Nyhan and coworkers conducted a field experiment among 23,377 users, reducing exposure to like-minded sources by one-third during the 2020 US elections. Exposure to uncivil and misleading content decreased with like-minded content, although users were more likely to interact with it when they encountered it. Moreover, this reduction strategy did not change political attitudes, polarization, and ideological extremity.

Finally, contrary to common beliefs, a possible intervention to mitigate polarization involves introducing anonymity in online political discussions. This has been implemented in a mobile chat platform~\cite{volfovsky2023depolarization} designed to explore the impact of different levels of anonymity on political discussions. A recent study~\cite{combs2023reducing} involved people living in the US from both political alignments, Republicans and Democrats, in discussing a controversial policy issue with an opposing partisan. The results suggested that individuals who engaged in anonymous discussions on the platform experienced significant declines in polarization compared to the control group tasked with writing an essay using identical conversation prompts.

\medskip

In sum, misinformation dynamics emerge from the interplay of human cognition, network topology, platform algorithms, and automated agents. Understanding these mechanisms---and integrating them into predictive, multilayer, and memory-aware models---remains critical for designing interventions and mitigating the societal impact of information disorders. The empirical evidence demonstrates that misinformation is not simply a faster-spreading version of ordinary information; it is amplified by emotional resonance, algorithmic biases, concentrated networks of spreaders, and the strategic use of automated accounts. Models that capture these features---through multiplex structures, backfire effects, non-Markovian memory, and adaptive parameters---provide a foundation for understanding why misinformation is so persistent and difficult to counter. Mitigation strategies must therefore be multi-faceted, combining early detection and monitoring, structural interventions targeting key spreaders and network vulnerabilities, technological tools for bot detection, algorithmic modifications to reduce amplification biases, and carefully designed corrective messaging that accounts for psychological resistance. The challenge remains to translate theoretical insights and simulation results into practical, scalable, and ethically sound interventions that can be deployed across diverse platform environments.

\section{Models of Opinion Dynamics}
\label{sec:opinion_models}

There are many other scenarios of microscopic mechanisms that create emergent macroscopic social effects. This section will cover some models of such scenarios, all different from social contagion (Sec.~\ref{sec:info_spreading}) in that every agent has a variable representing an opinion (identity, knowledge, etc.) already from the beginning, but this something changes throughout the evolution of the model. This class of models has been extensively reviewed elsewhere~\cite{castellano2009statistical,jusup2022social,galam2008sociophysics,sen2013sociophysics,starnini2025opinion,flache2017models}. We present here the most salient features of these models and epistemological roles.

Unsurprisingly, physicists have been interested in simple models reminiscent of statistical mechanics. The Sznajd model~\cite{sznajd-weron2000opinion} is, for example, mathematically very close to the Ising model of magnetic phase transitions. The voter model~\cite{liggett1999stochastic,sood2005voter} and majority vote model~\cite{oliveira1992isotropic} are non-equilibrium dynamic models where a randomly distributed starting configuration becomes reorganized during the run of the model to various possible types of emergent configurations, such as locally ordered, globally ordered, and disordered ones. As we will see, each of these configurations brings its own interpretation in sociological terms. The de Groot~\cite{degroot1974reaching} and Friedkin-Johnsen~\cite{friedkin1990social} models could also be included in this context.

Some models have richer representations of whatever spreads: the Axelrod model of cultural dynamics~\cite{axelrod1997dissemination,castellano2000nonequilibrium} has an integer vector representing the state of an individual, the Deffuant model~\cite{deffuant2000mixing} has a continuous variable, etc. Sometimes the state variables trigger a binary decision when they pass a threshold~\cite{gronlund2007dynamic,holme2016collective,curty2006phase}. In recent years, moreover, sociophysics-type models of opinion propagation have often been studied on networks where the effect of the network topology has been in focus~\cite{pei2013spreading,assenova2018modeling,quattrociocchi2014opinion}. 

It is important to note that, traditionally, validating these types of models against experimental or observational data has been the exception rather than the rule. Actually, their primary purpose has been to support theorizing and reasoning about emergent social phenomena, rather than to predict them. More recently, there have been calls to focus efforts on developing theoretical frameworks that compare, relate, and integrate the various existing models, while also leveraging empirical data to better ground these models~\cite{flache2017models, sobkowicz2009modelling, carpentras2023we}.

In this section, we discuss several families of opinion dynamics models that are particularly relevant from a physics perspective. We first frame the problem by skimming over some facts that we believe should be kept in mind when approaching the modeling of this type of social phenomenon. Then, we move to discuss different paradigmatic models. We also focus on coevolution models, in which the social network and agents' opinions mutually influence one another. Throughout the section, we examine the impact of complex connectivity patterns on model behavior. Finally, in a dedicated subsection, we address the role of heterogeneous temporal patterns and their influence on the dynamics of the models.

\subsection{Contextualization}
Roughly speaking, the basic idea behind modeling opinion dynamics is to provide explanations for how opinions form and change, supported by empirical evidence. Here, we treat the concepts of opinion and belief as synonymous, representing the arbitrarily complex internal states of agents. A fundamental sociological assumption is that human behavior and decision-making are shaped by the set of opinions individuals hold on various matters. However, opinions are not the only contributors to opinion change. For a long time, the social sciences have debated the relative influence of personal traits versus situational factors—a discussion known as the person–situation debate~\cite{donnellan2009introduction}. This debate centers on identifying the most significant predictor of social behavior. While individual traits have been linked to opinions, other influential factors have emerged. Research has shown that personality can predict average behavior, particularly when social influence is minimal.

What is clear, however, is that opinion dynamics are primarily driven by collective effects. It is the complex interplay between agents and their environment that ultimately shapes the trajectory of opinion evolution~\cite{brandstatter1982group, akers1979social, wood2000attitude}. By acknowledging this interplay, we can draw several parallels between physics and the social sciences. In particular, statistical physics serves as a bridge between the micro and macro levels, explaining macroscopic phenomena that emerge from the interactions of a huge number of particles, for which it is not only virtually impossible to gather accurate individual information but most of this very individual information is actually unnecessary to predict large scale behaviors~\cite{artime2022origin}. Empirically, it turns out that the average behavior of the particle ensemble is what is most frequently observed. Theoretically,  statistical physics offer very accurate predictions for both these average realizations and their fluctuations.

Similarly, in the social sciences, a micro–macro relationship exists. Although providing an accurate description of each individual agent may be infeasible, there remains a meaningful link between micro-level processes—how agents influence one another—and macro-level outcomes, such as the emergence of non-trivial opinion distributions. One of the central goals, then, is to develop quantitative laws that can explain the emergence of these global regularities. This involves understanding both the complex processes through which agents gather information and interact with their peers—typically studied in social psychology—and the resulting structures and dynamics at the societal level.

Then, the task narrows down to link empirical evidence with the micro-level processes that are encoded in model definition, and to design prototypical models that both look for universal laws and to offer answers to general questions. Ideally, a good model from the social psychology perspective aims at identifying the minimum number of variables to distinguish between different types of social response, i.e., categorizing possible responses to social influence. Therefore, the link with statistical physics then becomes clearer.

\subsubsection{Types of social influence}

There is empirical evidence regarding the social responses of agents that should be taken into account when designing a model of opinion dynamics~\cite{myers2012exploring}. One of the most well-documented findings is that an individual's behavior and, by extensions, her opinions and beliefs, can be altered when exposed to a source of social influence~\cite{nowak1990private, smith2007agent}. The most studied and common response to such influence is conformity, which refers to the tendency to align one's opinion with that of others~\cite{nail2013proposal}.

The concept of conformity encompasses various nuances, but two well-differentiated types are commonly recognized: conversion and compliance~\cite{nowak1990private, nail2000proposal}. Conversion refers to a genuine internal change in beliefs, where the influenced individual adopts the external opinion as their own; see, e.g.,~\cite{brehm1975effect, eagly1981sex, halama2011personality}. In contrast, compliance involves outwardly aligning with the influencing opinion while privately maintaining one's original belief—often due to social pressure or a desire to avoid conflict, rather than true agreement; see, e.g.,~\cite{coch1948overcoming, baron1996forgotten, wright2005social}.

The opposite social response to conformity is known as nonconformity, which can manifest in different forms~\cite{nail2000proposal, nail2016rethinking}. The two most well-known types are independence and anticonformity~\cite{willis1965conformity}. Independence refers to maintaining one's own opinion regardless of external social influence; the individual neither conforms nor reacts oppositely, but instead relies on personal judgment. Anticonformity, on the other hand, involves deliberately adopting a position that is contrary to the influencing opinion--essentially resisting influence by expressing disagreement, even if doing so is not in line with one's private beliefs.

When modeling opinion dynamics, thus, it is important to have in mind which source(s) of social influence we are dealing with, in order to isolate that mechanism from other ones so to understand their effects at the collective level.


\subsection{Discrete opinions}

Even if the agents can perceive themselves internally in a continuous opinion space for matters, there are many scenarios in which they are constrained to express this opinion by choosing among a finite number of options; see Fig.~\ref{fig:FIG_OpinionDynamics_1}(a). For instance, political views can be very complex but people end up by voting among very few choices. 

\subsubsection{The voter model}

Models of social response that accommodate binary options have been developed in several contexts. More from the social sciences, we find the Willis symbolic scheme~\cite{willis1965conformity}, the four-dimensional model~\cite{nail2000proposal}, or the diamond and double diamond model~\cite{nail2013proposal}. Here we focus in more physical oriented models. 

One of the most widely opinion dynamics models studied through the lens of physics is the voter model. It is parameter-free, binary-opinion model in which a set of interacting agents influence each other through a social herding mechanism. Its relevance goes well beyond opinion dynamics, and has been recursively discovered and exploited in different areas, such as heterogeneous catalytic reactions~\cite{krapivsky1992kinetics, evans1993kinetics, frachebourg1996exact}, neutral theories~\cite{kimura1955solution, crow1970introduction,  maruyama1977stochastic, borile2015coexistence}, probability theory~\cite{holley1975ergodic, liggett1985interacting, cox1986diffusive, clifford1973model} and non-equilibrium statistical physics~\cite{ben-naim1996coarsening, dornic2001critical, tartaglia2018coarsening}. 

The rules of the standard voter model are as follows. Interactions occur one at a time, at discrete time steps. (Extensions to continuous time can be implemented, for instance, by using first-reaction algorithms such as the Gillespie algorithm~\cite{gillespie1977exact, masuda2022gillespie}.) At each step, an agent is chosen at random and adopts the opinion of one of its neighbors, who is also selected uniformly at random; see Fig.~\ref{fig:FIG_OpinionDynamics_1}(b).

After many rounds of interactions, the long-term behavior of the system is examined. If all agents eventually hold the same opinion, the system reaches an absorbing state from which it cannot escape, thus breaking ergodicity. These configurations where all agents agree in their opinion are referred to as consensus states in the jargon of opinion dynamics. A central question in opinion dynamics is to unveil  the conditions under which the system will eventually reach a long-lived, ever-changing state or instead become trapped in a consensus configuration where no further opinion changes occur.

It turns out that the way agents are connected has a strong influence on the outcomes of voter dynamics; see Fig.~\ref{fig:FIG_OpinionDynamics_1}(c). When agents interact on a regular, low-dimensional lattice with dimension $d \leq 2$, the system exhibits coarsening, meaning that domains of agents sharing the same opinion grow steadily over time. A common way to quantify this process is by measuring the fraction of links that connect agents holding different opinions, denoted by $\rho(t)$, which decreases as the system coarsens. Clusters grow by taking over small clusters of the contrary opinion, leading to internal homogenization that makes agents agree and so $\rho(t)$ is reduced in time. $\rho(t) = 0$ in the consensus states. This cluster growth is not through a curvature-driven mechanism, as in the Ising model, but through the so-called interfacial noise~\cite{dornic2001critical}.

As far as we set in dimensions $d > 2$, the system does not coarsen. Instead, it reaches a partially ordered metastable state and remains there until a finite-size fluctuation drives the system to consensus. That translates in $\rho(t)$, on average, setting into a constant value until a fluctuation makes it decay to $0$ exponentially fast (see Fig.~\ref{fig:FIG_OpinionDynamics_1}(c)). The timescale for this fluctuation to appear scales with the number of agents $N$. Therefore, even if consensus is reached for any dimension in finite population of agents, the underlying physical mechanisms that drive the system there is completely different. Furthermore, opinion clusters will not grow and consensus will not be approached when dealing with infinitely large systems $N \to \infty $ in $d > 2$.

We refer to~\cite{krapivsky2010kinetic} for a detailed and pedagogical account of these behaviors in regular lattices. For analytical solutions of the voter model in the setting where all agents can interact with all others, i.e., on the so-called complete graph or mean-field limit, we refer to~\cite{slanina2003analytical, mckane2007singular}.

Note that non-trivial networked interaction patterns, in general, can be seen as high-dimensional structures. As a result, the qualitative behavior observed for dimensions $d > 2$ is also observed in such networks; see Fig.~\ref{fig:FIG_OpinionDynamics_1}(c). Quantitatively, however, there are notable differences compared to the complete graph scenario, where each agent interacts with all others in the network, and also belongs to the regime $d > 2$. First, the level of order at which the system stabilizes, measured by the value of $\rho(t)$, is generally lower in complex networks than in the complete graph. This has been analytically studied, and the development of approximation schemes for analyzing binary-state dynamics on complex networks has become an active research area in recent years. Notable approaches include the approximate master equation framework~\cite{gleeson2013binary} and various forms of pair approximations~\cite{vazquez2008analytical, pugliese2009heterogeneous, peralta2020binary}.

Additional effects investigated concern the conservation laws that broke in specific topologies~\cite{suchecki2004conservation, suchecki2005voter} as well as the time required to reach consensus. The mean time has been shown to depend on the first and second moments of the network's degree distribution~\cite{sood2005voter, sood2008voter}, while the entire first-passage distribution to consensus can display power-law behaviors depending on the initial configuration~\cite{artime2018first}.

\subsubsection{Group-size effects: the q-voter model and non-linear variations}

\begin{figure*}[h!]
    \includegraphics[width=0.95\textwidth]{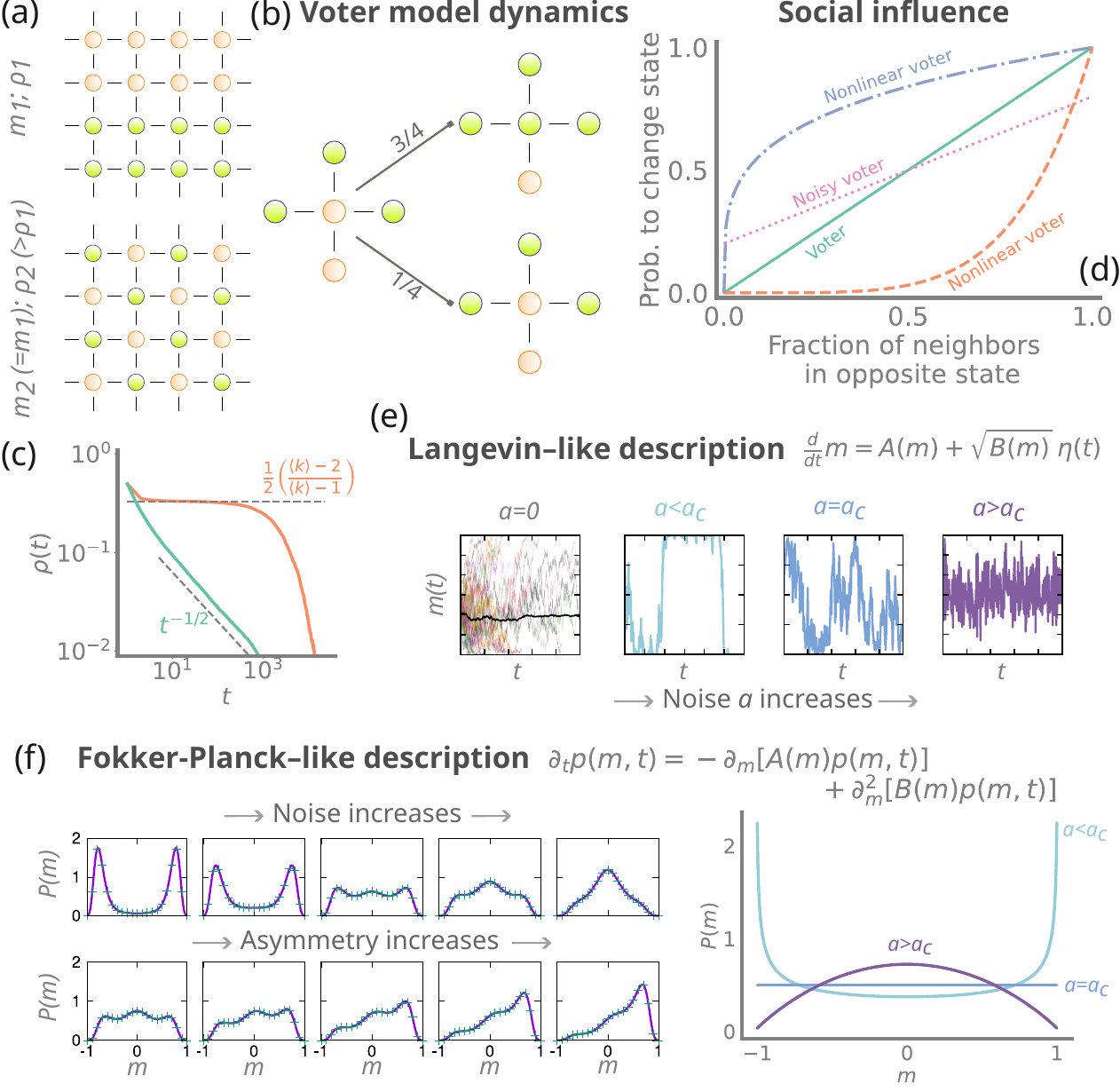}
    \caption{In (a), sketch representing the logic behind the physically-inspired metrics introduced in the main text to characterize opinion dynamics. In the two regular networks, nodes (agents) can be in any of two states $S_i=\pm 1$, here represented by the two different colors. Both networks have the same value for the magnetization $m = \sum_i S_i / N$, since the proportion of the two opinions is the same. On the contrary, the density of active links $\rho$ measures the level of order in the system, and it is lower in the above network because opinion clusters are larger there. 
    In (b), sketch representing a single update in the voter model. The central agent is chosen to attempt an opinion update and adopts the opinion of a neighbor selected uniformly at random. In this case, it becomes green and orange with probability $3/4$ and $1/4$, respectively. 
    In (c), we display how $\rho(t)$ captures different ways to reach the consensus state, i.e., $|m| = 1$ and $\rho = 0$: either through a coarsening process ($ \rho(t) \sim t^{-1/2} $; voter model in a one-dimensional lattice) or through a finite-size fluctuation after the stabilization in a metastable plateau $\rho^{\text{st}} = \frac{1}{2}\frac{\langle k \rangle - 2}{\langle k \rangle - 1} $ for a timescale that grows with $N$, where $\langle k \rangle$ is the mean degree of the complex network on top of which the voter model evolves. 
    In (d), we sketch the functional form of the probability for an agent to change state depending on the fraction of her neighbors in the opposite state, for different opinion dynamics models explained in the main text. 
    In (e), we show individual trajectories of the magnetization. For the standard voter model ($a=0$), 50 independent trajectories are presented, all departing from $m(0) = 1/3$. We can appreciate how the average magnetization is conserved over the ensemble of trajectories (solid line). In the other panels, we show a single trajectory of the noisy voter model in the bimodal ($a < a_c$) and unimodal ($a > a_c$) phase, and one at the critical value of the noise ($a = a_c$). This panel has been adapted from~\cite{artime2019herding}.
    In (f), stationary probability distribution for the magnetization $P(m)$ for the nonlinear noisy voter model (with exponent $\alpha = 6$) on the left and for the noisy voter model on the right. In the panels of the top row, the system is kept symmetric and, as noise is increased, we observe a transition from a bimodal to a trimodal distribution and from a trimodal to a unimodal one.  
    In the bottom panels, the noise is kept constant and the asymmetry is varied. We identify a transition from a trimodal to a bimodal distribution and a transition from bimodal to unimodal one.
    Markers come from numerical simulations, while solid lines are the analytical approximations. This panel has been adapted from~\cite{peralta2018analytical}.
    The rightmost plot shows the modality transition for the noisy voter model. Lines corresponds to the noise values used in the magnetization trajectories in (e). 
    }\label{fig:FIG_OpinionDynamics_1}
\end{figure*}

Mathematically, the voter model is relatively simple, as the probability of an agent changing her opinion increases linearly with the fraction of neighbors holding the opposite opinion; see Fig.\ref{fig:FIG_OpinionDynamics_1}(b, d). This simplicity, along with the parameter-free nature of the model, makes it a solid foundation upon which more sophisticated or realistic assumptions about social influence can be built. These modifications are typically encoded in the rules governing how agents update their internal state, or alternatively, how the influence exerted by neighbors leads an agent to change her opinion (see Fig.\ref{fig:FIG_OpinionDynamics_1}(d)).

One such effect relates to group size. As discussed earlier, compliance is a frequently observed in social interactions. Experimental research has shown that the impact of compliance is proportional to the size of a unanimous influence group, but only up to a certain threshold. Beyond a group size of approximately five individuals, the effect of additional members yields diminishing returns~\cite{bond2005group, asch1955opinions, asch1956studies}. Moreover, compliance is significantly reduced when the group of influence is not unanimous, highlighting the importance of perceived consensus in social influence~\cite{asch1955opinions, asch1956studies}.

One of the most paradigmatic models that include group effects is the $q$-voter model~\cite{castellano2009nonlinear}. It encapsulates the idea of unanimity, in which the social pressure on an agent is effective only if she confronts several times with her neighbors and finds the same opinion all the times, after which she adopts that opinion. Sometimes, the model is endowed with some noise, encoded in a constant probability $\epsilon$ to change state if the $q$ verifications did not result in the same opinion. The properties of the $q$-voter model have been extensively studied over the years, for example, its probability of reaching consensus~\cite{przybyla2011exit, timpanaro2014exit, timpanaro2015analytical, moretti2013mean, vieira2020pair}, the time needed to reach consensus~\cite{moretti2013mean}, and the effects of the network topology in the phase diagram~\cite{moretti2013mean}. For a review, see~\cite{jkedrzejewski2019statistical}.


\begin{table}[h!]
    \includegraphics[width=\linewidth]{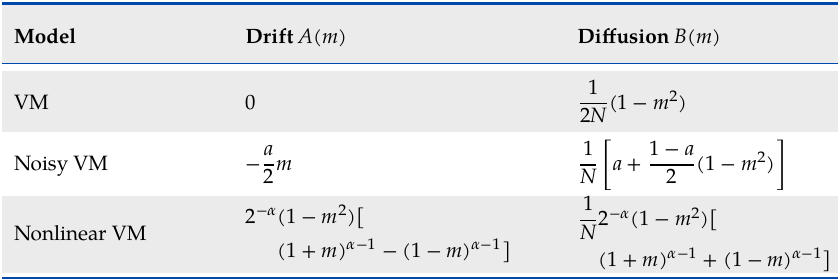}
    \caption{Drift $A(m)$ and diffusion $B(m)$ functions of three variants of the voter model as functions of magnetization $ m \in [-1, 1] $. $ a $ denotes the noise strength in the noisy voter model. $\alpha$ is the nonlinear flipping probability. In Ito sense, these drift and diffusion are related to the Fokker-Planck and Langevin equation such that $\partial_t p(m,t) = - \partial_m \left[ A(m) p(m,t) \right] + \partial_m^2 \left[ B(m) p(m,t) \right]$ and $\frac{d}{dt} m = A(m) + \sqrt{B(m)} \, \eta(t)$, respectively. Here $\eta(t)$ is a Gaussian white noise.}
    \label{tab:table_drif_diffusion}
\end{table}

From a more general perspective, the social influence mechanism of the $q$-voter model can be understood as a deviation from the linear peer influence assumed in the standard voter model. This deviation is formalized in the class of nonlinear voter models, which account for scenarios where the dependence on opinion frequency is not linear~\cite{schweitzer2009nonlinear, vazquez2008systems}.

More specifically, in nonlinear models, the probability that an agent switches opinion is proportional to a continuous power $\alpha$ of the fraction of neighbors holding the opposite opinion. This framework enables the modeling of various types of group influence on agent opinion changes: for instance, for $\alpha > 1$ results into a convex opinion-switching probability, meaning that agents are less influenced by their peers in opposite opinion or, alternatively, more resistant to change opinion than in the standard voter model case. On the other hand, $\alpha < 1$ models the opposite behavior: agents have a higher propensity and preference to change opinion upon interaction with contrary-minded agents; see Fig.~\ref{fig:FIG_OpinionDynamics_1}(d). Introducing such nonlinearity gives rise to a rich physical phenomenology. In particular, the stationary distribution of the system's magnetization (i.e., the net opinion balance) can undergo modality transitions, including the emergence of bimodal or unimodal distributions, and may exhibit a tricritical point, where the nature of the phase transition changes~\cite{peralta2018analytical}; see Fig.~\ref{fig:FIG_OpinionDynamics_1}(f).

Beyond opinion dynamics, this type of group-size-based nonlinearity has been applied in other areas of sociophysics, such as language dynamics, where similar mechanisms govern language competition and evolution~\cite{abrams2003modelling, vazquez2010agent}.

\subsubsection{Idiosyncratic choices: the noisy voter model}


Complementing group-based influences, there are mechanisms in decision-making and opinion formation that do not depend on the states of neighboring agents, yet have been extensively incorporated into opinion dynamics models. From a social perspective, such influences can be motivated by factors like idiosyncratic choices or mass media effects~\cite{eminente2022interplay}. From a physics-inspired viewpoint, these mechanisms are often referred to as noise or temperature, analogous to their role in thermal phase transitions. We adopt this latter nomenclature here.


The effects of noise in binary-state stochastic models have been widely studied. Introducing noise into the voter update rules results in a model that is isomorphic to several well-known systems in complex dynamics, such as the Kirman model~\cite{kirman1993ants, alfarano2008time, carro2015markets}, which has been used to explain heterogeneous trading strategies and bounded rationality in behavioral financial models, as well as in models of surface-catalytic reactions~\cite{fichthorn1989noise, considine1989comment}. In terms of the transition rates for opinion changes, these remain linear as in the standard voter model, but with a smaller slope, due to the nonzero probability of changing opinion when the agent agrees with all her neighbors and the less-than-one probability of changing opinion when all the neighbors are in the opposite state; see Fig.~\ref{fig:FIG_OpinionDynamics_1}(d).

Idiosyncratic opinion changes in opinion dynamics yields a phenomenology fundamentally different from that of the standard voter model. Most notably, absorbing states are eliminated and ergodicity is restored. This results in a non-trivial non-equilibrium steady state, whose properties depend strongly on the balance between social herding and individual idiosyncratic behavior.

When noisy (random) updates are rare, the system tends to remain near consensus states for long periods of time, with occasional excursions away from consensus. The time scale of these excursions is much smaller than the time the system spends near consensus. As the noise level increases, such excursions become more frequent, eventually leading to a situation in which all opinion configurations are approximately equally likely. Beyond a critical noise level, the system self-organizes around a state of perfect opinion coexistence. 
In Fig.~\ref{fig:FIG_OpinionDynamics_1}(e), we display individual trajectories of the magnetization in the different regimes, alongside the typical trajectories for the noiseless case $a=0$, i.e., the standard voter model. In the right panel of Fig.~\ref{fig:FIG_OpinionDynamics_1}(f), through the stationary distribution of the magnetization, the three regimes can be easily appreciated.

This behavior reflects a modality transition, from a bimodal regime (where probability concentrates near consensus states) to a unimodal regime (where probability peaks around zero magnetization). The transition point is inversely proportional to the number of agents $N$ in the system. Consequently, as $N$ increases, the likelihood of observing the bimodal phase, and hence the transition, decreases~\cite{carro2016noisy, granovsky1995noisy, peralta2018stochastic, kononovicius2020noisy}.

\subsubsection{Mathematical treatment}

The mathematical treatment of all the variations of the voter model described here, and others, can be approached with tools developed to study stochastic processes~\cite{van1992stochastic}. In particular, the most widespread approaches are the Langevin and Fokker-Planck descriptions, obtained by performing the proper approximations to the master equation. Let $S_i$ be the opinion of agent $i$, which we take $+1$ or $-1$ in the binary settings described in this section. In a single step $\delta t$ of the dynamics, three events can occur. If the chosen agent changes opinion, the magnetization can increase an amount $\delta m = 2/N$ in the case of the transition $S_i = -1 \to S_i = 1$ or decrease the same amount if $S_i = 1 \to S_i = -1$ occurs. If there are no opinion change, $S_i = \pm 1 \to S_i = \pm 1$, the magnetization remains constant. Denoting these first two processes by $R(m) \equiv \text{Prob} \left\{ m \to m + \delta m\right\}$ and $ L(m) \equiv \text{Prob} \left\{ m \to m - \delta m\right\}$, the last one reads $\text{Prob} \left\{ m\to m\right\} = 1 - R(m) - L(m)$, due to probability conservation. If $p(m,t)$ stands for the probability of observing the system with magnetization $m$ at time $t$,
we have~\cite{sood2008voter}
\begin{eqnarray}
\label{eq:discrete_me}
  p (m, t + \delta t)&=& R(m-\delta m)p(m-\delta m, t)\nonumber\\
&~&+ L(m + \delta m)p(m + \delta m, t)\nonumber\\
&~&+ [1- R(m) - L(m)]p(m, t), 
\end{eqnarray}
where $\delta t = 1/N$ is typically chosen so $N$ updates result into one Monte Carlo step. Expanding Eq.~\eqref{eq:discrete_me} to first order in time and to second order in magnetization, we obtain the Fokker-Planck equation:
\begin{equation}
  \frac{\partial p(m,t)}{\partial t} = \frac{\partial}{\partial m}
 [A(m)p(m,t)]+\frac{\partial^2}{\partial m^2} [B (m)p(m,t)], 
  \label{eq:fp_eq}
\end{equation}
where
\begin{equation}
A(m)\equiv \frac{\delta m}{\delta t} \left[R(m)-L(m) \right]    
\end{equation} 
is the drift term and
\begin{equation}
  B(m)\equiv \frac{1}{2} \frac{\delta m^2}{\delta t} \left[R(m) + L(m) \right], 
\end{equation}
is the diffusion term.

Using the well-known relation between the Fokker-Planck and the Langevin equation, we can write down
\begin{equation}
    \frac{d}{dt} m = A(m) + \sqrt{B (m)} \, \eta(t),
\end{equation}
where $\eta(t)$ is a Gaussian white noise. These stochastic descriptions of opinion dynamics models allows one to unveil where it is more probable to find the system in the opinion space (consensus, perfect coexistence of opinions, etc.) as well as to address the first passage statistics between different opinion configurations~\cite{artime2018first}.

In light of this, the mathematical description of binary-state opinion dynamics models boils down to identifying what the probabilities $R(m) $ and $  L(m)$ are. For instance, for the limit of large $N \gg 1 $, in the voter model we have~\cite{slanina2003analytical}
\begin{equation}
    \begin{aligned}
        & R(m) \equiv \text{Prob} \left\{ m \to m + \delta m\right\}=\frac{1}{4}\left(1-m^2\right) \nonumber\\
        & L(m) \equiv \text{Prob} \left\{ m \to m - \delta m\right\}=
\frac{1}{4}\left(1-m^2\right)
\label{eq:transitionprobabilitiesII}\\
    & \text{Prob} \left\{ m\to m\right\} = 1 - R(m) - L(m)= \frac{1}{2}\left(1 + m^2 \right). \nonumber 
    \end{aligned}
\end{equation}
A summary of the drift and diffusion terms for the models discussed here is given in Table~\ref{tab:table_drif_diffusion}.

\subsubsection{Other discrete opinion models}

Here, we have reviewed the voter model and two of its most prominent variations, in which modifications to the transition rules are implemented to capture different types of social influence. Other variations of the voter model, based on alternative forms of social influence that agents can experience, have been extensively discussed in the literature. Further details can be found in the reviews~\cite{redner2019reality, sirbu2016opinion} and the references therein.  


The unconditional social influence assumed in the voter model is a reasonable approximation in situations where agents blindly trust their peers or where there is no mechanism for evaluating the reliability or popularity of opinions. However, the voter model's blind copy mechanism is not universal and in many empirical scenarios other mechanisms of opinion change are at play. One alternative is the majority rule model, which was originally introduced as a model for hierarchical voting in societies~\cite{galam1986majority}, and later applied to opinion dynamics~\cite{galam2002minority} to explain the spread and eventual dominance of minority opinions under specific conditions.

In its original formulation, the majority rule model for opinion dynamics assumes a population of agents, each holding an opinion, say, without loss of generality, $+1$ or $-1$. Agents interact within so-called meeting cells, whose sizes, i.e., the number of agents participating in a discussion, follow a given distribution and are fixed at the beginning of each independent realization of the dynamics.

At each time step, agents are randomly assigned to meeting cells, and all agents within the same cell adopt the opinion of the local majority. This is a process reflecting assimilative influence, and can be interpreted as a deliberation in which the group must collectively voice a single, agreed-upon opinion, which is always that of the majority.

A key mechanism by which a minority opinion may spread is through social inertia~\cite{friedman1985tyranny}, namely, a bias that manifests as resistance to change or reform. In the event of a tie within a cell, one opinion is adopted by default; this is typically set arbitrarily, for instance, in favor of opinion $+1$.

A striking and counterintuitive result is that an initially minority opinion, i.e., when fewer than half of the agents hold opinion $+1$, can ultimately dominate the public discourse. Put otherwise, the initial minority becomes the consensus majority over time, leading to a bottom-up rejection of the opinion originally supported by the majority. This convergence to consensus occurs on a time scale proportional to $\ln N$, where $N$ is again the number of agents~\cite{tessone2004neighborhood}.


Another central model in opinion dynamics is the Sznajd model~\cite{sznajd-weron2000opinion}, whose dynamics are inspired by social impact theory. In this model, a group of neighboring agents is selected. In its original formulation, the model was studied on a one-dimensional lattice, where pairs of adjacent agents were chosen. Later extensions to two-dimensional lattices involved selecting groups of neighboring agents arranged in plaquettes~\cite{stauffer2000generalization}.

If the selected group of agents shares the same opinion, it is assumed that they are able to convince their immediate neighbors, who then update their opinions to align with that of the group. Conversely, if the selected group is not in agreement, their neighbors retain their original opinions.

Although both the majority rule model and the Sznajd model lead to a consensus in the long run, there are manifest differences between both of them, regarding the directionality of influence. In the majority rule model, influence flows inward, from the surrounding neighbors to the central agent or group. In contrast, in the Sznajd model, influence flows outward, spreading from a central, unified group to its neighbors.

\subsection{Continuous opinion}
As anticipated, there are scenarios in which a finite number of opinions cannot represent well certain social phenomena, and we need to allow agents to hold opinions in a continuous spectrum. At odds with discrete opinions, dealing with continuous ones opens the possibility that interacting agents reach compromises among them, resulting in asymptotic configurations that cannot be obtained in the discrete setting. Similarly, the mechanisms of social influence operate differently for the case of continuous opinions. These mechanisms are approached with the so-called models of assimilative and similarity-biased influence, that encompass agents gradually adjusting their opinion upon interaction. The particularity is that interactions are restricted under the bounded confidence condition, i.e. agents need to be close enough in the opinion space to influence each other. In these terms, bounded confidence is a sort of homophilic relation.

The basic assumptions to treat mathematically continuous opinion dynamics is to work with an opinion space $S \subset \mathbb{R}$ and with the agents' opinions $\textbf{x}(t) \in S^N$, where $N$ is the number of agents. Without loss of generality, the typical choice is $S=[0,1]$, i.e., opinion values are normalized to unity. Note that generalization to multidimensional continuous variables $S \subset \mathbb{R}^d$ and $S=[0,1]^d$ are also possible, yet one must taken into account how the different opinion dimensions influence each other~\cite{pedraza2021analytical}. Two alternative viewpoints for the dynamics of the opinions predominate in the literature~\cite{lorenz2007continuous}. The first one, the agent-based approach, provides a discrete map $\textbf{x}(t + 1) = f\left(\textbf{x}(t), t\right)$ that specifies how each agent $i$ updates her opinion $x_i$.  Complementarily, the density-based approach~\cite{ben-naim2003bifurcations} employs the density function $P(x, t)$, such that $P (x, t)dx$, is the fraction of agents that have opinions in the range $[x, x + dx]$ at time $t$. The ultimate goal is to find empirically-grounded rules behind the temporal evolution of $P(x, t)$ that are able to explain the experimental data and predict unknown scenarios to be tested in new experiments. 

There are two main models that have been studied in the literature that assume continuous opinions that evolve through the bounded confidence of the agents. Both of them were introduced at the turn of the century, yet in very different contexts. The Deffuant-Weisbuch (DW) model~\cite{deffuant2000mixing, weisbuch2002meet} was developed within an EU project for the improvement of agri-environmental policies. The other, the so-called Hegselmann-Krause (HK) model~\cite{hegselmann2002opinion}, was first introduced as a nonlinear version of older consensus models~\cite{degroot1974reaching}. Their main difference is rooted in the social influence mechanism. In the former, interactions are pairwise and symmetric, meaning that agents $i$ and $j$, upon interacting, influence each other and their opinions becomes closer. In the latter, agent $i$ takes the average opinion of all the agents that are in her confidence interval.  

\subsubsection{The Deffuant-Weisbuch model}

\begin{figure*}[h!]\label{fig:FIG_OpinionDynamics_2}
    \includegraphics[width=\textwidth]{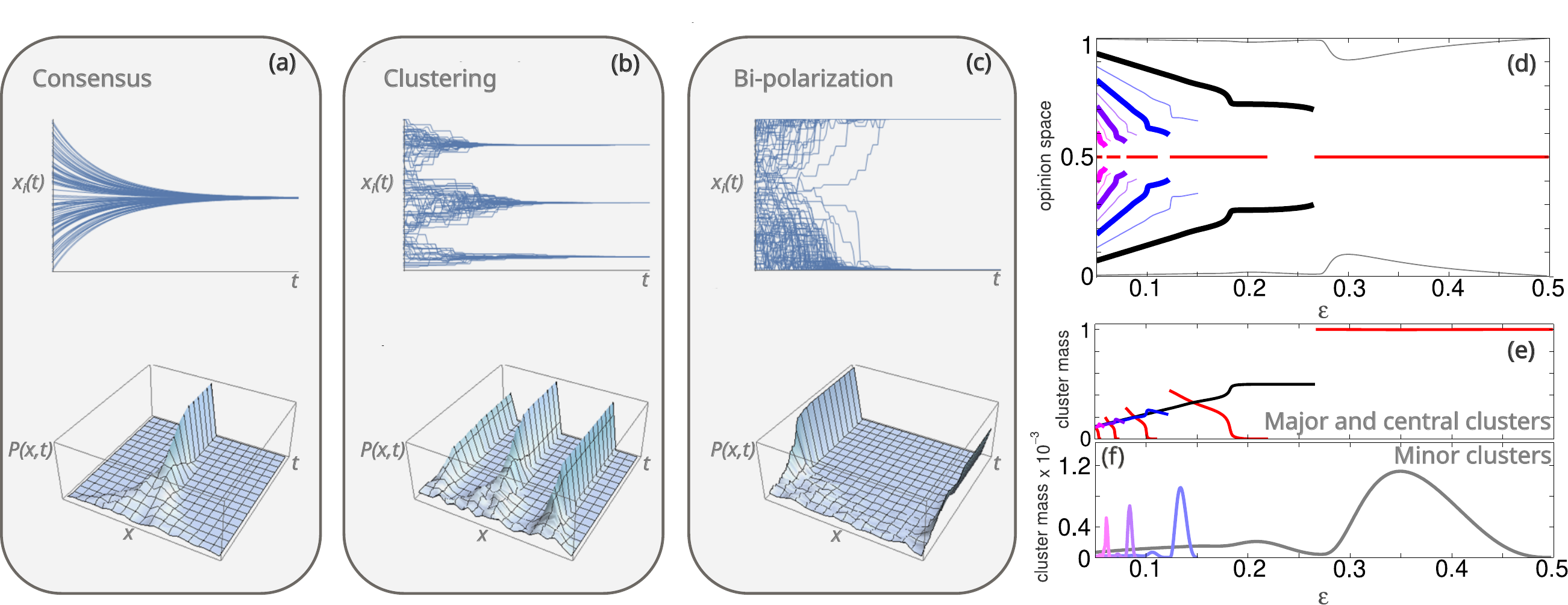}
    \caption{Possible outcomes for the Deffuant-Weisbuch continuous-opinion model as a function of the tolerance parameter $\epsilon$: (a) consensus, with all nodes sharing the same opinion; (b) clustering into several intermediate opinions; (c) polarization in two extremes of the opinion space. 
    On the right, panel (d) shows the bifurcation diagram for the DF model: the position of the opinion cluster(s) is presented as a function of the tolerance parameter. Thick lines corresponds to the clusters formed by the majority of agents, while thinner lines indicate the location of clusters clusters that, even with a finite number of agents, are not majority. 
    (e) and (f) show the masses of the related clusters. Notice the existence of minor clusters at the extreme opinions and between the location of major clusters. 
    Panels (a), (b) and (c) have been adapted from~\cite{flache2017models}, and (d), (e) and (f) from~\cite{lorenz2007continuous}.}
\end{figure*}

Let us focus first in the DW model. At each time step, two interacting agents $i$ and $j$ are selected. The discrete map for the DW reads
\begin{equation}
    \label{eq:df}
    \begin{aligned}
        x_i (t + 1) = & x_i (t) + \mu \left[ x_j (t) - x_i (t) \right], \\
        x_j (t + 1) = & x_j (t) + \mu \left[ x_i (t) - x_j (t) \right]
    \end{aligned}
\end{equation}
as far as $|x_i (t) - x_j (t)| \leq \epsilon $; otherwise $x_i (t)$ and $x_j (t)$ remain unchanged. Here, $\mu \in (0, 0.5]$ is sometimes called persuasiveness or persuasibility, and $\epsilon \in [0, 1]$ is the so-called confidence or tolerance parameter. Hence, $i$ and $j$'s opinion become simultaneously similar to each other if both agents hold an opinion that, in the first place, is close enough in the opinion space. Given an initial condition $\textbf{x}(0)$, the system evolves toward an absorbing state in which exists full consensus $x_i(t \to \infty) = x^*, \, \forall i$, or fragmented clusters with local consensus values whose separation in the opinion space is larger than $\epsilon$~\cite{lorenz2005stabilization}. The possible states to which $x_i(t)$ can tend are displayed in Fig.~\ref{fig:FIG_OpinionDynamics_2}(a)--(c). The goal, therefore, is to determine in which configuration the system ends up given the initial distribution of opinions $\textbf{x}(0)$, the network of interactions and the parameters of the model. 

The alternative density-based approach facilitates the analytical treatment, as well as reduces the computational effort as far as one is not interested in the effects of small-population sizes. In the DW model, one finds~\cite{fennell2021generalized}
\begin{equation}
    \begin{aligned}
    \frac{\partial P(x,t)}{\partial t} = \frac{1}{\mu} \int_{|x-y|<\epsilon\mu} P(y,t)&P\left(y+\frac{1}{\mu}(x-y),t\right) \,dy \\
& - \int_{|x-y|<\epsilon} P(x,t)P\left(y,t\right) \,dy.
    \end{aligned}
    \label{eq:originalME}
\end{equation}
This is simply a master equation, with a gain and a loss term, accounting for the expected number of agents whose opinions respectively enters in and leaves out the opinion interval $[x, x + dx)$ in $dt$. Alternative forms of the master equation for the DW model can be found in Refs.~\cite{ben-naim2003bifurcations, lorenz2007continuous}.

Note that Eq.~\eqref{eq:originalME} assumes a well-mixed population, where all agents can virtually interact with the entire network. However, it has been recently reported that the network of interactions of the agents impacts the dynamics of the DW model, leading to outcomes that are potentially different from those predicted by the mean-field equation~\cite{meng2018opinion}. For instance, on Erdos-Renyi graphs the number of stable opinion clusters grows as the confidence bound decreases. In contrast, on cycle graphs, the confidence bound has no influence, and the population ultimately reaches a unanimous opinion. The dynamics on other network types, such as lattices~\cite{deffuant2000mixing} or Barabasi-Albert~\cite{stauffer2004simulation} and Watts-Strogatz networks~\cite{gandica2010continuous}, also exhibit different behaviors compared to complete graphs. Notably, it was not until recently that the limiting, all-to-all assumption has been lifted from the analytical treatments. In Refs.~\cite{fennell2021generalized, dubovskaya2023analysis}, degree-based mean-field equations are proposed for the evolution of the probabilities $P_k(x,t)$, standing for the fraction of agents of degree $k$ with opinion $x$ at time $t$.

As can be readily seen from Eq.~\eqref{eq:df}, the opinion displacement is weighted by the persuasiveness $\mu$ of one agent into the other. (As a matter of fact, $\mu$ can be equivalently interpreted as the amount an agent is willing to change its opinion upon a successful interaction). It essentially determines the speed of convergence towards the absorbing state and, as a result, plays a role in influencing the final number of clusters~\cite{laguna2004minorities, porfiri2007decline}. Specifically, when $\mu$ takes on intermediate or large values, convergence occurs rapidly at the extremes of the opinion spectrum. This can lead to a few agents remaining isolated in those regions, unable to communicate with the rest of the population. Hence, they retain their initial opinion and form two small extremist clusters close to $x = 0$ and $1$. In contrast, when the persuasiveness is small, the convergence is slower, and all agents have the chance to interact and be influenced by others, gradually aligning their opinions with the larger clusters over time.

The other key parameter of the DW model is the confidence $\epsilon $, and it also controls the number of clusters at the absorbing state. When all agents share the same confidence parameter $\epsilon_i = \epsilon, \, \forall i$, we talk about the homogeneous scenario; otherwise we are in the so-called heterogeneous setting. Intuitively, we can expect that the more open-minded agents are, the less likely will be to find polarized configurations. Indeed, this expectation is verified, on average, for initial opinions that are distributed uniformly at random and agents with a confidence $\epsilon \geq 0.5$: full consensus is reached at the intermediate opinion value. For $\epsilon < 0.5$, a richer phenomenology appears, with several clusters of different size emerging. 

To capture the location of these clusters and to understand where the system of agents polarizes, bifurcation diagrams are used; see Fig.~\ref{fig:FIG_OpinionDynamics_2}(d)--(f). They mark the opinion of the clusters as a function of the confidence $\epsilon$, and help identify at which $\epsilon$-values clusters emerge or split. These bifurcations have been reported to occur in non-trivial ways~\cite{ben-naim2003bifurcations}. In particular, four fundamental types of bifurcation behavior repeat in decreasing order of $\epsilon$ and across progressively smaller $\epsilon$ intervals. For instance, as already hinted, for $\epsilon \geq 0.5$, a single large central cluster forms, encompassing most of the population. As $\epsilon$ drops below $0.5$, two minor clusters begin to emerge near the boundaries of the opinion spectrum. Around $\epsilon \approx 0.266$, the central cluster splits into two major clusters. As $\epsilon$ continues to decrease, the central cluster reappears as a minor cluster at approximately $\epsilon \approx 0.222$, and then rapidly gains mass around $\epsilon \approx 0.182$, pushing the two major clusters further apart. This behavior keeps repeating over and over as the confidence is decreased. A scaling of these intervals with $1 / \epsilon$ has deemed plausible, based on the numerical evidence. 

The location of the stationary clusters mentioned above is delta-peaked, namely, $P_{st} = \sum_{i=1}^r m_i \delta(x - x_i)$, where $r$ is the number of clusters, $x_i$ is the opinion of cluster $i$ and $m_i$ is the fraction of agents in cluster $i$, also known as its mass. As far as noise is introduced in the DW model, stationary clusters display a certain width with respect $x_i$, hence bifurcation diagrams are constructed by reporting the maximum opinion value within the cluster. Moreover, because of noise, the agreement between theory and simulations may be compromised~\cite{pineda2009noisy, carro2013role}. The initial opinion distribution has been also reported to play an important role in the final configuration of the clusters, which can force or prevent consensus~\cite{carro2013role}.

The intriguing simplicity of DW model has converted it into a reference model on top of which further influence mechanisms have been incorporated. One is algorithmic bias~\cite{sirbu2019algorithmic}, where agents are selected to interact more frequently if they are closer in the opinion space, mimicking recommendation systems of online media. Incorporating such an effect leads to an increased tendency to opinion fragmentation, increased polarization and, counterintuitively, longer convergence times to the steady state~\cite{sirbu2019algorithmic}. Adaptive and heterogeneous confidence bounds have been found to lead to fewer major opinion clusters and longer convergence times than the baseline DW model~\cite{li2025bounded}. Finally, longer times to achieve consensus, as well as more opinion fragmentation with respect to the baseline model, have been also reported if agents are assigned a heterogeneously distributed activity level that mediates the probability of being chosen for interaction~\cite{li2023bounded}.

Another context in which the DW model has been employed is when agents need to form a consensual product, in which, on top of the pairwise agent influence, a global (indirect) interaction is mediated through a medium that holds and modifies its opinion. Notably, this can display transitions between consensus and perpetual conflict, qualitatively agreeing with consensus formation in Wikipedia~\cite{torok2013opinions}.

\subsubsection{The Hegselmann-Krause model}

At odds with the pairwise influence of the DW model, the opinion influence in the Hegselmann-Krause model is interpreted as a higher-order effect among all the neighboring agents within the confidence intervals. Denoting by $\mathcal{N}_i$ the neighborhood of node $i$, the agent-based updates for the HG model are recursively written as 
\begin{equation}
    x_i(t+1) = \frac{1}{|\mathcal{I}_i(t)|} \sum_{j \in \mathcal{I}_i(t)} x_j(t),
\end{equation}
where $\mathcal{I}_i(t) = \lbrace  j \in \mathcal{N}_i \mid |x_j(t) - x_i(t)| < \epsilon \rbrace$ is the confidence set of agent $i$ at time $t$, i.e., the set of neighbors whose opinions lie within $\epsilon$ of agent $i$'s opinion. At each time step all agents update simultaneously: each agent $i$ sets its new opinion equal to the arithmetic mean of the opinions $x_j(t)$ of the agents in $\mathcal{I}_i(t)$. However, in~\cite{li2022mixed}, a version of the HK model has been proposed to interpolate between the fully synchronous  and asynchronous case by incorporating agent stubbornness. Note, moreover, that when the confidence parameter $\epsilon$ tends to 1, the bounded confidence models introduced here reduce to the well-known (unweighted) DeGroot model~\cite{degroot1974reaching}.  

A density-based description of the dynamics of the HK model can be obtained in a similar way to that of the DW model (Eq.~\eqref{eq:originalME}), i.e., through a master equation that individuates the gain-loss terms~\cite{lorenz2007continuous}. Let us define the vector of individual confidences $\bm{\epsilon} = (\epsilon_1, \ldots, \epsilon_N)$. In this setting of heterogeneous confidences, the confidence set $\mathcal{I}_i(t)$ includes all neighbors $j$ for which $|x_i(t) - x_j(t)| < \epsilon_i$. Let us also define the $\epsilon$-local mean
\begin{equation}
M_1(x, \bm{\epsilon}) = \frac{\int_{x-\epsilon}^{x+\epsilon} y \, P(y,t) \, dy}{\int_{x-\epsilon}^{x+\epsilon} P(y, t) \, dy}.
\end{equation}
This function gives the expected value of the fraction of agents whose opinion is within the interval $[x - \epsilon, x + \epsilon]$. For a homogeneous bound of confidence $\epsilon = [\epsilon_1, \epsilon_2]$, the time evolution of the density function reads~\cite{lorenz2007continuous}
\begin{equation}
    \label{eq:HK_density}
    \begin{aligned}
    \frac{\partial}{\partial t}P(x,t) = & \int_S dy \, \left[ \delta(M_1(y, \bm{\epsilon})  - x)P(y, t) \right. \\
    & \hspace{1.5cm} \left. - \delta(M_1(x,\bm{\epsilon}) - y)P(x, t) \right].  
\end{aligned}
\end{equation}
Again, this is a gain-loss equation, where the first and the second $\delta$-terms in the integral indicate, respectively, the fraction of agents that transition into, and leave from, opinion $x$.
The generalization of Eq.~\eqref{eq:HK_density} to heterogeneous confidence levels follows directly~\cite{lorenz2007continuous}.

The distribution of the opinion clusters in the HK model has been reported to be highly sensitive to the initial conditions and the interaction thresholds~\cite{hegselmann2002opinion, lorenz2007continuous, kou2012multi}. Some regularities emerge, though. For instance, in the all-to-all case, global consensus is reached for $\epsilon > \epsilon_c$, where $\epsilon_c \sim 0.2$. Surprisingly, the value of $\epsilon_c \sim 0.2$ is quite robust, as far as the mean degree of the network diverges when we take the limit $N \to \infty$. In networks where $\langle k \rangle$ remains finite, one gets $\epsilon_c \sim 1/2$ when the same limit is taken~\cite{fortunato2005consensus}. Cluster splitting occurs through bifurcations as a function of the confidence level. Notably, signatures of dynamical phase transitions, like critical slowing down, have been reported near these bifurcations~\cite{slanina2011dynamical}. However, when not all neighboring agents are treated equally and, instead, one weighs most on the opinion of the neighbors that is not shared by others, these opinion cluster bifurcations disappear and final consensus is always guaranteed~\cite{yang2014opinion}.

A myriad of generalizations to the Hegselmann-Krauss model have been explored in order account for a broader range of social situations it can model. One is to include multi-dimensional opinion spaces, even though this generalization leads to similar qualitative behavior as far as the higher-dimensional confidence area is symmetric, e.g., a circle or a square in two-dimensional opinion spaces~\cite{fortunato2005vector, pluchino2006compromise, lanchier2022consensus}. Another family of variations concerns in modifying the updates rules, for instance, to take into account external influences~\cite{hegselmann2006truth} and environmental or communication noise among agents~\cite{chen2020heterogeneous}. Fuzzy logic has been used to model interaction ambiguity among agents~\cite{zhao2021fuzzy}. Interactions mediated in time has been explored too, for instance, with the incorporation of decaying confidence~\cite{morarescu2010opinion} or distance-dependent interaction weight~\cite{motsch2014heterophilious}. The concept of bounded influence, in contraposition to the bounded confidence, has been explored in Ref.~\cite{mirtabatabaei2012opinion}. In such a model, interactions are mediated by the confidence parameter of the neighbor and not the focal node's one, i.e., $\mathcal{I}_i(t) = \lbrace  j \in \mathcal{N}_i \mid |x_j(t) - x_i(t)| < \epsilon_j \rbrace$. Finally, in~\cite{schawe2020collective} it has been explored the case where agents hold some resources that are used when opinion changes, hence modeling scenarios in which changing opinion require some effort. The inclusion of the cost of opinion change leads to interesting phenomenology, such as a re-entrant phase transition between global consensus and cluster fragmentation with well-defined critical exponents. 

\subsection{Flocking-inspired opinion dynamics models}

We close this section by emphasizing the parallels between opinion dynamics and the seemingly unrelated problem of flocking. In recent years, non-equilibrium statistical physics has sought to explain the phenomenology underlying the collective motion of animals and active matter~\cite{vicsek2012collective}. Flocking models are inherently spatial: agents move in physical space and align their headings by averaging those of neighboring agents. In flocking models, however, neighborhoods are defined by spatial proximity, typically via a fixed interaction radius or a set number of nearest neighbors.

The study of flocking can inform both continuous and discrete opinion dynamics models, and vice-versa. For instance, in the classical Vicsek model~\cite{vicsek1995novel}, agents move at constant speed and, at each time step, update their heading to the mean heading of neighbors within the interaction radius, subject to stochastic noise. This can be mapped to a noisy Hegselmann–Krause model, where the confidence parameter $\epsilon$ plays the role of the interaction radius. On the other hand, local averaging is not always observed empirically; instead, voter-like copying mechanisms are sometimes more appropriate~\cite{jhawar2020noise}. In this context, the multistate voter model with imperfect copying can be mapped to a discrete version of a flocking model; this has been studied for both finite~\cite{loscar2021noisy} and infinite~\cite{vazquez2019multistate} interaction radii. Overall, the transition from a disordered to an ordered flock is mediated by the level of noise in the copying mechanism. This mirrors results from consensus formation in purely opinion-based models and underscores, once again, the central role of idiosyncratic choices in the emergence of macroscopic patterns.

A common feature of most flocking models is a rapidly changing interaction topology induced by spatial motion. Spatial embedding both constrains interaction patterns and is itself altered by the evolving states, producing a bidirectional feedback absent from the opinion models discussed so far in this section. Depending on the social system under study, the assumption of a static topology may not always be empirically accurate, and important phenomenological features may be overlooked. In the next section, we examine this coupling between state dynamics and network structure in the so-called co-evolving network opinion models. For further details on the continuous opinion dynamics models discussed here and other variations, we refer to the reviews~\cite{lorenz2007continuous, starnini2025opinion, bernardo2024bounded, peralta2022opinion, hegselmann2019consensus, xia2011opinion} and references therein.

\subsection{Coevolution of opinions and social structure and fragmentation transitions}

\begin{figure*}[h!]\label{fig:FIG_OpinionDynamics_3}
    \includegraphics[width=\textwidth]{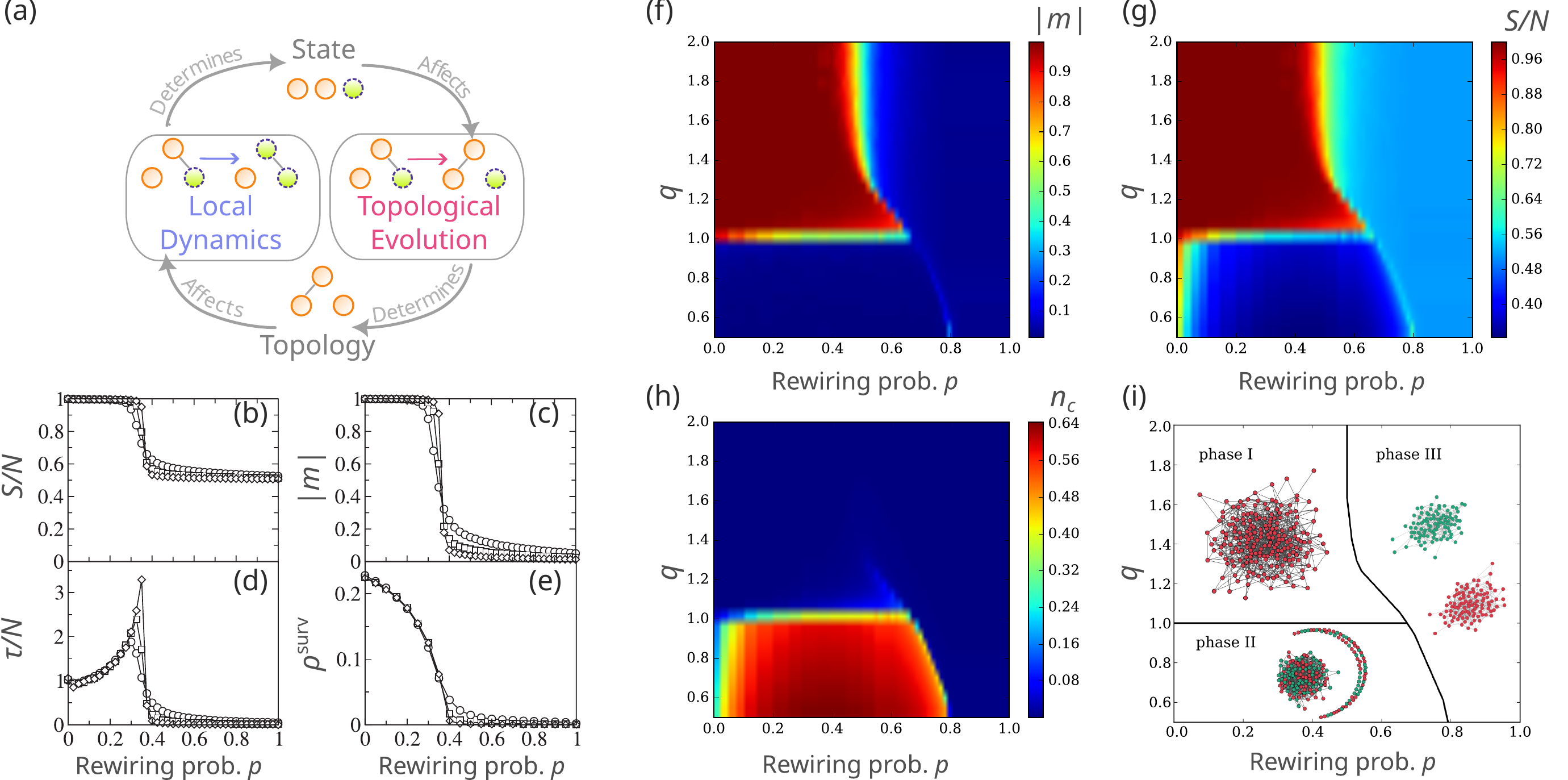}
    \caption{(a) Sketch showing the feedback logic between state and structure changes in coevolving adaptive processes.
    In (b)--(d), different metrics used to characterize the fragmentation transition as a function of the rewiring probability $p$ for the coevolving voter model introduced in the text: 
    normalized size of the largest connected component $S$ (b), 
    absolute value of the magnetization $|m|$ (c),
    normalized average convergence time to the absorbing state $\tau$ (d), and 
    stationary density of active links in surviving runs $\rho^{\text{surv}}$. Averages are taken over $10^4$ realizations of random regular network where all nodes have $\mu = 4$ connections, whose number of nodes are $N=250$ (circles), $1000$ (squares) and $4000$ (diamonds).
    In panels (f)--(i), characterization of the fragmentation transition in a nonlinear voter model with triadic closure in the rewiring events~\cite{raducha2018coevolving}, for an Erdos-Renyi network with mean degree $8$.
    It is shown the $(p, q)$-phase diagrams for the absolute value of the magnetization (f), for the normalized size of the largest connected component (g) and for the fraction of nodes belonging to isolated components in the fragmented state. In (i), a sketch of the phase diagrams with a network realization corresponding to each phase.
    Panel (a) adapted from~\cite{gross2008adaptive}, panels (b)--(h) adapted from~\cite{vazquez2008generic} and panel (i) adapted from~\cite{raducha2018coevolving}.
    }
\end{figure*}

A key element that has been shown to critically impact the behavior of opinion dynamics models is the structure of the interactions between agents~\cite{barrat2008dynamical}. Traditional models were proposed either in low-dimensional lattices, oftentimes drawing inspiration from statistical physics spin-like models~\cite{marro2005nonequilibrium}, or in the all-to-all regime, where all agents interact can interact among them. Yet, contact patterns both in online and offline social contexts, are more complex, displaying several features such as high levels of heterogeneity in the number of connections, homophilic and assortative relations and hierarchical community structures. In the previous sections, we have already hinted at the novel phenomenology that can emerge due to the non-trivial social structure, but we have assumed throughout that networks were static. When studying opinion dynamics, in certain situations this may be justified because the timescales for network rewiring can be much longer than those of the unfolding social process. At the other extreme, we have processes for which the focus is on how the network evolves but the information of the state of the nodes is not deemed relevant and completely disregarded. However, the conditions assumed by these two extreme cases are rarely met in empirical scenarios, and we need models that help us addressing the situation in which the timescales for network re-adaption are similar to the inner scales of the opinion dynamics. Furthermore, both processes can influence each other, leading to highly nontrivial outcomes. The study of coevolving dynamics is, at least, as old as the field of network science, yet in its origin it was mainly inspired by biological problems, e.g., see~\cite{sole1996extinction, bornholdt1998neutral, christensen1998evolution, bornholdt2000topological}, and the review~\cite{gross2008adaptive}. However, the topology-dynamics feedback mechanisms are general and can also be framed in the precise context of opinion dynamics (Fig.~\ref{fig:FIG_OpinionDynamics_3}).

Most coevolving opinion models assume an underlying social influence mechanism operating among agents, coupled to a probabilistically link rewiring when certain conditions are met. Through the combination of both elements, opinion-based network fragmentation is observed. Actually, the emergence and maintenance of fragmented outcomes normally occur through a nonequilibrium transition with nontrivial time scalings~\cite{holme2006nonequilibrium, nardini2008who}. Even if there several choices for the network restructuring rules and the opinion dynamics model has been proposed in the literature, the transition from full consensus inside a single connected component of agents to a fragmented society with isolated groups holding different opinions has been proved to be robust~\cite{vazquez2008generic}. 

An illustration of the phenomenology behind coevolving models is presented in Fig.~\ref{fig:FIG_OpinionDynamics_3}. Agents hold a binary opinion $S_i = \lbrace \pm 1 \rbrace$, and at each time step, a node $i$ and one of her neighbors $j$ are chosen. If both agents agree on the matter, $S_i = S_j$, nothing occurs. When they disagree, $S_i \neq S_j$, with probability $1-p$, agent $i$ adopts $j$'s opinion, thus $S_i = S_j$. But with probability $p$, $i$ breaks her relationship with $j$ and creates a new tie with a randomly chosen agent $a$ such that she is not already connected to $i$ and adopts her opinion, thus $S_i = S_a$. Clearly, $p$ is the rewiring probability and is used as a control parameter. It turns out that there exist a critical value $p_c$ separating various phases. When looking at the relative size of the largest network component $S$, it separates the region for which $S=1$, for small rewiring, from the region in which $S/N \neq 1$, occurring at $ p > p_c $. In the latter case, the network gets fragmented into two large components of size $S \lesssim N/2$ along with components whose size is much smaller than the number of agents $N$ in the population. The magnetization $|m| = \sum_i S_i / N$ also brings information about the nature of the transition and the final configuration of the system. Indeed, in the single-connected regime $ p \leq p_c $ all nodes share the same opinion, displaying full consensus $|m|=1$. When the network fragments, the magnetization tends to $|m|=0$, indicating that both opinion have a similar share of agents. Whether disconnected clusters are in local consensus or not is evaluated through the density of active links, which are the links that join nodes in different opinion. It is observed that stationary density of active links in surviving runs of the dynamics, $\rho^{\mbox{\scriptsize surv}}$, reaches a finite steady value, which, however decreases with the rewiring probability $p$ within the region $p \leq p_c$. It vanishes continuously at $p = p_c$, indicating a second-order phase transition between the active and the frozen configurations. Put otherwise, in the final fragmented configuration, there exist consensus within the isolated groups. Regarding the average time $\tau$ to reach an absorbing state, we observe its divergence when $p \to p_c$, a signature of the critical slowing down of phase transitions.
 
For scenarios with discrete opinions, voter models~\cite{vazquez2008generic, nardini2008who} and multistate variants of it~\cite{holme2006nonequilibrium, herrera2011general} have been largely employed as testbeds of opinion-topology coevolution. Yet, alternative social models with discrete node states have been also put forward~\cite{zanette2006opinion, gil2006coevolution}. In a similar fashion to the analyses in static networks, social influence mechanisms other than the voter model's copying one have been implemented into the models in order to understand their impact on the final configuration of the adaptive process. For instance, idiosyncratic opinion changes, modeled as noise, in the standard (linear) voter model~\cite{diakonova2015noise} and a nonlinear variation~\cite{raducha2020emergence}. Regarding the group influence in the agent's opinion change, the nonlinear voter has been extensively studied in this coevolving context~\cite{min2017fragmentation, raducha2018coevolving, jkedrzejewski2020spontaneous}.
Moreover, these family of models have also been approached from a more classical statistical physics perspective, with efforts directed toward unraveling conservation laws~\cite{toruniewska2017coupling} and symmetry breakings~\cite{jkedrzejewski2020spontaneous}.

Coevolutionary dynamics has been also explored in the context of continuous opinion models. For the Deffuant-Weisbuch model discussed above has been also studied under different types of link rewiring. For random rewiring~\cite{kozma2008consensus, kozma2008consensus1}, it is found that the convergence
time to the steady state is determined by the characteristic time of link rearrangement. Achieving global consensus, on the other hand, is harder than in the baseline model because the rewiring makes it easier for large connected clusters to be broken in smaller parts. In fact, the number of finite-size clusters significantly decreases, as agents can more easily find others to agree with. A rewiring protocol that accounts for transitive homophily and neighbor effects has been also proposed~\cite{kan2023adaptive, krishnagopal2024bounded}. It favors link reconnections towards agents that hold an average neighbor opinion that is closer to the opinion of the focal node that decided to rewire its link. When implemented, networks with a smaller degree assortativity, a smaller spectral gap, and fewer connected components than the baseline adaptive DW model are obtained. Other continuous opinion models have been explicitly laid down to be studied under the coevolving framework as well~\cite{iniguez2009opinion}. 

Liu et al. \cite{liu2023emergence} explored a coevolutionary percolation model in which the network structure itself adapts dynamically based on user opinions and homophily. Their model coupled Deffuant opinion updates, where opinions converge if they fall within a certain tolerance, with a rewiring mechanism. Specifically, if the opinions of two connected users differ by more than a tolerance $\epsilon$, they sever the tie with probability $h$ and then reconnect to a like-minded node. Their mean-field analysis focused on the cross-opinion edge density ($\rho_\times$) and revealed a critical rewiring probability $h_c$, above which the network fragments into polarized echo chambers. Numerical simulations consistently validated these analytical predictions, producing scaling laws for fragmentation time, $T_{\mathrm{frag}} \sim |h - h_c|^{-\alpha}$, with the exponent $\alpha$ being dependent on degree heterogeneity. This work compellingly demonstrates that even subtle changes in homophily-driven rewiring can dramatically impact polarization outcomes and, consequently, reshape information diffusion paths. This study effectively bridges opinion dynamics and network evolution, offering a dynamic perspective on how social structure is both shaped by and, in turn, influences the spread of ideas.

Kozitsin et al. \cite{kozitsin2022general} analyzed VKontakte data to fit a mean-field kinetic opinion model that explicitly incorporates the effects of algorithmic recommendations, or filter bubbles. In their model, each user's propensity to interact with another is weighted by a similarity function $w_{ij} \propto e^{-|x_i - x_j|/\sigma}$, where $\sigma$ reflects the platform's filtering strength. The kinetic equation governing opinion evolution yields polarized steady states for small values of $\sigma$. They found that the observed VKontakte polarization matched an empirical $\sigma_{\mathrm{emp}} \approx 0.4$, suggesting that recommendation algorithms effectively lower the tolerance for opinion differences ($\epsilon_c$), thereby accelerating the formation of echo chambers. This study is significant for bridging opinion dynamics, network science, and the real-world impact of algorithmic design on social polarization, offering a physics-inspired framework to understand how platform policies shape collective behavior and information exposure.

Coevolutinary dynamics with continuous opinions has not only been addressed in the context of pairwise interactions of the DW model, but also in the Hegselmann-Krause model; see, e.g.,Ref.~\cite{su2014coevolution}.

As a final remark, we note that coevolving adaptive models have been used to address socially-inspired phenomena other than opinion dynamics itself, such as role differentiation~\cite{eguiluz2005cooperation} and evolution of cooperation~\cite{zschaler2012adaptive} in game theory, language dynamics through majority rules~\cite{carro2014fragmentation}, cultural evolution~\cite{vazquez2007time}, collective motion~\cite{zschaler2012adaptive} and synchronization~\cite{ito2003spontaneous}. The so-called activity-driven time-varying networks framework has been also developed to integrate activation patterns of the agents with the network evolution, but it has been primarily explored in spreading (epidemiological, rumors) contexts and not so much in opinion dynamics~\cite{perra2012activity, liu2014controlling, karsai2014time}.

\subsection{Realistic features in the temporal interactions: memory effects, burstiness and non-Markovianity}

\begin{figure}[h!]\label{fig:FIG_OpinionDynamics_4}
    \includegraphics[width=\columnwidth]{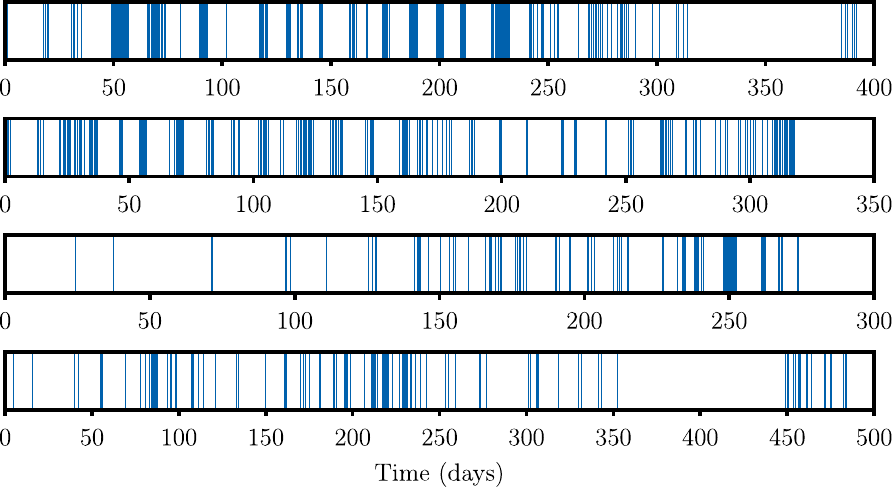}
    \caption{Activity patterns between 4 different pairs of Twitter users. Every vertical line corresponds to a directed public interaction among them, so interevent times are the distance among two consecutive ticks. The measure of time is relative to each pair of users, thus, $t = 0$ corresponds to their first recorded interaction. In the figures are displayed $1914$, $1198$, $695$ and $823$ messages, respectively.}
\end{figure}

In the last couple of decades, big data analyses have revealed that the temporal dimension of human activity patterns is fairly more complex than what traditional hypotheses supposed. This includes sociotechnical interactions that require both a direct and indirect interactions of people with other people or with technological devices, such as phone calls~\cite{miritello2011dynamical}, writing emails~\cite{eckmann2004entropy}, web browsing~\cite{gonccalves2008human}, or face-to-face encounters between school students~\cite{fournet2014contact}, among others. On top of that, evidence from psychological experiments abound regarding the fact that agents' opinion and beliefs are shaped by memory, e.g., in political voting~\cite{michelson2014memory}, in the evaluation of pleasurable experiences~\cite{do2008evaluations} and pain~\cite{kahneman1993when, redelmeier2003memories}. Normally, the working assumption has been that the underlying statistics of opinion dynamics, when it comes to model them, are Poissonian and memoryless, i.e., the social activities are performed at a constant rate and independently of other previous activities~\cite{haight1967handbook, greene1970production, anderson2003fixed}. As a matter of fact, both the mathematical approaches based on master, Langevin and Fokker-Planck equations, and the computational approaches based on standard Monte Carlo methods~\cite{newman1999monte}, the most widespread techniques to simulate statistical physics problems, present the aforementioned properties. Data find systematic and significant deviations from this behavior; see Fig.~\ref{fig:FIG_OpinionDynamics_4} for an example of the interaction patterns between 4 different pairs of Twitter users. Since opinion dynamics relies on the interaction among agents, it is fundamental to know how to identify and measure these nontrivial temporal patterns and to unravel the mechanisms that originate them, with the ultimate goal of incorporating them into the opinion dynamics models to understand the qualitative and quantitative role they play. This is the pipeline that we follow in this section. A wealth of research has been devoted to these aspects; for recent reviews see~\cite{karsai2018bursty, karsai2024measuring}. 

Temporal activation patterns can be seen as time series $ \{t_i\} $, $i=0, 1, 2, \ldots$, where $ t_i $ denotes the time at which an agent performs an activity that yields interactions with other agents. We disregard time series such that $ \{(t_i,z_i)\} $, where $ z_i $ is a measure associated to the event $i$, because most works on heterogeneous temporal patterns in opinion dynamics are primarily interested in scenarios in which the events do not have duration (e.g., receiving an email) or their duration is much smaller than the total time window under consideration (e.g., a phone call against a lifelong kinship) so they can be approximated as a point in the time dimension. We stick to this convention here. Based on this, a quantity of paramount importance to characterize temporal patterns is the interevent time (IET), the time elapsed between two consecutive events $ \tau_i = t_i - t_{i-1}$. $ P(\tau) $ denotes the probability density function of $\{ \tau_i \}$. 

A common feature present in most human interactions is the so-called burstiness~\cite{oliveira2005darwin, barabasi2010bursts}, i.e., periods of very intense activity within short time windows, the bursts, followed by long periods of inactivity; see Fig.~\ref{fig:FIG_OpinionDynamics_4} for examples of bursty time series. Generally, $ P(\tau) $ is fat-tailed and there is no dominance of a single timescale. 

To quantify the amount of burstiness of a time series, Goh and Barab{\'a}si have proposed the burstiness coefficient~\cite{goh2008burstiness}
\begin{equation}
    B = \frac{\sigma - \langle \tau \rangle}{\sigma + \langle \tau \rangle},    
\end{equation}
where $ \langle \tau \rangle $ and $ \sigma $ are the mean and variance of the $P(\tau)$. Clearly, $ B \in [-1,1] $. The extreme case $ B = - 1 $ corresponds to a completely regular time series $ P(\tau) = \delta(\tau - \langle \tau \rangle) $, while $ B \to 1 $  is approached for interevent time distributions with larger and larger variances, that is, very bursty signals. For a Poisson process, we have $ B = 0 $.

The interevent time distribution, and by extension, the burstiness coefficient, does not contain information about the order of the events. Therefore, these quantities are not suitable to obtain information about temporal correlations in social processes, i.e., how past events influence future ones. There are several proposals to quantify these correlations. One is the so-called memory coefficient~\cite{goh2008burstiness}
\begin{equation}
    M = \frac{1}{n-2} \sum_{i = 1}^{n-2} \frac{(\tau_i - \langle \tau \rangle _1)(\tau_{i+1} - \langle \tau \rangle _2)}{\sigma_1 \sigma_2},
\end{equation}
where $ n $ indicates the number of events in the time series, $ \langle \tau \rangle _1$ (resp. $ \langle \tau \rangle _2$) and $ \sigma_1 $ (resp. $\sigma_2$) are the mean and the standard deviation of the set of values $ \{ \tau_i \} $ (resp. $ \{ \tau_{i+1} \} $), with $ i = 1, \, \ldots, \, n - 2 $. This is nothing else than the Pearson correlation coefficient for each pair of consecutive events. The memory coefficient ranges from $ -1 $ to $ 1 $ and the value $ M = 0 $ corresponds to uncorrelated time series. $ M > 0 $ represents positively correlated activity patterns in which short (long) interevent times are followed by short (long) ones. For negative correlations, $ M < 0 $, the behavior is reversed: short (long) interevent times are followed by long (short) ones.

In a similar spirit, the so-called local variation parameter~\cite{shinomoto2003differences}
\begin{equation}
    LV = \frac{3}{n-2} \sum_{i=0}^{n-2} \left(\frac{\tau_i - \tau_{i+1}}{\tau_i + \tau_{i+1}} \right)^2,
\end{equation}
has been introduced in the context of neuroscience, which takes the values $ 0 $, $ 1 $ and $ 3 $ for regular, Poisson, and strongly bursty time series. The advantage of this measure over the memory coefficient is that it does not require the moments of the time series but only the values of consecutive interevent times, so the correlation can be dynamically computed. Finally, one can compute the conditional probability $ P(\tau|\tau') $, defined as the probability of observing an interevent time of value $ \tau $ given the anterior one was $ \tau' $~\cite{artime2017dynamics}. This is a quantitative measure but, at the same time, brings qualitative information. Indeed, by plotting $ P(\tau|\tau') $, for example as a heat map, one can know where the probability concentrates, and how the time series is positively or negatively correlated.

To measure correlations beyond consecutive events, we have the autocorrelation function~\cite{box2015time}
\begin{equation}
    A(t_d) = \frac{\langle x(t)x(t+t_d) \rangle_t - \langle x(t) \rangle_t^2}{\langle x(t)^2 \rangle_t - \langle x(t) \rangle_t^2},
\end{equation}
where $ t_d $ is the delay time, that sets the lag between two observations, and $ \langle \cdot \rangle_t $ is the time average over the observation period. When there are correlations in the signal, it decays as a power law $ A(t_d) \sim t_d^{-\gamma} $. The exponent $ \gamma $ can be related with other quantities, such as the exponent of the underlying fat-tailed interevent time distribution~\cite{vajna2013modelling} or the Hurst exponent~\cite{kantelhardt2001detecting}.

It may occur that the autocorrelation function decays as a power law if the time series is heterogeneous enough, yet with independent interevent times, as reported in~\cite{karsai2012universal}. This is an important point to remark, as sometimes it yields confusion: power-law interevent time distributions and power-law autocorrelation functions can be found even if the sequence of interevent times is uncorrelated. Aiming at finding a measure to correctly disentangle the correlations from the heterogeneity, \cite{karsai2012universal} proposed the bursty train size distribution $ P_{\Delta t}(E) $. It computes the number of consecutive events $ E $ that are separated from one to another by a time window smaller than $ \Delta t $. This distribution is exponentially distributed for uncorrelated sequences of interevent times. Hence, any deviation from an exponential $ P_{\Delta t}(E) $ indicates event-event correlations. 

Once burstiness and interevent correlations in human interacting patterns are reported, we need to unravel the mechanisms that make these properties emerge in order to later test their consequences in models of opinion dynamics. In the literature, we find several proposals that are apparently contradictory from one to another, hinting at the fact that there is not a universal mechanism responsible for the emergence of burstiness and correlations and one must be careful when transferring the conclusions from one scenario to another.

One possible explanation for the power-law decay of the interevent time distribution is based on queuing theory~\cite{barabasi2005origin}. Agents have a list of activities, each of which have a priority assigned. At each time step, the agent chooses between performing the highest priority task with probability $ p $ or performing a random task with the complementary probability. These probabilities can be justified by the finite cognitive capacities in the task handling of human beings. In this setting, the time a task spends in the queue follows a power law of exponent $ -1 $, which reproduced the empirical findings in an email dataset. When human-human interactions are taken into account in the queuing model, interevent time distributions with a larger variety of exponents are found~\cite{oliveira2009impact}, better encompassing the wealth of decays observed in different datasets~\cite{karsai2018bursty}. Other modifications in the particular rules of the model can induce other decay exponents that agree with other datasets~\cite{vazquez2006modeling, gonccalves2008human, masuda2009priority}. Analytical treatments of queuing models have been put forward~\cite{vazquez2005exact, gabrielli2007invasion, anteneodo2009exact}, as well as its study in nontrivial topologies~\cite{min2009waiting}.

Explanations of other nature consider coupled Poissonian models. \cite{hidalgo2006conditions} proposed a simple setting to obtain power-law interevent times by assuming a group of Poissonian agents with different characteristic timescales or a single nonstationary agent that change its rate. Other work in this direction is the one of \cite{malmgren2008poissonian}, which assumes that when performing activities, there are coupled two types of processes acting at different timescales. The primary one is a non-homogeneous Poisson process with periodic time-dependent rate $ \lambda(t) = \lambda(t+T) $, being $ T $ a period (the authors account for circadian and weekly patterns). The secondary process, that is performed when the primary one activates, is a homogeneous Poisson process that leads to a cascade of activity for a randomly selected number of events. The authors achieve an accurate reproduction of email activity patterns and they conclude that the circadian and weekly cycles are key elements to describe the heavy tails in agents' communication. Some years after, however, it has been proposed~\cite{jo2012circadian} a method to de-season a time series, that is, to remove the effects of the daily and weekly cycles from the data, finding that the heavy tails remain robust after this procedure. In other words, the consideration of circadian and weekly rhythms in a model can induce broad interevent time distributions and possibly has an impact on real task execution patterns, but these cycles may not be the fundamental reason for the observation of long tails in the distributions.

Although the phenomenon of broad interevent time distributions has been extensively studied and models of diverse nature have been proposed to reproduce them, the discussion on how to generate temporal series with interevent correlations to be incorporated in the opinion dynamics models has been much more scarce. On the one hand, some numerical algorithms to generate correlated time series with underlying power-law interevent distributions are known, such as~\cite{boguna2014simulating, masuda2018gillespie, jo2019copula}. On the other hand, an option for mechanistic models to generate the desired correlations is to endow the system with some sort of memory. For instance, in \cite{karsai2012universal} an agent is assumed to be in two possible states, the normal and the excited one, and there are transitions between these states. In the former, events are executed independently, while the events performed in the latter correspond to the bursts. The probability to excite the system is constant, but the de-excitation probability depends on the number of consecutive events in the excited state. Besides, the probability of performing the next event also depends on the time elapsed since the last event.

Models that endow agent with memory can also generate fat-tailed interevent time distributions from a bottom-up manner. This has been achieved through what has been dubbed aging. The age of an agent $\tau$ is not to be confused with her physical age, but it is taken as the time since the last change of her state/opinion. Aging is then implemented such that the probability of execution of an event (e.g. an interaction with a neighboring agent in order to adopt her opinion) is modified in accordance with a age-dependent probability $ P_a(\tau) $. This kind of long-term memory in the nodes and also in the links have been used to explain the emergence of heterogeneous activity observed in large-scale communication
networks~\cite{vestergaard2014how}.

To mimic the positive correlations observed in data, this probability must be a decaying function of the age, i.e., the longer an an interaction has not been taken place, the more difficult it is to do so. However, different choices of $ P_a(\tau) $ will modify in different manner the dynamics of the model on which aging is considered. For example, in the case of the voter model, it has found that the the formation of consensus can be accelerated by selecting a linear probability in the age~\cite{stark2008decelerating} or by including latency periods after opinion changes of the agents~\cite{wang2014freezing}. In the latter case, the voter model can also experience oscillatory behavior~\cite{palermo2024spontaneous}. Fernández-Gracia and coatuhors have shown that a decaying $ P_a(\tau) \sim \tau^{-1} $ not only yields power-law interevent time distributions between opinion changes but also drives the voter model to consensus through a coarsening process even if simulated in the regime $ d > 2 $~\cite{fernandez2011update}. In low-dimensional lattices, the coarsening properties of the voter model have been also modified through memory effects~\cite{dallasta2007effective}. Hence, the inclusion of memory in the actions of the agents leads to phenomena that are radically different from the one observed for the memory-less cases. This has been further verified by including aging in other models of opinion dynamics, such as in the noisy voter model~\cite{artime2018aging}. In there, it has been shown that aging is able to change the universality class of the dynamical model, and make the pseudo-critical size-dependent critical point of the modality transition becomes independent of $N$. This type of aging has been generalized and its consequences investigated recently, within the family of voter models~\cite{peralta2020ordering, baron2022analytical, llabres2024aging} and other models relevant to sociophysics, such as the Schelling model~\cite{abella2022aging}, threshold models~\cite{abella2023aging} and the majority rule model~\cite{chen2020non}, among others. For the $q$-voter model, a time- and memory-dependent noise have been also implemented, motivated by the fact that agents may possess their own memories of past experiences related to the social costs and benefits of being independent (noisy updates) or conformist (group influence)~\cite{jkedrzejewski2018impact}.

Moreover, we underscore that including aging not only offers new dynamical behaviors, which is interesting from a fundamental point of view, but also is capable of reproducing real-world features, such as global oscillations of the global dynamics in the voter model~\cite{perez2016competition} and in the spatial prisoner's dilemma game~\cite{szolnoki2009impact, szolnoki2010dynamically}.

Another aspect that has been largely explored under the lens of heterogeneous interaction patterns is the time it takes for a system to reach consensus. It has been found that consensus can be delayed with respect to the memoryless case~\cite{takaguchi2011voter}, but the interaction network plays a crucial role in determining the final amount. However, when burstiness and temporal correlations are intertwined with topological correlations, in the form of communities, it has been reported that the time to consensus may speed-up or slow-down with respect the uncorrelated case, depending on the intensity of each correlation~\cite{artime2017dynamics}. Similar results regarding the speed-up/slow-down due to the interplay between topology and temporal correlations are found in continuous-opinion bounded confidence models such as the Deffuant-Weisbuch and the Hegselmann-Krause discussed above~\cite{chu2024bounded, zarei2024bursts} and diffusive processes~\cite{scholtes2014causality}; see also~\cite{delvenne2015diffusion, rosvall2014memory}. For spreading models, it has been shown that prevalence decay times under long-tailed interevent time statistics make prevalence decay times considerably larger than what is predicted by the standard Poisson process based models~\cite{vazquez2007impact} and that the large heterogeneity found in the response time is responsible for the slow dynamics of information at the collective level~\cite{iribarren2009impact}. This is robust for other type of temporal correlations~\cite{karsai2011small}.


\section{Summary and Conclusions}
\label{sec:conclusions}
The advent of the Internet and social media has fundamentally reshaped the information landscape, blurring the boundaries between physical and social networks and creating a dynamic, complex adaptive information ecosystem. This landscape, where news, rumors, and opinions spread at unprecedented speed and scale, is rife with unreliable and competing content, making rigorous scientific analysis essential. The inherent complexity of these socio-technological systems—characterized by disordered connectivity patterns, nonlinear dynamics, and active, adaptive agents—poses significant challenges for traditional analytical approaches. This review has systematically demonstrated that the analytical power of statistical physics and network science provides the necessary, quantitative framework to clarify the fundamental principles governing these systems, moving beyond descriptive accounts to reveal the underlying mechanisms.

The study of collective social dynamics necessitates first characterizing the medium. We have shown how the analysis of social media platforms (particularly X, Facebook, and Reddit) reveals distinct topological and algorithmic features that shape information diffusion. Network models, including random graphs, preferential attachment frameworks, and fitness-based approaches, have proven essential in quantifying how information disorders emerge from platform architectures. Maximum entropy null models, in particular, provide a robust baseline for distinguishing organic information flow from artificially amplified campaigns. At the mesoscopic scale, structural features like echo chambers and polarized communities arise naturally from homophilic interactions and algorithmic reinforcement, while at the macroscopic level, cascading failures in information integrity often mirror critical phase transitions found in physical systems. The properties of these networks—being directed, weighted, and signed—encode how information flows and provide critical insights into the macroscopic spread and microscopic user dynamics.

The core of the review addressed how and how fast news, rumors, and opinions spread, showing that the dynamics of propagation are governed by both simple and complex contagion processes. While some false narratives spread through viral, broadcast-like mechanisms (simple contagion), others require repeated exposure or social reinforcement (complex contagion), particularly in ideologically insulated communities. Threshold models and bounded-confidence opinion dynamics further explain how misinformation becomes entrenched, with co-evolutionary network effects leading to fragmentation and sustained belief polarization. The temporal dimension adds further complexity, including phenomena such as burstiness, memory effects, and non-Markovian interactions. This requires temporally resolved models that capture the non-stationary dynamics of these ecosystems, reinforcing the need to investigate the principles at the basis of spreading patterns in disordered structures, as well as measuring the pervasiveness of different narratives in online media.

The question of under what conditions a population reaches consensus, polarization, or fragmentation was answered by reviewing discrete opinion models, like the Voter model and its variations, and continuous models, such as Deffuant-Weisbuch and Hegselmann-Krause. These frameworks reveal that even minor perturbations in network structure or information exposure can trigger large-scale shifts in collective belief systems. Human factors, including confirmation bias, motivated reasoning, and identity-protective cognition, play a critical role, as social psychological research underscores how these biases amplify susceptibility to false narratives. The interplay between individual decision-making and social dilemmas—such as the trade-off between sharing speed and accuracy—further complicates mitigation efforts, highlighting the necessity of bridging micro-level cognitive biases with macro-level societal divides through mesoscopic analysis.

The focused study of information disorder (misinformation, disinformation, and mal-information) represents a critical and timely application of this physics-based framework. Here, key structural and dynamical insights have been leveraged to examine the mechanisms through which false narratives propagate, persist, and evolve. Efforts to combat disinformation must account for the adaptive nature of both malicious actors and platform ecosystems. Source detection methods, including network-based attribution and machine learning classifiers, face persistent challenges from obfuscation tactics, such as sockpuppet networks and cross-platform hopping. Meanwhile, content moderation policies often struggle to balance censorship risks with the need for rapid intervention, particularly in encrypted or private spaces where disinformation thrives.

This leads to the practical question of whether we can design effective, physics-informed interventions to mitigate misinformation diffusion. Innovative approaches—such as probabilistic fact-checking, network inoculation strategies, and decentralized reputation systems—show promise but require scalable implementation. The integration of behavioral nudges (e.g., accuracy prompts) with structural interventions (e.g., algorithmic transparency) may offer a multi-layered defense. However, the rapid evolution of generative AI and synthetic media demands continuous adaptation of detection and response frameworks.

Ultimately, the physics of news, rumors, and opinions has firmly established itself as a unifying science, demonstrating that these phenomena are not merely social or technological issues but complex adaptive systems governed by fundamental physical principles. Future research should prioritize cross-platform analyses to track information migration and resilience, and develop policy-relevant simulations to stress-test interventions under realistic, dynamic conditions. By grounding solutions in empirical network science and theoretical advances in stochastic processes, phase transitions, and percolation theory, while respecting the complexities of human behavior, we can better navigate the challenges of our hyperconnected information age, fostering a more informed and resilient digital society.

\section*{Acknowledgements}
GC and AN acknowledge support from the EU project NODES ``Narratives Observatory Combatting Disinformation in Europe Systemically''. 
O. A. acknowledges support from the Spanish Grants No. PID2021-128005NB-C22 and No. PID2024-158120NB-C22, funded by MCIN/AEI/10.13039/501100011033 and ``ERDF A way of making Europe'', from Generalitat de Catalunya (2021SGR00856).
SG and FS acknowledge support from the project “CODE – Coupling Opinion Dynamics with Epidemics”, funded under PNRR Mission 4 "Education and Research" - Component C2 - Investment 1.1 - Next Generation EU "Fund for National Research Program and Projects of Significant National Interest" PRIN 2022 PNRR, grant code P2022AKRZ9, CUP B53D23026080001. 

\bibliographystyle{elsarticle-num}

\begin{thebibliography}{100}
\expandafter\ifx\csname url\endcsname\relax
  \def\url#1{\texttt{#1}}\fi
\expandafter\ifx\csname urlprefix\endcsname\relax\def\urlprefix{URL }\fi
\expandafter\ifx\csname href\endcsname\relax
  \def\href#1#2{#2} \def\path#1{#1}\fi

\bibitem{artime2022origin}
O.~Artime, M.~De~Domenico, From the origin of life to pandemics: Emergent
  phenomena in complex systems (2022).

\bibitem{caldarelli2002scale}
G.~Caldarelli, A.~Capocci, P.~{De Los Rios}, M.~A. Mu{\~n}oz,
  \href{https://link.aps.org/doi/10.1103/PhysRevLett.89.258702}{{Scale-Free
  Networks from Varying Vertex Intrinsic Fitness}}, Physical Review Letters
  89~(25) (2002) 258702.
\newblock \href {https://doi.org/10.1103/PhysRevLett.89.258702}
  {\path{doi:10.1103/PhysRevLett.89.258702}}.
\newline\urlprefix\url{https://link.aps.org/doi/10.1103/PhysRevLett.89.258702}

\bibitem{barabasi2005origin}
A.-L. Barabasi, The origin of bursts and heavy tails in human dynamics, Nature
  435~(7039) (2005) 207--211.

\bibitem{lazer2009social}
D.~Lazer, A.~Pentland, L.~Adamic, S.~Aral, A.-L. Barabasi, D.~Brewer,
  N.~Christakis, N.~Contractor, J.~Fowler, M.~Gutmann, et~al., Social science.
  computational social science., Science (New York, NY) 323~(5915) (2009)
  721--723.

\bibitem{simini2012universal}
F.~Simini, M.~C. Gonz{\'a}lez, A.~Maritan, A.-L. Barab{\'a}si, A universal
  model for mobility and migration patterns, Nature 484~(7392) (2012) 96--100.

\bibitem{pappalardo2015returners}
L.~Pappalardo, F.~Simini, S.~Rinzivillo, D.~Pedreschi, F.~Giannotti, A.-L.
  Barab{\'a}si, Returners and explorers dichotomy in human mobility, Nature
  communications 6~(1) (2015) 8166.

\bibitem{lazer2021meaningful}
D.~Lazer, E.~Hargittai, D.~Freelon, S.~Gonzalez-Bailon, K.~Munger,
  K.~Ognyanova, J.~Radford, Meaningful measures of human society in the
  twenty-first century, Nature 595~(7866) (2021) 189--196.

\bibitem{caldarelli2018physics}
G.~Caldarelli, S.~Wolf, Y.~Moreno,
  \href{https://www.nature.com/articles/s41567-018-0266-x}{{Physics of humans,
  physics for society}} (sep 2018).
\newblock \href {https://doi.org/10.1038/s41567-018-0266-x}
  {\path{doi:10.1038/s41567-018-0266-x}}.
\newline\urlprefix\url{https://www.nature.com/articles/s41567-018-0266-x}

\bibitem{nadal2012manifesto}
J.~P. Nadal, et~al., Manifesto of computational social science, The European
  Physical Journal Special Topics 214 (2012) 325--346.

\bibitem{pareto1935mind}
V.~Pareto, The mind and society, Vol.~1, New York: Harcourt, Brace and Company,
  1935.

\bibitem{zipf1949human}
G.~K. Zipf, Human behavior and the principle of least effort: An introduction
  to human ecology, Addison Wesley, 1949.

\bibitem{onnela2007analysis}
J.-P. Onnela, J.~Saram{\"a}ki, J.~Hyv{\"o}nen, G.~Szab{\'o}, M.~A. De~Menezes,
  K.~Kaski, A.-L. Barab{\'a}si, J.~Kert{\'e}sz, Analysis of a large-scale
  weighted network of one-to-one human communication, New journal of physics
  9~(6) (2007) 179.

\bibitem{song2010modelling}
C.~Song, T.~Koren, P.~Wang, A.-L. Barab{\'a}si, Modelling the scaling
  properties of human mobility, Nature physics 6~(10) (2010) 818--823.

\bibitem{jiang2013calling}
Z.-Q. Jiang, W.-J. Xie, M.-X. Li, B.~Podobnik, W.-X. Zhou, H.~E. Stanley,
  Calling patterns in human communication dynamics, Proceedings of the National
  Academy of Sciences 110~(5) (2013) 1600--1605.

\bibitem{schlapfer2014scaling}
M.~Schl{\"a}pfer, L.~M. Bettencourt, S.~Grauwin, M.~Raschke, R.~Claxton,
  Z.~Smoreda, G.~B. West, C.~Ratti, The scaling of human interactions with city
  size, Journal of the Royal Society Interface 11~(98) (2014) 20130789.

\bibitem{alessandretti2020scales}
L.~Alessandretti, U.~Aslak, S.~Lehmann, The scales of human mobility, Nature
  587~(7834) (2020) 402--407.

\bibitem{xu2021emergence}
F.~Xu, Y.~Li, D.~Jin, J.~Lu, C.~Song, Emergence of urban growth patterns from
  human mobility behavior, Nature Computational Science 1~(12) (2021) 791--800.

\bibitem{haken1975cooperative}
H.~Haken, Cooperative phenomena in systems far from thermal equilibrium and in
  nonphysical systems, Reviews of modern physics 47~(1) (1975) 67.

\bibitem{castellano2009statistical}
C.~Castellano, S.~Fortunato, V.~Loreto,
  \href{https://link.aps.org/doi/10.1103/RevModPhys.81.591}{{Statistical
  physics of social dynamics}}, Reviews of Modern Physics 81~(2) (2009)
  591--646.
\newblock \href {http://arxiv.org/abs/0710.3256} {\path{arXiv:0710.3256}},
  \href {https://doi.org/10.1103/RevModPhys.81.591}
  {\path{doi:10.1103/RevModPhys.81.591}}.
\newline\urlprefix\url{https://link.aps.org/doi/10.1103/RevModPhys.81.591}

\bibitem{montanari2010spread}
A.~Montanari, A.~Saberi, The spread of innovations in social networks,
  Proceedings of the National Academy of Sciences 107~(47) (2010) 20196--20201.

\bibitem{dedomenico2013anatomy}
M.~{De Domenico}, A.~Lima, P.~Mougel, M.~Musolesi,
  \href{https://www.nature.com/articles/srep02980}{{The anatomy of a scientific
  rumor}}, Scientific Reports 3~(1) (2013) 2980.
\newblock \href {http://arxiv.org/abs/1301.2952} {\path{arXiv:1301.2952}},
  \href {https://doi.org/10.1038/srep02980} {\path{doi:10.1038/srep02980}}.
\newline\urlprefix\url{https://www.nature.com/articles/srep02980}

\bibitem{gallotti2016stochastic}
R.~Gallotti, A.~Bazzani, S.~Rambaldi, M.~Barthelemy, A stochastic model of
  randomly accelerated walkers for human mobility, Nature communications 7~(1)
  (2016) 12600.

\bibitem{hauser2019social}
O.~P. Hauser, C.~Hilbe, K.~Chatterjee, M.~A. Nowak, Social dilemmas among
  unequals, Nature 572~(7770) (2019) 524--527.

\bibitem{cimini2019statistical}
G.~Cimini, T.~Squartini, F.~Saracco, D.~Garlaschelli, A.~Gabrielli,
  G.~Caldarelli, \href{http://www.nature.com/articles/s42254-018-0002-6}{{The
  statistical physics of real-world networks}} (jan 2019).
\newblock \href {http://arxiv.org/abs/1810.05095} {\path{arXiv:1810.05095}},
  \href {https://doi.org/10.1038/s42254-018-0002-6}
  {\path{doi:10.1038/s42254-018-0002-6}}.
\newline\urlprefix\url{http://www.nature.com/articles/s42254-018-0002-6}

\bibitem{anderson1972more}
P.~W. Anderson, More is different: Broken symmetry and the nature of the
  hierarchical structure of science., Science 177~(4047) (1972) 393--396.

\bibitem{ladyman2013what}
J.~Ladyman, J.~Lambert, K.~Wiesner, What is a complex system?, European Journal
  for Philosophy of Science 3 (2013) 33--67.

\bibitem{buckley2017society}
W.~Buckley, Society as a complex adaptive system, in: Systems Research for
  Behavioral Science, Routledge, 2017, pp. 490--513.

\bibitem{caldarelli2020perspective}
G.~Caldarelli,
  \href{https://iopscience.iop.org/article/10.1088/2632-072X/ab9a24}{{A
  perspective on complexity and networks science}} (sep 2020).
\newblock \href {https://doi.org/10.1088/2632-072X/ab9a24}
  {\path{doi:10.1088/2632-072X/ab9a24}}.
\newline\urlprefix\url{https://iopscience.iop.org/article/10.1088/2632-072X/ab9a24}

\bibitem{newman2018networks}
M.~Newman, Networks, Oxford university press, 2018.

\bibitem{kitano2004biological}
H.~Kitano, Biological robustness, Nature Reviews Genetics 5~(11) (2004)
  826--837.

\bibitem{gosak2018network}
M.~Gosak, R.~Markovi{\v{c}}, J.~Dolen{\v{s}}ek, M.~S. Rupnik, M.~Marhl,
  A.~Sto{\v{z}}er, M.~Perc, Network science of biological systems at different
  scales: A review, Physics of life reviews 24 (2018) 118--135.

\bibitem{barucca2018tackling}
P.~Barucca, G.~Caldarelli, T.~Squartini,
  \href{http://link.springer.com/10.1007/s10955-018-2076-z}{{Tackling
  Information Asymmetry in Networks: A New Entropy-Based Ranking Index}},
  Journal of Statistical Physics 173~(3-4) (2018) 1028--1044.
\newblock \href {http://arxiv.org/abs/1710.09656} {\path{arXiv:1710.09656}},
  \href {https://doi.org/10.1007/s10955-018-2076-z}
  {\path{doi:10.1007/s10955-018-2076-z}}.
\newline\urlprefix\url{http://link.springer.com/10.1007/s10955-018-2076-z}

\bibitem{bardoscia2021physics}
M.~Bardoscia, P.~Barucca, S.~Battiston, F.~Caccioli, G.~Cimini,
  D.~Garlaschelli, F.~Saracco, T.~Squartini, G.~Caldarelli, The physics of
  financial networks, Nature Reviews Physics 3~(7) (2021) 490--507.

\bibitem{caldarelli2000fractal}
G.~Caldarelli, R.~Marchetti, L.~Pietronero, The fractal properties of internet,
  Europhysics letters 52~(4) (2000) 386.

\bibitem{pastor-satorras2001dynamical}
R.~Pastor-Satorras, A.~V{\'a}zquez, A.~Vespignani, Dynamical and correlation
  properties of the internet, Physical review letters 87~(25) (2001) 258701.

\bibitem{watts1998collective}
D.~J. Watts, S.~H. Strogatz, Collective dynamics of 'small-world' networks,
  Nature 393 (1998) 440--442.

\bibitem{barabasi1999emergence}
A.-L. Barab{\'a}si, R.~Albert, Emergence of scaling in random networks, science
  286~(5439) (1999) 509--512.

\bibitem{caldarelli2007scale}
G.~Caldarelli,
  \href{https://books.google.it/books?hl=en\&lr=\&id=6pwVDAAAQBAJ\&oi=fnd\&pg=PR7\&dq=info:LRDBgmL06e8J:scholar.google.com\&ots=X8\_u32S3TV\&sig=r1uw5MOkLDgEFXQZSuqaMbat3RA\&redir\_esc=y\#v=onepage\&q\&f=false}{{Scale-Free
  Networks: Complex Webs in Nature and Technology}}, Vol. 9780199211, Oxford
  University Press, UK, 2007.
\newblock \href {https://doi.org/10.1093/acprof:oso/9780199211517.001.0001}
  {\path{doi:10.1093/acprof:oso/9780199211517.001.0001}}.
\newline\urlprefix\url{https://books.google.it/books?hl=en\&lr=\&id=6pwVDAAAQBAJ\&oi=fnd\&pg=PR7\&dq=info:LRDBgmL06e8J:scholar.google.com\&ots=X8\_u32S3TV\&sig=r1uw5MOkLDgEFXQZSuqaMbat3RA\&redir\_esc=y\#v=onepage\&q\&f=false}

\bibitem{mucha2010community}
P.~J. Mucha, T.~Richardson, K.~Macon, M.~A. Porter, J.-P. Onnela, Community
  structure in time-dependent, multiscale, and multiplex networks, science
  328~(5980) (2010) 876--878.

\bibitem{de2013mathematical}
M.~De~Domenico, A.~Sol{\'e}-Ribalta, E.~Cozzo, M.~Kivel{\"a}, Y.~Moreno, M.~A.
  Porter, S.~G{\'o}mez, A.~Arenas, Mathematical formulation of multilayer
  networks, Physical Review X 3~(4) (2013) 041022.

\bibitem{nicosia2013growing}
V.~Nicosia, G.~Bianconi, V.~Latora, M.~Barthelemy, Growing multiplex networks,
  Physical review letters 111~(5) (2013) 058701.

\bibitem{wang2015evolutionary}
Z.~Wang, L.~Wang, A.~Szolnoki, M.~Perc, Evolutionary games on multilayer
  networks: a colloquium, The European physical journal B 88~(5) (2015) 124.

\bibitem{buldyrev2010catastrophic}
S.~V. Buldyrev, R.~Parshani, G.~Paul, H.~E. Stanley, S.~Havlin, Catastrophic
  cascade of failures in interdependent networks, Nature 464~(7291) (2010)
  1025--1028.

\bibitem{gao2012networks}
J.~Gao, S.~V. Buldyrev, H.~E. Stanley, S.~Havlin, Networks formed from
  interdependent networks, Nature physics 8~(1) (2012) 40--48.

\bibitem{artime2022multilayer}
O.~Artime, B.~Benigni, G.~Bertagnolli, V.~d'Andrea, R.~Gallotti, A.~Ghavasieh,
  S.~Raimondo, M.~De~Domenico, Multilayer network science: from cells to
  societies, Cambridge University Press, 2022.

\bibitem{de2023more}
M.~De~Domenico, More is different in real-world multilayer networks, Nature
  Physics 19~(9) (2023) 1247--1262.

\bibitem{robins2001random}
G.~Robins, P.~Pattison, Random graph models for temporal processes in social
  networks, Journal of Mathematical Sociology 25~(1) (2001) 5--41.

\bibitem{holme2012temporal}
P.~Holme, J.~Saram{\"a}ki, Temporal networks, Physics reports 519~(3) (2012)
  97--125.

\bibitem{perra2012activity}
N.~Perra, B.~Gon{\c{c}}alves, R.~Pastor-Satorras, A.~Vespignani, Activity
  driven modeling of time varying networks, Scientific reports 2~(1) (2012)
  469.

\bibitem{borge-holthoefer2016dynamics}
J.~Borge-Holthoefer, N.~Perra, B.~Gon{\c{c}}alves, S.~Gonz{\'a}lez-Bail{\'o}n,
  A.~Arenas, Y.~Moreno, A.~Vespignani, The dynamics of information-driven
  coordination phenomena: A transfer entropy analysis, Science advances 2~(4)
  (2016) e1501158.

\bibitem{matamalas2016assessing}
J.~T. Matamalas, M.~De~Domenico, A.~Arenas, Assessing reliable human mobility
  patterns from higher order memory in mobile communications, Journal of The
  Royal Society Interface 13~(121) (2016) 20160203.

\bibitem{williams2022shape}
O.~E. Williams, L.~Lacasa, A.~P. Mill{\'a}n, V.~Latora, The shape of memory in
  temporal networks, Nature communications 13~(1) (2022) 499.

\bibitem{lehmann2012dynamical}
J.~Lehmann, B.~Gon\c{c}alves, J.~J. Ramasco, C.~Cattuto,
  \href{http://dx.doi.org/10.1145/2187836.2187871}{Dynamical classes of
  collective attention in twitter}, in: Proceedings of the 21st international
  conference on World Wide Web, WWW 2012, ACM, 2012.
\newblock \href {https://doi.org/10.1145/2187836.2187871}
  {\path{doi:10.1145/2187836.2187871}}.
\newline\urlprefix\url{http://dx.doi.org/10.1145/2187836.2187871}

\bibitem{mocanu2015collective}
D.~Mocanu, L.~Rossi, Q.~Zhang, M.~Karsai, W.~Quattrociocchi,
  \href{http://dx.doi.org/10.1016/j.chb.2015.01.024}{Collective attention in
  the age of (mis)information}, Computers in Human Behavior 51 (2015)
  1198–1204.
\newblock \href {https://doi.org/10.1016/j.chb.2015.01.024}
  {\path{doi:10.1016/j.chb.2015.01.024}}.
\newline\urlprefix\url{http://dx.doi.org/10.1016/j.chb.2015.01.024}

\bibitem{lorenz-spreen2019accelerating}
P.~Lorenz-Spreen, B.~M. M{\o}nsted, P.~H{\"{o}}vel, S.~Lehmann,
  \href{https://www.nature.com/articles/s41467-019-09311-w}{{Accelerating
  dynamics of collective attention}}, Nature Communications 10~(1) (2019) 1759.
\newblock \href {https://doi.org/10.1038/s41467-019-09311-w}
  {\path{doi:10.1038/s41467-019-09311-w}}.
\newline\urlprefix\url{https://www.nature.com/articles/s41467-019-09311-w}

\bibitem{dedomenico2020unraveling}
M.~{De Domenico}, E.~G. Altmann,
  \href{https://www.nature.com/articles/s41598-020-61523-z}{{Unraveling the
  Origin of Social Bursts in Collective Attention}}, Scientific Reports 10~(1)
  (2020) 4629.
\newblock \href {http://arxiv.org/abs/1903.06588} {\path{arXiv:1903.06588}},
  \href {https://doi.org/10.1038/s41598-020-61523-z}
  {\path{doi:10.1038/s41598-020-61523-z}}.
\newline\urlprefix\url{https://www.nature.com/articles/s41598-020-61523-z}

\bibitem{vicsek1995novel}
T.~Vicsek, A.~Czir{\'o}k, E.~Ben-Jacob, I.~Cohen, O.~Shochet, Novel type of
  phase transition in a system of self-driven particles, Physical review
  letters 75~(6) (1995) 1226.

\bibitem{cavagna2019dynamical}
A.~Cavagna, L.~Di~Carlo, I.~Giardina, L.~Grandinetti, T.~S. Grigera,
  G.~Pisegna, Dynamical renormalization group approach to the collective
  behavior of swarms, Physical Review Letters 123~(26) (2019) 268001.

\bibitem{delvenne2015diffusion}
J.~C. Delvenne, R.~Lambiotte, L.~E. Rocha,
  \href{https://www.nature.com/articles/ncomms8366}{{Diffusion on networked
  systems is a question of time or structure}}, Nature Communications 6~(1)
  (2015) 7366.
\newblock \href {http://arxiv.org/abs/1309.4155} {\path{arXiv:1309.4155}},
  \href {https://doi.org/10.1038/ncomms8366} {\path{doi:10.1038/ncomms8366}}.
\newline\urlprefix\url{https://www.nature.com/articles/ncomms8366}

\bibitem{budak2024misunderstanding}
C.~Budak, B.~Nyhan, D.~M. Rothschild, E.~Thorson, D.~J. Watts,
  \href{https://doi.org/10.1038/s41586-024-07417-w}{Misunderstanding the harms
  of online misinformation}, Nature 630~(8015) (2024) 45--53.
\newblock \href {https://doi.org/10.1038/s41586-024-07417-w}
  {\path{doi:10.1038/s41586-024-07417-w}}.
\newline\urlprefix\url{https://doi.org/10.1038/s41586-024-07417-w}

\bibitem{falkenberg2025towards}
M.~Falkenberg, M.~Cinelli, A.~Galeazzi, C.~A. Bail, R.~M. Benito, A.~Bruns,
  A.~Gruzd, D.~Lazer, J.~K. Lee, J.~McCoy, et~al., Towards global equity in
  political polarization research, arXiv preprint arXiv:2504.11090 (2025).

\bibitem{urman2020context}
A.~Urman, \href{https://doi.org/10.1177/0163443719876541}{Context matters:
  political polarization on twitter from a comparative perspective}, Media,
  Culture \& Society 42~(6) (2020) 857--879.
\newblock \href {http://arxiv.org/abs/https://doi.org/10.1177/0163443719876541}
  {\path{arXiv:https://doi.org/10.1177/0163443719876541}}, \href
  {https://doi.org/10.1177/0163443719876541}
  {\path{doi:10.1177/0163443719876541}}.
\newline\urlprefix\url{https://doi.org/10.1177/0163443719876541}

\bibitem{vanvliet2021political}
L.~Van~Vliet, P.~Törnberg, J.~Uitermark, Political systems and political
  networks: The structure of parliamentarians’ retweet networks in 19
  countries, International Journal of Communication 15 (2021) 2156 – 2176.

\bibitem{broido2019scale}
A.~D. Broido, A.~Clauset, Scale-free networks are rare, Nature communications
  10~(1) (2019) 1--10.

\bibitem{clauset2009power}
A.~Clauset, C.~R. Shalizi, M.~E.~J. Newman,
  \href{https://doi.org/10.1137/070710111}{Power-law distributions in empirical
  data}, SIAM Review 51~(4) (2009) 661--703.
\newblock \href {http://arxiv.org/abs/https://doi.org/10.1137/070710111}
  {\path{arXiv:https://doi.org/10.1137/070710111}}, \href
  {https://doi.org/10.1137/070710111} {\path{doi:10.1137/070710111}}.
\newline\urlprefix\url{https://doi.org/10.1137/070710111}

\bibitem{ravasz2003hierarchical}
E.~Ravasz, A.-L. Barab{\'a}si, Hierarchical organization in complex networks,
  Physical review E 67~(2) (2003) 026112.

\bibitem{fortunato2010community}
S.~Fortunato, Community detection in graphs, Physics reports 486~(3-5) (2010)
  75--174.

\bibitem{erdos1959on}
P.~Erd\"os, A.~R\'enyi, On random graphs i, Publicationes Mathematicae Debrecen
  6 (1959) 290--297.

\bibitem{erdos1960on}
P.~Erd\H{o}s, A.~R\'{e}nyi, On the evolution of random graphs, Publications of
  the Mathematical Institute of the Hungarian Academy of Sciences 5 (1960)
  17--60.

\bibitem{bollobas2011random}
B.~Bollob{\'a}s, Random graphs, in: Modern graph theory, Springer, 2011, pp.
  215--252.

\bibitem{bianconi2001competition}
G.~Bianconi, A.-L. Barab{\'a}si, Competition and multiscaling in evolving
  networks, Europhysics letters 54~(4) (2001) 436.

\bibitem{holland1981exponential}
P.~W. Holland, S.~Leinhardt, An exponential family of probability distributions
  for directed graphs, Journal of the american Statistical association 76~(373)
  (1981) 33--50.

\bibitem{frank1986markov}
O.~Frank, D.~Strauss, Markov graphs, Journal of the american Statistical
  association 81~(395) (1986) 832--842.

\bibitem{holland1983stochastic}
P.~W. Holland, K.~B. Laskey, S.~Leinhardt, Stochastic blockmodels: First steps,
  Social networks 5~(2) (1983) 109--137.

\bibitem{peixoto2014hierarchical}
T.~P. Peixoto, Hierarchical block structures and high-resolution model
  selection in large networks, Physical Review X 4~(1) (2014) 011047.

\bibitem{papadopoulos2012popularity}
F.~Papadopoulos, M.~Kitsak, M.~{\'A}. Serrano, M.~Bogun{\'a}, D.~Krioukov,
  Popularity versus similarity in growing networks, Nature 489~(7417) (2012)
  537--540.

\bibitem{molloy1995critical}
M.~Molloy, B.~Reed,
  \href{https://biblioproxy.cnr.it:2270/doi/abs/10.1002/rsa.3240060204}{A
  critical point for random graphs with a given degree sequence}, Random
  Structures \& Algorithms 6~(2-3) (1995) 161--180.
\newblock \href
  {http://arxiv.org/abs/https://biblioproxy.cnr.it:2270/doi/pdf/10.1002/rsa.3240060204}
  {\path{arXiv:https://biblioproxy.cnr.it:2270/doi/pdf/10.1002/rsa.3240060204}},
  \href
  {https://doi.org/https://biblioproxy.cnr.it:2481/10.1002/rsa.3240060204}
  {\path{doi:https://biblioproxy.cnr.it:2481/10.1002/rsa.3240060204}}.
\newline\urlprefix\url{https://biblioproxy.cnr.it:2270/doi/abs/10.1002/rsa.3240060204}

\bibitem{hoff2002latent}
P.~D. Hoff, A.~E. Raftery, M.~S. Handcock, Latent space approaches to social
  network analysis, Journal of the American Statistical Association 97~(460)
  (2002) 1090--1098.

\bibitem{karrer2011stochastic}
B.~Karrer, M.~E.~J. Newman, Stochastic blockmodels and community structure in
  networks, Physical Review E 83~(1) (2011) 016107.

\bibitem{peel2017ground}
L.~Peel, D.~B. Larremore, A.~Clauset, Ground truth for network community
  detection: A survey, ACM Computing Surveys 50~(1) (2017) 1--40.

\bibitem{park2004statistical}
J.~Park, M.~E. Newman, Statistical mechanics of networks, Physical Review
  E—Statistical, Nonlinear, and Soft Matter Physics 70~(6) (2004) 066117.

\bibitem{squartini2011analytical}
T.~Squartini, D.~Garlaschelli, Analytical maximum-likelihood method to detect
  patterns in real networks, New Journal of Physics 13~(8) (2011) 083001.

\bibitem{bianconi2009entropy}
G.~Bianconi, Entropy of network ensembles, Physical Review E 79~(3) (2009)
  036114.

\bibitem{krioukov2010hyperbolic}
D.~Krioukov, F.~Papadopoulos, M.~Kitsak, A.~Vahdat, M.~Bogu{\~n}{\'a},
  Hyperbolic geometry of complex networks, Physical Review E 82~(3) (2010)
  036106.

\bibitem{boguna2010sustaining}
M.~Bogu{\~n}{\'a}, F.~Papadopoulos, D.~Krioukov, Sustaining the internet with
  hyperbolic mapping, Nature Communications 1 (2010) 62.

\bibitem{muscoloni2018machine}
A.~Muscoloni, C.~V. Cannistraci, Machine learning meets complex networks via
  coalescent embedding in the hyperbolic space, Nature Communications 9 (2018)
  3969.

\bibitem{boguna2009navigability}
M.~Bogu{\~n}{\'a}, D.~Krioukov, K.~Claffy, Navigability of complex networks,
  Nature Physics 5~(1) (2009) 74--80.

\bibitem{krapivsky2000connectivity}
P.~L. Krapivsky, S.~Redner, F.~Leyvraz, Connectivity of growing random
  networks, Physical review letters 85~(21) (2000) 4629.

\bibitem{bianconi2001bose}
G.~Bianconi, A.-L. Barab{\'a}si, Bose-einstein condensation in complex
  networks, Physical review letters 86~(24) (2001) 5632.

\bibitem{medo2011temporal}
M.~c.~v. Medo, G.~Cimini, S.~Gualdi,
  \href{https://link.aps.org/doi/10.1103/PhysRevLett.107.238701}{Temporal
  effects in the growth of networks}, Phys. Rev. Lett. 107 (2011) 238701.
\newblock \href {https://doi.org/10.1103/PhysRevLett.107.238701}
  {\path{doi:10.1103/PhysRevLett.107.238701}}.
\newline\urlprefix\url{https://link.aps.org/doi/10.1103/PhysRevLett.107.238701}

\bibitem{gronlund2004networking}
A.~Gr\"onlund, P.~Holme,
  \href{https://link.aps.org/doi/10.1103/PhysRevE.70.036108}{Networking the
  seceder model: Group formation in social and economic systems}, Phys. Rev. E
  70 (2004) 036108.
\newblock \href {https://doi.org/10.1103/PhysRevE.70.036108}
  {\path{doi:10.1103/PhysRevE.70.036108}}.
\newline\urlprefix\url{https://link.aps.org/doi/10.1103/PhysRevE.70.036108}

\bibitem{jin2001structure}
E.~M. Jin, M.~Girvan, M.~E.~J. Newman, Structure of growing social networks,
  Phys. Rev. E 64 (2001) 046132.

\bibitem{mcpherson2001birds}
M.~McPherson, L.~Smith-Lovin, J.~M. Cook, Birds of a feather: Homophily in
  social networks, Annual review of sociology 27~(1) (2001) 415--444.

\bibitem{myers2014bursty}
S.~A. Myers, J.~Leskovec, The bursty dynamics of the {T}witter information
  network, in: Proceedings of the 23rd International Conference on World Wide
  Web, Association for Computing Machinery, New York, NY, USA, 2014, p.
  913–924.

\bibitem{aad2013evidence}
G.~Aad, T.~Abajyan, B.~Abbott, J.~Abdallah, S.~A. Khalek, R.~Aben, B.~Abi,
  M.~Abolins, O.~AbouZeid, H.~Abramowicz, et~al., Evidence for the spin-0
  nature of the higgs boson using atlas data, Physics Letters B 726~(1-3)
  (2013) 120--144.

\bibitem{jaynes1957information}
E.~T. Jaynes, Information theory and statistical mechanics, Physical review
  106~(4) (1957) 620.

\bibitem{garlaschelli2007self}
D.~Garlaschelli, A.~Capocci, G.~Caldarelli, Self-organized network evolution
  coupled to extremal dynamics, Nature Physics 3~(11) (2007) 813--817.

\bibitem{garlaschelli2008maximum}
D.~Garlaschelli, M.~I. Loffredo, Maximum likelihood: Extracting unbiased
  information from complex networks, Physical Review E - Statistical,
  Nonlinear, and Soft Matter Physics 78 (2008) 1--5.
\newblock \href {https://doi.org/10.1103/PhysRevE.78.015101}
  {\path{doi:10.1103/PhysRevE.78.015101}}.

\bibitem{huang2009introduction}
K.~Huang, Introduction to statistical physics, Chapman and Hall/CRC, 2009.

\bibitem{bianconi2007entropy}
G.~Bianconi, \href{https://dx.doi.org/10.1209/0295-5075/81/28005}{The entropy
  of randomized network ensembles}, Europhysics Letters 81~(2) (2007) 28005.
\newblock \href {https://doi.org/10.1209/0295-5075/81/28005}
  {\path{doi:10.1209/0295-5075/81/28005}}.
\newline\urlprefix\url{https://dx.doi.org/10.1209/0295-5075/81/28005}

\bibitem{conover2011political}
M.~Conover, J.~Ratkiewicz, M.~Francisco, B.~Gon{\c{c}}alves, F.~Menczer,
  A.~Flammini, Political polarization on twitter, in: Proceedings of the
  international aaai conference on web and social media, Vol.~5, 2011, pp.
  89--96.

\bibitem{wu2011who}
S.~Wu, J.~M. Hofman, W.~A. Mason, D.~J. Watts,
  \href{https://doi.org/10.1145/1963405.1963504}{Who says what to whom on
  twitter}, in: Proceedings of the 20th International Conference on World Wide
  Web, WWW '11, Association for Computing Machinery, New York, NY, USA, 2011,
  p. 705–714.
\newblock \href {https://doi.org/10.1145/1963405.1963504}
  {\path{doi:10.1145/1963405.1963504}}.
\newline\urlprefix\url{https://doi.org/10.1145/1963405.1963504}

\bibitem{grimmett2012percolation}
G.~Grimmett, What is percolation?, in: Percolation, Springer, 2012, pp. 1--31.

\bibitem{stauffer2018introduction}
D.~Stauffer, A.~Aharony, Introduction to percolation theory, Taylor \& Francis,
  2018.

\bibitem{newman2001random}
M.~E. Newman, S.~H. Strogatz, D.~J. Watts, Random graphs with arbitrary degree
  distributions and their applications, Physical review E 64~(2) (2001) 026118.

\bibitem{callaway2000network}
D.~S. Callaway, M.~E. Newman, S.~H. Strogatz, D.~J. Watts, Network robustness
  and fragility: Percolation on random graphs, Physical review letters 85~(25)
  (2000) 5468.

\bibitem{watson1875probability}
H.~W. Watson, F.~Galton, On the probability of the extinction of families, The
  Journal of the Anthropological Institute of Great Britain and Ireland 4
  (1875) 138--144.

\bibitem{harris1963theory}
T.~E. Harris, et~al., The theory of branching processes, Vol.~6, Springer
  Berlin, 1963.

\bibitem{athreya2012branching}
K.~B. Athreya, P.~E. Ney, Branching processes, Vol. 196, Springer Science \&
  Business Media, 2012.

\bibitem{hawkes1971spectra}
A.~G. Hawkes, Spectra of some self-exciting and mutually exciting point
  processes, Biometrika 58~(1) (1971) 83--90.

\bibitem{crane2008robust}
R.~Crane, D.~Sornette, Robust dynamic classes revealed by measuring the
  response function of a social system, Proceedings of the National Academy of
  Sciences 105~(41) (2008) 15649--15653.

\bibitem{watts2002simple}
D.~J. Watts, \href{http://www.ncbi.nlm.nih.gov/pubmed/16578874
  http://www.pubmedcentral.nih.gov/articlerender.fcgi?artid=PMC122850
  https://pnas.org/doi/full/10.1073/pnas.082090499}{{A simple model of global
  cascades on random networks}}, Proceedings of the National Academy of
  Sciences of the United States of America 99~(9) (2002) 5766--5771.
\newblock \href {https://doi.org/10.1073/pnas.082090499}
  {\path{doi:10.1073/pnas.082090499}}.
\newline\urlprefix\url{http://www.ncbi.nlm.nih.gov/pubmed/16578874
  http://www.pubmedcentral.nih.gov/articlerender.fcgi?artid=PMC122850
  https://pnas.org/doi/full/10.1073/pnas.082090499}

\bibitem{kempe2003maximizing}
D.~Kempe, J.~Kleinberg, {\'E}.~Tardos, Maximizing the spread of influence
  through a social network, in: Proceedings of the ninth ACM SIGKDD
  international conference on Knowledge discovery and data mining, 2003, pp.
  137--146.

\bibitem{pastor2002epidemic}
R.~Pastor-Satorras, A.~Vespignani, Epidemic dynamics in finite size scale-free
  networks, Physical Review E 65~(3) (2002) 035108.

\bibitem{pastor2001epidemic}
R.~Pastor-Satorras, A.~Vespignani, Epidemic spreading in scale-free networks,
  Physical review letters 86~(14) (2001) 3200.

\bibitem{wang2003epidemic}
Y.~Wang, D.~Chakrabarti, C.~Wang, C.~Faloutsos, Epidemic spreading in real
  networks: An eigenvalue viewpoint, in: 22nd International Symposium on
  Reliable Distributed Systems, 2003. Proceedings., IEEE, 2003, pp. 25--34.

\bibitem{chakrabarti2008epidemic}
D.~Chakrabarti, Y.~Wang, C.~Wang, J.~Leskovec, C.~Faloutsos, Epidemic
  thresholds in real networks, ACM Transactions on Information and System
  Security (TISSEC) 10~(4) (2008) 1--26.

\bibitem{gomez2010discrete}
S.~G{\'o}mez, A.~Arenas, J.~Borge-Holthoefer, S.~Meloni, Y.~Moreno,
  Discrete-time markov chain approach to contact-based disease spreading in
  complex networks, Europhysics Letters 89~(3) (2010) 38009.

\bibitem{pastor-satorras2015epidemic}
R.~Pastor-Satorras, C.~Castellano, P.~Van~Mieghem, A.~Vespignani, Epidemic
  processes in complex networks, Reviews of modern physics 87~(3) (2015) 925.

\bibitem{kiss2017mathematics}
I.~Z. Kiss, J.~C. Miller, P.~L. Simon, Mathematics of Epidemics on Networks:
  From Exact to Approximate Models, Vol.~46 of Interdisciplinary Applied
  Mathematics, Springer, 2017.

\bibitem{britton2019stochastic}
T.~Britton, E.~Pardoux (Eds.),
  \href{https://link.springer.com/book/10.1007/978-3-030-30900-8}{Stochastic
  Epidemic Models with Inference}, Vol. 2255 of Lecture Notes in Mathematics,
  Springer, Cham, 2019.
\newblock \href {https://doi.org/10.1007/978-3-030-30900-8}
  {\path{doi:10.1007/978-3-030-30900-8}}.
\newline\urlprefix\url{https://link.springer.com/book/10.1007/978-3-030-30900-8}

\bibitem{masuda2013predicting}
N.~Masuda, P.~Holme, Predicting and controlling infectious disease epidemics
  using temporal networks, F1000Prime Rep. 5 (2013) 6.
\newblock \href {https://doi.org/10.12703/P5-6} {\path{doi:10.12703/P5-6}}.

\bibitem{wang2024epidemic}
W.~Wang, Y.~Nie, W.~Li, T.~Lin, M.-S. Shang, S.~Su, Y.~Tang, Y.-C. Zhang, G.-Q.
  Sun,
  \href{https://www.sciencedirect.com/science/article/pii/S0370157324000176}{Epidemic
  spreading on higher-order networks}, Physics Reports 1056 (2024) 1--70,
  epidemic spreading on higher-order networks.
\newblock \href {https://doi.org/https://doi.org/10.1016/j.physrep.2024.01.003}
  {\path{doi:https://doi.org/10.1016/j.physrep.2024.01.003}}.
\newline\urlprefix\url{https://www.sciencedirect.com/science/article/pii/S0370157324000176}

\bibitem{masuda2017random}
N.~Masuda, M.~A. Porter, R.~Lambiotte, Random walks and diffusion on networks,
  Phys. Rep. 716-717 (2017) 1--58.

\bibitem{cohen2003scale}
R.~Cohen, S.~Havlin, Scale-free networks are ultrasmall, Physical review
  letters 90~(5) (2003) 058701.

\bibitem{allen-perkins2019markov}
A.~Allen-Perkins, A.~B. Serrano, T.~A. De~Assis, J.~M. Pastor, R.~F.~S.
  Andrade, Markov chain approach to anomalous diffusion on newman--watts
  networks, Journal of Statistical Mechanics: Theory and Experiment 2019~(4)
  (2019) 043301.

\bibitem{ji2023signal}
P.~Ji, J.~Ye, Y.~Mu, W.~Lin, Y.~Tian, C.~Hens, M.~Perc, Y.~Tang, J.~Sun,
  J.~Kurths, Signal propagation in complex networks, Physics reports 1017
  (2023) 1--96.

\bibitem{watts2002identity}
D.~J. Watts, P.~S. Dodds, M.~E.~J. Newman, Identity and search in social
  networks, Science 296~(5571) (2002) 1302--1305.

\bibitem{kleinberg2000navigation}
J.~M. Kleinberg, Navigation in a small world, Nature 406~(6798) (2000) 845.

\bibitem{milgram1967small}
S.~Milgram, The small-world problem, Psychology Today 1~(1) (1967) 61--67.

\bibitem{travers1969experimental}
J.~Travers, S.~Milgram, \href{https://www.jstor.org/stable/2786545}{An
  experimental study of the small world problem}, Sociometry 32~(4) (1969)
  425--443.
\newblock \href {https://doi.org/10.2307/2786545} {\path{doi:10.2307/2786545}}.
\newline\urlprefix\url{https://www.jstor.org/stable/2786545}

\bibitem{boissevain1974friends}
J.~Boissevain, Friends of Friends, Basil Blackwell, Oxford, 1974.

\bibitem{granovetter1973strength}
M.~S. Granovetter, The strength of weak ties, Am. J. Sociol. 78~(6) (1973)
  1360--1380.

\bibitem{lee2011pathlength}
S.~H. Lee, P.~Holme, Pathlength scaling in graphs with incomplete navigational
  information, Physica A 390~(21-22) (2011) 3996--4001.
\newblock \href {https://doi.org/10.1016/j.physa.2011.06.012}
  {\path{doi:10.1016/j.physa.2011.06.012}}.

\bibitem{benevenuto2012characterizing}
F.~Benevenuto, T.~Rodrigues, M.~Cha, V.~Almeida,
  \href{https://www.sciencedirect.com/science/article/pii/S0020025511006372}{Characterizing
  user navigation and interactions in online social networks}, Information
  Sciences 195 (2012) 1--24.
\newblock \href {https://doi.org/https://doi.org/10.1016/j.ins.2011.12.009}
  {\path{doi:https://doi.org/10.1016/j.ins.2011.12.009}}.
\newline\urlprefix\url{https://www.sciencedirect.com/science/article/pii/S0020025511006372}

\bibitem{yan2006efficient}
G.~Yan, T.~Zhou, B.~Hu, Z.-Q. Fu, B.-H. Wang, Efficient routing on complex
  networks, Physical Review E 73~(4) (2006) 046108.

\bibitem{danila2006optimal}
B.~Danila, Y.~Yu, J.~A. Marsh, K.~E. Bassler, Optimal transport on complex
  networks, Physical Review E—Statistical, Nonlinear, and Soft Matter Physics
  74~(4) (2006) 046106.

\bibitem{fraigniaud2014greedy}
P.~Fraigniaud, G.~Giakkoupis, Greedy routing in small-world networks with
  power-law degrees, in: Proceedings of the 2014 ACM symposium on Principles of
  distributed computing, 2014, pp. 311--320.

\bibitem{kleinberg2000small}
J.~Kleinberg, The small-world phenomenon: An algorithmic perspective, in:
  Proceedings of the thirty-second annual ACM symposium on Theory of computing,
  2000, pp. 163--170.

\bibitem{kleinberg2004small}
J.~Kleinberg, The small-world phenomenon and decentralized search, SiAM News
  37~(3) (2004) 1--2.

\bibitem{kleinrock1977hierarchical}
L.~Kleinrock, F.~Kamoun, Hierarchical routing for large networks performance
  evaluation and optimization, Computer Networks (1976) 1~(3) (1977) 155--174.

\bibitem{corominas2002hierarchical}
B.~Corominas-Murtra, R.~V. Sol{\'e}, Hierarchical social networks and
  information flow, Physica A: Statistical mechanics and its applications
  316~(1-4) (2002) 695--708.

\bibitem{granovetter1978threshold}
M.~Granovetter, \href{https://www.jstor.org/stable/2778111}{Threshold models of
  collective behavior}, Am. J. Sociol. 83~(6) (1978) 1420--1443.
\newblock \href {https://doi.org/10.1086/226707} {\path{doi:10.1086/226707}}.
\newline\urlprefix\url{https://www.jstor.org/stable/2778111}

\bibitem{schelling2006micromotives}
T.~C. Schelling, Micromotives and macrobehavior, WW Norton \& Company, 2006.

\bibitem{bikhchandani1992theory}
S.~Bikhchandani, D.~Hirshleifer, I.~Welch, A theory of fads, fashion, custom,
  and cultural change as informational cascades, Journal of political Economy
  100~(5) (1992) 992--1026.

\bibitem{banerjee1992simple}
A.~V. Banerjee, A simple model of herd behavior, The Quarterly Journal of
  Economics 107~(3) (1992) 797--817.

\bibitem{acemoglu2011opinion}
D.~Acemoglu, A.~Ozdaglar, Opinion dynamics and learning in social networks,
  Dynamic Games and Applications 1~(1) (2011) 3--49.

\bibitem{bharathi2007competitive}
S.~Bharathi, D.~Kempe, M.~Salek, Competitive influence maximization in social
  networks, in: International workshop on web and internet economics, Springer,
  2007, pp. 306--311.

\bibitem{gleeson2007seed}
J.~P. Gleeson, D.~J. Cahalane, Seed size strongly affects cascades on random
  networks, Physical Review E—Statistical, Nonlinear, and Soft Matter Physics
  75~(5) (2007) 056103.

\bibitem{leskovec2007cost}
J.~Leskovec, A.~Krause, C.~Guestrin, C.~Faloutsos, J.~VanBriesen, N.~Glance,
  Cost-effective outbreak detection in networks, in: Proceedings of the 13th
  ACM SIGKDD international conference on Knowledge discovery and data mining,
  2007, pp. 420--429.

\bibitem{cohen2014sketch}
E.~Cohen, D.~Delling, T.~Pajor, R.~F. Werneck, Sketch-based influence
  maximization and computation: Scaling up with guarantees, in: Proceedings of
  the 23rd ACM international conference on conference on information and
  knowledge management, 2014, pp. 629--638.

\bibitem{borgs2014maximizing}
C.~Borgs, M.~Brautbar, J.~Chayes, B.~Lucier, Maximizing social influence in
  nearly optimal time, in: Proceedings of the twenty-fifth annual ACM-SIAM
  symposium on Discrete algorithms, SIAM, 2014, pp. 946--957.

\bibitem{chen2010scalable}
W.~Chen, C.~Wang, Y.~Wang, Scalable influence maximization for prevalent viral
  marketing in large-scale social networks, in: Proceedings of the 16th ACM
  SIGKDD international conference on Knowledge discovery and data mining, 2010,
  pp. 1029--1038.

\bibitem{tang2014influence}
Y.~Tang, X.~Xiao, Y.~Shi, Influence maximization: Near-optimal time complexity
  meets practical efficiency, in: Proceedings of the 2014 ACM SIGMOD
  international conference on Management of data, 2014, pp. 75--86.

\bibitem{jackson2010social}
M.~O. Jackson, Social and Economic Networks, Princeton University Press,
  Princeton, NJ, 2010.

\bibitem{easley2010networks}
D.~Easley, J.~Kleinberg, et~al., Networks, crowds, and markets: Reasoning about
  a highly connected world, Vol.~1, Cambridge university press Cambridge, 2010.

\bibitem{ising1925beitrag}
E.~Ising, Beitrag zur theorie des ferromagnetismus, Zeitschrift f{\"u}r Physik
  A Hadrons and Nuclei 31~(1) (1925) 253--258.

\bibitem{stanley1971phase}
H.~E. Stanley, Phase transitions and critical phenomena, Clarendon Press,
  Oxford, 1971.

\bibitem{baxter2016exactly}
R.~J. Baxter, Exactly solved models in statistical mechanics, Elsevier, 2016.

\bibitem{stanley1999scaling}
H.~E. Stanley, Scaling, universality, and renormalization: Three pillars of
  modern critical phenomena, Reviews of modern physics 71~(2) (1999) S358.

\bibitem{beggs2003neuronal}
J.~M. Beggs, D.~Plenz, Neuronal avalanches in neocortical circuits, Journal of
  neuroscience 23~(35) (2003) 11167--11177.

\bibitem{kadanoff1966scaling}
L.~P. Kadanoff, Scaling laws for ising models near t c, Physics Physique Fizika
  2~(6) (1966) 263.

\bibitem{fisher1974renormalization}
M.~E. Fisher, The renormalization group in the theory of critical behavior,
  Reviews of Modern Physics 46~(4) (1974) 597.

\bibitem{galam1991towards}
S.~Galam, S.~Moscovici, Towards a theory of collective phenomena: Consensus and
  attitude changes in groups, European Journal of Social Psychology 21~(1)
  (1991) 49--74.

\bibitem{weidlich2006sociodynamics}
W.~Weidlich, Sociodynamics: A systematic approach to mathematical modelling in
  the social sciences, Courier Corporation, 2006.

\bibitem{suchecki2005voter}
K.~Suchecki, V.~M. Egu{\'\i}luz, M.~San~Miguel, Voter model dynamics in complex
  networks: Role of dimensionality, disorder, and degree distribution, Physical
  Review E 72~(3) (2005) 036132.

\bibitem{gleeson2013binary}
J.~P. Gleeson, {Binary-state dynamics on complex networks: Pair approximation
  and beyond}, Physical Review X 3~(2) (2013) 021004.

\bibitem{strogatz2000kuramoto}
S.~H. Strogatz, From kuramoto to crawford: exploring the onset of
  synchronization in populations of coupled oscillators, Physica D: Nonlinear
  Phenomena 143~(1-4) (2000) 1--20.

\bibitem{rodrigues2016kuramoto}
F.~A. Rodrigues, T.~K.~D. Peron, P.~Ji, J.~Kurths, The kuramoto model in
  complex networks, Physics Reports 610 (2016) 1--98.

\bibitem{hong2011kuramoto}
H.~Hong, S.~H. Strogatz, Kuramoto model of coupled oscillators with positive
  and negative coupling parameters: an example of conformist and contrarian
  oscillators, Physical Review Letters 106~(5) (2011) 054102.

\bibitem{agcom2020journalism}
{AGCOM}, \href{https://www.agcom.it/osservatorio-sul-giornalismo}{Osservatorio
  sul giornalismo}, report nazionale, Accessed: 2025-09-21 (2020).
\newline\urlprefix\url{https://www.agcom.it/osservatorio-sul-giornalismo}

\bibitem{reuters2025dnr}
R.~I. for the Study~of Journalism,
  \href{https://reutersinstitute.politics.ox.ac.uk/digital-news-report/2025}{Digital
  news report 2024}, accessed: 2025-09-21 (2025).
\newline\urlprefix\url{https://reutersinstitute.politics.ox.ac.uk/digital-news-report/2025}

\bibitem{x_api_about}
{X Developer Documentation},
  \href{https://docs.x.com/x-api/getting-started/about-x-api}{About the x api},
  Online, accessed: 2025-09-21 (2025).
\newline\urlprefix\url{https://docs.x.com/x-api/getting-started/about-x-api}

\bibitem{EU_DSA_2022_Article40}
{European Union}, Regulation (eu) 2022/2065: Digital services act, article 40
  – data access and scrutiny,
  \url{https://eur-lex.europa.eu/eli/reg/2022/2065/oj/eng}, see Article 40 of
  Regulation (EU) 2022/2065 for obligations on providers of very large online
  platforms to share data with vetted researchers under conditions defined by
  the DSA. (October 2022).

\bibitem{romero2011differences}
D.~M. Romero, B.~Meeder, J.~Kleinberg, Differences in the mechanics of
  information diffusion across topics: Idioms, political hashtags, and complex
  contagion on twitter, in: Proceedings of the 20th International Conference on
  World Wide Web (WWW '11), 2011, pp. 695--704.

\bibitem{kwak2010what}
H.~Kwak, C.~Lee, H.~Park, S.~Moon,
  \href{https://dl.acm.org/doi/10.1145/1772690.1772751}{{What is Twitter, a
  social network or a news media?}}, in: Proceedings of the 19th International
  Conference on World Wide Web, WWW '10, ACM, New York, NY, USA, 2010, pp.
  591--600.
\newblock \href {https://doi.org/10.1145/1772690.1772751}
  {\path{doi:10.1145/1772690.1772751}}.
\newline\urlprefix\url{https://dl.acm.org/doi/10.1145/1772690.1772751}

\bibitem{goel2015structural}
S.~Goel, A.~Anderson, J.~Hofman, D.~J. Watts, The structural virality of online
  diffusion, Management Science 62~(1) (2015) 180--196.

\bibitem{zhao2018trendburst}
C.~Zhao, Y.~Chen, Y.~Wang, N.~Li, Trendburst: Detecting and summarizing bursts
  in microblog streams, Journal of the Association for Information Science and
  Technology 69~(7) (2018) 903--920.

\bibitem{kobayashi2021tideh}
R.~Kobayashi, R.~Lambiotte,
  \href{http://dx.doi.org/10.1609/icwsm.v10i1.14717}{Tideh: Time-dependent
  hawkes process for predicting retweet dynamics}, Proceedings of the
  International AAAI Conference on Web and Social Media 10~(1) (2021)
  191–200.
\newblock \href {https://doi.org/10.1609/icwsm.v10i1.14717}
  {\path{doi:10.1609/icwsm.v10i1.14717}}.
\newline\urlprefix\url{http://dx.doi.org/10.1609/icwsm.v10i1.14717}

\bibitem{Lerman2009Computational}
K.~Lerman, R.~Ghosh, Computational social dynamics: the technical perspective,
  EPJ Data Science 2~(1) (2009) 4.

\bibitem{Morstatter2013Sample}
F.~Morstatter, J.~Pfeffer, H.~Liu, K.~M. Carley, Is the sample good enough?
  comparing data from twitter's streaming api with twitter's firehose, in:
  Proceedings of the 7th International AAAI Conference on Weblogs and Social
  Media (ICWSM '13), 2013.

\bibitem{meta_content_library_2025}
{Meta}, Meta content library and api,
  \url{https://transparency.meta.com/en-gb/researchtools/meta-content-library/},
  transparency Center (2025).

\bibitem{icpsr_2025}
{ICPSR}, Inter-university consortium for political and social research,
  \url{https://www.icpsr.umich.edu/sites/icpsr/home}, institute for Social
  Research, University of Michigan (2025).

\bibitem{SocialMediaArchive_FAQ}
S.~M. Archive, Faqs,
  \url{https://socialmediaarchive.org/pages/?page=FAQs&ln=en}, accessed:
  2025-09-21.

\bibitem{bachrach2012personality}
Y.~Bachrach, M.~Kosinski, T.~Graepel, P.~Kohli, D.~Stillwell, Personality and
  patterns of facebook usage, in: Proceedings of the 4th annual ACM web science
  conference, 2012, pp. 24--32.

\bibitem{quercia2012personality}
D.~Quercia, R.~Lambiotte, D.~Stillwell, M.~Kosinski, J.~Crowcroft, The
  personality of popular facebook users, in: Proceedings of the ACM 2012
  conference on computer supported cooperative work, 2012, pp. 955--964.

\bibitem{cadwalladr2017great}
C.~Cadwalladr, The great british brexit robbery: how our democracy was
  hijacked, The Guardian 7 (2017).

\bibitem{ugander2011anatomy}
J.~Ugander, B.~Karrer, L.~Backstrom, C.~Marlow,
  \href{http://arxiv.org/abs/1111.4503}{{The Anatomy of the Facebook Social
  Graph}} (nov 2011).
\newblock \href {http://arxiv.org/abs/1111.4503} {\path{arXiv:1111.4503}}.
\newline\urlprefix\url{http://arxiv.org/abs/1111.4503}

\bibitem{backstrom2012four}
L.~Backstrom, P.~Boldi, M.~Rosa, J.~Ugander, S.~Vigna, Four degrees of
  separation, in: Proceedings of the 4th annual ACM Web science conference,
  2012, pp. 33--42.

\bibitem{traud2012social}
A.~L. Traud, P.~J. Mucha, M.~A. Porter,
  \href{https://www.sciencedirect.com/science/article/abs/pii/S0378437111009186}{{Social
  structure of Facebook networks}}, Physica A: Statistical Mechanics and its
  Applications 391~(16) (2012) 4165--4180.
\newblock \href {http://arxiv.org/abs/1102.2166} {\path{arXiv:1102.2166}},
  \href {https://doi.org/10.1016/j.physa.2011.12.021}
  {\path{doi:10.1016/j.physa.2011.12.021}}.
\newline\urlprefix\url{https://www.sciencedirect.com/science/article/abs/pii/S0378437111009186}

\bibitem{centola2010spread}
D.~Centola, The spread of behavior in an online social network experiment,
  Science 329~(5996) (2010) 1194--1197.

\bibitem{bakshy2015exposure}
E.~Bakshy, S.~Messing, L.~A. Adamic,
  \href{https://www.science.org/doi/10.1126/science.aaa1160}{{Exposure to
  ideologically diverse news and opinion on Facebook}}, Science 348~(6239)
  (2015) 1130--1132.
\newblock \href {https://doi.org/10.1126/science.aaa1160}
  {\path{doi:10.1126/science.aaa1160}}.
\newline\urlprefix\url{https://www.science.org/doi/10.1126/science.aaa1160}

\bibitem{preston2021detecting}
S.~Preston, A.~Anderson, D.~J. Robertson, M.~P. Shephard, N.~Huhe, Detecting
  fake news on facebook: The role of emotional intelligence, Plos one 16~(3)
  (2021) e0246757.

\bibitem{gonzalez-bailon2014social}
S.~González-Bailón, N.~Shao, J.~Dunbar, G.~Cui, Social science in the age of
  social media, PeerJ Computer Science 2014 (2014) 1--17.

\bibitem{reddit_q2_2025}
{Reddit, Inc.},
  \href{https://s203.q4cdn.com/380862485/files/doc_financials/2025/q2/Q2-25-Earnings-Press-Release.pdf}{Reddit
  announces second quarter 2025 results}, Quarterly earnings report, Reddit,
  Inc., q2 2025 Financial Results and Key Operating Metrics (July 2025).
\newline\urlprefix\url{https://s203.q4cdn.com/380862485/files/doc_financials/2025/q2/Q2-25-Earnings-Press-Release.pdf}

\bibitem{weninger2012exploration}
T.~Weninger, D.~Farag, R.~Choudhury, J.~Han, An exploration of discussion
  cascades in social news websites, in: International AAAI Conference on
  Weblogs and Social Media (ICWSM), 2012.

\bibitem{Janetschek2021Understanding}
H.~Janetschek, M.~Salathé, Understanding misinformation dynamics in reddit
  communities, EPJ Data Science 10~(15) (2021).

\bibitem{Kumar2018Community}
S.~Kumar, F.~Speicher, V.~Subrahmanian, K.~Lerman, Community interaction and
  conflict on the web, Proceedings of the National Academy of Sciences 116~(6)
  (2018) 185--190.

\bibitem{Hu2014What}
Y.~Hu, L.~Manikonda, S.~Kambhampati, What we instagram: A first analysis of
  instagram photo content and user types, in: International AAAI Conference on
  Weblogs and Social Media (ICWSM), 2014.

\bibitem{highfield2016instagrammatics}
T.~Highfield, T.~Leaver, Instagrammatics and digital methods: Studying visual
  social media, from selfies and gifs to memes and emoji, Communication
  Research and Practice 2~(1) (2016) 47--62.

\bibitem{Zhou2017Rumor}
X.~Zhou, Q.~Guo, H.~Lin, Rumor spreading model with reader node influence on
  microblogging networks, Physica A: Statistical Mechanics and its Applications
  467 (2017) 154--168.

\bibitem{Yang2012Rumor}
S.~Yang, C.~Zhou, L.~Shi, Rumor tracking on social media: An ai-based approach,
  in: IEEE International Conference on Data Mining, 2012.

\bibitem{Kumar2020Characterizing}
S.~Kumar, R.~West, J.~Leskovec, Characterizing misinformation propagation on
  whatsapp, in: IEEE International Conference on Data Mining, 2020.

\bibitem{lerman2010information}
K.~Lerman, R.~Ghosh, Information contagion: An empirical study of the spread of
  news on digg and twitter social networks, in: Proceedings of the
  International AAAI Conference on Web and Social Media, Vol.~4, 2010, pp.
  90--97.

\bibitem{Altay2018Media}
S.~Altay, J.-Y. Choi, G.~Bauer, Media sharing and cross-platform contagion in
  news ecosystems, Journal of Communication 68~(3) (2018) 583--605.

\bibitem{wardle2017information}
C.~Wardle, H.~Derakhshan,
  \href{https://rm.coe.int/information-disorder-toward-an-interdisciplinary-framework-for-researc/168076277c}{Information
  disorder: Toward an interdisciplinary framework for research and policy
  making}, Council of Europe Policy Report DGI(2017)09, Council of Europe
  (2017).
\newline\urlprefix\url{https://rm.coe.int/information-disorder-toward-an-interdisciplinary-framework-for-researc/168076277c}

\bibitem{lazer2018science}
D.~M. Lazer, M.~A. Baum, Y.~Benkler, A.~J. Berinsky, K.~M. Greenhill,
  F.~Menczer, M.~J. Metzger, B.~Nyhan, G.~Pennycook, D.~Rothschild,
  M.~Schudson, S.~A. Sloman, C.~R. Sunstein, E.~A. Thorson, D.~J. Watts, J.~L.
  Zittrain, \href{http://www.ncbi.nlm.nih.gov/pubmed/29590025}{{The science of
  fake news: Addressing fake news requires a multidisciplinary effort}},
  Science 359~(6380) (2018) 1094--1096.
\newblock \href {https://doi.org/10.1126/science.aao2998}
  {\path{doi:10.1126/science.aao2998}}.
\newline\urlprefix\url{http://www.ncbi.nlm.nih.gov/pubmed/29590025}

\bibitem{vosoughi2018spread}
S.~Vosoughi, D.~Roy, S.~Aral,
  \href{https://www.science.org/doi/10.1126/science.aap9559}{{The spread of
  true and false news online}}, Science 359~(6380) (2018) 1146--1151.
\newblock \href {https://doi.org/10.1126/science.aap9559}
  {\path{doi:10.1126/science.aap9559}}.
\newline\urlprefix\url{https://www.science.org/doi/10.1126/science.aap9559}

\bibitem{grinberg2019political}
N.~Grinberg, K.~Joseph, L.~Friedland, B.~Swire-Thompson, D.~Lazer,
  \href{https://www.science.org/doi/10.1126/science.aau2706}{{Political
  science: Fake news on Twitter during the 2016 U.S. presidential election}},
  Science 363~(6425) (2019) 374--378.
\newblock \href {https://doi.org/10.1126/science.aau2706}
  {\path{doi:10.1126/science.aau2706}}.
\newline\urlprefix\url{https://www.science.org/doi/10.1126/science.aau2706}

\bibitem{juul2021comparing}
J.~L. Juul, J.~Ugander,
  \href{https://pnas.org/doi/full/10.1073/pnas.2100786118}{{Comparing
  information diffusion mechanisms by matching on cascade size}}, Proceedings
  of the National Academy of Sciences of the United States of America 118~(46)
  (2021) e2100786118.
\newblock \href {https://doi.org/10.1073/pnas.2100786118}
  {\path{doi:10.1073/pnas.2100786118}}.
\newline\urlprefix\url{https://pnas.org/doi/full/10.1073/pnas.2100786118}

\bibitem{vincent2025measuring}
E.~M. Vincent, D.~Crisan, B.~Carniel,
  \href{https://science.feedback.org/wp-content/uploads/2025/09/SIMODS-Report-1.pdf}{Measuring
  the state of online disinformation in europe on very large online platforms:
  First report of the simods project (structural indicators to monitor online
  disinformation scientifically)}, Technical report, Science Feedback (SIMODS
  consortium), how to cite (inside PDF): "Vincent EM, Crisan D, Carniel B
  (2025)". Accessed: 2025-10-03 (sep 2025).
\newline\urlprefix\url{https://science.feedback.org/wp-content/uploads/2025/09/SIMODS-Report-1.pdf}

\bibitem{keller2020political}
F.~B. Keller, D.~Schoch, S.~Stier, J.~Yang, Political astroturfing on twitter:
  How to coordinate a disinformation campaign, Political communication 37~(2)
  (2020) 256--280.

\bibitem{delvicario2016spreading}
M.~D. Vicario, A.~Bessi, F.~Zollo, F.~Petroni, A.~Scala, G.~Caldarelli, H.~E.
  Stanley, W.~Quattrociocchi,
  \href{https://pnas.org/doi/full/10.1073/pnas.1517441113}{{The spreading of
  misinformation online}}, Proceedings of the National Academy of Sciences of
  the United States of America 113~(3) (2016) 554--559.
\newblock \href {https://doi.org/10.1073/pnas.1517441113}
  {\path{doi:10.1073/pnas.1517441113}}.
\newline\urlprefix\url{https://pnas.org/doi/full/10.1073/pnas.1517441113}

\bibitem{weedon2017information}
J.~Weedon, W.~Nuland, A.~Stamos, Information operations and facebook, Retrieved
  from Facebook: https://fbnewsroomus. files. wordpress.
  com/2017/04/facebook-and-information-operations-v1. pdf (2017).

\bibitem{scheufele2019science}
D.~A. Scheufele, N.~M. Krause,
  \href{http://dx.doi.org/10.1073/pnas.1805871115}{Science audiences,
  misinformation, and fake news}, Proceedings of the National Academy of
  Sciences 116~(16) (2019) 7662–7669.
\newblock \href {https://doi.org/10.1073/pnas.1805871115}
  {\path{doi:10.1073/pnas.1805871115}}.
\newline\urlprefix\url{http://dx.doi.org/10.1073/pnas.1805871115}

\bibitem{davis2016botornot}
C.~A. Davis, O.~Varol, E.~Ferrara, A.~Flammini, F.~Menczer,
  \href{http://dl.acm.org/citation.cfm?doid=2872518.2889302}{{BotOrNot: A
  System to Evaluate Social Bots}}, in: WWW 2016 Companion - Proceedings of the
  25th International Conference on World Wide Web, ACM Press, New York, New
  York, USA, 2016, pp. 273--274.
\newblock \href {http://arxiv.org/abs/1602.00975} {\path{arXiv:1602.00975}},
  \href {https://doi.org/10.1145/2872518.2889302}
  {\path{doi:10.1145/2872518.2889302}}.
\newline\urlprefix\url{http://dl.acm.org/citation.cfm?doid=2872518.2889302}

\bibitem{tangcharoensathien2020framework}
V.~Tangcharoensathien, N.~Calleja, T.~Nguyen, T.~Purnat, M.~D’Agostino,
  S.~Garcia-Saiso, M.~Landry, A.~Rashidian, C.~Hamilton, A.~AbdAllah, I.~Ghiga,
  A.~Hill, D.~Hougendobler, J.~van Andel, M.~Nunn, I.~Brooks, P.~L. Sacco,
  M.~De~Domenico, P.~Mai, A.~Gruzd, A.~Alaphilippe, S.~Briand,
  \href{http://dx.doi.org/10.2196/19659}{Framework for managing the covid-19
  infodemic: Methods and results of an online, crowdsourced who technical
  consultation}, Journal of Medical Internet Research 22~(6) (2020) e19659.
\newblock \href {https://doi.org/10.2196/19659} {\path{doi:10.2196/19659}}.
\newline\urlprefix\url{http://dx.doi.org/10.2196/19659}

\bibitem{calleja2021public}
N.~Calleja, A.~AbdAllah, N.~Abad, N.~Ahmed, D.~Albarracin, E.~Altieri, J.~N.
  Anoko, R.~Arcos, A.~A. Azlan, J.~Bayer, A.~Bechmann, S.~Bezbaruah, S.~C.
  Briand, I.~Brooks, L.~M. Bucci, S.~Burzo, C.~Czerniak, M.~De~Domenico, A.~G.
  Dunn, U.~K.~H. Ecker, L.~Espinosa, C.~Francois, K.~Gradon, A.~Gruzd, B.~S.
  G\"{u}lg\"{u}n, R.~Haydarov, C.~Hurley, S.~I. Astuti, A.~Ishizumi,
  N.~Johnson, D.~Johnson~Restrepo, M.~Kajimoto, A.~Koyuncu, S.~Kulkarni,
  J.~Lamichhane, R.~Lewis, A.~Mahajan, A.~Mandil, E.~McAweeney, M.~Messer,
  W.~Moy, P.~Ndumbi~Ngamala, T.~Nguyen, M.~Nunn, S.~B. Omer, C.~Pagliari,
  P.~Patel, L.~Phuong, D.~Prybylski, A.~Rashidian, E.~Rempel, S.~Rubinelli,
  P.~Sacco, A.~Schneider, K.~Shu, M.~Smith, H.~Sufehmi, V.~Tangcharoensathien,
  R.~Terry, N.~Thacker, T.~Trewinnard, S.~Turner, H.~Tworek, S.~Uakkas,
  E.~Vraga, C.~Wardle, H.~Wasserman, E.~Wilhelm, A.~W\"{u}rz, B.~Yau, L.~Zhou,
  T.~D. Purnat, \href{http://dx.doi.org/10.2196/30979}{A public health research
  agenda for managing infodemics: Methods and results of the first who
  infodemiology conference}, JMIR Infodemiology 1~(1) (2021) e30979.
\newblock \href {https://doi.org/10.2196/30979} {\path{doi:10.2196/30979}}.
\newline\urlprefix\url{http://dx.doi.org/10.2196/30979}

\bibitem{jordan1982human}
T.~G. Jordan (Ed.), The Human Mosaic, Harper Row, New York, 1982.

\bibitem{lazarsfeld1944people}
P.~F. Lazarsfeld, B.~Berelson, H.~Gaudet, The People's Choice, Columbia
  University Press, New York, 1944.

\bibitem{katz1955personal}
E.~Katz, P.~F. Lazarsfeld, Personal Influence: The Part Played by People in the
  Flow of Mass Communications, The Free Press, Glencoe IL, 1955.

\bibitem{cha2012world}
M.~Cha, F.~Benevenuto, H.~Haddadi, K.~Gummadi, The world of connections and
  information flow in {T}witter, IEEE Trans. Syst. Man Cybern. Syst. 42~(4)
  (2012) 991--998.
\newblock \href {https://doi.org/10.1109/TSMCA.2012.2183359}
  {\path{doi:10.1109/TSMCA.2012.2183359}}.

\bibitem{harrison1993kinetic}
L.~G. Harrison, Kinetic Theory of Living Pattern, Cambridge University Press,
  Cambridge, 1993.

\bibitem{rosen2002anatomy}
E.~Rosen, The anatomy of buzz: How to create word of mouth marketing, Crown
  Currency, New York, 2002.

\bibitem{rogers1983diffusion}
E.~M. Rogers, Diffusion of Innovations, 3rd Edition, The Free Press, New York,
  1983.

\bibitem{hagerstrand1953innovationsf}
T.~H\"agerstrand, Innovationsf\"orloppet ur korologisk synpunkt, Ph.D. thesis,
  Lund University, Lund (1953).

\bibitem{valente1996network}
T.~W. Valente, \href{https://doi.org/10.1007/BF00240425}{Network models of the
  diffusion of innovations}, Comput. Math. Organ. Theory 2~(2) (1996) 163--164.
\newblock \href {https://doi.org/10.1007/BF00240425}
  {\path{doi:10.1007/BF00240425}}.
\newline\urlprefix\url{https://doi.org/10.1007/BF00240425}

\bibitem{valente1996social}
T.~W. Valente, Social network thresholds in the diffusion of innovations, Soc.
  Netw. 18~(1) (1996) 69--89.

\bibitem{toole2012modeling}
J.~L. Toole, M.~Cha, M.~C. Gonz\'alez, Modeling the adoption of innovations in
  the presence of geographic and media influences, PLOS ONE 7 (2012) 0029528.

\bibitem{arieli2020speed}
I.~Arieli, Y.~Babichenko, R.~Peretz, H.~P. Young, The speed of innovation
  diffusion in social networks, Econometrica 88~(2) (2020) 569--594.

\bibitem{kooti2021emergence}
F.~Kooti, H.~Yang, M.~Cha, K.~Gummadi, W.~Mason,
  \href{https://ojs.aaai.org/index.php/ICWSM/article/view/14267}{The emergence
  of conventions in online social networks}, Proceedings of the International
  AAAI Conference on Web and Social Media 6~(1) (2021) 194--201.
\newblock \href {https://doi.org/10.1609/icwsm.v6i1.14267}
  {\path{doi:10.1609/icwsm.v6i1.14267}}.
\newline\urlprefix\url{https://ojs.aaai.org/index.php/ICWSM/article/view/14267}

\bibitem{eisenstein2014diffusion}
J.~Eisenstein, B.~O'Connor, N.~A. Smith, E.~P. Xing, Diffusion of lexical
  change in social media, PloS one 9~(11) (2014) e113114.

\bibitem{zajonc1968attitudinal}
R.~B. Zajonc, Attitudinal effects of mere exposure., Journal of personality and
  social psychology 9~(2p2) (1968) 1.

\bibitem{hassan2021effects}
A.~Hassan, S.~J. Barber, The effects of repetition frequency on the illusory
  truth effect, Cognitive research: principles and implications 6~(1) (2021)
  38.

\bibitem{nickerson1998confirmation}
R.~S. Nickerson, Confirmation bias: A ubiquitous phenomenon in many guises,
  Review of general psychology 2~(2) (1998) 175--220.

\bibitem{berger2012what}
J.~Berger, K.~L. Milkman, What makes online content viral?, Journal of
  marketing research 49~(2) (2012) 192--205.

\bibitem{fodor2016how}
{\'E}.~Fodor, C.~Nardini, M.~E. Cates, J.~Tailleur, P.~Visco, F.~Van~Wijland,
  How far from equilibrium is active matter?, Physical review letters 117~(3)
  (2016) 038103.

\bibitem{shaebani2020computational}
M.~R. Shaebani, A.~Wysocki, R.~G. Winkler, G.~Gompper, H.~Rieger, Computational
  models for active matter, Nature Reviews Physics 2~(4) (2020) 181--199.

\bibitem{liben-nowell2008tracing}
D.~Liben-Nowell, J.~Kleinberg, Tracing information flow on a global scale using
  internet chain-letter data, Proc. Natl. Acad. Sci. USA 105~(12) (2008)
  4633--4638.

\bibitem{cha2008characterizing}
M.~Cha, A.~Mislove, B.~Adams, K.~P. Gummadi,
  \href{https://doi.org/10.1145/1397735.1397739}{Characterizing social cascades
  in {Flickr}}, in: Proceedings of the First Workshop on Online Social
  Networks, WOSN '08, Association for Computing Machinery, New York, NY, USA,
  2008, p. 13–18.
\newblock \href {https://doi.org/10.1145/1397735.1397739}
  {\path{doi:10.1145/1397735.1397739}}.
\newline\urlprefix\url{https://doi.org/10.1145/1397735.1397739}

\bibitem{centola2018how}
D.~Centola, How Behavior Spreads, Princeton University Press, Princeton NJ,
  2018.

\bibitem{daley1964epidemics}
D.~Daley, D.~Kendall, Epidemics and rumours, Nature 204 (1964) 1118.

\bibitem{daley1965stochastic}
D.~J. Daley, D.~G. Kendall, Stochastic rumours, IMA Journal of Applied
  Mathematics 1~(1) (1965) 42--55.

\bibitem{maki1973mathematical}
D.~P. Maki, M.~Thompson, Mathematical Models and Applications: With Emphasis on
  the Social, Life, and Management Sciences, Prentice-Hall, Englewood Cliffs,
  NJ, 1973.

\bibitem{nekovee2007theory}
M.~Nekovee, Y.~Moreno, G.~Bianconi, M.~Marsili, Theory of rumour spreading in
  complex social networks, Physica A: Statistical Mechanics and its
  Applications 374~(1) (2007) 457--470.

\bibitem{kandhway2014optimal}
K.~Kandhway, J.~Kuri, Optimal control of information epidemics modeled as maki
  thompson rumors, Communications in Nonlinear Science and Numerical Simulation
  19~(12) (2014) 4135--4147.

\bibitem{gani2000maki}
J.~Gani, The maki--thompson rumour model: a detailed analysis, Environmental
  Modelling \& Software 15~(8) (2000) 721--725.

\bibitem{belen2011classical}
S.~Belen, E.~Kropat, G.-W. Weber, On the classical maki--thompson rumour model
  in continuous time, Central European Journal of Operations Research 19~(1)
  (2011) 1--17.

\bibitem{ferraz2022subcritical}
G.~Ferraz~de Arruda, L.~G. Jeub, A.~S. Mata, F.~A. Rodrigues, Y.~Moreno, From
  subcritical behavior to a correlation-induced transition in rumor models,
  Nature Communications 13~(1) (2022) 3049.

\bibitem{aral2017exercise}
S.~Aral, C.~Nicolaides,
  \href{https://www.nature.com/articles/ncomms14753}{Exercise contagion in a
  global social network}, Nat. Commun. 8 (2017) 14753.
\newblock \href {https://doi.org/10.1038/ncomms14753}
  {\path{doi:10.1038/ncomms14753}}.
\newline\urlprefix\url{https://www.nature.com/articles/ncomms14753}

\bibitem{berk1974gaming}
R.~A. Berk, A gaming approach to crowd behavior, Am. Sociol. Rev. (1974)
  355--373.

\bibitem{karimi2013threshold}
F.~Karimi, P.~Holme, Threshold model of cascades in empirical temporal
  networks, Physica A: Statistical Mechanics and its Applications 392~(16)
  (2013) 3476--3483.

\bibitem{takaguchi2013bursty}
T.~Takaguchi, N.~Masuda, P.~Holme, Bursty communication patterns facilitate
  spreading in a threshold-based epidemic dynamics, PLOS ONE 8~(7) (2013)
  e68629.

\bibitem{watts2007influentials}
D.~J. Watts, P.~S. Dodds,
  \href{https://academic.oup.com/jcr/article-lookup/doi/10.1086/518527}{{Influentials,
  networks, and public opinion formation}}, Journal of Consumer Research 34~(4)
  (2007) 441--458.
\newblock \href {https://doi.org/10.1086/518527} {\path{doi:10.1086/518527}}.
\newline\urlprefix\url{https://academic.oup.com/jcr/article-lookup/doi/10.1086/518527}

\bibitem{centola2007cascade}
D.~Centola, V.~M. Egu\'iluz, M.~W. Macy,
  \href{https://www.sciencedirect.com/science/article/pii/S0378437106007679}{Cascade
  dynamics of complex propagation}, Physica A: Statistical Mechanics and its
  Applications 374~(1) (2007) 449--456.
\newblock \href {https://doi.org/https://doi.org/10.1016/j.physa.2006.06.018}
  {\path{doi:https://doi.org/10.1016/j.physa.2006.06.018}}.
\newline\urlprefix\url{https://www.sciencedirect.com/science/article/pii/S0378437106007679}

\bibitem{gronlund2005network}
A.~Grönlund, P.~Holme, A network-based threshold model for the spreading of
  fads in society and markets, Adv. Complex Syst. 8~(2) (2005) 261--273.
\newblock \href {https://doi.org/10.1142/S0219525905000439}
  {\path{doi:10.1142/S0219525905000439}}.

\bibitem{lee2017social}
E.~Lee, P.~Holme, Social contagion with degree-dependent thresholds, Phys. Rev.
  E 96~(1) (2017) 012315.
\newblock \href {https://doi.org/10.1103/PhysRevE.96.012315}
  {\path{doi:10.1103/PhysRevE.96.012315}}.

\bibitem{nishioka2022cascading}
S.~Nishioka, T.~Hasegawa, Cascading behavior of an extended watts model on
  networks, J. Phys. Soc. Jpn. 91~(12) (2022) 124801.

\bibitem{gleeson2008cascades}
J.~P. Gleeson,
  \href{https://link.aps.org/doi/10.1103/PhysRevE.77.046117}{{Cascades on
  correlated and modular random networks}}, Physical Review E - Statistical,
  Nonlinear, and Soft Matter Physics 77~(4) (2008) 046117.
\newblock \href {https://doi.org/10.1103/PhysRevE.77.046117}
  {\path{doi:10.1103/PhysRevE.77.046117}}.
\newline\urlprefix\url{https://link.aps.org/doi/10.1103/PhysRevE.77.046117}

\bibitem{centola2007complex}
D.~Centola, M.~Macy, Complex contagions and the weakness of long ties, American
  journal of Sociology 113~(3) (2007) 702--734.

\bibitem{dodds2004universal}
P.~S. Dodds, D.~J. Watts,
  \href{https://link.aps.org/doi/10.1103/PhysRevLett.92.218701}{{Universal
  behavior in a generalized model of contagion}}, Physical Review Letters
  92~(21) (2004) 218701.
\newblock \href {http://arxiv.org/abs/0403699} {\path{arXiv:0403699}}, \href
  {https://doi.org/10.1103/PhysRevLett.92.218701}
  {\path{doi:10.1103/PhysRevLett.92.218701}}.
\newline\urlprefix\url{https://link.aps.org/doi/10.1103/PhysRevLett.92.218701}

\bibitem{bianconi2021higher}
G.~Bianconi, Higher-Order Networks, Elements in the Structure and Dynamics of
  Complex Networks, Cambridge University Press, 2021.

\bibitem{majhi2022dynamics}
S.~Majhi, M.~Perc, D.~Ghosh, Dynamics on higher-order networks: A review, J.
  Roy. Soc. Interface 19~(188) (2022) 20220043.

\bibitem{iacopini2019simplicial}
I.~Iacopini, G.~Petri, A.~Barrat, V.~Latora, Simplicial models of social
  contagion, Nat. Commun. 10~(1) (2019) 2485.

\bibitem{cencetti2023distinguishing}
G.~Cencetti, D.~A. Contreras, M.~Mancastroppa, A.~Barrat, Distinguishing simple
  and complex contagion processes on networks, Phys. Rev. Lett. 130~(24) (2023)
  247401.

\bibitem{lee2017modeling}
E.~Lee, P.~Holme, S.~H. Lee, Modeling the dynamics of dissent, Physica A 486
  (2017) 262--272.
\newblock \href {https://doi.org/10.1016/j.physa.2017.05.047}
  {\path{doi:10.1016/j.physa.2017.05.047}}.

\bibitem{holme2005modelling}
P.~Holme, A.~Gr\"onlund, \href{https://www.jasss.org/8/3/3.html}{Modelling the
  dynamics of youth subcultures}, J. Artif. Soc. Soc. Simul. 8~(3) (2005) 3.
\newline\urlprefix\url{https://www.jasss.org/8/3/3.html}

\bibitem{barrett2014sensitivity}
S.~Barrett, A.~Dannenberg,
  \href{https://www.nature.com/articles/nclimate2059}{Sensitivity of collective
  action to uncertainty about climate tipping points}, Nat. Clim. Change 4~(1)
  (2014) 36--39.
\newblock \href {https://doi.org/10.1038/nclimate2059}
  {\path{doi:10.1038/nclimate2059}}.
\newline\urlprefix\url{https://www.nature.com/articles/nclimate2059}

\bibitem{goffman1964generalization}
W.~Goffman, V.~Newill, Generalization of epidemic theory, Nature 204~(4955)
  (1964) 225--228.

\bibitem{ruan2015kinetics}
Z.~Ruan, G.~I{\~n}iguez, M.~Karsai, J.~Kert\'esz,
  \href{https://link.aps.org/doi/10.1103/PhysRevLett.115.218702}{Kinetics of
  social contagion}, Phys. Rev. Lett. 115 (2015) 218702.
\newblock \href {https://doi.org/10.1103/PhysRevLett.115.218702}
  {\path{doi:10.1103/PhysRevLett.115.218702}}.
\newline\urlprefix\url{https://link.aps.org/doi/10.1103/PhysRevLett.115.218702}

\bibitem{castiello2023using}
M.~Castiello, D.~Conte, S.~Iscaro,
  \href{http://dx.doi.org/10.3390/a16080391}{Using epidemiological models to
  predict the spread of information on twitter}, Algorithms 16~(8) (2023) 391.
\newblock \href {https://doi.org/10.3390/a16080391}
  {\path{doi:10.3390/a16080391}}.
\newline\urlprefix\url{http://dx.doi.org/10.3390/a16080391}

\bibitem{borgeholthoefer2012emergence}
J.~Borge-Holthoefer, S.~Meloni, B.~Gon\c{c}alves, Y.~Moreno,
  \href{http://dx.doi.org/10.1007/s10955-012-0595-6}{Emergence of influential
  spreaders in modified rumor models}, Journal of Statistical Physics
  151~(1–2) (2012) 383–393.
\newblock \href {https://doi.org/10.1007/s10955-012-0595-6}
  {\path{doi:10.1007/s10955-012-0595-6}}.
\newline\urlprefix\url{http://dx.doi.org/10.1007/s10955-012-0595-6}

\bibitem{sprague2017evidence}
D.~A. Sprague, T.~House, Evidence for complex contagion models of social
  contagion from observational data, PloS ONE 12~(7) (2017) e0180802.

\bibitem{hodas2014simple}
N.~O. Hodas, K.~Lerman, The simple rules of social contagion, Sci. Rep. 4~(1)
  (2014) 4343.

\bibitem{ugander2012structural}
J.~Ugander, L.~Backstrom, C.~Marlow, J.~Kleinberg,
  \href{https://www.pnas.org/doi/10.1073/pnas.1116502109}{Structural diversity
  in social contagion}, Proc. Natl. Acad. Sci. USA 109~(16) (2012) 5962--5966.
\newblock \href {https://doi.org/10.1073/pnas.1116502109}
  {\path{doi:10.1073/pnas.1116502109}}.
\newline\urlprefix\url{https://www.pnas.org/doi/10.1073/pnas.1116502109}

\bibitem{mason2012collaborative}
W.~Mason, D.~J. Watts, Collaborative learning in networks, Proc. Natl. Acad.
  Sci. USA 109~(3) (2012) 764--769.

\bibitem{feng2015competing}
L.~Feng, Y.~Hu, B.~Li, H.~E. Stanley, S.~Havlin, L.~A. Braunstein,
  \href{https://doi.org/10.1371/journal.pone.0126090}{Competing for attention
  in social media under information overload conditions}, PLOS ONE 10~(7)
  (2015) 1--13.
\newblock \href {https://doi.org/10.1371/journal.pone.0126090}
  {\path{doi:10.1371/journal.pone.0126090}}.
\newline\urlprefix\url{https://doi.org/10.1371/journal.pone.0126090}

\bibitem{xie2021detecting}
J.~Xie, F.~Meng, J.~Sun, X.~Ma, G.~Yan, Y.~Hu, Detecting and modelling real
  percolation and phase transitions of information on social media, Nature
  Human Behaviour 5~(9) (2021) 1161--1168.

\bibitem{sobkowicz2018opinion}
P.~Sobkowicz, Opinion dynamics model based on cognitive biases of complex
  agents, Journal of Artificial Societies and Social Simulation (JASSS) 21~(4)
  (2018) 8.

\bibitem{battiston2020networks}
F.~Battiston, G.~Cencetti, I.~Iacopini, V.~Latora, M.~Lucas, A.~Patania, J.~D.
  Young, G.~Petri, Networks beyond pairwise interactions: Structure and
  dynamics, Physics Reports 874 (2020) 1--92.

\bibitem{rimmon-kenan2003narrative}
S.~Rimmon-Kenan, Narrative fiction: Contemporary poetics, Routledge, 2003.

\bibitem{murray2015narrative}
M.~Murray, et~al., Narrative psychology, Qualitative psychology: A practical
  guide to research methods (2015) 85--107.

\bibitem{schank2013scripts}
R.~C. Schank, R.~P. Abelson, Scripts, plans, goals, and understanding: An
  inquiry into human knowledge structures, Psychology press, 2013.

\bibitem{bruner2009actual}
J.~S. Bruner, Actual minds, possible worlds, Harvard university press, 2009.

\bibitem{tversky1973availability}
A.~Tversky, D.~Kahneman, Availability: A heuristic for judging frequency and
  probability, Cognitive psychology 5~(2) (1973) 207--232.

\bibitem{nowak2023integration}
A.~Nowak, R.~R. Vallacher, W.~Bartkowski, L.~Olson, Integration and expression:
  The complementary functions of self-reflection, Journal of Personality 91~(4)
  (2023) 947--962.

\bibitem{wason1960failure}
P.~C. Wason, On the failure to eliminate hypotheses in a conceptual task,
  Quarterly journal of experimental psychology 12~(3) (1960) 129--140.

\bibitem{zollo2018social}
F.~Zollo, W.~Quattrociocchi, et~al., Social dynamics in the age of credulity:
  the misinformation risk and its fallout, in: Digital Dominance. The Power of
  Google, Amazon, Facebook, and Apple, Oxford University Press, 2018.

\bibitem{peruzzi2019confirmation}
A.~Peruzzi, F.~Zollo, A.~L. Schmidt, W.~Quattrociocchi, From confirmation bias
  to echo-chambers: A data-driven approach, Sociologia e Politiche
  Sociali~(2018/3) (2019).

\bibitem{stieglitz2013emotions}
S.~Stieglitz, L.~Dang-Xuan, Emotions and information diffusion in social
  media—sentiment of microblogs and sharing behavior, Journal of management
  information systems 29~(4) (2013) 217--248.

\bibitem{cotter2008influence}
E.~M. Cotter, Influence of emotional content and perceived relevance on spread
  of urban legends: A pilot study, Psychological reports 102~(2) (2008)
  623--629.

\bibitem{heath2001emotional}
C.~Heath, C.~Bell, E.~Sternberg, Emotional selection in memes: the case of
  urban legends., Journal of personality and social psychology 81~(6) (2001)
  1028.

\bibitem{van2019storytelling}
T.~Van~Laer, S.~Feiereisen, L.~M. Visconti, Storytelling in the digital era: A
  meta-analysis of relevant moderators of the narrative transportation effect,
  Journal of Business Research 96 (2019) 135--146.

\bibitem{green2000role}
M.~C. Green, T.~C. Brock, The role of transportation in the persuasiveness of
  public narratives., Journal of personality and social psychology 79~(5)
  (2000) 701.

\bibitem{sangalang2019potential}
A.~Sangalang, Y.~Ophir, J.~N. Cappella, The potential for narrative correctives
  to combat misinformation, Journal of communication 69~(3) (2019) 298--319.

\bibitem{eom2015twitter}
Y.~H. Eom, M.~Puliga, J.~Smailovic, I.~Mozetic, G.~Caldarelli,
  \href{https://dx.plos.org/10.1371/journal.pone.0131184}{{Twitter-based
  analysis of the dynamics of collective attention to political parties}}, PLoS
  ONE 10~(7) (2015) e0131184.
\newblock \href {http://arxiv.org/abs/1504.06861} {\path{arXiv:1504.06861}},
  \href {https://doi.org/10.1371/journal.pone.0131184}
  {\path{doi:10.1371/journal.pone.0131184}}.
\newline\urlprefix\url{https://dx.plos.org/10.1371/journal.pone.0131184}

\bibitem{li2017deepcas}
C.~Li, J.~Ma, X.~Guo, Q.~Mei, Deepcas: An end-to-end predictor of information
  cascades, in: Proceedings of the 26th international conference on World Wide
  Web, 2017, pp. 577--586.

\bibitem{gao2021public}
L.~Gao, Y.~Liu, H.~Zhuang, H.~Wang, B.~Zhou, A.~Li, Public opinion early
  warning agent model: A deep learning cascade virality prediction model based
  on multi-feature fusion, Frontiers in Neurorobotics 15 (2021) 674322.

\bibitem{qiu2018deepinf}
J.~Qiu, J.~Tang, H.~Ma, Y.~Dong, K.~Wang, J.~Tang, Deepinf: Social influence
  prediction with deep learning, in: Proceedings of the 24th ACM SIGKDD
  international conference on knowledge discovery \& data mining, 2018, pp.
  2110--2119.

\bibitem{xu2020casgcn}
Z.~Xu, M.~Qian, X.~Huang, J.~Meng, Casgcn: Predicting future cascade growth
  based on information diffusion graph, arXiv preprint arXiv:2009.05152 (2020).

\bibitem{wang2021dydiff}
R.~Wang, Z.~Huang, S.~Liu, H.~Shao, D.~Liu, J.~Li, T.~Wang, D.~Sun, S.~Yao,
  T.~Abdelzaher, Dydiff-vae: A dynamic variational framework for information
  diffusion prediction, in: Proceedings of the 44th International ACM SIGIR
  Conference on Research and Development in Information Retrieval, 2021, pp.
  163--172.

\bibitem{wang2020modeling}
H.~Wang, F.~Wang, K.~Xu, Modeling information diffusion in online social
  networks with partial differential equations, Vol.~7, Springer Nature, 2020.

\bibitem{cheng2024information}
Z.~Cheng, F.~Zhou, X.~Xu, K.~Zhang, G.~Trajcevski, T.~Zhong, P.~S. Yu,
  Information cascade popularity prediction via probabilistic diffusion, IEEE
  Transactions on Knowledge and Data Engineering (2024).

\bibitem{foroozani2021nonlinear}
A.~Foroozani, M.~Ebrahimi, Nonlinear anomalous information diffusion model in
  social networks, Communications in Nonlinear Science and Numerical Simulation
  103 (2021) 106019.

\bibitem{xian2019misinformation}
J.~Xian, D.~Yang, L.~Pan, W.~Wang, Z.~Wang, Misinformation spreading on
  correlated multiplex networks, Chaos: An Interdisciplinary Journal of
  Nonlinear Science 29~(11) (2019).

\bibitem{liu2018epidemic}
Q.-H. Liu, X.~Xiong, Q.~Zhang, N.~Perra, Epidemic spreading on time-varying
  multiplex networks, Physical review E 98~(6) (2018) 062303.

\bibitem{chang2021analytical}
X.~Chang, C.-R. Cai, J.-Q. Zhang, C.-Y. Wang, Analytical solution of epidemic
  threshold for coupled information-epidemic dynamics on multiplex networks
  with alterable heterogeneity, Physical Review E 104~(4) (2021) 044303.

\bibitem{granell2013dynamical}
C.~Granell, S.~G\'omez, A.~Arenas,
  \href{https://link.aps.org/doi/10.1103/PhysRevLett.111.128701}{Dynamical
  interplay between awareness and epidemic spreading in multiplex networks},
  Phys. Rev. Lett. 111 (2013) 128701.
\newblock \href {https://doi.org/10.1103/PhysRevLett.111.128701}
  {\path{doi:10.1103/PhysRevLett.111.128701}}.
\newline\urlprefix\url{https://link.aps.org/doi/10.1103/PhysRevLett.111.128701}

\bibitem{cellai2013percolation}
D.~Cellai, E.~L{\'o}pez, J.~Zhou, J.~P. Gleeson, G.~Bianconi, Percolation in
  multiplex networks with overlap, Physical Review E—Statistical, Nonlinear,
  and Soft Matter Physics 88~(5) (2013) 052811.

\bibitem{santoro2020optimal}
A.~Santoro, V.~Nicosia, Optimal percolation in correlated multilayer networks
  with overlap, Physical Review Research 2~(3) (2020) 033122.

\bibitem{wu2020spreading}
Q.~Wu, S.~Chen, Spreading of two interacting diseases in multiplex networks,
  Chaos 30 (2020) 093113.
\newblock \href {https://doi.org/10.1063/5.0011878}
  {\path{doi:10.1063/5.0011878}}.

\bibitem{obrien2019spreading}
J.~D. O’Brien, I.~K. Dassios, J.~P. Gleeson,
  \href{https://dx.doi.org/10.1088/1367-2630/ab05ef}{Spreading of memes on
  multiplex networks}, New Journal of Physics 21~(2) (2019) 025001.
\newblock \href {https://doi.org/10.1088/1367-2630/ab05ef}
  {\path{doi:10.1088/1367-2630/ab05ef}}.
\newline\urlprefix\url{https://dx.doi.org/10.1088/1367-2630/ab05ef}

\bibitem{notarmuzi2022universality}
D.~Notarmuzi, C.~Castellano, A.~Flammini, D.~Mazzilli, F.~Radicchi,
  \href{http://dx.doi.org/10.1038/s41467-022-28964-8}{Universality, criticality
  and complexity of information propagation in social media}, Nature
  Communications 13~(1) (March 2022).
\newblock \href {https://doi.org/10.1038/s41467-022-28964-8}
  {\path{doi:10.1038/s41467-022-28964-8}}.
\newline\urlprefix\url{http://dx.doi.org/10.1038/s41467-022-28964-8}

\bibitem{chen2018optimal}
X.~Chen, W.~Wang, S.~Cai, H.~E. Stanley, L.~A. Braunstein, Optimal resource
  diffusion for suppressing disease spreading in multiplex networks, Journal of
  Statistical Mechanics: Theory and Experiment 2018~(5) (2018) 053501.

\bibitem{obrien2021hierarchical}
J.~D. O'Brien, K.~A. Oliveira, J.~P. Gleeson, M.~Asllani,
  \href{https://link.aps.org/doi/10.1103/PhysRevResearch.3.023117}{{Hierarchical
  route to the emergence of leader nodes in real-world networks}}, Physical
  Review Research 3~(2) (2021) 023117.
\newblock \href {http://arxiv.org/abs/2012.01098} {\path{arXiv:2012.01098}},
  \href {https://doi.org/10.1103/PhysRevResearch.3.023117}
  {\path{doi:10.1103/PhysRevResearch.3.023117}}.
\newline\urlprefix\url{https://link.aps.org/doi/10.1103/PhysRevResearch.3.023117}

\bibitem{sacco2021emergence}
P.~L. Sacco, R.~Gallotti, F.~Pilati, N.~Castaldo, M.~De~Domenico, Emergence of
  knowledge communities and information centralization during the covid-19
  pandemic, Social Science \& Medicine 285 (2021) 114215.

\bibitem{flamino2023political}
J.~Flamino, A.~Galeazzi, S.~Feldman, M.~W. Macy, B.~Cross, Z.~Zhou,
  M.~Serafino, A.~Bovet, H.~A. Makse, B.~K. Szymanski,
  \href{https://www.nature.com/articles/s41562-023-01550-8}{{Political
  polarization of news media and influencers on Twitter in the 2016 and 2020 US
  presidential elections}}, Nature Human Behaviour 7~(6) (2023) 904--916.
\newblock \href {https://doi.org/10.1038/s41562-023-01550-8}
  {\path{doi:10.1038/s41562-023-01550-8}}.
\newline\urlprefix\url{https://www.nature.com/articles/s41562-023-01550-8}

\bibitem{pei2013spreading}
S.~Pei, H.~A. Makse, Spreading dynamics in complex networks, J. Stat. Mech.
  Theor. Exp. 2013~(12) (2013) P12002.

\bibitem{teng2016collective}
X.~Teng, S.~Pei, F.~Morone, H.~A. Makse,
  \href{http://dx.doi.org/10.1038/srep36043}{Collective influence of multiple
  spreaders evaluated by tracing real information flow in large-scale social
  networks}, Scientific Reports 6~(1) (October 2016).
\newblock \href {https://doi.org/10.1038/srep36043}
  {\path{doi:10.1038/srep36043}}.
\newline\urlprefix\url{http://dx.doi.org/10.1038/srep36043}

\bibitem{bontorin2023multi}
S.~Bontorin, M.~De~Domenico,
  \href{http://dx.doi.org/10.1038/s42005-023-01204-1}{Multi pathways temporal
  distance unravels the hidden geometry of network-driven processes},
  Communications Physics 6~(1) (June 2023).
\newblock \href {https://doi.org/10.1038/s42005-023-01204-1}
  {\path{doi:10.1038/s42005-023-01204-1}}.
\newline\urlprefix\url{http://dx.doi.org/10.1038/s42005-023-01204-1}

\bibitem{altarelli2013optimizing}
F.~Altarelli, A.~Braunstein, L.~Dall’Asta, R.~Zecchina,
  \href{https://dx.doi.org/10.1088/1742-5468/2013/09/P09011}{Optimizing spread
  dynamics on graphs by message passing}, Journal of Statistical Mechanics:
  Theory and Experiment 2013~(09) (2013) P09011.
\newblock \href {https://doi.org/10.1088/1742-5468/2013/09/P09011}
  {\path{doi:10.1088/1742-5468/2013/09/P09011}}.
\newline\urlprefix\url{https://dx.doi.org/10.1088/1742-5468/2013/09/P09011}

\bibitem{osullivan2015mathematical}
D.~J.~P. O'Sullivan, G.~J. O'Keeffe, P.~G. Fennell, J.~P. Gleeson,
  \href{https://www.frontiersin.org/journals/physics/articles/10.3389/fphy.2015.00071}{Mathematical
  modeling of complex contagion on clustered networks}, Frontiers in Physics
  Volume 3 - 2015 (2015).
\newblock \href {https://doi.org/10.3389/fphy.2015.00071}
  {\path{doi:10.3389/fphy.2015.00071}}.
\newline\urlprefix\url{https://www.frontiersin.org/journals/physics/articles/10.3389/fphy.2015.00071}

\bibitem{liu2003propagation}
Z.~Liu, Y.-C. Lai, N.~Ye,
  \href{https://link.aps.org/doi/10.1103/PhysRevE.67.031911}{Propagation and
  immunization of infection on general networks with both homogeneous and
  heterogeneous components}, Phys. Rev. E 67 (2003) 031911.
\newblock \href {https://doi.org/10.1103/PhysRevE.67.031911}
  {\path{doi:10.1103/PhysRevE.67.031911}}.
\newline\urlprefix\url{https://link.aps.org/doi/10.1103/PhysRevE.67.031911}

\bibitem{moreno2004dynamics}
Y.~Moreno, M.~Nekovee, A.~F. Pacheco,
  \href{https://link.aps.org/doi/10.1103/PhysRevE.69.066130}{Dynamics of rumor
  spreading in complex networks}, Phys. Rev. E 69 (2004) 066130.
\newblock \href {https://doi.org/10.1103/PhysRevE.69.066130}
  {\path{doi:10.1103/PhysRevE.69.066130}}.
\newline\urlprefix\url{https://link.aps.org/doi/10.1103/PhysRevE.69.066130}

\bibitem{zanette2001critical}
D.~H. Zanette,
  \href{https://link.aps.org/doi/10.1103/PhysRevE.64.050901}{Critical behavior
  of propagation on small-world networks}, Phys. Rev. E 64 (2001) 050901.
\newblock \href {https://doi.org/10.1103/PhysRevE.64.050901}
  {\path{doi:10.1103/PhysRevE.64.050901}}.
\newline\urlprefix\url{https://link.aps.org/doi/10.1103/PhysRevE.64.050901}

\bibitem{zanette2002dynamics}
D.~H. Zanette,
  \href{https://link.aps.org/doi/10.1103/PhysRevE.65.041908}{Dynamics of rumor
  propagation on small-world networks}, Phys. Rev. E 65 (2002) 041908.
\newblock \href {https://doi.org/10.1103/PhysRevE.65.041908}
  {\path{doi:10.1103/PhysRevE.65.041908}}.
\newline\urlprefix\url{https://link.aps.org/doi/10.1103/PhysRevE.65.041908}

\bibitem{davis2020phase}
J.~T. Davis, N.~Perra, Q.~Zhang, Y.~Moreno, A.~Vespignani,
  \href{http://dx.doi.org/10.1038/s41567-020-0810-3}{Phase transitions in
  information spreading on structured populations}, Nature Physics 16~(5)
  (2020) 590–596.
\newblock \href {https://doi.org/10.1038/s41567-020-0810-3}
  {\path{doi:10.1038/s41567-020-0810-3}}.
\newline\urlprefix\url{http://dx.doi.org/10.1038/s41567-020-0810-3}

\bibitem{banos2013role}
R.~A. Baños, J.~Borge-Holthoefer, Y.~Moreno,
  \href{http://dx.doi.org/10.1140/epjds18}{The role of hidden influentials in
  the diffusion of online information cascades}, EPJ Data Science 2~(1) (July
  2013).
\newblock \href {https://doi.org/10.1140/epjds18} {\path{doi:10.1140/epjds18}}.
\newline\urlprefix\url{http://dx.doi.org/10.1140/epjds18}

\bibitem{su2018optimal}
Z.~Su, W.~Wang, L.~Li, H.~E. Stanley, L.~A. Braunstein,
  \href{https://dx.doi.org/10.1088/1367-2630/aac0c9}{Optimal community
  structure for social contagions}, New Journal of Physics 20~(5) (2018)
  053053.
\newblock \href {https://doi.org/10.1088/1367-2630/aac0c9}
  {\path{doi:10.1088/1367-2630/aac0c9}}.
\newline\urlprefix\url{https://dx.doi.org/10.1088/1367-2630/aac0c9}

\bibitem{shaw2012politics}
F.~Shaw, \href{https://doi.org/10.1177/1329878X1214200106}{The politics of
  blogs: Theories of discursive activism online}, Media International Australia
  142~(1) (2012) 41--49.
\newblock \href
  {http://arxiv.org/abs/https://doi.org/10.1177/1329878X1214200106}
  {\path{arXiv:https://doi.org/10.1177/1329878X1214200106}}, \href
  {https://doi.org/10.1177/1329878X1214200106}
  {\path{doi:10.1177/1329878X1214200106}}.
\newline\urlprefix\url{https://doi.org/10.1177/1329878X1214200106}

\bibitem{radicioni2021analysing}
T.~Radicioni, F.~Saracco, E.~Pavan, T.~Squartini,
  \href{https://www.nature.com/articles/s41598-021-92337-2}{Analysing twitter
  semantic networks: the case of 2018 italian elections}, Scientific Reports
  2021 11:1 11 (2021) 1--22.
\newblock \href {https://doi.org/10.1038/s41598-021-92337-2}
  {\path{doi:10.1038/s41598-021-92337-2}}.
\newline\urlprefix\url{https://www.nature.com/articles/s41598-021-92337-2}

\bibitem{guarino2024verified}
S.~Guarino, A.~Mounim, G.~Caldarelli, F.~Saracco,
  \href{https://arxiv.org/abs/2405.04896}{Verified authors shape x/twitter
  discursive communities} (2024).
\newblock \href {http://arxiv.org/abs/2405.04896} {\path{arXiv:2405.04896}}.
\newline\urlprefix\url{https://arxiv.org/abs/2405.04896}

\bibitem{becatti2019extracting}
C.~Becatti, G.~Caldarelli, R.~Lambiotte, F.~Saracco,
  \href{https://www.nature.com/articles/s41599-019-0300-3}{{Extracting
  significant signal of news consumption from social networks: the case of
  Twitter in Italian political elections}}, Palgrave Communications 5~(1)
  (2019) 91.
\newblock \href {http://arxiv.org/abs/1901.07933} {\path{arXiv:1901.07933}},
  \href {https://doi.org/10.1057/s41599-019-0300-3}
  {\path{doi:10.1057/s41599-019-0300-3}}.
\newline\urlprefix\url{https://www.nature.com/articles/s41599-019-0300-3}

\bibitem{Wiki_verified}
\href{https://en.wikipedia.org/wiki/Twitter\_verification}{Wikipedia.org:
  Twitter verification}.
\newline\urlprefix\url{https://en.wikipedia.org/wiki/Twitter\_verification}

\bibitem{X2009}
\href{https://blog.x.com/official/en\_us/a/2009/not-playing-ball.html}{X blog:
  Not playing ball} (2009).
\newline\urlprefix\url{https://blog.x.com/official/en\_us/a/2009/not-playing-ball.html}

\bibitem{X2016}
\href{https://web.archive.org/web/20160719090643/https://support.twitter.com/groups/31-twitter-basics/topics/111-features/articles/119135-about-verified-accounts}{X
  blog: Faqs about verified accounts} (2016).
\newline\urlprefix\url{https://web.archive.org/web/20160719090643/https://support.twitter.com/groups/31-twitter-basics/topics/111-features/articles/119135-about-verified-accounts}

\bibitem{X2022}
\href{https://help.x.com/en/using-x/x-premium}{X help center: About x premium}
  (2022).
\newline\urlprefix\url{https://help.x.com/en/using-x/x-premium}

\bibitem{caldarelli2020role}
G.~Caldarelli, R.~{De Nicola}, F.~{Del Vigna}, M.~Petrocchi, F.~Saracco, {The
  role of bot squads in the political propaganda on Twitter}, Communication
  Physics 3 (2020) 81.

\bibitem{caldarelli2021flow}
G.~Caldarelli, R.~{De Nicola}, M.~Petrocchi, M.~Pratelli, F.~Saracco,
  \href{https://epjdatascience.springeropen.com/articles/10.1140/epjds/s13688-021-00289-4}{{Flow
  of online misinformation during the peak of the COVID-19 pandemic in Italy}},
  EPJ Data Science 10~(1) (2021) 34.
\newblock \href {http://arxiv.org/abs/2010.01913} {\path{arXiv:2010.01913}},
  \href {https://doi.org/10.1140/epjds/s13688-021-00289-4}
  {\path{doi:10.1140/epjds/s13688-021-00289-4}}.
\newline\urlprefix\url{https://epjdatascience.springeropen.com/articles/10.1140/epjds/s13688-021-00289-4}

\bibitem{mattei2022bow}
M.~Mattei, M.~Pratelli, G.~Caldarelli, M.~Petrocchi, F.~Saracco,
  \href{https://www.nature.com/articles/s41598-022-16603-7}{{Bow-tie structures
  of twitter discursive communities}}, Scientific Reports 12~(1) (2022) 12944.
\newblock \href {http://arxiv.org/abs/2202.03316} {\path{arXiv:2202.03316}},
  \href {https://doi.org/10.1038/s41598-022-16603-7}
  {\path{doi:10.1038/s41598-022-16603-7}}.
\newline\urlprefix\url{https://www.nature.com/articles/s41598-022-16603-7}

\bibitem{pratelli2024entropy}
M.~Pratelli, F.~Saracco, M.~Petrocchi, {Entropy-based detection of Twitter echo
  chambers}, PNAS Nexus 3~(5) (2024) pgae177.
\newblock \href
  {http://arxiv.org/abs/https://academic.oup.com/pnasnexus/article-pdf/3/5/pgae177/57505528/pgae177.pdf}
  {\path{arXiv:https://academic.oup.com/pnasnexus/article-pdf/3/5/pgae177/57505528/pgae177.pdf}},
  \href {https://doi.org/10.1093/pnasnexus/pgae177}
  {\path{doi:10.1093/pnasnexus/pgae177}}.

\bibitem{han2024modelling}
Y.~Han, M.~Bazzi, P.~Turrini, Modelling and predicting online vaccination views
  using bow-tie decomposition, Royal Society Open Science 11~(2) (2024) 231792.

\bibitem{weeks2015online}
B.~E. Weeks, A.~Ardèvol-Abreu, H.~Gil~de Zúñiga,
  \href{https://doi.org/10.1093/ijpor/edv050}{{Online Influence? Social Media
  Use, Opinion Leadership, and Political Persuasion}}, International Journal of
  Public Opinion Research 29~(2) (2015) 214--239.
\newblock \href
  {http://arxiv.org/abs/https://academic.oup.com/ijpor/article-pdf/29/2/214/17694136/edv050.pdf}
  {\path{arXiv:https://academic.oup.com/ijpor/article-pdf/29/2/214/17694136/edv050.pdf}},
  \href {https://doi.org/10.1093/ijpor/edv050}
  {\path{doi:10.1093/ijpor/edv050}}.
\newline\urlprefix\url{https://doi.org/10.1093/ijpor/edv050}

\bibitem{choi2015two}
S.~Choi, \href{https://doi.org/10.1177/0894439314556599}{The two-step flow of
  communication in twitter-based public forums}, Social Science Computer Review
  33~(6) (2015) 696--711.
\newblock \href {http://arxiv.org/abs/https://doi.org/10.1177/0894439314556599}
  {\path{arXiv:https://doi.org/10.1177/0894439314556599}}, \href
  {https://doi.org/10.1177/0894439314556599}
  {\path{doi:10.1177/0894439314556599}}.
\newline\urlprefix\url{https://doi.org/10.1177/0894439314556599}

\bibitem{hilbert2017one}
M.~Hilbert, J.~Vásquez, D.~Halpern, S.~Valenzuela, E.~Arriagada,
  \href{https://doi.org/10.1177/0894439316639561}{One step, two step, network
  step? complementary perspectives on communication flows in twittered citizen
  protests}, Social Science Computer Review 35~(4) (2017) 444--461.
\newblock \href {http://arxiv.org/abs/https://doi.org/10.1177/0894439316639561}
  {\path{arXiv:https://doi.org/10.1177/0894439316639561}}, \href
  {https://doi.org/10.1177/0894439316639561}
  {\path{doi:10.1177/0894439316639561}}.
\newline\urlprefix\url{https://doi.org/10.1177/0894439316639561}

\bibitem{dubois2020who}
E.~Dubois, S.~Minaeian, A.~Paquet-Labelle, S.~Beaudry,
  \href{https://doi.org/10.1177/2056305120913993}{Who to trust on social media:
  How opinion leaders and seekers avoid disinformation and echo chambers},
  Social Media + Society 6~(2) (2020) 2056305120913993.
\newblock \href {http://arxiv.org/abs/https://doi.org/10.1177/2056305120913993}
  {\path{arXiv:https://doi.org/10.1177/2056305120913993}}, \href
  {https://doi.org/10.1177/2056305120913993}
  {\path{doi:10.1177/2056305120913993}}.
\newline\urlprefix\url{https://doi.org/10.1177/2056305120913993}

\bibitem{bennett2006one}
W.~L. Bennett, J.~B. Manheim,
  \href{https://doi.org/10.1177/0002716206292266}{The one-step flow of
  communication}, The ANNALS of the American Academy of Political and Social
  Science 608~(1) (2006) 213--232.
\newblock \href {http://arxiv.org/abs/https://doi.org/10.1177/0002716206292266}
  {\path{arXiv:https://doi.org/10.1177/0002716206292266}}, \href
  {https://doi.org/10.1177/0002716206292266}
  {\path{doi:10.1177/0002716206292266}}.
\newline\urlprefix\url{https://doi.org/10.1177/0002716206292266}

\bibitem{broder2000graph}
A.~Broder, R.~Kumar, F.~Maghoul, P.~Raghavan, S.~Rajagopalan, R.~Stata,
  A.~Tomkins, J.~Wiener, Graph structure in the web, Computer Networks (2000).
\newblock \href {https://doi.org/10.1016/S1389-1286(00)00083-9}
  {\path{doi:10.1016/S1389-1286(00)00083-9}}.

\bibitem{yang2011bow}
R.~Yang, L.~Zhuhadar, O.~Nasraoui, Bow-tie decomposition in directed graphs,
  2011.

\bibitem{bernaschi2022onion}
M.~Bernaschi, A.~Celestini, M.~Cianfriglia, S.~Guarino, F.~Lombardi,
  E.~Mastrostefano, \href{http://dx.doi.org/10.1007/s11280-022-01044-z}{Onion
  under microscope: An in-depth analysis of the tor web}, World Wide Web 25~(3)
  (2022) 1287–1313.
\newblock \href {https://doi.org/10.1007/s11280-022-01044-z}
  {\path{doi:10.1007/s11280-022-01044-z}}.
\newline\urlprefix\url{http://dx.doi.org/10.1007/s11280-022-01044-z}

\bibitem{vitali2011network}
S.~Vitali, J.~B. Glattfelder, S.~Battiston,
  \href{https://doi.org/10.1371/journal.pone.0025995}{The network of global
  corporate control}, PLOS ONE 6~(10) (2011) 1--6.
\newblock \href {https://doi.org/10.1371/journal.pone.0025995}
  {\path{doi:10.1371/journal.pone.0025995}}.
\newline\urlprefix\url{https://doi.org/10.1371/journal.pone.0025995}

\bibitem{jamieson2008echo}
K.~Jamieson, J.~Cappella, Echo Chamber: Rush Limbaugh and the Conservative
  Media Establishment, Oxford University Press, 2008.

\bibitem{garrett2009echo}
R.~K. Garrett, Echo chambers online?: Politically motivated selective exposure
  among internet news users, Journal of Computer-Mediated Communication 14
  (2009) 265--285.
\newblock \href {https://doi.org/10.1111/J.1083-6101.2009.01440.X}
  {\path{doi:10.1111/J.1083-6101.2009.01440.X}}.

\bibitem{bail2018exposure}
C.~A. Bail, L.~P. Argyle, T.~W. Brown, J.~P. Bumpus, H.~Chen, M.~B.~F.
  Hunzaker, J.~Lee, M.~Mann, F.~Merhout, A.~Volfovsky,
  \href{http://dx.doi.org/10.1073/pnas.1804840115}{Exposure to opposing views
  on social media can increase political polarization}, Proceedings of the
  National Academy of Sciences 115~(37) (2018) 9216–9221.
\newblock \href {https://doi.org/10.1073/pnas.1804840115}
  {\path{doi:10.1073/pnas.1804840115}}.
\newline\urlprefix\url{http://dx.doi.org/10.1073/pnas.1804840115}

\bibitem{cinelli2021echo}
M.~Cinelli, G.~De~Francisci~Morales, A.~Galeazzi, W.~Quattrociocchi,
  M.~Starnini, The echo chamber effect on social media, Proceedings of the
  national academy of sciences 118~(9) (2021) e2023301118.

\bibitem{defrancisci2021noecho}
G.~De~Francisci~Morales, C.~Monti, M.~Starnini,
  \href{https://doi.org/10.1038/s41598-021-81531-x}{No echo in the chambers of
  political interactions on reddit}, Scientific Reports 11~(1) (2021) 2818.
\newblock \href {https://doi.org/10.1038/s41598-021-81531-x}
  {\path{doi:10.1038/s41598-021-81531-x}}.
\newline\urlprefix\url{https://doi.org/10.1038/s41598-021-81531-x}

\bibitem{monti2023evidence}
C.~Monti, J.~D’Ignazi, M.~Starnini, G.~De~Francisci~Morales,
  \href{http://dx.doi.org/10.1145/3543507.3583468}{Evidence of demographic
  rather than ideological segregation in news discussion on reddit}, in:
  Proceedings of the ACM Web Conference 2023, WWW ’23, ACM, 2023, p.
  2777–2786.
\newblock \href {https://doi.org/10.1145/3543507.3583468}
  {\path{doi:10.1145/3543507.3583468}}.
\newline\urlprefix\url{http://dx.doi.org/10.1145/3543507.3583468}

\bibitem{guarino2021information}
S.~Guarino, F.~Pierri, M.~D. Giovanni, A.~Celestini, Information disorders
  during the covid-19 infodemic: The case of italian facebook, Online Social
  Networks and Media 22 (2021) 100124.
\newblock \href {https://doi.org/10.1016/J.OSNEM.2021.100124}
  {\path{doi:10.1016/J.OSNEM.2021.100124}}.

\bibitem{berner2023adaptive}
R.~Berner, T.~Gross, C.~Kuehn, J.~Kurths, S.~Yanchuk, Adaptive dynamical
  networks, arXiv preprint arXiv:2304.05652 (2023).

\bibitem{ghosh2022synchronized}
D.~Ghosh, M.~Frasca, A.~Rizzo, S.~Majhi, S.~Rakshit, K.~Alfaro-Bittner,
  S.~Boccaletti, The synchronized dynamics of time-varying networks, Physics
  Reports 949 (2022) 1--63.

\bibitem{baumann2020modeling}
F.~Baumann, P.~Lorenz-Spreen, I.~M. Sokolov, M.~Starnini, Modeling echo
  chambers and polarization dynamics in social networks, Physical Review
  Letters 124~(4) (2020) 048301.

\bibitem{ayi2021mean}
N.~Ayi, N.~P. Duteil, Mean-field and graph limits for collective dynamics
  models with time-varying weights, Journal of Differential Equations 299
  (2021) 65--110.

\bibitem{duteil2021mean}
N.~P. Duteil, Mean-field limit of collective dynamics with time-varying
  weights, arXiv preprint arXiv:2103.06527 (2021).

\bibitem{gkogkas2022mean}
M.~A. Gkogkas, C.~Kuehn, C.~Xu, Mean field limits of co-evolutionary
  heterogeneous networks, arXiv preprint arXiv:2202.01742 (2022).

\bibitem{gkogkas2021continuum}
M.~A. Gkogkas, C.~Kuehn, C.~Xu, Continuum limits for adaptive network dynamics,
  arXiv preprint arXiv:2109.05898 (2021).

\bibitem{unicomb2021dynamics}
S.~Unicomb, G.~Iñiguez, J.~P. Gleeson, M.~Karsai,
  \href{http://dx.doi.org/10.1038/s41467-020-20398-4}{Dynamics of cascades on
  burstiness-controlled temporal networks}, Nature Communications 12~(1)
  (January 2021).
\newblock \href {https://doi.org/10.1038/s41467-020-20398-4}
  {\path{doi:10.1038/s41467-020-20398-4}}.
\newline\urlprefix\url{http://dx.doi.org/10.1038/s41467-020-20398-4}

\bibitem{masoomy2023impact}
H.~Masoomy, T.~Chou, L.~Böttcher,
  \href{https://doi.org/10.1063/5.0139844}{Impact of random and targeted
  disruptions on information diffusion during outbreaks}, Chaos: An
  Interdisciplinary Journal of Nonlinear Science 33~(3) (2023) 033145.
\newblock \href {https://doi.org/10.1063/5.0139844}
  {\path{doi:10.1063/5.0139844}}.
\newline\urlprefix\url{https://doi.org/10.1063/5.0139844}

\bibitem{sethna2001crackling}
J.~P. Sethna, K.~A. Dahmen, C.~R. Myers, Crackling noise, nature 410~(6825)
  (2001) 242--250.

\bibitem{goldenfeld2018lectures}
N.~Goldenfeld, Lectures on phase transitions and the renormalization group, CRC
  Press, 2018.

\bibitem{hall2019statistical}
G.~Hall, W.~Bialek, \href{https://dx.doi.org/10.1088/1742-5468/ab3af0}{The
  statistical mechanics of twitter communities}, Journal of Statistical
  Mechanics: Theory and Experiment 2019~(9) (2019) 093406.
\newblock \href {https://doi.org/10.1088/1742-5468/ab3af0}
  {\path{doi:10.1088/1742-5468/ab3af0}}.
\newline\urlprefix\url{https://dx.doi.org/10.1088/1742-5468/ab3af0}

\bibitem{alodjants2022mean}
A.~P. Alodjants, A.~Y. Bazhenov, A.~Y. Khrennikov, A.~V. Bukhanovsky,
  \href{http://dx.doi.org/10.1038/s41598-022-12327-w}{Mean-field theory of
  social laser}, Scientific Reports 12~(1) (May 2022).
\newblock \href {https://doi.org/10.1038/s41598-022-12327-w}
  {\path{doi:10.1038/s41598-022-12327-w}}.
\newline\urlprefix\url{http://dx.doi.org/10.1038/s41598-022-12327-w}

\bibitem{pluchino2006opinion}
A.~Pluchino, S.~Boccaletti, V.~Latora, A.~Rapisarda, Opinion dynamics and
  synchronization in a network of scientific collaborations, Physica A 372
  (2006) 316--325.

\bibitem{ren2013symmetry}
Q.~Ren, Q.~Long, J.~Zhao, Symmetry and symmetry breaking in a kuramoto model
  induced on a {M\"obius} strip, Physical Review E 87~(2) (2013) 022811.

\bibitem{zhang2025generalized}
Z.~Zhang, S.~Al-Abri, F.~Zhang, A generalized kuramoto model for opinion
  dynamics on the unit sphere, Automatica 171 (2025) 111957.

\bibitem{cinelli2020covid}
M.~Cinelli, W.~Quattrociocchi, A.~Galeazzi, C.~M. Valensise, E.~Brugnoli, A.~L.
  Schmidt, P.~Zola, F.~Zollo, A.~Scala,
  \href{http://dx.doi.org/10.1038/s41598-020-73510-5}{The covid-19 social media
  infodemic}, Scientific Reports 10~(1) (October 2020).
\newblock \href {https://doi.org/10.1038/s41598-020-73510-5}
  {\path{doi:10.1038/s41598-020-73510-5}}.
\newline\urlprefix\url{http://dx.doi.org/10.1038/s41598-020-73510-5}

\bibitem{zarocostas2020how}
J.~Zarocostas, How to fight an infodemic, Lancet 395~(10225) (2020) 676.

\bibitem{chiou2022future}
H.~Chiou, C.~Voegeli, E.~Wilhelm, J.~Kolis, K.~Brookmeyer, D.~Prybylski, The
  future of infodemic surveillance as public health surveillance, Emerging
  Infectious Diseases 28~(Suppl 1) (2022) S121.

\bibitem{singh2022misinformation}
K.~Singh, G.~Lima, M.~Cha, C.~Cha, J.~Kulshrestha, Y.-Y. Ahn, O.~Varol,
  Misinformation, believability, and vaccine acceptance over 40 countries:
  Takeaways from the initial phase of the {COVID}-19 infodemic, PLOS ONE 17~(2)
  (2022) 0263381.

\bibitem{simon2023autopsy}
F.~M. Simon, C.~Q. Camargo, Autopsy of a metaphor: The origins, use and blind
  spots of the `infodemic', New Media \& Society 25~(8) (2023) 2219--2240.

\bibitem{bovet2019influence}
A.~Bovet, H.~A. Makse,
  \href{https://www.nature.com/articles/s41467-018-07761-2}{{Influence of fake
  news in Twitter during the 2016 US presidential election}}, Nature
  Communications 10~(1) (2019) 7.
\newblock \href {http://arxiv.org/abs/1803.08491} {\path{arXiv:1803.08491}},
  \href {https://doi.org/10.1038/s41467-018-07761-2}
  {\path{doi:10.1038/s41467-018-07761-2}}.
\newline\urlprefix\url{https://www.nature.com/articles/s41467-018-07761-2}

\bibitem{cresci2015fame}
S.~Cresci, R.~Di~Pietro, M.~Petrocchi, A.~Spognardi, M.~Tesconi,
  \href{https://doi.org/10.1016/j.dss.2015.09.003}{Fame for sale}, Decis.
  Support Syst. 80~(C) (2015) 56–71.
\newblock \href {https://doi.org/10.1016/j.dss.2015.09.003}
  {\path{doi:10.1016/j.dss.2015.09.003}}.
\newline\urlprefix\url{https://doi.org/10.1016/j.dss.2015.09.003}

\bibitem{ferrara2016rise}
E.~Ferrara, O.~Varol, C.~Davis, F.~Menczer, A.~Flammini,
  \href{https://dl.acm.org/doi/10.1145/2818717}{{The rise of social bots}},
  Communications of the ACM 59~(7) (2016) 96--104.
\newblock \href {http://arxiv.org/abs/1407.5225} {\path{arXiv:1407.5225}},
  \href {https://doi.org/10.1145/2818717} {\path{doi:10.1145/2818717}}.
\newline\urlprefix\url{https://dl.acm.org/doi/10.1145/2818717}

\bibitem{cresci2020decade}
S.~Cresci, \href{https://doi.org/10.1145/3409116}{A decade of social bot
  detection}, Commun. ACM 63~(10) (2020) 72–83.
\newblock \href {https://doi.org/10.1145/3409116} {\path{doi:10.1145/3409116}}.
\newline\urlprefix\url{https://doi.org/10.1145/3409116}

\bibitem{lopez-joya2025dissecting}
S.~Lopez-Joya, J.~A. Diaz-Garcia, M.~D. Ruiz, M.~J. Martin-Bautista,
  \href{http://dx.doi.org/10.1007/s13278-025-01410-5}{Dissecting a social bot
  powered by generative ai: anatomy, new trends and challenges}, Social Network
  Analysis and Mining 15~(1) (March 2025).
\newblock \href {https://doi.org/10.1007/s13278-025-01410-5}
  {\path{doi:10.1007/s13278-025-01410-5}}.
\newline\urlprefix\url{http://dx.doi.org/10.1007/s13278-025-01410-5}

\bibitem{ferrara2016detection}
E.~Ferrara, O.~Varol, F.~Menczer, A.~Flammini, Detection of promoted social
  media campaigns, in: Proceedings of the international aaai conference on web
  and social media, Vol.~10, 2016, pp. 563--566.

\bibitem{ferrara2017disinformation}
E.~Ferrara,
  \href{https://firstmonday.org/ojs/index.php/fm/article/view/8005}{{Disinformation
  and social bot operations in the run up to the 2017 French presidential
  election}}, First Monday 22~(8) (jul 2017).
\newblock \href {http://arxiv.org/abs/1707.00086} {\path{arXiv:1707.00086}},
  \href {https://doi.org/10.5210/fm.v22i8.8005}
  {\path{doi:10.5210/fm.v22i8.8005}}.
\newline\urlprefix\url{https://firstmonday.org/ojs/index.php/fm/article/view/8005}

\bibitem{stella2018bots}
M.~Stella, E.~Ferrara, M.~{De Domenico},
  \href{https://pnas.org/doi/full/10.1073/pnas.1803470115}{{Bots increase
  exposure to negative and inflammatory content in online social systems}},
  Proceedings of the National Academy of Sciences of the United States of
  America 115~(49) (2018) 12435--12440.
\newblock \href {http://arxiv.org/abs/1802.07292} {\path{arXiv:1802.07292}},
  \href {https://doi.org/10.1073/pnas.1803470115}
  {\path{doi:10.1073/pnas.1803470115}}.
\newline\urlprefix\url{https://pnas.org/doi/full/10.1073/pnas.1803470115}

\bibitem{bruno2022brexit}
M.~Bruno, R.~Lambiotte, F.~Saracco,
  \href{http://dx.doi.org/10.1140/epjds/s13688-022-00330-0}{Brexit and bots:
  characterizing the behaviour of automated accounts on twitter during the uk
  election}, EPJ Data Science 11~(1) (Mar. 2022).
\newblock \href {https://doi.org/10.1140/epjds/s13688-022-00330-0}
  {\path{doi:10.1140/epjds/s13688-022-00330-0}}.
\newline\urlprefix\url{http://dx.doi.org/10.1140/epjds/s13688-022-00330-0}

\bibitem{badawy2018analyzing}
A.~Badawy, E.~Ferrara, K.~Lerman,
  \href{http://arxiv.org/abs/1802.04291}{{Analyzing the Digital Traces of
  Political Manipulation: The 2016 Russian Interference Twitter Campaign}}
  (2018).
\newblock \href {http://arxiv.org/abs/1802.04291} {\path{arXiv:1802.04291}},
  \href {https://doi.org/10.1145/nnnnnnn.nnnnnnn}
  {\path{doi:10.1145/nnnnnnn.nnnnnnn}}.
\newline\urlprefix\url{http://arxiv.org/abs/1802.04291}

\bibitem{gonzalez-bailon2021bots}
S.~Gonz{\'{a}}lez-Bail{\'{o}}n, M.~{De Domenico},
  \href{https://pnas.org/doi/full/10.1073/pnas.2013443118}{{Bots are less
  central than verified accounts during contentious political events}},
  Proceedings of the National Academy of Sciences of the United States of
  America 118~(11) (2021) e2013443118.
\newblock \href {https://doi.org/10.1073/pnas.2013443118}
  {\path{doi:10.1073/pnas.2013443118}}.
\newline\urlprefix\url{https://pnas.org/doi/full/10.1073/pnas.2013443118}

\bibitem{stella2019influence}
M.~Stella, M.~Cristoforetti, M.~{De Domenico},
  \href{https://dx.plos.org/10.1371/journal.pone.0214210}{{Influence of
  augmented humans in online interactions during voting events}}, PLoS ONE
  14~(5) (2019) e0214210.
\newblock \href {http://arxiv.org/abs/1803.08086} {\path{arXiv:1803.08086}},
  \href {https://doi.org/10.1371/journal.pone.0214210}
  {\path{doi:10.1371/journal.pone.0214210}}.
\newline\urlprefix\url{https://dx.plos.org/10.1371/journal.pone.0214210}

\bibitem{shao2018spread}
C.~Shao, G.~L. Ciampaglia, O.~Varol, K.~C. Yang, A.~Flammini, F.~Menczer,
  \href{https://www.nature.com/articles/s41467-018-06930-7}{{The spread of
  low-credibility content by social bots}}, Nature Communications 9~(1) (2018)
  4787.
\newblock \href {http://arxiv.org/abs/1707.07592} {\path{arXiv:1707.07592}},
  \href {https://doi.org/10.1038/s41467-018-06930-7}
  {\path{doi:10.1038/s41467-018-06930-7}}.
\newline\urlprefix\url{https://www.nature.com/articles/s41467-018-06930-7}

\bibitem{cresci2023demystifying}
S.~Cresci, K.-C. Yang, A.~Spognardi, R.~Di~Pietro, F.~Menczer, M.~Petrocchi,
  Demystifying misconceptions in social bots research, Social Science Computer
  Review (2023) 08944393251376707.

\bibitem{chesney2019deepfakes}
R.~Chesney, D.~Citron, Deepfakes and the new disinformation war: The coming age
  of post-truth geopolitics, Foreign Aff. 98 (2019) 147.

\bibitem{westerlund2019emergence}
M.~Westerlund, The emergence of deepfake technology: A review, Technology
  Innovation Management Review 9~(11) (2019) 39--52.
\newblock \href {https://doi.org/10.22215/timreview/1282}
  {\path{doi:10.22215/timreview/1282}}.

\bibitem{lee2024deepfakes}
H.-P. Lee, Y.-J. Yang, T.~S. Von~Davier, J.~Forlizzi, S.~Das, Deepfakes,
  phrenology, surveillance, and more! a taxonomy of ai privacy risks, in:
  Proceedings of the 2024 CHI Conference on Human Factors in Computing Systems,
  2024, pp. 1--19.

\bibitem{di2020reversible}
B.~A.~M. Di~Muro, S.~V. Buldyrev, L.~A. Braunstein, Reversible bootstrap
  percolation: Fake news and fact checking, Physical Review E 101 (2020)
  042307.
\newblock \href {https://doi.org/10.1103/PhysRevE.101.042307}
  {\path{doi:10.1103/PhysRevE.101.042307}}.

\bibitem{spearing2025countering}
E.~R. Spearing, C.~I. Gile, A.~L. Fogwill, T.~Prike, B.~Swire-Thompson,
  S.~Lewandowsky, U.~K. Ecker, Countering ai-generated misinformation with
  pre-emptive source discreditation and debunking, Royal Society Open Science
  12~(6) (2025) 242148.

\bibitem{lovato2024diverse}
J.~Lovato, L.~Hebert-Dufresne, J.~St-Onge, R.~Harp, G.~Salazar~Lopez,
  S.~Rogers, I.~Ul~Haq, J.~Onaolapo, Diverse misinformation: Impacts of human
  biases on detection of deepfakes on networks, npj Complex SystemsPreprint,
  see arXiv:2210.10026 (2024).

\bibitem{castioni2022voice}
P.~Castioni, G.~Andrighetto, R.~Gallotti, E.~Polizzi, M.~{De Domenico},
  \href{https://royalsocietypublishing.org/doi/10.1098/rsos.220716}{{The voice
  of few, the opinions of many: Evidence of social biases in Twitter COVID-19
  fake news sharing}}, Royal Society Open Science 9~(10) (oct 2022).
\newblock \href {http://arxiv.org/abs/2112.01304} {\path{arXiv:2112.01304}},
  \href {https://doi.org/10.1098/rsos.220716} {\path{doi:10.1098/rsos.220716}}.
\newline\urlprefix\url{https://royalsocietypublishing.org/doi/10.1098/rsos.220716}

\bibitem{gonzalez-bailon2023asymmetric}
S.~Gonz{\'{a}}lez-Bail{\'{o}}n, D.~Lazer, P.~Barber{\'{a}}, M.~Zhang,
  H.~Allcott, T.~Brown, A.~Crespo-Tenorio, D.~Freelon, M.~Gentzkow, A.~M.
  Guess, S.~Iyengar, Y.~M. Kim, N.~Malhotra, D.~Moehler, B.~Nyhan, J.~Pan,
  C.~V. Rivera, J.~Settle, E.~Thorson, R.~Tromble, A.~Wilkins, M.~Wojcieszak,
  C.~K. de~Jonge, A.~Franco, W.~Mason, N.~J. Stroud, J.~A. Tucker,
  \href{https://www.science.org/doi/10.1126/science.ade7138}{{Asymmetric
  ideological segregation in exposure to political news on Facebook}}, Science
  (New York, N.Y.) 381~(6656) (2023) 392--398.
\newblock \href {https://doi.org/10.1126/science.ade7138}
  {\path{doi:10.1126/science.ade7138}}.
\newline\urlprefix\url{https://www.science.org/doi/10.1126/science.ade7138}

\bibitem{gonzalez-bailon2022advantage}
S.~Gonz{\'{a}}lez-Bail{\'{o}}n, V.~D'Andrea, D.~Freelon, M.~{De Domenico},
  \href{https://academic.oup.com/pnasnexus/article/doi/10.1093/pnasnexus/pgac137/6651695}{{The
  advantage of the right in social media news sharing}}, PNAS Nexus 1~(3) (jul
  2022).
\newblock \href {https://doi.org/10.1093/pnasnexus/pgac137}
  {\path{doi:10.1093/pnasnexus/pgac137}}.
\newline\urlprefix\url{https://academic.oup.com/pnasnexus/article/doi/10.1093/pnasnexus/pgac137/6651695}

\bibitem{proferes2021studying}
N.~Proferes, N.~Jones, S.~Gilbert, C.~Fiesler, M.~Zimmer, Studying reddit: A
  systematic overview of disciplines, approaches, methods, and ethics, Social
  Media+ Society 7~(2) (2021) 20563051211019004.

\bibitem{peralta2021effect}
A.~F. Peralta, M.~Neri, J.~Kert{\'e}sz, G.~I{\~n}iguez, Effect of algorithmic
  bias and network structure on coexistence, consensus, and polarization of
  opinions, Physical Review E 104~(4) (2021) 044312.

\bibitem{corsi2024evaluating}
E.~Corsi, Evaluating twitter’s algorithmic amplification of low-credibility
  content: An observational study, EPJ Data Science 13 (2024) 18.
\newblock \href {https://doi.org/10.1140/epjds/s13688-024-00456-3}
  {\path{doi:10.1140/epjds/s13688-024-00456-3}}.

\bibitem{metaxas2012social}
P.~T. Metaxas, E.~Mustafaraj,
  \href{https://www.science.org/doi/10.1126/science.1230456}{{Social media and
  the elections}} (oct 2012).
\newblock \href {https://doi.org/10.1126/science.1230456}
  {\path{doi:10.1126/science.1230456}}.
\newline\urlprefix\url{https://www.science.org/doi/10.1126/science.1230456}

\bibitem{babaei2021analyzing}
M.~Babaei, J.~Kulshrestha, A.~Chakraborty, E.~M. Redmiles, M.~Cha, K.~P.
  Gummadi, Analyzing biases in perception of truth in news stories and their
  implications for fact checking, IEEE Trans. Comput. Soc. 9~(3) (2021)
  839--850.

\bibitem{vargo2017agenda}
C.~J. Vargo, L.~Guo, M.~A. Amazeen,
  \href{http://journals.sagepub.com/doi/10.1177/1461444817712086}{{The
  agenda-setting power of fake news: A big data analysis of the online media
  landscape from 2014 to 2016}}, New Media \& Society 20~(5) (2017)
  146144481771208.
\newblock \href {https://doi.org/10.1177/1461444817712086}
  {\path{doi:10.1177/1461444817712086}}.
\newline\urlprefix\url{http://journals.sagepub.com/doi/10.1177/1461444817712086}

\bibitem{doshi2018impact}
A.~R. Doshi, S.~Raghavan, R.~Weiss, E.~Petitt,
  \href{https://www.ssrn.com/abstract=3093397}{{The Impact of the Supply of
  Fake News on Consumer Behavior During the 2016 US Election}} (jun 2018).
\newblock \href {https://doi.org/10.2139/ssrn.3093397}
  {\path{doi:10.2139/ssrn.3093397}}.
\newline\urlprefix\url{https://www.ssrn.com/abstract=3093397}

\bibitem{gallotti2020assessing}
R.~Gallotti, F.~Valle, N.~Castaldo, P.~Sacco, M.~{De Domenico},
  \href{https://www.nature.com/articles/s41562-020-00994-6}{{Assessing the
  risks of ‘infodemics' in response to COVID-19 epidemics}}, Nature Human
  Behaviour 4~(12) (2020) 1285--1293.
\newblock \href {http://arxiv.org/abs/2004.03997} {\path{arXiv:2004.03997}},
  \href {https://doi.org/10.1038/s41562-020-00994-6}
  {\path{doi:10.1038/s41562-020-00994-6}}.
\newline\urlprefix\url{https://www.nature.com/articles/s41562-020-00994-6}

\bibitem{tambuscio2018network}
M.~Tambuscio, D.~F.~M. Oliveira, G.~L. Ciampaglia, G.~Ruffo,
  \href{http://dx.doi.org/10.1007/s42001-018-0018-9}{Network segregation in a
  model of misinformation and fact-checking}, Journal of Computational Social
  Science 1~(2) (2018) 261–275.
\newblock \href {https://doi.org/10.1007/s42001-018-0018-9}
  {\path{doi:10.1007/s42001-018-0018-9}}.
\newline\urlprefix\url{http://dx.doi.org/10.1007/s42001-018-0018-9}

\bibitem{han2013competing}
W.~Han, A.~Rakhlin, K.~Sridharan,
  \href{https://proceedings.mlr.press/v30/Han13.html}{Competing with
  strategies}, in: S.~Shalev-Shwartz, I.~Steinwart (Eds.), Proceedings of the
  26th Annual Conference on Learning Theory, Vol.~30 of Proceedings of Machine
  Learning Research, PMLR, Princeton, NJ, USA, 2013, pp. 966--992.
\newline\urlprefix\url{https://proceedings.mlr.press/v30/Han13.html}

\bibitem{bak-coleman2022combining}
J.~B. Bak-Coleman, I.~Kennedy, M.~Wack, A.~Beers, J.~S. Schafer, E.~S. Spiro,
  K.~Starbird, J.~D. West, Combining interventions to reduce the spread of
  viral misinformation., Nat Hum Behav 6~(10) (2022) 1372--1380.

\bibitem{morone2015influence}
F.~Morone, H.~A. Makse,
  \href{https://www.nature.com/articles/nature14604}{{Influence maximization in
  complex networks through optimal percolation}}, Nature 524~(7563) (2015)
  65--68.
\newblock \href {http://arxiv.org/abs/1506.08326} {\path{arXiv:1506.08326}},
  \href {https://doi.org/10.1038/nature14604} {\path{doi:10.1038/nature14604}}.
\newline\urlprefix\url{https://www.nature.com/articles/nature14604}

\bibitem{nwala2023language}
A.~C. Nwala, A.~Flammini, F.~Menczer,
  \href{https://epjdatascience.springeropen.com/articles/10.1140/epjds/s13688-023-00410-9}{{A
  language framework for modeling social media account behavior}}, EPJ Data
  Science 12~(1) (2023) 33.
\newblock \href {https://doi.org/10.1140/epjds/s13688-023-00410-9}
  {\path{doi:10.1140/epjds/s13688-023-00410-9}}.
\newline\urlprefix\url{https://epjdatascience.springeropen.com/articles/10.1140/epjds/s13688-023-00410-9}

\bibitem{varol2017online}
O.~Varol, E.~Ferrara, C.~A. Davis, F.~Menczer, A.~Flammini,
  \href{https://ojs.aaai.org/index.php/ICWSM/article/view/14871}{{Online
  human-bot interactions: Detection, estimation, and characterization}}, in:
  Proceedings of the 11th International Conference on Web and Social Media,
  ICWSM 2017, Vol.~11, 2017, pp. 280--289.
\newblock \href {http://arxiv.org/abs/1703.03107} {\path{arXiv:1703.03107}},
  \href {https://doi.org/10.1609/icwsm.v11i1.14871}
  {\path{doi:10.1609/icwsm.v11i1.14871}}.
\newline\urlprefix\url{https://ojs.aaai.org/index.php/ICWSM/article/view/14871}

\bibitem{sayyadiharikandeh2020detection}
M.~Sayyadiharikandeh, O.~Varol, K.-C. Yang, A.~Flammini, F.~Menczer, Detection
  of novel social bots by ensembles of specialized classifiers, in: Proceedings
  of the 29th ACM international conference on information \& knowledge
  management, 2020, pp. 2725--2732.

\bibitem{yang2019arming}
K.-C. Yang, O.~Varol, C.~A. Davis, E.~Ferrara, A.~Flammini, F.~Menczer, Arming
  the public with artificial intelligence to counter social bots, Human
  Behavior and Emerging Technologies 1~(1) (2019) 48--61.

\bibitem{yang2020scalable}
K.-C. Yang, O.~Varol, P.-M. Hui, F.~Menczer,
  \href{https://ojs.aaai.org/index.php/AAAI/article/view/5460}{Scalable and
  generalizable social bot detection through data selection}, Proceedings of
  the AAAI Conference on Artificial Intelligence 34~(01) (2020) 1096--1103.
\newblock \href {https://doi.org/10.1609/aaai.v34i01.5460}
  {\path{doi:10.1609/aaai.v34i01.5460}}.
\newline\urlprefix\url{https://ojs.aaai.org/index.php/AAAI/article/view/5460}

\bibitem{guess2023reshares}
A.~M. Guess, N.~Malhotra, J.~Pan, P.~Barber{\'{a}}, H.~Allcott, T.~Brown,
  A.~Crespo-Tenorio, D.~Dimmery, D.~Freelon, M.~Gentzkow,
  S.~Gonz{\'{a}}lez-Bail{\'{o}}n, E.~Kennedy, Y.~M. Kim, D.~Lazer, D.~Moehler,
  B.~Nyhan, C.~V. Rivera, J.~Settle, D.~R. Thomas, E.~Thorson, R.~Tromble,
  A.~Wilkins, M.~Wojcieszak, B.~Xiong, C.~K. de~Jonge, A.~Franco, W.~Mason,
  N.~J. Stroud, J.~A. Tucker,
  \href{https://www.science.org/doi/10.1126/science.add8424}{{Reshares on
  social media amplify political news but do not detectably affect beliefs or
  opinions}}, Science (New York, N.Y.) 381~(6656) (2023) 404--408.
\newblock \href {https://doi.org/10.1126/science.add8424}
  {\path{doi:10.1126/science.add8424}}.
\newline\urlprefix\url{https://www.science.org/doi/10.1126/science.add8424}

\bibitem{guess2023how}
A.~M. Guess, N.~Malhotra, J.~Pan, P.~Barber{\'{a}}, H.~Allcott, T.~Brown,
  A.~Crespo-Tenorio, D.~Dimmery, D.~Freelon, M.~Gentzkow,
  S.~Gonz{\'{a}}lez-Bail{\'{o}}n, E.~Kennedy, Y.~M. Kim, D.~Lazer, D.~Moehler,
  B.~Nyhan, C.~V. Rivera, J.~Settle, D.~R. Thomas, E.~Thorson, R.~Tromble,
  A.~Wilkins, M.~Wojcieszak, B.~Xiong, C.~K. de~Jonge, A.~Franco, W.~Mason,
  N.~J. Stroud, J.~A. Tucker,
  \href{https://www.science.org/doi/10.1126/science.abp9364}{{How do social
  media feed algorithms affect attitudes and behavior in an election
  campaign?}}, Science (New York, N.Y.) 381~(6656) (2023) 398--404.
\newblock \href {https://doi.org/10.1126/science.abp9364}
  {\path{doi:10.1126/science.abp9364}}.
\newline\urlprefix\url{https://www.science.org/doi/10.1126/science.abp9364}

\bibitem{nyhan2023like}
B.~Nyhan, J.~Settle, E.~Thorson, M.~Wojcieszak, P.~Barber{\'{a}}, A.~Y. Chen,
  H.~Allcott, T.~Brown, A.~Crespo-Tenorio, D.~Dimmery, D.~Freelon, M.~Gentzkow,
  S.~Gonz{\'{a}}lez-Bail{\'{o}}n, A.~M. Guess, E.~Kennedy, Y.~M. Kim, D.~Lazer,
  N.~Malhotra, D.~Moehler, J.~Pan, D.~R. Thomas, R.~Tromble, C.~V. Rivera,
  A.~Wilkins, B.~Xiong, C.~K. de~Jonge, A.~Franco, W.~Mason, N.~J. Stroud,
  J.~A. Tucker,
  \href{https://www.nature.com/articles/s41586-023-06297-w}{{Like-minded
  sources on Facebook are prevalent but not polarizing}}, Nature 620~(7972)
  (2023) 137--144.
\newblock \href {https://doi.org/10.1038/s41586-023-06297-w}
  {\path{doi:10.1038/s41586-023-06297-w}}.
\newline\urlprefix\url{https://www.nature.com/articles/s41586-023-06297-w}

\bibitem{volfovsky2023depolarization}
A.~Volfovsky, C.~Bail, Depolarization via anonymous mobile online communication
  (2023).

\bibitem{combs2023reducing}
A.~Combs, G.~Tierney, B.~Guay, F.~Merhout, C.~A. Bail, D.~S. Hillygus,
  A.~Volfovsky,
  \href{https://www.nature.com/articles/s41562-023-01655-0}{{Reducing political
  polarization in the United States with a mobile chat platform}}, Nature Human
  Behaviour 7~(9) (2023) 1454--1461.
\newblock \href {https://doi.org/10.1038/s41562-023-01655-0}
  {\path{doi:10.1038/s41562-023-01655-0}}.
\newline\urlprefix\url{https://www.nature.com/articles/s41562-023-01655-0}

\bibitem{jusup2022social}
M.~Jusup, P.~Holme, K.~Kanazawa, M.~Takayasu, I.~Romic, Z.~Wang, S.~Gecek,
  T.~Lipic, B.~Podobnik, L.~Wang, W.~Luo, T.~Klanjscek, J.~Fan, S.~Boccaletti,
  M.~Perc, Social physics, Phys. Rep. 948 (2022) 1--148.
\newblock \href {https://doi.org/10.1016/j.physrep.2021.10.005}
  {\path{doi:10.1016/j.physrep.2021.10.005}}.

\bibitem{galam2008sociophysics}
S.~Galam, Sociophysics: A review of {G}alam models, Int. J. Mod. Phys. C 19~(3)
  (2008) 409--440.

\bibitem{sen2013sociophysics}
P.~Sen, B.~K. Chakrabarti, Sociophysics: An Introduction, Oxford University
  Press, Oxford, 2013.

\bibitem{starnini2025opinion}
M.~Starnini, F.~Baumann, T.~Galla, D.~Garcia, G.~I{\~n}iguez, M.~Karsai,
  J.~Lorenz, K.~Sznajd-Weron, Opinion dynamics: Statistical physics and beyond,
  arXiv preprint arXiv:2507.11521 (2025).

\bibitem{flache2017models}
A.~Flache, M.~M{\"a}s, T.~Feliciani, E.~Chattoe-Brown, G.~Deffuant, S.~Huet,
  J.~Lorenz, Models of social influence: Towards the next frontiers, {JASSS-The
  Journal of Artificial Societies and Social} Simulation 20~(4) (2017) 2.

\bibitem{sznajd-weron2000opinion}
K.~SZNAJD-WERON, J.~SZNAJD, Opinion evolution in closed community, Int. J. Mod.
  Phys. C 11~(06) (2000) 1157--1165.

\bibitem{liggett1999stochastic}
T.~M. Liggett, Stochastic Interacting Systems: Contact, Voter, and Exclusion
  Processes, Springer-Verlag, 1999.

\bibitem{sood2005voter}
V.~Sood, S.~Redner, Voter model on heterogeneous graphs, Physical review
  letters 94~(17) (2005) 178701.

\bibitem{oliveira1992isotropic}
M.~J. de~Oliveira, Isotropic majority-vote model on a square lattice, J. Stat.
  Phys. 66 (1992) 273--281.

\bibitem{degroot1974reaching}
M.~H. {DeGroot}, Reaching a consensus, J. Am. Stat. Assoc. 69~(345) (1974)
  118--121.

\bibitem{friedkin1990social}
N.~E. Friedkin, E.~C. Johnsen, Social influence and opinions, J. Math. Sociol.
  15~(3-4) (1990) 193--206.

\bibitem{axelrod1997dissemination}
R.~Axelrod, \href{https://www.jstor.org/stable/174371}{The dissemination of
  culture: A model with local convergence and global polarization}, J. Confl.
  Resolut. 41~(2) (1997) 203--226.
\newblock \href {https://doi.org/10.1177/0022002797041002001}
  {\path{doi:10.1177/0022002797041002001}}.
\newline\urlprefix\url{https://www.jstor.org/stable/174371}

\bibitem{castellano2000nonequilibrium}
C.~Castellano, M.~Marsili, A.~Vespignani,
  \href{https://link.aps.org/doi/10.1103/PhysRevLett.85.3536}{Nonequilibrium
  phase transition in a model for social influence}, Phys. Rev. Lett. 85~(16)
  (2000) 3536--3539.
\newblock \href {https://doi.org/10.1103/PhysRevLett.85.3536}
  {\path{doi:10.1103/PhysRevLett.85.3536}}.
\newline\urlprefix\url{https://link.aps.org/doi/10.1103/PhysRevLett.85.3536}

\bibitem{deffuant2000mixing}
G.~Deffuant, D.~Neau, F.~Amblard, G.~Weisbuch, Mixing beliefs among interacting
  agents, Adv. Complex Syst. 3 (2000) 87--98.
\newblock \href {https://doi.org/10.1142/S0219525900000078}
  {\path{doi:10.1142/S0219525900000078}}.

\bibitem{gronlund2007dynamic}
A.~Gr\"onlund, P.~Holme, P.~Minnhagen, Dynamic scaling regimes of collective
  decision making, Europhys. Lett. 81~(2) (2007) 28003.

\bibitem{holme2016collective}
P.~Holme, H.-H. Jo, Collective decision making with a mix of majority and
  minority seekers, Phys. Rev. E 93~(5) (2016) 052308.
\newblock \href {https://doi.org/10.1103/PhysRevE.93.052308}
  {\path{doi:10.1103/PhysRevE.93.052308}}.

\bibitem{curty2006phase}
P.~Curty, M.~Marsili, Phase coexistence in a forecasting game, J. Stat. Mech.
  2006~(03) (2006) P03013.

\bibitem{assenova2018modeling}
V.~A. Assenova, Modeling the diffusion of complex innovations as a process of
  opinion formation through social networks, PloS ONE 13~(5) (2018) e0196699.

\bibitem{quattrociocchi2014opinion}
W.~Quattrociocchi, G.~Caldarelli, A.~Scala,
  \href{https://www.nature.com/articles/srep04938}{Opinion dynamics on
  interacting networks: Media competition and social influence}, Scientific
  Reports 4 (2014) 4938.
\newblock \href {https://doi.org/10.1038/srep04938}
  {\path{doi:10.1038/srep04938}}.
\newline\urlprefix\url{https://www.nature.com/articles/srep04938}

\bibitem{sobkowicz2009modelling}
P.~Sobkowicz, Modelling opinion formation with physics tools: Call for closer
  link with reality, Journal of Artificial Societies and Social Simulation
  12~(1) (2009) 11.

\bibitem{carpentras2023we}
D.~Carpentras, Why we are failing at connecting opinion dynamics to the
  empirical world, Review of Artificial Societies and Social Simulations
  (2023).

\bibitem{donnellan2009introduction}
M.~B. Donnellan, R.~E. Lucas, W.~Fleeson, Introduction to personality and
  assessment at age 40: Reflections on the legacy of the person--situation
  debate and the future of person--situation integration, Journal of Research
  in Personality 43~(2) (2009) 117--119.

\bibitem{brandstatter1982group}
H.~Brandst{\"a}tter, J.~H. Davis, G.~Stocker-Kreichgauer, Group decision
  making, no.~25, Academic Press London, 1982.

\bibitem{akers1979social}
R.~L. Akers, M.~D. Krohn, L.~Lanza-Kaduce, M.~Radosevich, Social learning and
  deviant behavior: A specific test of a general theory, American Sociological
  Review (1979) 636--655.

\bibitem{wood2000attitude}
W.~Wood, Attitude change: Persuasion and social influence, Annual Review of
  Psychology 51~(1) (2000) 539--570.

\bibitem{myers2012exploring}
D.~G. Myers, J.~M. Twenge, Exploring social psychology, McGraw-Hill New York,
  2012.

\bibitem{nowak1990private}
A.~Nowak, J.~Szamrej, B.~Latan{\'e}, From private attitude to public opinion: A
  dynamic theory of social impact., Psychological Review 97~(3) (1990) 362.

\bibitem{smith2007agent}
E.~R. Smith, F.~R. Conrey, Agent-based modeling: A new approach for theory
  building in social psychology, Personality and Social Psychology Review
  11~(1) (2007) 87--104.

\bibitem{nail2013proposal}
P.~R. Nail, S.~I. Di~Domenico, G.~MacDonald, Proposal of a double diamond model
  of social response, Review of General Psychology 17~(1) (2013) 1--19.

\bibitem{nail2000proposal}
P.~R. Nail, G.~MacDonald, D.~A. Levy, Proposal of a four-dimensional model of
  social response., Psychological Bulletin 126~(3) (2000) 454.

\bibitem{brehm1975effect}
J.~W. Brehm, M.~Mann, Effect of importance of freedom and attraction to group
  members on influence produced by group pressure, Journal of Personality and
  Social Psychology 31~(5) (1975) 816.

\bibitem{eagly1981sex}
A.~H. Eagly, W.~Wood, L.~Fishbaugh, Sex differences in conformity: Surveillance
  by the group as a determinant of male nonconformity, Journal of Personality
  and Social Psychology 40~(2) (1981) 384.

\bibitem{halama2011personality}
P.~Halama, M.~La{\v{c}}n{\'a}, Personality change following religious
  conversion: Perceptions of converts and their close acquaintances, Mental
  Health, Religion \& Culture 14~(8) (2011) 757--768.

\bibitem{coch1948overcoming}
L.~Coch, J.~R.~P. French~Jr, Overcoming resistance to change, Human relations
  1~(4) (1948) 512--532.

\bibitem{baron1996forgotten}
R.~S. Baron, J.~A. Vandello, B.~Brunsman, The forgotten variable in conformity
  research: Impact of task importance on social influence., Journal of
  personality and social psychology 71~(5) (1996) 915.

\bibitem{wright2005social}
D.~B. Wright, S.~A. Mathews, E.~M. Skagerberg, Social recognition memory: the
  effect of other people's responses for previously seen and unseen items.,
  Journal of Experimental Psychology: Applied 11~(3) (2005) 200.

\bibitem{nail2016rethinking}
P.~R. Nail, K.~Sznajd-Weron, Rethinking the diamond model: Theory and research
  support self-anticonformity as a basic response and influence process, The
  psychology of consumer and social influence: Theory and research (2016)
  99--136.

\bibitem{willis1965conformity}
R.~H. Willis, Conformity, independence, and anticonformity, Human Relations
  18~(4) (1965) 373--388.

\bibitem{krapivsky1992kinetics}
P.~L. Krapivsky, Kinetics of monomer-monomer surface catalytic reactions,
  Physical Review A 45~(2) (1992) 1067.

\bibitem{evans1993kinetics}
J.~W. Evans, T.~Ray, Kinetics of the monomer-monomer surface reaction model,
  Physical Review E 47~(2) (1993) 1018.

\bibitem{frachebourg1996exact}
L.~Frachebourg, P.~Krapivsky, Exact results for kinetics of catalytic
  reactions, Physical Review E 53~(4) (1996) R3009.

\bibitem{kimura1955solution}
M.~Kimura, Solution of a process of random genetic drift with a continuous
  model, Proceedings of the National Academy of Sciences 41~(3) (1955)
  144--150.

\bibitem{crow1970introduction}
J.~F. Crow, M.~Kimura, An introduction to population genetics theory, An
  introduction to population genetics theory. (1970).

\bibitem{maruyama1977stochastic}
T.~Maruyama, Stochastic problems in population genetics, Springer, 1977.

\bibitem{borile2015coexistence}
C.~Borile, D.~Molina-Garcia, A.~Maritan, M.~A. Munoz, Coexistence in neutral
  theories: interplay of criticality and mild local preferences, Journal of
  Statistical Mechanics: Theory and Experiment 2015~(1) (2015) P01030.

\bibitem{holley1975ergodic}
R.~A. Holley, T.~M. Liggett, Ergodic theorems for weakly interacting infinite
  systems and the voter model, The Annals of Probability (1975) 643--663.

\bibitem{liggett1985interacting}
T.~M. Liggett, Interacting particle systems, Springer Science \& Business
  Media, 1985.

\bibitem{cox1986diffusive}
J.~T. Cox, D.~Griffeath, Diffusive clustering in the two dimensional voter
  model, The Annals of Probability (1986) 347--370.

\bibitem{clifford1973model}
P.~Clifford, A.~Sudbury, A model for spatial conflict, Biometrika 60~(3) (1973)
  581--588.

\bibitem{ben-naim1996coarsening}
E.~Ben-Naim, L.~Frachebourg, P.~L. Krapivsky, Coarsening and persistence in the
  voter model, Physical Review E 53~(4) (1996) 3078.

\bibitem{dornic2001critical}
I.~Dornic, H.~Chat{\'e}, J.~Chave, H.~Hinrichsen, {Critical coarsening without
  surface tension: The universality class of the voter model}, Physical Review
  Letters 87~(4) (2001) 045701.

\bibitem{tartaglia2018coarsening}
A.~Tartaglia, L.~F. Cugliandolo, M.~Picco, Coarsening and percolation in the
  kinetic $2d$ ising model with spin exchange updates and the voter model,
  Journal of Statistical Mechanics: Theory and Experiment 2018~(8) (2018)
  083202.

\bibitem{gillespie1977exact}
D.~T. Gillespie, Exact stochastic simulation of coupled chemical reactions, The
  Journal of Physical Chemistry 81~(25) (1977) 2340--2361.

\bibitem{masuda2022gillespie}
N.~Masuda, C.~L. Vestergaard, Gillespie algorithms for stochastic multiagent
  dynamics in populations and networks, Cambridge University Press, 2022.

\bibitem{krapivsky2010kinetic}
P.~L. Krapivsky, S.~Redner, E.~Ben-Naim, A kinetic view of statistical physics,
  Cambridge University Press, 2010.

\bibitem{slanina2003analytical}
F.~Slanina, H.~Lavicka, Analytical results for the sznajd model of opinion
  formation, The European Physical Journal B-Condensed Matter and Complex
  Systems 35~(2) (2003) 279--288.

\bibitem{mckane2007singular}
A.~J. McKane, D.~Waxman, Singular solutions of the diffusion equation of
  population genetics, Journal of Theoretical Biology 247~(4) (2007) 849--858.

\bibitem{vazquez2008analytical}
F.~V{\'a}zquez, V.~M. Egu{\'\i}luz, Analytical solution of the voter model on
  uncorrelated networks, New Journal of Physics 10~(6) (2008) 063011.

\bibitem{pugliese2009heterogeneous}
E.~Pugliese, C.~Castellano, Heterogeneous pair approximation for voter models
  on networks, EPL (Europhysics Letters) 88~(5) (2009) 58004.

\bibitem{peralta2020binary}
A.~F. Peralta, R.~Toral, Binary-state dynamics on complex networks: Stochastic
  pair approximation and beyond, Physical Review Research 2~(4) (2020) 043370.

\bibitem{suchecki2004conservation}
K.~Suchecki, V.~M. Egu{\'\i}luz, M.~San~Miguel, Conservation laws for the voter
  model in complex networks, Europhysics Letters 69~(2) (2004) 228.

\bibitem{sood2008voter}
V.~Sood, T.~Antal, S.~Redner, Voter models on heterogeneous networks, Physical
  Review E 77~(4) (2008) 041121.

\bibitem{artime2018first}
O.~Artime, N.~Khalil, R.~Toral, M.~San~Miguel, First-passage distributions for
  the one-dimensional fokker-planck equation, Physical Review E 98~(4) (2018)
  042143.

\bibitem{bond2005group}
R.~Bond, Group size and conformity, Group Processes \& Intergroup Relations
  8~(4) (2005) 331--354.

\bibitem{asch1955opinions}
S.~E. Asch, Opinions and social pressure, Scientific American 193~(5) (1955)
  31--35.

\bibitem{asch1956studies}
S.~E. Asch, Studies of independence and conformity: I. a minority of one
  against a unanimous majority., Psychological Monographs: General and Applied
  70~(9) (1956) 1.

\bibitem{artime2019herding}
O.~Artime, A.~Carro, A.~F. Peralta, J.~J. Ramasco, M.~San~Miguel, R.~Toral,
  Herding and idiosyncratic choices: Nonlinearity and aging-induced transitions
  in the noisy voter model, Comptes Rendus Physique 20~(4) (2019) 262--274.

\bibitem{peralta2018analytical}
A.~F. Peralta, A.~Carro, M.~San~Miguel, R.~Toral, Analytical and numerical
  study of the non-linear noisy voter model on complex networks, Chaos: An
  Interdisciplinary Journal of Nonlinear Science 28~(7) (2018).

\bibitem{castellano2009nonlinear}
C.~Castellano, M.~A. Mu{\~n}oz, R.~Pastor-Satorras, Nonlinear $q$-voter model,
  Physical Review E 80~(4) (2009) 041129.

\bibitem{przybyla2011exit}
P.~Przyby{\l}a, K.~Sznajd-Weron, M.~Tabiszewski, Exit probability in a
  one-dimensional nonlinear $q$-voter model, Physical Review E 84~(3) (2011)
  031117.

\bibitem{timpanaro2014exit}
A.~M. Timpanaro, C.~P. Prado, Exit probability of the one-dimensional $q$-voter
  model: Analytical results and simulations for large networks, Physical Review
  E 89~(5) (2014) 052808.

\bibitem{timpanaro2015analytical}
A.~M. Timpanaro, S.~Galam, Analytical expression for the exit probability of
  the $q$-voter model in one dimension, Physical Review E 92~(1) (2015) 012807.

\bibitem{moretti2013mean}
P.~Moretti, S.~Liu, C.~Castellano, R.~Pastor-Satorras, Mean-field analysis of
  the $q$-voter model on networks, Journal of Statistical Physics 151 (2013)
  113--130.

\bibitem{vieira2020pair}
A.~R. Vieira, A.~F. Peralta, R.~Toral, M.~S. Miguel, C.~Anteneodo, Pair
  approximation for the noisy threshold $q$-voter model, Physical Review E
  101~(5) (2020) 052131.

\bibitem{jkedrzejewski2019statistical}
A.~J{\k{e}}drzejewski, K.~Sznajd-Weron, Statistical physics of opinion
  formation: Is it a spoof?, Comptes Rendus Physique 20~(4) (2019) 244--261.

\bibitem{schweitzer2009nonlinear}
F.~Schweitzer, L.~Behera, Nonlinear voter models: the transition from invasion
  to coexistence, The European Physical Journal B 67 (2009) 301--318.

\bibitem{vazquez2008systems}
F.~Vazquez, C.~L{\'o}pez, Systems with two symmetric absorbing states: relating
  the microscopic dynamics with the macroscopic behavior, Physical Review E
  78~(6) (2008) 061127.

\bibitem{abrams2003modelling}
D.~M. Abrams, S.~H. Strogatz, Modelling the dynamics of language death, Nature
  424~(6951) (2003) 900--900.

\bibitem{vazquez2010agent}
F.~Vazquez, X.~Castell{\'o}, M.~San~Miguel, Agent based models of language
  competition: macroscopic descriptions and order--disordertransitions, Journal
  of Statistical Mechanics: Theory and Experiment 2010~(04) (2010) P04007.

\bibitem{eminente2022interplay}
C.~Eminente, O.~Artime, M.~{De Domenico},
  \href{https://www.sciencedirect.com/science/article/abs/pii/S0960077922009389}{{Interplay
  between exogenous triggers and endogenous behavioral changes in contagion
  processes on social networks}}, Chaos, Solitons and Fractals 165 (2022)
  112759.
\newblock \href {https://doi.org/10.1016/j.chaos.2022.112759}
  {\path{doi:10.1016/j.chaos.2022.112759}}.
\newline\urlprefix\url{https://www.sciencedirect.com/science/article/abs/pii/S0960077922009389}

\bibitem{kirman1993ants}
A.~Kirman, Ants, rationality, and recruitment, The Quarterly Journal of
  Economics 108~(1) (1993) 137--156.

\bibitem{alfarano2008time}
S.~Alfarano, T.~Lux, F.~Wagner, Time variation of higher moments in a financial
  market with heterogeneous agents: An analytical approach, Journal of Economic
  Dynamics and Control 32~(1) (2008) 101--136.

\bibitem{carro2015markets}
A.~Carro, R.~Toral, M.~San~Miguel, Markets, herding and response to external
  information, PloS one 10~(7) (2015) e0133287.

\bibitem{fichthorn1989noise}
K.~Fichthorn, E.~Gulari, R.~Ziff, Noise-induced bistability in a monte carlo
  surface-reaction model, Physical Review Letters 63~(14) (1989) 1527.

\bibitem{considine1989comment}
D.~Considine, S.~Redner, H.~Takayasu, Comment on ‘‘noise-induced
  bistability in a monte carlo surface-reaction model’’, Physical Review
  Letters 63~(26) (1989) 2857.

\bibitem{carro2016noisy}
A.~Carro, R.~Toral, M.~San~Miguel, The noisy voter model on complex networks,
  Scientific Reports 6~(1) (2016) 24775.

\bibitem{granovsky1995noisy}
B.~L. Granovsky, N.~Madras, The noisy voter model, Stochastic Processes and
  their applications 55~(1) (1995) 23--43.

\bibitem{peralta2018stochastic}
A.~F. Peralta, A.~Carro, M.~S. Miguel, R.~Toral, Stochastic pair approximation
  treatment of the noisy voter model, New Journal of Physics 20~(10) (2018)
  103045.

\bibitem{kononovicius2020noisy}
A.~Kononovicius, Noisy voter model for the anomalous diffusion of parliamentary
  presence, Journal of Statistical Mechanics: Theory and Experiment 2020~(6)
  (2020) 063405.

\bibitem{van1992stochastic}
N.~G. Van~Kampen, Stochastic processes in physics and chemistry, Vol.~1,
  Elsevier, 1992.

\bibitem{redner2019reality}
S.~Redner, Reality-inspired voter models: A mini-review, Comptes Rendus
  Physique 20~(4) (2019) 275--292.

\bibitem{sirbu2016opinion}
A.~S{\^\i}rbu, V.~Loreto, V.~D. Servedio, F.~Tria, Opinion dynamics: models,
  extensions and external effects, in: Participatory Sensing, Opinions and
  Collective Awareness, Springer, 2016, pp. 363--401.

\bibitem{galam1986majority}
S.~Galam, Majority rule, hierarchical structures, and democratic
  totalitarianism: A statistical approach, Journal of Mathematical Psychology
  30~(4) (1986) 426--434.

\bibitem{galam2002minority}
S.~Galam, Minority opinion spreading in random geometry, The European Physical
  Journal B-Condensed Matter and Complex Systems 25 (2002) 403--406.

\bibitem{friedman1985tyranny}
M.~Friedman, R.~D. Friedman, Tyranny of the status quo, Penguin Harmondsworth,
  1985.

\bibitem{tessone2004neighborhood}
C.~J. Tessone, R.~Toral, P.~Amengual, H.~S. Wio, M.~San~Miguel, Neighborhood
  models of minority opinion spreading, The European Physical Journal
  B-Condensed Matter and Complex Systems 39~(4) (2004) 535--544.

\bibitem{stauffer2000generalization}
D.~Stauffer, A.~O. Sousa, S.~M. De~Oliveira, Generalization to square lattice
  of sznajd sociophysics model, International Journal of Modern Physics C
  11~(06) (2000) 1239--1245.

\bibitem{pedraza2021analytical}
L.~Pedraza, J.~P. Pinasco, N.~Saintier, P.~Balenzuela, An analytical
  formulation for multidimensional continuous opinion models, Chaos, Solitons
  \& Fractals 152 (2021) 111368.

\bibitem{lorenz2007continuous}
J.~Lorenz, Continuous opinion dynamics under bounded confidence: A survey,
  International Journal of Modern Physics C 18~(12) (2007) 1819--1838.

\bibitem{ben-naim2003bifurcations}
E.~Ben-Naim, P.~L. Krapivsky, S.~Redner, Bifurcations and patterns in
  compromise processes, Physica D: Nonlinear Phenomena 183~(3-4) (2003)
  190--204.

\bibitem{weisbuch2002meet}
G.~Weisbuch, G.~Deffuant, F.~Amblard, J.-P. Nadal, Meet, discuss, and
  segregate!, Complexity 7~(3) (2002) 55--63.

\bibitem{hegselmann2002opinion}
R.~Hegselmann, K.~Ulrich, Opinion dynamics and bounded confidence: Models,
  analysis and simulation, Journal of Artificial Societies and Social
  Simulation 5~(3) (2002).

\bibitem{lorenz2005stabilization}
J.~Lorenz, A stabilization theorem for dynamics of continuous opinions, Physica
  A: Statistical Mechanics and its Applications 355~(1) (2005) 217--223.

\bibitem{fennell2021generalized}
S.~C. Fennell, K.~Burke, M.~Quayle, J.~P. Gleeson, Generalized mean-field
  approximation for the deffuant opinion dynamics model on networks, Physical
  Review E 103~(1) (2021) 012314.

\bibitem{meng2018opinion}
X.~F. Meng, R.~A. Van~Gorder, M.~A. Porter, Opinion formation and distribution
  in a bounded-confidence model on various networks, Physical Review E 97~(2)
  (2018) 022312.

\bibitem{stauffer2004simulation}
D.~Stauffer, H.~Meyer-Ortmanns, Simulation of consensus model of deffuant et
  al. on a barabasi--albert network, International Journal of Modern Physics C
  15~(02) (2004) 241--246.

\bibitem{gandica2010continuous}
Y.~Gandica, M.~del Castillo-Mussot, G.~J. V{\'a}zquez, S.~Rojas, Continuous
  opinion model in small-world directed networks, Physica A: Statistical
  Mechanics and its Applications 389~(24) (2010) 5864--5870.

\bibitem{dubovskaya2023analysis}
A.~Dubovskaya, S.~C. Fennell, K.~Burke, J.~P. Gleeson, D.~O’Kiely, Analysis
  of mean-field approximation for deffuant opinion dynamics on networks, SIAM
  Journal on Applied Mathematics 83~(2) (2023) 436--459.

\bibitem{laguna2004minorities}
M.~F. Laguna, G.~Abramson, D.~H. Zanette, Minorities in a model for opinion
  formation, Complexity 9~(4) (2004) 31--36.

\bibitem{porfiri2007decline}
M.~Porfiri, E.~Bollt, D.~Stilwell, Decline of minorities in stubborn societies,
  The European Physical Journal B 57 (2007) 481--486.

\bibitem{pineda2009noisy}
M.~Pineda, R.~Toral, E.~Hernandez-Garcia, Noisy continuous-opinion dynamics,
  Journal of Statistical Mechanics: Theory and Experiment 2009~(08) (2009)
  P08001.

\bibitem{carro2013role}
A.~Carro, R.~Toral, M.~San~Miguel, The role of noise and initial conditions in
  the asymptotic solution of a bounded confidence, continuous-opinion model,
  Journal of Statistical Physics 151 (2013) 131--149.

\bibitem{sirbu2019algorithmic}
A.~S{\^i}rbu, D.~Pedreschi, F.~Giannotti, J.~Kert{\'e}sz, Algorithmic bias
  amplifies opinion fragmentation and polarization: A bounded confidence model,
  PLOS ONE 14~(3) (2019) e0213246.
\newblock \href {https://doi.org/10.1371/journal.pone.0213246}
  {\path{doi:10.1371/journal.pone.0213246}}.

\bibitem{li2025bounded}
G.~J. Li, J.~Luo, M.~A. Porter, Bounded-confidence models of opinion dynamics
  with adaptive confidence bounds, SIAM Journal on Applied Dynamical Systems
  24~(2) (2025) 994--1041.

\bibitem{li2023bounded}
G.~J. Li, M.~A. Porter, Bounded-confidence model of opinion dynamics with
  heterogeneous node-activity levels, Physical Review Research 5~(2) (2023)
  023179.

\bibitem{torok2013opinions}
J.~T{\"o}r{\"o}k, G.~I{\~n}iguez, T.~Yasseri, M.~San~Miguel, K.~Kaski,
  J.~Kert{\'e}sz, Opinions, conflicts, and consensus: modeling social dynamics
  in a collaborative environment, Physical Review Letters 110~(8) (2013)
  088701.

\bibitem{li2022mixed}
H.-L. Li, Mixed hegselmann-krause dynamics, Discrete and Continuous Dynamical
  Systems - B 27~(2) (2022) 1149--1162.

\bibitem{kou2012multi}
G.~Kou, Y.~Zhao, Y.~Peng, Y.~Shi, Multi-level opinion dynamics under bounded
  confidence (2012).

\bibitem{fortunato2005consensus}
S.~Fortunato, On the consensus threshold for the opinion dynamics of
  krause--hegselmann, International Journal of Modern Physics C 16~(02) (2005)
  259--270.

\bibitem{slanina2011dynamical}
F.~Slanina, Dynamical phase transitions in hegselmann-krause model of opinion
  dynamics and consensus, The European Physical Journal B 79~(1) (2011)
  99--106.

\bibitem{yang2014opinion}
Y.~Yang, D.~V. Dimarogonas, X.~Hu, Opinion consensus of modified
  hegselmann--krause models, Automatica 50~(2) (2014) 622--627.

\bibitem{fortunato2005vector}
S.~Fortunato, V.~Latora, A.~Pluchino, A.~Rapisarda, Vector opinion dynamics in
  a bounded confidence consensus model, International Journal of Modern Physics
  C 16~(10) (2005) 1535--1551.

\bibitem{pluchino2006compromise}
A.~Pluchino, V.~Latora, A.~Rapisarda, Compromise and synchronization in opinion
  dynamics, The European Physical Journal B-Condensed Matter and Complex
  Systems 50~(1) (2006) 169--176.

\bibitem{lanchier2022consensus}
N.~Lanchier, H.-L. Li, Consensus in the hegselmann--krause model, Journal of
  Statistical Physics 187~(3) (2022) 20.

\bibitem{hegselmann2006truth}
R.~Hegselmann, U.~Krause, et~al., Truth and cognitive division of labor: First
  steps towards a computer aided social epistemology, Journal of Artificial
  Societies and Social Simulation 9~(3) (2006) 10.

\bibitem{chen2020heterogeneous}
G.~Chen, W.~Su, S.~Ding, Y.~Hong, Heterogeneous hegselmann--krause dynamics
  with environment and communication noise, IEEE Transactions on Automatic
  Control 65~(8) (2020) 3409--3424.

\bibitem{zhao2021fuzzy}
Y.~Zhao, M.~Xu, Y.~Dong, Y.~Peng, Fuzzy inference based hegselmann--krause
  opinion dynamics for group decision-making under ambiguity, Information
  Processing \& Management 58~(5) (2021) 102671.

\bibitem{morarescu2010opinion}
I.-C. Morarescu, A.~Girard, Opinion dynamics with decaying confidence:
  Application to community detection in graphs, IEEE Transactions on Automatic
  Control 56~(8) (2010) 1862--1873.

\bibitem{motsch2014heterophilious}
S.~Motsch, E.~Tadmor, Heterophilious dynamics enhances consensus, SIAM Review
  56~(4) (2014) 577--621.

\bibitem{mirtabatabaei2012opinion}
A.~Mirtabatabaei, F.~Bullo, Opinion dynamics in heterogeneous networks:
  Convergence conjectures and theorems, SIAM Journal on Control and
  Optimization 50~(5) (2012) 2763--2785.

\bibitem{schawe2020collective}
H.~Schawe, L.~Hern{\'a}ndez, Collective effects of the cost of opinion change,
  Scientific Reports 10~(1) (2020) 13825.

\bibitem{vicsek2012collective}
T.~Vicsek, A.~Zafeiris, Collective motion, Physics Reports 517~(3-4) (2012)
  71--140.

\bibitem{jhawar2020noise}
J.~Jhawar, R.~G. Morris, U.~Amith-Kumar, M.~Danny~Raj, T.~Rogers, H.~Rajendran,
  V.~Guttal, Noise-induced schooling of fish, Nature Physics 16~(4) (2020)
  488--493.

\bibitem{loscar2021noisy}
E.~S. Loscar, G.~Baglietto, F.~Vazquez, Noisy multistate voter model for
  flocking in finite dimensions, Physical Review E 104~(3) (2021) 034111.

\bibitem{vazquez2019multistate}
F.~Vazquez, E.~S. Loscar, G.~Baglietto, Multistate voter model with imperfect
  copying, Physical Review E 100~(4) (2019) 042301.

\bibitem{bernardo2024bounded}
C.~Bernardo, C.~Altafini, A.~Proskurnikov, F.~Vasca, Bounded confidence opinion
  dynamics: A survey, Automatica 159 (2024) 111302.

\bibitem{peralta2022opinion}
A.~F. Peralta, J.~Kert{\'e}sz, G.~I{\~n}iguez, Opinion dynamics in social
  networks: From models to data, arXiv preprint arXiv:2201.01322 (2022).

\bibitem{hegselmann2019consensus}
R.~Hegselmann, U.~Krause, Consensus and fragmentation of opinions with a focus
  on bounded confidence, The American Mathematical Monthly 126~(8) (2019)
  700--716.

\bibitem{xia2011opinion}
H.~Xia, H.~Wang, Z.~Xuan, Opinion dynamics: A multidisciplinary review and
  perspective on future research, International Journal of Knowledge and
  Systems Science (IJKSS) 2~(4) (2011) 72--91.

\bibitem{barrat2008dynamical}
A.~Barrat, M.~Barthelemy, A.~Vespignani, Dynamical processes on complex
  networks, Cambridge University Press, 2008.

\bibitem{marro2005nonequilibrium}
J.~Marro, R.~Dickman, Nonequilibrium phase transitions in lattice models,
  Cambridge University Press, 2005.

\bibitem{sole1996extinction}
R.~V. Sol{\'e}, S.~C. Manrubia, Extinction and self-organized criticality in a
  model of large-scale evolution, Physical Review E 54~(1) (1996) R42.

\bibitem{bornholdt1998neutral}
S.~Bornholdt, K.~Sneppen, Neutral mutations and punctuated equilibrium in
  evolving genetic networks, Physical Review Letters 81~(1) (1998) 236.

\bibitem{christensen1998evolution}
K.~Christensen, R.~Donangelo, B.~Koiller, K.~Sneppen, Evolution of random
  networks, Physical Review Letters 81~(11) (1998) 2380.

\bibitem{bornholdt2000topological}
S.~Bornholdt, T.~Rohlf, Topological evolution of dynamical networks: Global
  criticality from local dynamics, Physical Review Letters 84~(26) (2000) 6114.

\bibitem{gross2008adaptive}
T.~Gross, B.~Blasius, Adaptive coevolutionary networks: a review, Journal of
  the Royal Society Interface 5~(20) (2008) 259--271.

\bibitem{holme2006nonequilibrium}
P.~Holme, M.~E.~J. Newman, Nonequilibrium phase transition in the coevolution
  of networks and opinions, Phys. Rev. E 74 (2006) 056108.

\bibitem{nardini2008who}
C.~Nardini, B.~Kozma, A.~Barrat, Who's talking first? consensus or lack thereof
  in coevolving opinion formation models, Physical Review Letters 100 (2008)
  158701.

\bibitem{vazquez2008generic}
F.~Vazquez, V.~M. Egu{\'\i}luz, M.~S. Miguel, Generic absorbing transition in
  coevolution dynamics, Physical Review Letters 100~(10) (2008) 108702.

\bibitem{herrera2011general}
J.~L. Herrera, M.~G. Cosenza, K.~Tucci, J.~C. Gonz{\'a}lez-Avella, General
  coevolution of topology and dynamics in networks, Europhysics Letters 95~(5)
  (2011) 58006.

\bibitem{zanette2006opinion}
D.~H. Zanette, S.~Gil, Opinion spreading and agent segregation on evolving
  networks, Physica D: Nonlinear Phenomena 224~(1-2) (2006) 156--165.

\bibitem{gil2006coevolution}
S.~Gil, D.~H. Zanette, Coevolution of agents and networks: Opinion spreading
  and community disconnection, Physics Letters A 356~(2) (2006) 89--94.

\bibitem{diakonova2015noise}
M.~Diakonova, V.~M. Egu{\'\i}luz, M.~San~Miguel, Noise in coevolving networks,
  Physical Review E 92~(3) (2015) 032803.

\bibitem{raducha2020emergence}
T.~Raducha, M.~San~Miguel, Emergence of complex structures from nonlinear
  interactions and noise in coevolving networks, Scientific Reports 10~(1)
  (2020) 15660.

\bibitem{min2017fragmentation}
B.~Min, M.~S. Miguel, Fragmentation transitions in a coevolving nonlinear voter
  model, Scientific Reports 7~(1) (2017) 12864.

\bibitem{raducha2018coevolving}
T.~Raducha, B.~Min, M.~San~Miguel, Coevolving nonlinear voter model with
  triadic closure, Europhysics Letters 124~(3) (2018) 30001.

\bibitem{jkedrzejewski2020spontaneous}
A.~J{\k{e}}drzejewski, J.~Toruniewska, K.~Suchecki, O.~Zaikin, J.~A. Ho{\l}yst,
  Spontaneous symmetry breaking of active phase in coevolving nonlinear voter
  model, Physical Review E 102~(4) (2020) 042313.

\bibitem{toruniewska2017coupling}
J.~Toruniewska, K.~Ku{\l}akowski, K.~Suchecki, J.~A. Ho{\l}yst, Coupling of
  link-and node-ordering in the coevolving voter model, Physical Review E
  96~(4) (2017) 042306.

\bibitem{kozma2008consensus}
B.~Kozma, A.~Barrat,
  \href{https://link.aps.org/doi/10.1103/PhysRevE.77.016102}{Consensus
  formation on adaptive networks}, Phys. Rev. E 77 (2008) 016102.
\newblock \href {https://doi.org/10.1103/PhysRevE.77.016102}
  {\path{doi:10.1103/PhysRevE.77.016102}}.
\newline\urlprefix\url{https://link.aps.org/doi/10.1103/PhysRevE.77.016102}

\bibitem{kozma2008consensus1}
B.~Kozma, A.~Barrat, Consensus formation on coevolving networks: groups'
  formation and structure, Journal of Physics A: Mathematical and Theoretical
  41~(22) (2008) 224020.

\bibitem{kan2023adaptive}
U.~Kan, M.~Feng, M.~A. Porter, An adaptive bounded-confidence model of opinion
  dynamics on networks, Journal of Complex Networks 11~(1) (2023) 415--444.

\bibitem{krishnagopal2024bounded}
S.~Krishnagopal, M.~A. Porter, Bounded-confidence models of opinion dynamics
  with neighborhood effects, arXiv preprint arXiv:2402.05368 (2024).

\bibitem{iniguez2009opinion}
G.~I{\~n}iguez, J.~Kert\'esz, K.~K. Kaski, R.~A. Barrio,
  \href{https://link.aps.org/doi/10.1103/PhysRevE.80.066119}{Opinion and
  community formation in coevolving networks}, Phys. Rev. E 80 (2009) 066119.
\newblock \href {https://doi.org/10.1103/PhysRevE.80.066119}
  {\path{doi:10.1103/PhysRevE.80.066119}}.
\newline\urlprefix\url{https://link.aps.org/doi/10.1103/PhysRevE.80.066119}

\bibitem{liu2023emergence}
J.~Liu, S.~Huang, N.~M. Aden, N.~F. Johnson, C.~Song, Emergence of polarization
  in coevolving networks, Physical Review Letters 130~(3) (2023) 037401.

\bibitem{kozitsin2022general}
I.~V. Kozitsin, \href{http://dx.doi.org/10.1038/s41598-022-09468-3}{A general
  framework to link theory and empirics in opinion formation models},
  Scientific Reports 12~(1) (April 2022).
\newblock \href {https://doi.org/10.1038/s41598-022-09468-3}
  {\path{doi:10.1038/s41598-022-09468-3}}.
\newline\urlprefix\url{http://dx.doi.org/10.1038/s41598-022-09468-3}

\bibitem{su2014coevolution}
J.~Su, B.~Liu, Q.~Li, H.~Ma, Coevolution of opinions and directed adaptive
  networks in a social group, Journal of Artificial Societies and Social
  Simulation 17~(2) (2014) 4.

\bibitem{eguiluz2005cooperation}
V.~M. Egu{\'\i}luz, M.~G. Zimmermann, C.~J. Cela-Conde, M.~S. Miguel,
  Cooperation and the emergence of role differentiation in the dynamics of
  social networks, American Journal of Sociology 110~(4) (2005) 977--1008.

\bibitem{zschaler2012adaptive}
G.~Zschaler, Adaptive-network models of collective dynamics, The European
  Physical Journal Special Topics 211~(1) (2012) 1--101.

\bibitem{carro2014fragmentation}
A.~Carro, F.~Vazquez, R.~Toral, M.~San~Miguel, Fragmentation transition in a
  coevolving network with link-state dynamics, Physical Review E 89~(6) (2014)
  062802.

\bibitem{vazquez2007time}
F.~Vazquez, J.~C. Gonz\'alez-Avella, V.~M. Egu\'{\i}luz, M.~San~Miguel,
  Time-scale competition leading to fragmentation and recombination transitions
  in the coevolution of network and states, Phys. Rev. E 76 (2007) 046120.

\bibitem{ito2003spontaneous}
J.~Ito, K.~Kaneko, Spontaneous structure formation in a network of dynamic
  elements, Physical Review E 67~(4) (2003) 046226.

\bibitem{liu2014controlling}
S.~Liu, N.~Perra, M.~Karsai, A.~Vespignani, Controlling contagion processes in
  activity driven networks, Physical Review Letters 112~(11) (2014) 118702.

\bibitem{karsai2014time}
M.~Karsai, N.~Perra, A.~Vespignani, Time varying networks and the weakness of
  strong ties, Scientific Reports 4~(1) (2014) 4001.

\bibitem{miritello2011dynamical}
G.~Miritello, E.~Moro, R.~Lara, Dynamical strength of social ties in
  information spreading, Physical Review E 83~(4) (2011) 045102.

\bibitem{eckmann2004entropy}
J.-P. Eckmann, E.~Moses, D.~Sergi, Entropy of dialogues creates coherent
  structures in e-mail traffic, Proceedings of the National Academy of Sciences
  101~(40) (2004) 14333--14337.

\bibitem{gonccalves2008human}
B.~Gon{\c{c}}alves, J.~J. Ramasco, Human dynamics revealed through web
  analytics, Physical Review E 78~(2) (2008) 026123.

\bibitem{fournet2014contact}
J.~Fournet, A.~Barrat, Contact patterns among high school students, PloS one
  9~(9) (2014) e107878.

\bibitem{michelson2014memory}
M.~R. Michelson, Memory and voter mobilization, Polity 46~(4) (2014) 591--610.

\bibitem{do2008evaluations}
A.~M. Do, A.~V. Rupert, G.~Wolford, Evaluations of pleasurable experiences: The
  peak-end rule, Psychonomic bulletin \& review 15~(1) (2008) 96--98.

\bibitem{kahneman1993when}
D.~Kahneman, B.~L. Fredrickson, C.~A. Schreiber, D.~A. Redelmeier, When more
  pain is preferred to less: Adding a better end, Psychological science 4~(6)
  (1993) 401--405.

\bibitem{redelmeier2003memories}
D.~A. Redelmeier, J.~Katz, D.~Kahneman, Memories of colonoscopy: a randomized
  trial, Pain 104~(1-2) (2003) 187--194.

\bibitem{haight1967handbook}
F.~A. Haight, {Handbook of the Poisson distribution}, Wiley, 1967.

\bibitem{greene1970production}
J.~H. Greene, Production and inventory control handbook, Vol.~1, McGraw-Hill,
  1970.

\bibitem{anderson2003fixed}
H.~R. Anderson, Fixed broadband wireless system design, John Wiley \& Sons,
  2003.

\bibitem{newman1999monte}
M.~Newman, G.~Barkema, {Monte Carlo methods in statistical physics}, Oxford
  University Press: New York, USA, 1999.

\bibitem{karsai2018bursty}
M.~Karsai, H.-H. Jo, K.~Kaski, Bursty human dynamics, Springer, 2018.

\bibitem{karsai2024measuring}
M.~Karsai, H.-H. Jo, Measuring and modeling bursty human phenomena, arXiv
  preprint arXiv:2412.13617 (2024).

\bibitem{oliveira2005darwin}
J.~G. Oliveira, A.-L. Barab{\'a}si, Darwin and einstein correspondence
  patterns, Nature 437~(7063) (2005) 1251--1251.

\bibitem{barabasi2010bursts}
A.-L. Barab{\'a}si, Bursts: the hidden patterns behind everything we do, from
  your e-mail to bloody crusades, Penguin, 2010.

\bibitem{goh2008burstiness}
K.-I. Goh, A.-L. Barab{\'a}si, Burstiness and memory in complex systems,
  Europhysics Letters 81~(4) (2008) 48002.

\bibitem{shinomoto2003differences}
S.~Shinomoto, K.~Shima, J.~Tanji, Differences in spiking patterns among
  cortical neurons, Neural Computation 15~(12) (2003) 2823--2842.

\bibitem{artime2017dynamics}
O.~Artime, J.~J. Ramasco, M.~San~Miguel, Dynamics on networks: competition of
  temporal and topological correlations, Scientific Reports 7 (2017) 41627.

\bibitem{box2015time}
G.~E. Box, G.~M. Jenkins, G.~C. Reinsel, G.~M. Ljung, Time series analysis:
  forecasting and control, 5th Edition, John Wiley \& Sons, 2015.

\bibitem{vajna2013modelling}
S.~Vajna, B.~T{\'o}th, J.~Kert{\'e}sz, Modelling bursty time series, New
  Journal of Physics 15~(10) (2013) 103023.

\bibitem{kantelhardt2001detecting}
J.~W. Kantelhardt, E.~Koscielny-Bunde, H.~H. Rego, S.~Havlin, A.~Bunde,
  Detecting long-range correlations with detrended fluctuation analysis,
  Physica A: Statistical Mechanics and its Applications 295~(3-4) (2001)
  441--454.

\bibitem{karsai2012universal}
M.~Karsai, K.~Kaski, A.-L. Barab{\'a}si, J.~Kert{\'e}sz, Universal features of
  correlated bursty behaviour, Scientific Reports 2 (2012) 397.

\bibitem{oliveira2009impact}
J.~G. Oliveira, A.~Vazquez, Impact of interactions on human dynamics, Physica
  A: Statistical Mechanics and its Applications 388~(2-3) (2009) 187--192.

\bibitem{vazquez2006modeling}
A.~V{\'a}zquez, J.~G. Oliveira, Z.~Dezs{\"o}, K.-I. Goh, I.~Kondor, A.-L.
  Barab{\'a}si, Modeling bursts and heavy tails in human dynamics, Physical
  Review E 73~(3) (2006) 036127.

\bibitem{masuda2009priority}
N.~Masuda, J.~Kim, B.~Kahng, Priority queues with bursty arrivals of incoming
  tasks, Physical Review E 79~(3) (2009) 036106.

\bibitem{vazquez2005exact}
A.~V{\'a}zquez, {Exact results for the Barab{\'a}si model of human dynamics},
  Physical Review Letters 95~(24) (2005) 248701.

\bibitem{gabrielli2007invasion}
A.~Gabrielli, G.~Caldarelli, {Invasion percolation and critical transient in
  the Barab{\'a}si model of human dynamics}, Physical Review Letters 98~(20)
  (2007) 208701.

\bibitem{anteneodo2009exact}
C.~Anteneodo, {Exact results for the Barab{\'a}si queuing model}, Physical
  Review E 80~(4) (2009) 041131.

\bibitem{min2009waiting}
B.~Min, K.-I. Goh, I.-M. Kim, Waiting time dynamics of priority-queue networks,
  Physical Review E 79~(5) (2009) 056110.

\bibitem{hidalgo2006conditions}
C.~A. Hidalgo, Conditions for the emergence of scaling in the inter-event time
  of uncorrelated and seasonal systems, Physica A: Statistical Mechanics and
  its Applications 369~(2) (2006) 877--883.

\bibitem{malmgren2008poissonian}
R.~D. Malmgren, D.~B. Stouffer, A.~E. Motter, L.~A. Amaral, {A Poissonian
  explanation for heavy tails in e-mail communication}, Proceedings of the
  National Academy of Sciences 105~(47) (2008) 18153--18158.

\bibitem{jo2012circadian}
H.-H. Jo, M.~Karsai, J.~Kert{\'e}sz, K.~Kaski, Circadian pattern and burstiness
  in mobile phone communication, New Journal of Physics 14~(1) (2012) 013055.

\bibitem{boguna2014simulating}
M.~Bogu{\~n}{\'a}, L.~F. Lafuerza, R.~Toral, M.~{\'A}. Serrano, {Simulating
  non-Markovian stochastic processes}, Physical Review E 90~(4) (2014) 042108.

\bibitem{masuda2018gillespie}
N.~Masuda, L.~E. Rocha, {A Gillespie algorithm for non-Markovian stochastic
  processes}, SIAM Review 60~(1) (2018) 95--115.

\bibitem{jo2019copula}
H.-H. Jo, B.-H. Lee, T.~Hiraoka, W.-S. Jung, Copula-based algorithm for
  generating bursty time series, Physical Review E 100~(2) (2019) 022307.

\bibitem{vestergaard2014how}
C.~L. Vestergaard, M.~G{\'e}nois, A.~Barrat, How memory generates heterogeneous
  dynamics in temporal networks, Physical Review E 90~(4) (2014) 042805.

\bibitem{stark2008decelerating}
H.-U. Stark, C.~J. Tessone, F.~Schweitzer, Decelerating microdynamics can
  accelerate macrodynamics in the voter model, Physical Review Letters 101~(1)
  (2008) 018701.

\bibitem{wang2014freezing}
Z.~Wang, Y.~Liu, L.~Wang, Y.~Zhang, Z.~Wang, Freezing period strongly impacts
  the emergence of a global consensus in the voter model, Scientific Reports
  4~(1) (2014) 3597.

\bibitem{palermo2024spontaneous}
G.~Palermo, A.~Mancini, A.~Desiderio, R.~Di~Clemente, G.~Cimini, Spontaneous
  opinion swings in the voter model with latency, Physical Review E 110~(2)
  (2024) 024313.

\bibitem{fernandez2011update}
J.~Fern{\'a}ndez-Gracia, V.~M. Egu{\'\i}luz, M.~San~Miguel, Update rules and
  interevent time distributions: Slow ordering versus no ordering in the voter
  model, Physical Review E 84~(1) (2011) 015103.

\bibitem{dallasta2007effective}
L.~Dall'Asta, C.~Castellano, Effective surface-tension in the noise-reduced
  voter model, Europhysics Letters 77~(6) (2007) 60005.

\bibitem{artime2018aging}
O.~Artime, A.~F. Peralta, R.~Toral, J.~J. Ramasco, M.~San~Miguel, Aging-induced
  continuous phase transition, Physical Review E 98~(3) (2018) 032104.

\bibitem{peralta2020ordering}
A.~F. Peralta, N.~Khalil, R.~Toral, Ordering dynamics in the voter model with
  aging, Physica A: Statistical Mechanics and its Applications 552 (2020)
  122475.

\bibitem{baron2022analytical}
J.~W. Baron, A.~F. Peralta, T.~Galla, R.~Toral, Analytical and numerical
  treatment of continuous ageing in the voter model, Entropy 24~(10) (2022)
  1331.

\bibitem{llabres2024aging}
J.~Llabr{\'e}s, S.~Oliver-Bonafoux, C.~Anteneodo, R.~Toral, Aging in some
  opinion formation models: A comparative study, Physics 6~(2) (2024) 515--528.

\bibitem{abella2022aging}
D.~Abella, M.~San~Miguel, J.~J. Ramasco, Aging effects in schelling segregation
  model, Scientific Reports 12~(1) (2022) 19376.

\bibitem{abella2023aging}
D.~Abella, M.~San~Miguel, J.~J. Ramasco, Aging in binary-state models: The
  threshold model for complex contagion, Physical Review E 107~(2) (2023)
  024101.

\bibitem{chen2020non}
H.~Chen, S.~Wang, C.~Shen, H.~Zhang, G.~Bianconi, Non-markovian majority-vote
  model, Physical Review E 102~(6) (2020) 062311.

\bibitem{jkedrzejewski2018impact}
A.~J{\k{e}}drzejewski, K.~Sznajd-Weron, Impact of memory on opinion dynamics,
  Physica A: Statistical Mechanics and its Applications 505 (2018) 306--315.

\bibitem{perez2016competition}
T.~P{\'e}rez, K.~Klemm, V.~M. Egu{\'\i}luz, Competition in the presence of
  aging: dominance, coexistence, and alternation between states, Scientific
  Reports 6 (2016) 21128.

\bibitem{szolnoki2009impact}
A.~Szolnoki, M.~Perc, G.~Szab{\'o}, H.-U. Stark, Impact of aging on the
  evolution of cooperation in the spatial prisoner’s dilemma game, Physical
  Review E 80~(2) (2009) 021901.

\bibitem{szolnoki2010dynamically}
A.~Szolnoki, Z.~Wang, J.~Wang, X.~Zhu, Dynamically generated cyclic dominance
  in spatial prisoner’s dilemma games, Physical Review E 82~(3) (2010)
  036110.

\bibitem{takaguchi2011voter}
T.~Takaguchi, N.~Masuda, Voter model with non-poissonian interevent intervals,
  Physical Review E 84~(3) (2011) 036115.

\bibitem{chu2024bounded}
W.~Chu, M.~A. Porter, Bounded-confidence opinion models with random-time
  interactions, arXiv preprint arXiv:2409.15148 (2024).

\bibitem{zarei2024bursts}
F.~Zarei, Y.~Gandica, L.~E. Rocha, Bursts of communication increase opinion
  diversity in the temporal deffuant model, Scientific Reports 14~(1) (2024)
  2222.

\bibitem{scholtes2014causality}
I.~Scholtes, N.~Wider, R.~Pfitzner, A.~Garas, C.~J. Tessone, F.~Schweitzer,
  Causality-driven slow-down and speed-up of diffusion in non-markovian
  temporal networks, Nature Communications 5~(1) (2014) 5024.

\bibitem{rosvall2014memory}
M.~Rosvall, A.~V. Esquivel, A.~Lancichinetti, J.~D. West, R.~Lambiotte, Memory
  in network flows and its effects on spreading dynamics and community
  detection, Nature Communications 5~(1) (2014) 4630.

\bibitem{vazquez2007impact}
A.~Vazquez, B.~Racz, A.~Lukacs, A.-L. Barabasi, Impact of non-poissonian
  activity patterns on spreading processes, Physical Review Letters 98~(15)
  (2007) 158702.

\bibitem{iribarren2009impact}
J.~L. Iribarren, E.~Moro, Impact of human activity patterns on the dynamics of
  information diffusion, Physical Review Letters 103~(3) (2009) 038702.

\bibitem{karsai2011small}
M.~Karsai, M.~Kivel{\"a}, R.~K. Pan, K.~Kaski, J.~Kert{\'e}sz, A.-L.
  Barab{\'a}si, J.~Saram{\"a}ki, Small but slow world: How network topology and
  burstiness slow down spreading, Physical Review E 83~(2) (2011) 025102.

\end{thebibliography}

\end{document}